\documentclass[aps,nofootinbib,floatfix]{revtex4}
\usepackage{bm,longtable}
\usepackage{graphicx,tabularx,natbib}
\usepackage{latexsym,color}
\usepackage{physics}
\usepackage{centernot}
\usepackage{relsize}
\usepackage{mathtools}
\usepackage{stmaryrd}
\usepackage{rotating}
\usepackage{amssymb}
\usepackage[T1]{fontenc}
\usepackage[latin1]{inputenc}
\usepackage{natbib}
\usepackage{amsmath,amssymb,amsthm,mathrsfs,amsfonts,dsfont}
\usepackage{color}
\usepackage[dvipsnames]{xcolor}
\definecolor{LinkBlue}{RGB}{6,69,173}
\definecolor{DarkBlue}{RGB}{11,0,128}
\usepackage[colorlinks=true,linkcolor=blue,urlcolor=magenta,
	citecolor=green,hyperfootnotes=true]{hyperref}

\usepackage{enumitem}

\def\be{\begin{equation}}
\def\ee{\end{equation}}
\def\bea{\begin{eqnarray}}
\newcommand{\cc}{\mathrm{C}}

\newcommand{\apjs}{Astrophys.\ J.\ Suppl.}
\newcommand{\mnras}{MNRAS}
\newcommand{\actaa}{Acta Astronomica}

\newcommand{\aap}{A\&A}
\newcommand{\Qa}{\mathcal{Q}}
\newcommand{\Sa}{\mathcal{S}}
\newcommand{\la}{\mathcal{A}}

\def\eea{\end{eqnarray}}

\newcommand{\mso}{\mathrm{mso}}

\newcommand{\il}{~}

\begin{document}
\title{{ Embedded \textbf{BHs} and multipole  globules: \\\emph{ Clustered   misaligned  thick accretion disks around  static \textbf{SMBHs}.} }}

\author{D. Pugliese\&Z. Stuchl\'{\i}k}

\affiliation{
Research Centre for Theoretical Physics and Astrophysics,Institute of Physics,
  Silesian University in Opava,
 Bezru\v{c}ovo n\'{a}m\v{e}st\'{i} 13, CZ-74601 Opava, Czech Republic \\
 \email{d.pugliese.physics@gmail.com;zdenek.stuchlik@physics.cz} }

\date{\today}
  \begin{abstract}
We investigate clusters of  misaligned (inclined) tori orbiting a central static Schwarzschild black hole. To this purpose we considered  a set of  geometrically thick, pressure supported,  perfect fluid tori   analyzing   purely hydrodynamic models. We study the  tori collision  emergence and, consequently,  the stability properties of the  aggregates  composed by tori with  different inclination angles relative to a fixed distant observer.
The  aggregate of tilted tori is modeled as a single   orbiting configuration, by  introducing  a leading function governing the distribution of toroids around  the  black hole attractor. {Eventually the tori agglomerate  can be seen, depending on the     tori thickness, as a (multipole) gobules of orbiting matter, with different toroidal   spin orientations , covering the  embedded central black hole.}
These systems are shown to include   tori with emerging instability phase  related to accretion onto the central black hole.  Therefore we    provide  an  evaluation  of  quantities  related  to tori energetics such as the  mass-flux,  the  enthalpy-flux,
and  the flux thickness depending on the model parameters for
polytropic fluids.
 Consequently this analysis  places constraints on the existence and properties of  tilted tori and aggregate of misaligned disks.
Some notes are included
on  aggregates  including  proto-jets, represented by open cusped solutions  associated to the geometrically thick   tori.
\end{abstract}
\date{\today}
%
%
%
%
%
\maketitle
\section{Introduction}\label{Sec:intro}
%
%
 Accretion
disks misaligned with respect to the spin of a central black hole
can be formed  in the  active galaxy nuclei (\textbf{AGNs});
in such \textbf{AGNs}, corotating and counterrotating tori, and strongly misaligned disks may be consequential to periods of chaotical accretion.
These structures  can constitute a basis for interpretation of   the mass accretion rates of supermassive black holes (\textbf{SMBHs}) in \textbf{AGNs}, and several other phenomena connected with energetics of the accretion disks, but also the evolution  of the central attractor with particular respect to its spin.
In the context of the misaligned tori, a  Kerr black  hole plays a relevant  role both  in the disk structure (there could be a warped-tilted relation) and for the effects that the disk  evolution has  on the evolution of the  central \textbf{BH} spin.
In the context of the misaligned tori, a \textbf{BH} warped torus evolves together with its attractor changing its mass, the magnitude of its spin, and the spin orientation.  The impact of the \textbf{BHs} spin on the misaligned disks reflects in the Bardeen--Petterson effect, causing shift of the tilted accretion disk into the \textbf{BH} equatorial plane  due to the combined effect of the disk inclination and the frame dragging of the Kerr spacetime \cite{BP}.
{ Misaligned disks  have been studied  for example in \cite{Martin:2014wja,King:2018mgw,2012ApJ...757L..24N,2012MNRAS.422.2547N,2015MNRAS.448.1526N,
 2006MNRAS.368.1196L,Feiler,King2005}
where  the misaligned tori investigation is grounded on various  analytical and numerical  methods, often based on a simplified set of equations. For time-dependent systems,  the interaction between the central \textbf{BH} and the tori  is investigated, which results in multi tori system formation and the \textbf{BH} spin evolution. This analysis   emphasizes   the  time  scales of the processes of the interacting systems composed by the jets, disks and central attractor. Compared to the analysis presented here, for the most part,  the investigation lies on  various details of the formation of the tilted tori,  for example by evaluating the effects of the disk viscosity  in conjunction with the Lense--Thirring effect induced by the central spinning  attractor.
The evolution of misaligned accretion disks  is affected by the disk torque induced by the central spin attractor and, viceversa, the torque can empower the  \textbf{BH} spin-down (or eventually a  \textbf{BH} spin-up). This analysis thus  focuses on the
  propagation of warping
modification in the accretion disks,   considering location of
warping radius    and the  interrelation with  dissipation and  accretion rate during these processes.
 Further  important  aspects  are  the role of counterrotating fluids, usually  within  the hypothesis of thin disks, the evaluation of the \textbf{BH}   spin and the obscuration and absorption of X--ray emission, tracing back the
 history of black holes  to include also
the process of \textbf{BH} spin alignment.
The disk  location  varies  with   radius and with time,
inducing an alignment torque between  the  spinning \textbf{BH}  and a tilted  accretion
disk,  leading, due to  the Bardeen--Petterson effect to the tearing up  of the disk evolving  in  many distinct planes.
}

 In this article  we  study  misaligned  tori within  appropriate  modifications of the   Ringed Accretion Disks (\textbf{RADs}) framework  introduced in \cite{pugtot,ringed}, and then developed  as aggregates of axisymmetric toroidal configurations, coplanar and centered on the equatorial plane of the central  Kerr attractor   in \textbf{AGNs}.
 \textbf{RAD} structures are governed mainly by geometry of the Kerr \textbf{SMBH} attractors  as shown in \cite{ringed,open,dsystem,proto-jet,long,Letter,mnras,Multy}.  {First introduced in\cite{pugtot}, \textbf{RADs} were  detailed as fully general relativistic modela of (equatorial) tori, shortly the \textbf{eRAD}  \cite{ringed}. The possibility of  instabilities typical of the ringed structure including open configurations related to jet emission is discussed in \cite{open,long}. Constrains  on double accreting configurations (an \textbf{eRAD} of the order 2) were considered in \cite{dsystem}, where  observational evidences were considered. Energetics of couple of tori corotating and counterrotating around a Kerr central super-massive \textbf{BH}  and  \textbf{eRADs}  tori collisions  are  the focus of   \cite{Letter}.  Proto-jet configurations in \textbf{eRADs} orbiting a Kerr \textbf{SMBH}  are considered in \cite{proto-jet}. In \cite{Multy}  Kerr \textbf{SMBHs} in active galactic nuclei are related to  \textbf{RADs} configurations, binding the fluids and \textbf{BH} characteristics and providing indications on the situations where to search for \textbf{RADs} observational evidences.
Kerr black holes  are classified  due to their dimensionless spin (regulated by 14 characteristic values), according to possible combinations of corotating and counterrotating equilibrium or unstable (accreting) tori composing the \textbf{RADs}. It is proved that the number of accreting tori in \textbf{RADs} cannot exceed $n=2$. One of the critical predictions states that a \textbf{RAD} tori couple formed by an outer accreting corotating  torus and an inner accreting counterrotating torus is expected to be observed only around slowly spinning ($a<0.46M$) \textbf{BHs}.  In \cite{mnras}  the effects of a toroidal magnetic field are  analyzed in the formation of several magnetized accretion tori aggregated as \textbf{eRAD} orbiting around one central Kerr \textbf{SMBH}. The central \textbf{BH} spin-mass ratio, the magnetic field and the relative fluid rotation and tori rotation with respect the central \textbf{BH}  play a  significant role in  determining the accretion tori features, providing ultimately evidence of a strict correlation between \textbf{SMBH} spin, the fluid rotation, and magnetic fields in \textbf{RADs} formation and evolution. The \textbf{RAD} with tilted disks is considered in \cite{next} where limiting effects in clusters of misaligned toroids orbiting static \textbf{SMBHs} were explored. The possibility that the twin peak high-frequency quasi-periodic oscillations (HF-QPOs) could be related to the agglomerate inner ringed structure,  has been discussed  considering several oscillation geodesic models associated to the toroids inner edges.}

{The majority of this analysis examines  aggregates on a single plane of symmetry but, in accordance with the observational data  related to accretion onto \textbf{SMBH} attractors, it is  probable  that at least in the first and transient phases of the life of the  \textbf{BH}--disks system this special set  of symmetries is actually not finalized. This grounds the   need to consider, in the ringed disk analysis, the further complication of a cluster of orbiting tilted disks. Moreover, as mentioned above, the inclusion of  tilted disks can enter a very large number of aspects of the attractor characteristics  such as the mass accretion rates of \textbf{SMBHs} at high red-shift,  or even the spin-down or spin-up processes, associated  with extraction of rotational  energy from the central \textbf{BH} due to the  interaction with the surrounding matter\cite{2018MNRAS.478L..89D,2020arXiv200407907T,2020Univ....6...26S}.
However, in the presence of a very strong attractor  it is clear that there are constraints  mainly due to the curvature of the background. We evaluate these limits also testing collateral hypotheses such as the presence  of globules.
The \textbf{GRHD} analysis   is considered as preliminary test for   more complex  \textbf{GRMHD}   models, usually these \textbf{GRHD} models of tori serve as initial surfaces  for numerical simulations of \textbf{GRMHD}  tori.}

{
With regard to the  data supporting the misalignment  hypothesis, there are many observational evidences  concerning  the existence of  different periods of accretion of \textbf{SMBHs} hosted in \textbf{\textbf{AGNs}}, which are characterized by multi-accreting periods leaving  traces in counterrotating and even misaligned structures orbiting around the \textbf{SMBHs}.
Chaotical, discontinuous   accretion episodes  can  produce sequences of orbiting toroidal structures  with strongly   different features as different rotation orientations \cite{Dyda:2014pia,Aligetal(2013),Carmona-Loaiza:2015fqa,Blanchard:2017zfe,Lovelace:1996kx,Gafton:2015jja,natures,Nixon:2013qfa,Dogan:2015ida,
Bonnerot:2015ara,Bonnell,WA,Aly:2015vqa}.}

{Ringed accretion disks, in the {\textbf{eRAD}} and more general \textbf{RAD} formulation, are essentially a constraining   models, based on the  use of Euler equations for each toroidal  component of the aggregate, boundary conditions for the construction of the ringed inner-structures (often via effective potential approach)  of a leading function  governing  the distribution of  tori around the central attractor. It is developed as a full general relativistic hydrodynamic model, although  more terms can be included in the force balance equations. Each torus,  based on Boyer theory of equilibrium of  rigid surfaces in GR, satisfies the  von Zeipel, being  based on the assumption of a barotropic equation of state for the fluid. Especially in the \textbf{eRAD}  case (including the case of \textbf{RAD} on spherically symmetric background) there is no evolution of the systems, more precisely the  torus model is stationary, i.e. the fluid four velocity  has only a toroidal and time component  in the frame adapted to the  torus symmetry plane (and it  can be set in correspondence with definition of stationary  observers). As consequences of the assumptions on the symmetries, the continuity (evolution) equations for the  tori density are always satisfied, similarly it can be proved that the disk verticality (defined by the polar gradient of the pressure and density in each adapted frame of the torus) can be determined by the  pressure and density radial gradient.
For all these reasons, the ringed disk is particularly relevant   where the influence of the strong central attractor is predominant in determining  the tori construction, for example in the case of  thick tori in \textbf{SMBHs}. The model aims to provide constrains on tori  location with respect to the attractor and relative tori location  and emerging  tori collisions. The configurations provided by the set of equations of the \textbf{RAD} frame and the constrains are intended  to be initial data for the  evolutive models, or constrain-configurations for later stages of evolution.
In this work we include in the \textbf{RAD} set-up, as further ingredient the  disk tilt angle, providing constrains on the misaligned tori, and eventually the hypothesis of \textbf{BH} embedded a in multipolar orbiting structure.}

In order to treat in a simple analytical form  aggregates of toroidal structures, tilted under various inclination angles with respect to a fixed distant observer, we focus our attention to the Schwarzschild \textbf{SMBHs} using an appropriately adapted \textbf{RADs} framework. Only due to the spacetime  spherical symmetry, each toroidal structure can be centered around  its own central plane, keeping the axial symmetry  for its fully general relativistic description in the approximation  of test toroidal structures, i.e., structures that have negligible gravitational influence on the spacetime geometry and the other tori of the aggregate. Of course, instabilities due to accretion or collision between the tori have to be taken into account  in this simplified model.

Our simplified Schwarzschild version of the adapted \textbf{RADs} can be to some extend applied to the case of Kerr black holes, namely for tori located far enough from the central attractor, at distances where the rotational effect (frame dragging) of the Kerr metric is negligible; exact analytical models of the tilted toroidal aggregates in the Kerr metric are challenging  due to its axial symmetry fixing the only equatorial plane for the toroidal structures. In the vicinity of the Kerr \textbf{BHs}, the \textbf{RADs} are expected in the equatorial plane due to the Bardeen-Petterson  effect \cite{BP}.
  In this context, to distinguish the case of  misaligned tori, considered in this article, by the case where all the tori are located on the fixed equatorial plane  of the central Kerr attractor, described in  \cite{ringed,open,dsystem}, we denote the later case as \emph{equatorial-\textbf{RAD}} or \textbf{eRAD}. Note that
  off-equatorial configurations are possible even around the Kerr \textbf{BHs}, if electromagnetic phenomena enter the game--\cite{Kovar:2011uh,Stuchlik:2004wk,Kovar:2010ty,Slany:2013rml,Kovar:2014tla,2013ApJS..209...15C,Trova:2016ton,Kovar:2016kqh,Trova:2018bsf,Schroven:2018agz}.
For a static, spherically symmetric Schwarzschild black hole,  each central plane  can be considered as an  equatorial plane and  symmetry plane for the \textbf{RAD} toroidal component.
 We take full advantage of this special symmetry
 in the spherically symmetric  spacetime where
all the tori, regardless of the reciprocal rotation orientation,  can be considered as \textbf{RAD} model (in the sense of stability properties, morphology, model description).
 Here we can consider unlimited \textbf{RADs}, as in the Schwarzschild spacetime there is no outer limit  on stable geodesics governing center of the tori. In the accelerating universe with non-zero cosmological constant, we have to use the Schwarzschild-de Sitter spacetime giving  so called static radius \cite{1983BAICz..34..129S,1999PhRvD..60d4006S}, limiting the existence of stable circular geodesics from above; moreover in this case also the toroidal structures cannot exceed this radius \cite{2000A&A...363..425S,2005MPLA...20..561S}, and even  self-gravitating structures cannot cross this limiting static radius \cite{2016PhRvD..94j3513S}. We plan to study the  role of cosmological constant in some of our future papers.

We can consider, in a given central symmetry  plane, all configurations as  $\ell$corotating sequence of orbiting tori,  {generally, in  the fixed symmetry plane,  we can define the $\ell$corotanting ($\ell$counterrotating) pair of tori if  there is   $\ell_i\ell_o>0$ ($\ell_i\ell_o<0$), with   $(\ell_i,\ell_o)$ being the fluid  specific angular momenta \cite{dsystem}}. In the \textbf{eRAD}  this constraint implied  that  the two tori could corotate, $a\ell_i>0$ and $a\ell_o>0$,  with the central Kerr \textbf{BH} of spin $a>0$ or, viceversa, be  both counterrotating i.e. $a\ell_i<0$ and $a\ell_o<0$.
In the \textbf{eRAD} a  state with  only the inner accreting (cusped) torus  is possible, keeping the   stability of the \textbf{RAD}  structure--we verify this and other characteristics inherited by the  \textbf{RAD}  model containing misaligned tori. The misalignment  of the toroidal structures  allows to reconsider in some extend the possibility of the presence of multi accreting  tori on different planes, enlightening interesting situations and phenomenologies   which were   not allowed for the \textbf{eRAD}. The analysis of the \textbf{eRAD} in \cite{ringed,open,dsystem}, can be for many aspects transferred to  the modified \textbf{RAD} case  provided that  we consider now  the spherical  radius $r$ as the  radius of {stability spheres}  centered in the \textbf{BH} singularity.
Particularly we explore the possibility that   the presence of more accreting \textbf{RAD} tori  (including collisional regions) could increase the accretion rate of the central \textbf{BH}.
 We test the hypothesis of a quasi-complete covering of the \textbf{BH} horizon with a \textbf{BH} embedded in a accreting ball made by a composition of \textbf{RAD} tori  having  different orientations of the fluids specific angular momentum (tori spins). (This object could be seen as a variation of the models for  non-self-gravitating shell model centered on a central attractor discussed in \cite{Shell1,Shell2,Shell3,Shell4}).

 A further development of this model may  consist in the possibility that very thick tori, as those foreseen by this model of perfect fluid may  in fact constitute a globus of cluster  tori  i.e. a multipoles orbiting matter embedding surrounding the central \textbf{BH} converging  the \textbf{BH} horizon at different view angles  for the observer.
 In the \textbf{RAD}  context,   under  special conditions  (depending on number of toroidal components and tori geometrical thickness),   the system of  \textbf{BH}  entangled tori  could be considered as a sort of matter embedding,  covering the central \textbf{BH} from a distant observer at different angles. This case occurs  especially  for  static attractors or, eventually,  in the low \textbf{BH} spin regime  more precisely according to the conditions  $ R=r/a\rightarrow\infty $ verified for disks located far away from the  central \textbf{BH} (center of disk as seen here on $R$ variable, $r$ being radial distance from the central attractor) or very small \textbf{BH}    spin $a$. These     \emph{globuli} of orbiting  matter have several  interesting properties. At this stage of model  the \textbf{RAD} tori are not (predominantly) self-gravitating as composed by  not-self gravitating disks. These   globuli, being  constituted  by orbiting matter   with different  spin orientations constitute in this sense  a multipole   orbiting configuration with a central  \textbf{BH} object that may have significant role in different epochs of the \textbf{BH} life.

\medskip

Before detailing  this article plan, we summarize   some  aspects  of the methodology,   the results  and  possible observational aspects.
\begin{itemize}
\item
The analysis  results eventually  in  the  characterization of   the \textbf{RAD} macrostructure determining the  \emph{set} of tori   and indicating the   possible  correlated observational properties.   We stress that  considering a generic tilt angle  we  characterize the
 \emph{clusters} of tilted tori, rather than one single torus,
discussing the  limitations on \textbf{RAD} existence and stability in Sec.\il(\ref{Sec:Misal}) and Sec.\il(\ref{Sec:doc-ready}), the energetic of the systems  in  Sec.\il(\ref{Sec:energ-RAD-poli}) and an overview of the possible phenomena   associated with  these structures as  the proto-jets. On  methodological view point, the novelty  of our approach, with respect to the  current studies of analogue systems in the context of multi orbiting disks, consists primarily in the fact that other   these studies  foresee a  strong numerical effort (often within a  dynamical frame)  with different very specific assumptions on the tori models, for example considering dust,  while  our analysis focus on pressure supported perfect fluid disks with any barotropic equation of state.  Our final configuration can  indeed provide specific initial data on tori configurations as we discuss  constraints on general classes of tori which can be considered for application  in very diversified scenarios,  including  GRMHD setup. Our results completely constraint the possible initial configurations with multiple tori considering both the possibility of tori collision and accretion emergence, or  their morphological characteristics.
This analysis  proposes a methodology and conceptual setup that constitutes the \textbf{RAD} frame pursuing the existence of a \emph{leading function} representing (the constrains on) the tori distribution around the central attractor.
Consequently we identified an energy function $K(r)$  providing indications on  stability, being  related to the  energetics of \textbf{BH}-accretion disks systems,   and defining relevant quantities as the  mass accretion rate and cusp luminosity.
    This scenario has   consequences and
ramifications  on several possible phenomena connected to the \textbf{RAD}  structure  and proceeding from having preferred the analytical and global approach. The global and structural aspect of this approach  has to be intended in the sense of  constraints on the \textbf{RAD} as a whole rather than focus on the details of each component constituting the aggregate, thus leaving the field free for a very large number of different applications .
\item
  A key  element  for  the observation, relevant   for   the  \textbf{RAD} recognition, is  the establishment of the \textbf{RAD}  morphologic characteristics. Specifically  we mention the tori distance from the central attractor, here considered in details in Sec.\il(\ref{Sec:doc-ready}) and Appendix\il(\ref{Sec:app}), especially the  characteristics of  the outer  and the inner  toroids (respectively the farthest and closest to the central attractor), considering firstly  the torus extension on  its symmetric plane, torus thickness  and  the  conditions for the  cusp emergence, and the tori energetics considered in Sec.\il(\ref{Sec:energ-RAD-poli}).
\item
The characterization  of the stability of these structures is the  second crucial point for the \textbf{RAD} analysis. The system stability is dramatically   different   for static or not static central attractor and more in generally  where the condition  of very large $R=r/a$ is not satisfied (in this case there is also a clear dependence  on the tori angle of inclination and the fluid rotation). In  the context considered in this work where the central \textbf{BH} spin is neglected  and   the spacetime is spherical symmetric, the  possible globulus  stability  analysis means essentially  analysis of occurrence of tori collisions, and
conditions for tori accretion into \textbf{BH}.
Because of the symmetries of the  \textbf{BH-RAD} system, there are several  similarities  between the \textbf{eRAD} and \textbf{RAD} cases, useful to consider many  of these globulis aspects    in the \textbf{ RAD} analysis.
For a discussion  of a perturbation analysis we mention \cite{ringed}.
However occurrence of tori collision is analyzed in Sec.\il(\ref{Sec:doc-ready})  providing constraints on the attractor distance from the central attractor, the torus dimensions and the specific fluid  angular momentum.
Instability associated to cusped emergence is entirely treatable analytically. A careful study of  the instability conditions  can be found in Sec.\il(\ref{Sec:mirpj}).  While  in Sec.\il(\ref{Sec:er})  we consider the energetics of  tori-\textbf{RAD} with for example the  mass accretion rates.
It is therefore convenient to sum-up here the main expected  \textbf{RAD} instabilities.
The \textbf{RAD} inherits the typical instabilities of the \textbf{eRAD}, which are defined in \cite{ringed} and \cite{open,dsystem,Multy}.
We must  also consider the system instability   originating from the presence of a  torus   tilt angle. This type of instability has  an extremely relevant  role in the presence of a  spinning central attractor, where  the   Lense--Thirring effect triggers  different outcomes according to different tori misalignments and rotations,
provoking eventually  also a torus break, inducing a rupture of the  torus with the possible  formation of two equatorial disks (the inner torus would finally be   corotating with respect to the central \textbf{BH}),
fostering, especially  for viscous disks, the so called Bardeen\&Petterson effect--\cite{BD75,Nealon:2015jya}. For more details on these effects in the \textbf{RAD} context we refer to \cite{Multy,Letter}.
Even the presence of magnetic fields  can influence the stability of the  \textbf{RAD}  composed by   plasma disks  (especially in the presence of disks with a significant counterrotating component  of the momentum  with respect to the central \textbf{BH} spin).
We can therefore distinguish   four main\textbf{RAD} instabilities:
 \begin{enumerate}\item First instability is inherited by  each  toroidal  component instability. This is the hydrodynamic (HD) mechanical Paczynski instability,  connected to the presence of the "cusp" of the toroid surface.
The existence of a  minimum of the   hydrostatic pressure  implies  the existence of a  critical topology for the fluid configuration, solution of the Euler equation, reducing  to the  a presence of cusp for the related toroid surface. The cusp corresponds to the  violation of the conditions for the  mechanical equilibrium in the orbiting fluid-- \cite{Paczynski:2000tz,P-W,abrafra}.  There are two  types of cusped solutions, according to the range of values of the fluid specific angular momentum  encoding  the centrifugal forces regulating together with other forces the torus stability. If these are  sufficiently large the cusp is associated to open configurations (proto-jets), or otherwise to  closed configurations , this last case is  related  to   the emergence to the accretion phase for the  torus, therefore shortly we refer to these cusped closed tori as accreting  tori. The cups, also known as  Paczynski  instability points, are   maxima of the effective potential of the fluid encoding in the model used here the centrifugal and gravitational components of the force balance equations.  Moreover, we should note that the Paczy\'nski accretion mechanics from  a  Roche lobe overflow   through the cusp induces   a mass loss  from tori being an important
 local stabilizing mechanism  against thermal and viscous instabilities, and globally   against the Papaloizou-Pringle instability which is in fact a typical instability emerging for geometrically thick tori, and not irrelevant in the eventual combination with typical MRI-instabilities in the correspondent MHD models\cite{mnrasB,gelli}.
\item Second instability consists  in the emergence of  tori collision  which here we thoroughly consider.
\item Third instability is a combination  of the first instability and the second one and consists in tori collision induced by emergence of the first kind of instability, the accretion, in one torus of the configuration generally the outer of the couple. Different outcomes of these instabilities are possible as well as different modalities  for the third kind of instabilities  to occur-- \cite{dsystem}. Consequently we can  identify two successive instability phases  of the \textbf{RAD}: the {first}  where  there is  formation of one or more  points of instabilities   involving eventually more toroids. The second phase is in fact  a \textbf{RAD}  global instability of the ringed disk      following   the first phase. The  combinations of  many processes may also  result in   possible destabilization of the entire  structure, especially  for tori collision. Therefore here we study carefully the occurrence of this situation.
 \item A further interesting mechanism of instability typical of geometrically thick disks orbiting  around a central \textbf{BH} in the scenario of a \textbf{RAD} system is called \emph{runaway-runaway} instability, which  consists in the combination of runaway instability from the inner edge of the inner accreting torus of the \textbf{RAD} with the consequent destabilization of the aggregate, induced by both a change of the inner torus morphology,  due to the  onset of unstable phase, and by  the change of background geometry, arising from a shift in mass $M$ (and eventually spin $a$) of the central \textbf{BH}. This instability,  consequent to the accretion, affects  therefore  both  the \textbf{BH} spacetime structure  and the-inner disk. The whole  \textbf{BH} spacetime-disk system  changes in a sort of  "breathing"  mode, in a recursive process both the  disks and geometry properties, combined with the interaction of further (inert) tori of the aggregate-- \cite{dsystem,Letter,Multy}.
Runaway instability mechanism is  expected to be relevant in the case of a thick torus-- \cite{Font02}. During the accretion the mass loss through the cusp of the inner  tori is transformed into a shift in  the \textbf{BH} parameters    consequently the spacetime   geometry is  modified and this,  in turn,  affects the accreting material  changing also  the location of the  disk  cusp,  for the change of disk conditions and also the  \textbf{BH} geometry.
Clearly the picturing  of all the possible situations arising in the \textbf{RAD} of misaligned tori  induced by a runaway mechanism  is a complex task.
 However, among all the possible outcomes coming from   the establishment of the runaway instability
we  mention  that in the \textbf{eRAD},  the runaway mechanism   could trigger a  sort of "drying-feeding" process characterized by  the occurrence  of several    stages  of instabilities for the \textbf{RAD} (a sort of   ``clumpy'' episodic accretion  process).
 Typically  geometrically thick  tori composing the   \textbf{RAD} have  very high accretion rates,  in Sec.\il(\ref{Sec:energ-RAD-poli})  considered for different tori with general polytropics, where however  the \textbf{BH} mass parameter $M$    is considered as scale parameter for the distances, whose  variation has  therefore to be considered in the evaluations of distances $r/M$.
\end{enumerate}
\item
In this work we therefore propose that \textbf{BH} could be  embedded a  multipole  shell of orbiting, not-self-gravitating  tori having  different orientations  at different distances from  the attractor, composed by matter with very diversified characteristics, which can be more or less thick (a globulus) or spherical--Sec.\il(\ref{Sec:sfer.J}), would be typical objects of periods of low activity,  (cold-globuli) to then be reactivated due to a change of external environmental conditions giving rise possibly laos to a catastrophic event with a outburst of energy and matter. It is therefore essential to understand before and after this state,  the limitations of the globulus/ \textbf{RAD} in terms of the RAD radius which is defined from the more external  torus of the RAD- we consider this special problem in Sec.\il(\ref{Sec:limiting})  and Sec.\il(\ref{Sec:zeipel}). In the eventual \textbf{RAD}  destruction  after instability the \textbf{RAD} is  distinguished by an huge  release of energy and matter,  resulting ultimately into a   \textbf{SMBH} with different mass and spin from the starting  attractor of the \textbf{RAD}. Otherwise the matter, accreting towards from  these pressure supported tori   the central \textbf{BH}  with high  accretion rates   could find  out the formation of   orbiting structures quite different from the starting setup.
 Here we include some notes on the assumptions adopted in this work. We imposed for this first analysis a spherically symmetric and static central \textbf{BH}--this analysis can be compared with similar studies in  \cite{Martin:2014wja,King:2018mgw,2012ApJ...757L..24N,2012MNRAS.422.2547N,2015MNRAS.448.1526N,2006MNRAS.368.1196L,Feiler,King2005}..
The assumption on the spherical symmetry  allows to  focus  on  a non-dynamical structure. (On the other hand we note that we could  follow the evolution of the torus considering as in \cite{pugtot} sequences of tori at differences stages  characterized by different  values of the model parameters as in \cite{pugtot} where such  evolutive parameter was the specific  angular momentum).  This assumption moreover has the remarkable    advantage to  allow  the adoption of  a metric frame adapted to each torus  of the \textbf{RAD}, preferring a rotated frame  adapted to the symmetry plane of each torus  to characterize  the    ("magnetic multipole"-like) structure of the \textbf{RAD}. Clearly there is no loss of generality within this frame assumption. With respect to other studies, this has the advantage  to be an  exact  analysis of the collision conditions, useful especially according to the task to provide the initial configurations of dynamic simulations of more complex situations.
\item
Regarding the confrontation of   our analysis  and its outcomes  with the dynamical simulations on static \textbf{BH} background,  which is planned as future work, we want to emphasize some  fundamental points both on the goals  of such  analysis, which are also the objectives of other  similar  studies in literature, and the details of procedures of  to be considered when focusing on the \textbf{RAD} structure.  In the numerical analysis it  is necessary to fix a very specific setup, fixing the Schwarzschild attractor, consisting  in the  number of tori components, the  tori inclination angles  and  each torus  model. Tori we expect to have rather  different characteristics for example varying  in the  additional parameters like  the viscosity the  eventual resistivity. The \textbf{RAD} tori  can be  formed consequent to  different periods of accretion of the \textbf{BH} for interactions with different companions  which explain  the  different initial conditions such as angle of inclination and distance from the  attractor. Tori can  also be   formed by break of a single disk, especially  in the case of a Kerr \textbf{BH}, in this case the two tori formed after this occurrence will have at least in each phases of they formation  similar characteristics,  included the disk fluid   rotation ($\ell$corotating) which can then change  because of  the frame dragging on the inner torus of the couple (or other factors as the presence of the magnetic  fields etc).
Here, by overcoming  this  aspect , we  provided the classes of   disks which can be used to the  match with the  subsequent phases of development of the  fluid dynamics.
 One main objective of this analysis, in our opinion,   would consist in providing the typical time scales for the emergence of structural instability and  the emerging modes of instabilities, considering the first three kinds of \textbf{RAD} instabilities induced by    tori collision/accretion. The inclusion of   runaway instability and subsequent runaway-runaway  phase  requires a more refined and complex approach. The establishment of these two aspects it is clear will shed light into the issue  of the  formation   of \textbf{RAD}.
\end{itemize}

\medskip

 The plan of this article is as follows:
 In section (\ref{Sec:Misal}) we give
the basic equations for the description of the misaligned perfect fluid tori orbiting a central Schwarzschild black hole.
Geometry of the modified  \textbf{RAD} accreting tori, their stability and collision emergence are studied  in Sec.\il(\ref{Sec:doc-ready}). Particularly in  Sec.\il(\ref{Sec:morph})
 we discuss the relations between the tori morphological characteristics and  tori stability, while the
{limiting surfaces in the \textbf{RAD}} are the focus of Sec.\il(\ref{Sec:limiting}).
Conditions on the quasi-sphericity of the torus and of the globulus  are discussed in Sec.\il(\ref{Sec:sfer.J}).
The formation of the outer torus of the  \textbf{RAD} is addressed in
Sec.\il(\ref{Sec:ount})

 In Sec.\il(\ref{Sec:er}) we   provide  also evaluations  of  quantities  related  to tori energetics such as the  mass-flux,  the  enthalpy-flux (evaluating also the temperature parameter),
and  the flux thickness depending on the model parameters for
polytropic fluids.
Finally in section (\ref{Sec:conclu}) we discuss  our results and summarize the  conclusions of this analysis.
An appendix section (\ref{Sec:app}) follows where we
include further notes on  tori construction and \textbf{RAD} limiting configurations:
Sec.\il(\ref{Sec:four-b}) provides details on the  \textbf{RAD} rotational function   $\ell(r)$ and  the energy function $K(r)$,
Sec.\il(\ref{Sec:fasc-eur}) explicit the {toroidal surfaces},
Sec.\il(\ref{Sec:zeipel}) concerns the upper limit on the \textbf{RAD} (globule) radius.
We included also Table (\ref{Table:pol-cy})  presenting a list and a description of  the main notation used throughout this article.

\begin{table*}
\centering
\resizebox{.95\textwidth}{!}{%
\begin{tabular}{lll}
 \hline \hline
 $r_{\gamma}=3M$& Marginally circular orbit  (last circular photon  orbit) & Eq.\il(\ref{Eq:swan})--Figs\il(\ref{Fig:SIGNS},\ref{Fig:KKNOK})
\\
$r_{mbo}=4M$& Marginally bounded circular orbits& Eq.\il(\ref{Eq:swan})-Figs\il(\ref{Fig:SIGNS},\ref{Fig:KKNOK})
 \\
$r_{mso}=6M$& Marginally stable  circular orbit (ISCO)& Eq.\il(\ref{Eq:swan})--Figs\il(\ref{Fig:SIGNS},\ref{Fig:KKNOK})
\\
 $\ell(r)$& \textbf{RAD} rotational law-- \textbf{RAD} specific  angular momentum  distribution &
Eq.\il(\ref{Eq:lqkp})
\\
 $K(r)$ & \textbf{RAD} energy function--distributions &Eq.\il(\ref{Eq:sincer-Spee})
 \\
 & of \textbf{RAD}  maximum and minimum density/pressure points
  &
\\
$r_{cent}$& Torus center, maximum density and pressure point in a torus & Eq.\il(\ref{Eq:rcentro})--Figs\il(\ref{Fig:DopoDra})
\\
$r_{cent}(\ell)$&
Torus center of \emph{accreting} torus as function  of  $\ell$&Eq.\il(\ref{Eq:rcentro})--
Figs\il(\ref{Fig:VClose},\ref{Fig:SMGerm})
\\
 $r_{crit}=r_{cusp}=\{r_{\times},r_j\}$&  Location of effective potential maximum points, minimum of pressure&   Eqs\il(\ref{Eq:swan},\ref{Eq:swan1})
\\
$r_{\times}$ & (Accreting) Torus cusp (minimum density and pressure point) & Eqs\il(\ref{Eq:swan1},\ref{Eq:schaubl})
\\
 $r_{\times}(\ell)$&  Inner edge of \emph{accreting} torus as function  of  $\ell$&Eq.\il(\ref{Eq:rcentro})--
Figs\il(\ref{Fig:VClose},\ref{Fig:SMGerm})
\\
$r_{j}$&  Proto-jet (open) configuration cusp &  Eq.\il(\ref{Eq:swan})
\\
 $r_p^{\ell}(r)$&  Solution of  $\ell(r)=\ell(r^{\ell}_p)$, relates $r_{crit}$ and $r_{cent}$ at equal $\ell$& Eq.\il(\ref{Eq:schaubl})
  \\
$r_{mbo}^b\approx10.4721M$& Solution of $\ell(r)=\ell_{mbo}\equiv \ell(r_{mbo})$  & Eq.\il(\ref{Eq:swan1})
\\
 $r_{\gamma}^{b}=22.3923M$& Solution of $\ell(r)=\ell_{\gamma}\equiv \ell(r_{\gamma})$  & Eq.\il(\ref{Eq:swan1})
 \\
 $(r_{inner},r_{out})$& Torus inner and outer edges & Eqs\il(\ref{Eq:over-top})
\\
($r_{inner}(\ell,K)$, $r_{out}(\ell,K))$& Inner  and outer torus edges,  as function of $(\ell,K)$ & Eq.\il(\ref{Eq:mer-panto-ex-resul})--
Figs\il(\ref{Fig:VClose},\ref{Fig:conicap}.)
\\
$(r_{out}^{\times}(r_{\times}),r_{inner}^{\times}(r_{\times}))$&
Outer and inner edges,  function of the cusps & Eq.\il(\ref{Eq:over-top})--Figs\il(\ref{Fig:woersigna})
\\
 & combine solutions
$r=r_{\times}$ and $r=r_{out}$ &
\\
 $r_{inner}^{BH}(\ell,K)$ &Radius of the { \emph{innermost configuration}}, function of $(\ell,K)$& Eq.\il(\ref{Eq:mer-panto-ex-resul})--
Figs\il(\ref{Fig:VClose},\ref{Fig:conicap})
\\
$r_{out}^{\times}(\ell)$ & The outer edge of the cusped torus ($\ell=\ell_{\times}\in]\ell_{mso},\ell_{mbo}[$)  & Eqs\il(\ref{Eq:dan-aga-mich},\ref{Eq:vast-maj})-- Figs\il(\ref{Fig:tookP},\ref{Fig:lom-b-emi},\ref{Fig:SMGerm})
\\
$K_{cent}(\ell)$ &
$K$-Parameter  at the  torus center, function of $\ell$&
Eqs\il(\ref{Eq:grod-lock})
\\
$K_{\times}(\ell)$&
$K$-Parameter  at
 the  inner edge of accreting torus  as function of $\ell$&
Eqs\il(\ref{Eq:us-doi})
\\
$r_p(r)\in[4M,6M]$&  Solution of $K(r)=K(r_p)$,
  relates tori  ($T_1$, $T_2$) with  $K_{cent}(T_1)=K_{crit}(T_2)$&Eq.\il(\ref{Eq:Krrp})--Figs\il(\ref{Fig:solidy})
\\
 $r_{mbo}^k\approx4.61803M$& Solution of  $K(r_{mbo}^k)=K(r_{mbo}^b)$ & Sec.\il(\ref{Sec:conK-r-con})--Figs\il(\ref{Fig:KKNOK},\ref{Fig:solidy})
\\
   $r_{\gamma}^k\approx4.21748M$&  Solution of  $K(r_{\gamma}^k)=K(r_{\gamma}^b)$&Sec.\il(\ref{Sec:conK-r-con})--Figs\il(\ref{Fig:KKNOK},\ref{Fig:solidy})
\\
 $\lambda\equiv r_{out}-r_{inner}$& Torus elongation on its symmetry plane & Eqs\il(\ref{Eq:lam})--Figs\il(\ref{Fig:SMGerm},\ref{Fig:VClose},\ref{Fig:conicap},\ref{Fig:lom-b-emi},\ref{Fig:woersigna})
  \\
 $\lambda(\ell,K)$& Torus elongation
 function of $(\ell,K)$&
 Eq.\il(\ref{Eq:lam})--Figs\il(\ref{Fig:conicap})
 \\
$r_{max}\equiv (x_{\max},y_{\max})$&  Location of torus  (surface) geometric maximum point&  Eqs\il(\ref{Eqs:rssrcitt}).
\\
$r_{\max}\equiv x_{\max}$ &  Location of the  \textbf{RAD} tori geometric  maximum&Figs\il(\ref{Fig:woersigna},\ref{Fig:SMGerm},\ref{Fig:lom-b-emi})
  \\
$r_{\max}^o(K,\ell)$ &  Location  of torus maximum of the torus surface, function of $(K,\ell)$ &  Eq.\il(\ref{Eq:mat-ye})
\\
 $r_{\max}^o(r_{crit})$&  Location  of torus maximum, function of $r_{crit}$& Eq.\il(\ref{Eqs:rssrcitt})
\\
$r_{\max}^i(K,\ell)$&
Geometric maximum  radius of the innermost surface&  Eq.\il(\ref{Eq:mat-ye})
 \\
  $r_{\max}^i(r_{crit})$& Location of
innermost surface maximum  radius function of $r_{crit}$& Eq.\il(\ref{Eq:pos-spe})
  \\
$h=h_{\max}\equiv y_{\max}$& Torus height (surface maximum)&Figs\il(\ref{Fig:woersigna},\ref{Fig:SMGerm},\ref{Fig:lom-b-emi})
  \\
$h_{\times}$ & Accreting torus   height (surface maximum)& Fig.\il(\ref{Fig:woersigna})
\\
$h_{\max}^o(K,\ell)$& Maximum  of the torus surface as function of $K$ and $\ell$ &
Eq.\il(\ref{Eq:xit-graci})
\\
 $h_{\max}^o(r_{\times})$&Torus height, function of the  cusp & Eq.\il(\ref{Eqs:rssrcitt})
  \\
$\Sa=2h_/\lambda$& Torus geometrical  thickness& Sec.\il(\ref{Sec:sfer.J})--Figs\il(\ref{Fig:torithick},\ref{Fig:SMGermMs})
 \\
($\lambda_{\times}$, $\Sa_{\times}=2 h_{\times}/\lambda_{\times}$)&Elongation and thickness of the cusped tori&Figs\il(\ref{Fig:woersigna})
 \\
$r_{\mathcal{M}}=12.9282M$ & Solution$\in]r_{mbo}^b,r_\gamma^b[$ of  $\partial_r^2 \ell(r)=0$,    &Figs\il\ref{Fig:KKNOK}--Sec.\il(\ref{Sec:dig})
 \\
 &radius   of maximum density of tori&
\\
 $r_\mathcal{M}^K=8.079M$ &  Solution of  $\partial_r^2 K(r)=0$ with  $K_\mathcal{M}^K=0.948996M$,&Figs\il\ref{Fig:KKNOK}--Sec.\il(\ref{Sec:dig})
   \\
   &maximum point of   $\partial_r K(r)$
&
\\
$(\ell_{crit}^o(K),  \ell_{crit}^i(K))$&Momentum $\ell$  function of $K$-parameter& Eq.\il(\ref{Eq:crititLdeval})-- Figs\il(\ref{Fig:leaduniUR})
 \\
 &($\ell_{crit}^o(K)> \ell_{crit}^i(K)>\ell_{mso}$,  $\ell_{crit}^i(K)\in[\ell_{mso},\ell_{\gamma}[$)&
 \\
$r_{crit}(K)$& Tori critical radii      as a function of $K_{crit}$--$r_{crit}^i(K_{\times})=r_{inner}^{\times}$& Eqs\il(\ref{Eq:nicergerplto})--
Figs\il(\ref{Fig:leaduniUR})
\\
  &(inner edge of accreting torus),    $r_{crit}^o(K_{cent})=r_{cent}^{\times}$ (center of cusped configurations)&
\\
$(r_K^{in}\equiv r_K^-,  r_K^{out}\equiv r_K^+)$&Radii from condition  $K=1$ on  the potential,& Eq.\il(\ref{Eq:polaplot})--Sec.\il(\ref{Sec:ount}).
 \\
 &set limits for location of the inner and outer edges of quiescent tori with $\ell>\ell_{mbo}=4$&
\\
$\ell_{\lim}^{\texttt{\textbf{couple}}}$&  momentum $\ell$ of the outer torus such that $r_{inner}(\ell_{\lim}^{\texttt{\textbf{couple}}})=r_{out}^{\times}(\ell)$ & Eqs\il(\ref{Eq:vast-maj})-- Figs\il(\ref{Fig:tookP},\ref{Fig:lom-b-emi})
\\
$K_{r_{\wp}}(r_{\wp})\in\{K_{cent},{K_{\times}}\}$&
 $K$-Parameter  at the center of torus, $K_{cent}$,  or the value $K_{\times}$ for  cusped  tori &
Eq.\il(\ref{Eq:rac.tes.tec})
 \\
 & function of   $r_{\wp}\in\{r_{out}^{\times},r_{cent}\}$,  or the outer edge of  cusped torus&
\\
\hline\hline
\end{tabular}}
\caption{{Main symbols and relevant notation  used throughout the article. HD in the table   is for hydrodynamic. Table includes the link to related sections, equations and figures. General notation convention includes the following rules: $\mathcal{\mathbf{i.}}$ For any quantity $\mathcal{Q}$ evaluated  on  a general radius $r_{\bullet}$, we  adopt notation $\mathcal{Q}_{\bullet}\equiv \mathcal{Q}(r_{\bullet})$; $\mathcal{\mathcal{\mathbf{ii.}}}$ Any quantity $\mathcal{Q}_{\times}$ generally it is intended related to a  cusped tori. {$\mathcal{\mathbf{iii.}}$ In general $r_{crit}$  is  understood  as minimum pressure and density points, unstable points as specified in the table. However, in particulary aspects of this analysis it has been in fact convenient to generally intend more  the  minimum  and \emph{maximum}   points in other word  the critical points of the effective potential in Eq.\il(\ref{Eq:sha-conf}), of course this case is  clearly stated in the text.
$\mathcal{\mathbf{iv.}}$  In general, when not otherwise specified, the  superscript or subscript $(i)$ and $(o)$ stands for inner  and outer torus $(T^i,T^o)$ respectively (and any quantity related to the two tori) according to location of the torus center (maximum pressure and density point in the disk), thus there is $r_{cent}^i<r_{cent}^o$  and $T^i$ is the torus closest to the central attractor, consequently we write
$T^i<T^o$.  The innermost configuration we refer in the table is the circled toroidal structure closely located (embracing) the \textbf{BH} horizon, disconnected from the outer configuration, solution of the same Euler equation with equal boundary conditions and parameters values, being the "inner lobe" at the cusp emergence--see Fig.\il(\ref{Fig:Quadr}) and for a discussion on the significance of this  configuration \cite{pugtot}. }}}
\label{Table:pol-cy}
\end{table*}
%
\section{Misaligned perfect fluid tori on a static background}\label{Sec:Misal}
The aggregates of misaligned test tori (not influencing gravitationally each other and the spacetime background) can be treated in fully analytical way in the spherically symmetric background where for  any  toroid we can choose a  central plane that  can be considered its symmetry plane. Such construction is not possible in the rotating Kerr spacetimes, as their axial symmetry fixes the equatorial symmetry plane that can be symmetry plane of the toroidal configurations.
In order to describe   misaligned (inclined)  perfect  fluid  tori orbiting a central Schwarzschild \textbf{BH}
we adopt the
  Euler  equation:
\bea\label{Eq:Eulerif0}
&&
(p+\rho)u^a\nabla_au^c+ \ h^{bc}\nabla_b p=0,
\eea
for one-species  (simple) fluid toroid
where $h_{ab}=g_{ab}+ u_a u_b$ and $g_{a b}$ is
the Schwarzschild metric tensor, $M$ is the \textbf{BH} mass, { in the following  we set  for simplicity  $M=1$  in the main functions}, however we generally  keep the explicit dependence on the  parameter $M$ in the evaluations of the distances scales of the problem. The time-like flow vector field  $u$  denotes the fluid
four-velocity\footnote{The fluid  four-velocity  satisfies the normalization condition $u^a u_a=-1$. We adopt the
geometric  units $c=1=G$ and  the $(-,+,+,+)$ signature.  The radius $r$ has unit of
mass $[M]$, and the angular momentum  units of $[M]^2$, the velocities  $[u^t]=[u^r]=1$
and $[u^{\varphi}]=[u^{\vartheta}]=[M]^{-1}$ with $[u^{\varphi}/u^{t}]=[M]^{-1}$ and
$[u_{\varphi}/u_{t}]=[M]$. For the seek of convenience, we always consider the
dimensionless  energy and effective potential $[V_{eff}]=1$ and an angular momentum per
unit of mass $[L]/[M]=[M]$.},
%
 $\rho$ and $p$ are  the total energy density and
pressure, respectively, as measured by  observers moving with the fluid, we consider here a barotropic equation of state.
The continuity equation,
  $u^a\nabla_a\rho+(p+\rho)\nabla^au_a=0\, $ (where $\nabla_a g_{bc}=0$) is identically satisfied because of the symmetries: all the quantities $\Qa$ satisfy the conditions $\partial_i \Qa=0$ where $i\in\{t, \phi\}$, in the   standard spherical Schwarzschild  coordinate  system $\{t,r,\theta,\phi\}$.
  Its general integral  reads:
\bea\label{Eq:sha-conf}\int\frac{dp}{\rho+p}=-W\;
\left(\int_0^{p_{in}}\frac{dp}{\rho+p}=-(W-W_{in})\right),
\quad\mbox{where} \quad W\equiv\ln\left[\sqrt{\frac{(r-2) r^2}{r^3-\ell^2  (r-2)}}\right]\equiv\ln V_{{eff}}.
\eea
The effective potential $V_{eff}(r)$ governs the interplay of the gravitational  and inertial forces, and it is given by the spacetime geometry and the distribution of the specific angular momentum of the orbiting matter (in the following assumed uniform).
Function $V_{eff}(r)$ is defined on each torus symmetry plane, which is equivalent  to properly chosen central plane of the Schwarzschild geometry, and    $W_{in}$ denotes the values at the inner edge of the torus.

The disk fluid configuration described by the Euler equations,  Eq.\il(\ref{Eq:Eulerif0}),  (or modifications including other components for the force balance equations,   for example  due to  the magnetic field)  has been widely studied by
many authors, in particular we refer to the general  review \cite{abrafra,Font} and to  \cite{mnras,epl} for an in-depth study of the Schwarzschild case.
This model is essentially based on the original  framework envisioning
 the boundary of any stationary, barotropic, perfect fluid body as the
equipotential surface $W(\ell,\theta)=\mbox{constant}$ (known also as   ``Boyer's condition'' for the
analytic theory of equilibrium configurations of rotating perfect fluid bodies,
initially developed by \cite{Boy:1965:PCPS:}). For a barotropic fluid the
surfaces of constant pressure are given by the equipotential surfaces of the potential
defined by the relation in Eq.\il(\ref{Eq:Eulerif0}). 

We emphasize that while each torus of the agglomerate is on its symmetry  (equatorial) plane regulated by Eq.\il(\ref{Eq:Eulerif0}) and therefore, because of the spacetime symmetries, is independent of the inclination angle $\theta$, the boundary conditions defining  the inner structure of  the macro-structure composed of several  tilted tori described in  Eq.\il(\ref{Eq:Eulerif0}) depend on  the tori  relative inclination angle   $\vartheta_{ij}$ which we shall consider in defining the model.
All the main features of the equipotential
surfaces for a generic rotation law $\Omega=\Omega(\ell)$, where $\Omega$ is  fluid relativistic angular velocity
 are described here by the
equipotential surface of the simplest configuration with uniform distribution of the
angular momentum density $\ell\equiv L/E$ (specific fluid angular momentum), where for the individual matter elements,
 $(E, L)$   are two constants of motion, the  energy and angular
momentum per unit of rest mass as seen by infinity, respectively. The equipotential surfaces of the marginally stable
configurations orbiting in a Schwarzschild spacetime are defined by the constant $\ell$
\footnote{
The surfaces known as the  {von Zeipel's cylinders}, are defined by the conditions:
$\ell=\mbox{constant}$ and $\Omega=\mbox{constant}$. More precisely, the von Zeipel condition states: the surfaces of constant
pressure coincide with the surfaces of  constant density  (i.e. the isobar surfaces  are
also isocore) \emph{if and only if} the surfaces with the  angular momentum $\ell=
\mbox{constant}$ coincide with the surfaces with constant angular  velocity. In the static spacetimes, the family of von
Zeipel's surfaces  does not depend on the particular rotation law of the fluid,
$\Omega=\Omega(\ell)$, but on the background
spacetime only. In the case of a barotropic fluid,  the von Zeipel's theorem guarantees that
the surfaces $\Omega=\mbox{constant}$ coincide with the surfaces  $\ell=\mbox{constant}$. We address more specifically the von Zeipel surfaces in Sec.\il(\ref{Sec:zeipel}). }

\begin{figure}
    \includegraphics[width=5cm]{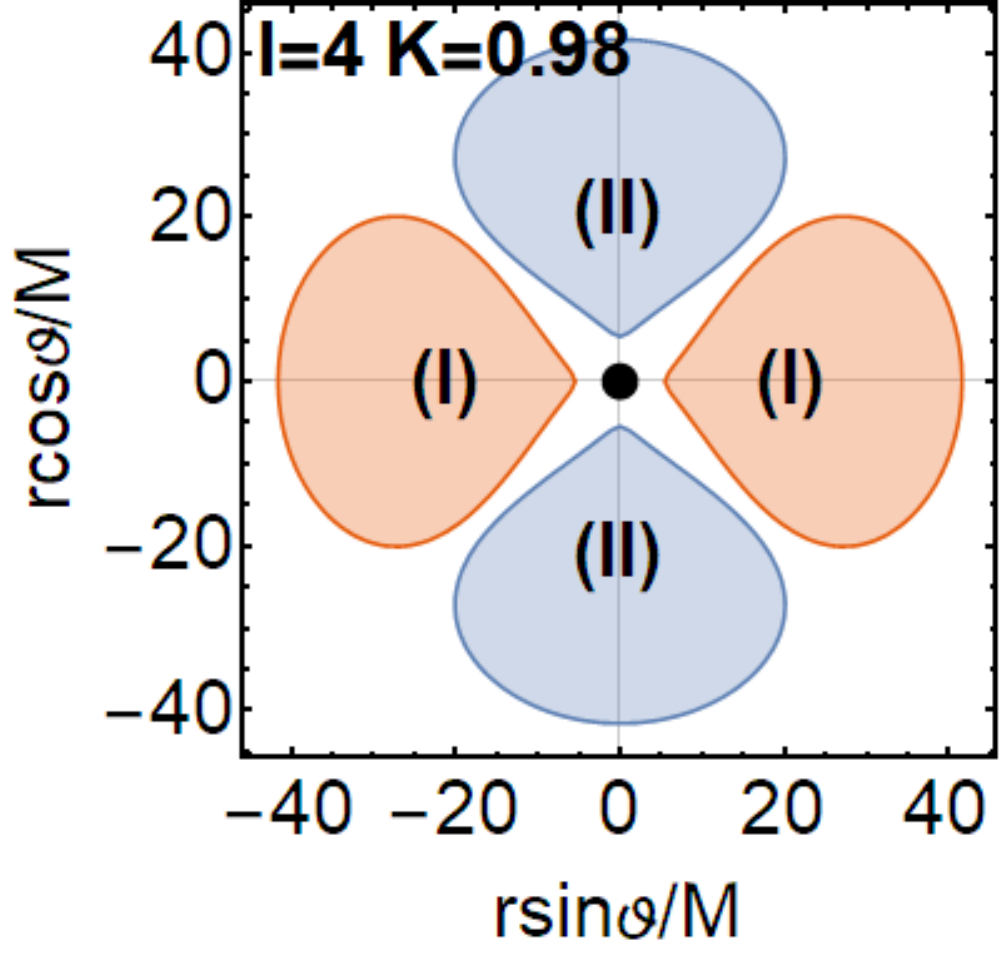}
  \includegraphics[width=5cm]{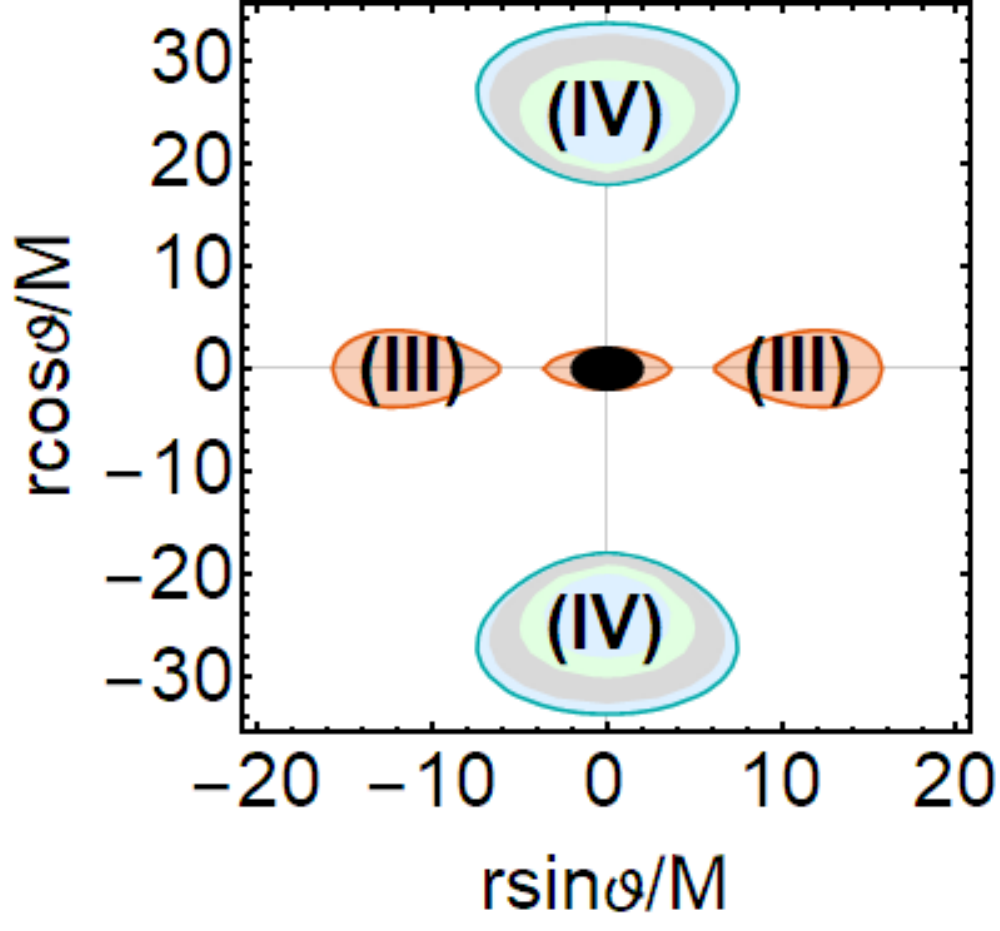}
  \caption{ Closed, not cusped (quiescent)    tori sections, solutions of the Euler equation (\ref{Eq:Eulerif0}). The system parameters are  the fluid specific angular momentum $\ell$, and   $K$  related to the energy function$K(r)$  Eq.\il(\ref{Eq:sincer-Spee}). Each torus \textbf{(I)},\textbf{ (II)}, \textbf{(III)} and \textbf{(IV)} represents a (magnetic like) dipole solution.  Right panel: couple  formed by  \textbf{(I)} and\textbf{ (II) } tori represents  a tori quadrupole configuration. Each component has parameters   $\ell= 4$, $K=0.98$. This configuration is  not possible  due to the  collision emergence.  Left panel shows the couple  of tori: \textbf{(III) } with parameters $\ell=\sqrt{15}$ and $K=0.96$, and  \textbf{(IV)}  with parameters $\ell=\sqrt{28}$ and $K=0.9813$,  constituting an observable  quadrupole tori. It is clear the presence of the innermost configuration embracing the central \textbf{BH}, in disks \textbf{(III)}. In the case of accreting  torus the innermost configuration merges  with the torus (outer Roche lobe) at the cusp (a Lagrangian point).}\label{Fig:Quadr}
\end{figure}

There are  three classes of solutions: {closed}, {open},
and {with a cusp} (self-crossing surfaces, which can be either closed or open). The
closed equipotential surfaces determine stationary equilibrium (quiescent) configurations: the fluid
can fill any closed surface. The open equipotential surfaces have been associated to
dynamical situations for example related to  the formation of  proto-jets \cite{open,proto-jet}.
 The critical, self-crossing and closed equipotential surfaces are relevant in the theory
of thick accretion disks, since the accretion onto the black hole can occur through the
cusp of the equipotential surface. In accordance with the original general idea in the development of the model,  the accretion is thus driven by a violation of the hydrostatic equilibrium (Paczy\'nski mechanism). The disk surface exceeds
the critical equipotential surface $W_{cusp}$ giving rise to a mechanical
non-equilibrium process that allows the  matter inflow  into the black hole. The
accretion onto the  \textbf{BH} is driven through the vicinity of the cusp due to
a little overcoming of the critical equipotential surface $W_{cusp}$ by the surface of
the disk.
Therefore, in this
accretion model the cusp of this equipotential surface corresponds to the inner edge of
the disk. We have to  study the equipotential surfaces,  defined by the condition
$W=\rm{constant}\equiv K$,  under assumption of uniform distribution of specific angular momentum ($\ell=\mbox{constant}$) corresponding to marginally stable perfect fluid configurations  \cite{Lei:2008ui,Abramowicz:2008bk,Abr-Jar-Sik:1978:ASTRA:}.
In the following we adopt the \textbf{RADs} framework developed in
 \cite{ringed,open,dsystem} for the  \textbf{eRADs} composed by tori orbiting on the equatorial plane of a Kerr attractor.
These systems, as confirmed by other analysis, have several  restrictions on the possibility  of formation,  their evolutions and  related  observational characteristics. They  are likely associated with transient  periods of the life of the attractor, especially  in the case of a  non-spherically symmetric attractor (for example undergoing a  dynamical phase  converging to a system of one  disk or with disks on the equatorial plane, implying moreover a change the \textbf{BH} characteristic parameters of mass and spins)  due to the possibility of  tori collision or tori accretion. The occurrence of both these conditions are here regulated by the model parameters.  Figs\il(\ref{Fig:Quadr})--\emph{right panel} shows an example of  quadrupole  configuration that cannot be observed,   while in  Figs\il(\ref{Fig:Quadr})--\emph{left panel}   is an example of quadrupole that can be observed. Furthermore in this quadrupole are possible all the solutions correspondent to minor values of $K$, therefore smaller tori at equal $\ell$ and $r_{cent}$ (equal maximum pressure and density points) at other  inclination angles. In this work particulary in Sec.\il(\ref{Sec:doc-ready}) we  provide the conditions determining these cases and the   observable characteristics  of the \textbf{RAD}.
\subsection{RAD rotational law}\label{Sec:RADloq}
\textbf{RADs}, and individual  tori of \textbf{RADs}, are governed by the Keplerian specific angular momentum  ($L/E$)  distribution.
In the Schwarzschild geometry it takes the form
\bea\label{Eq:lqkp}
\ell(r)\equiv \sqrt{\frac{r^3}{(r-2)^2}};
\eea
in order to fully reflect properties  of the inclined  toroidal structure, we introduce also an "energetic" Keplerian function
\bea\label{Eq:sincer-Spee}
K(r)\equiv V_{eff}(r,\ell(r))=\sqrt{\frac{(r-2)^2}{(r-3) r}},
\eea
 governing the local extrema of the effective potential (\ref{Eq:Eulerif0}):
$\ell=\ell(r)$ is thus the  magnitude of the specific angular momentum of a toroid centered at distance $r$ from the central \textbf{BH} on a general  central plane (polar angle $\theta=$constant).  Each  toroidal component can have   different relative  orientation (any inclination angle $\vartheta_{ij}=$constant).
We provide  further  notes on the derivation of   $\ell(r)$ and $K(r)$ and their significance in the Schwarzschild background in Sec.\il(\ref{Sec:four-b}).

\begin{figure}
 \includegraphics[width=9.7cm]{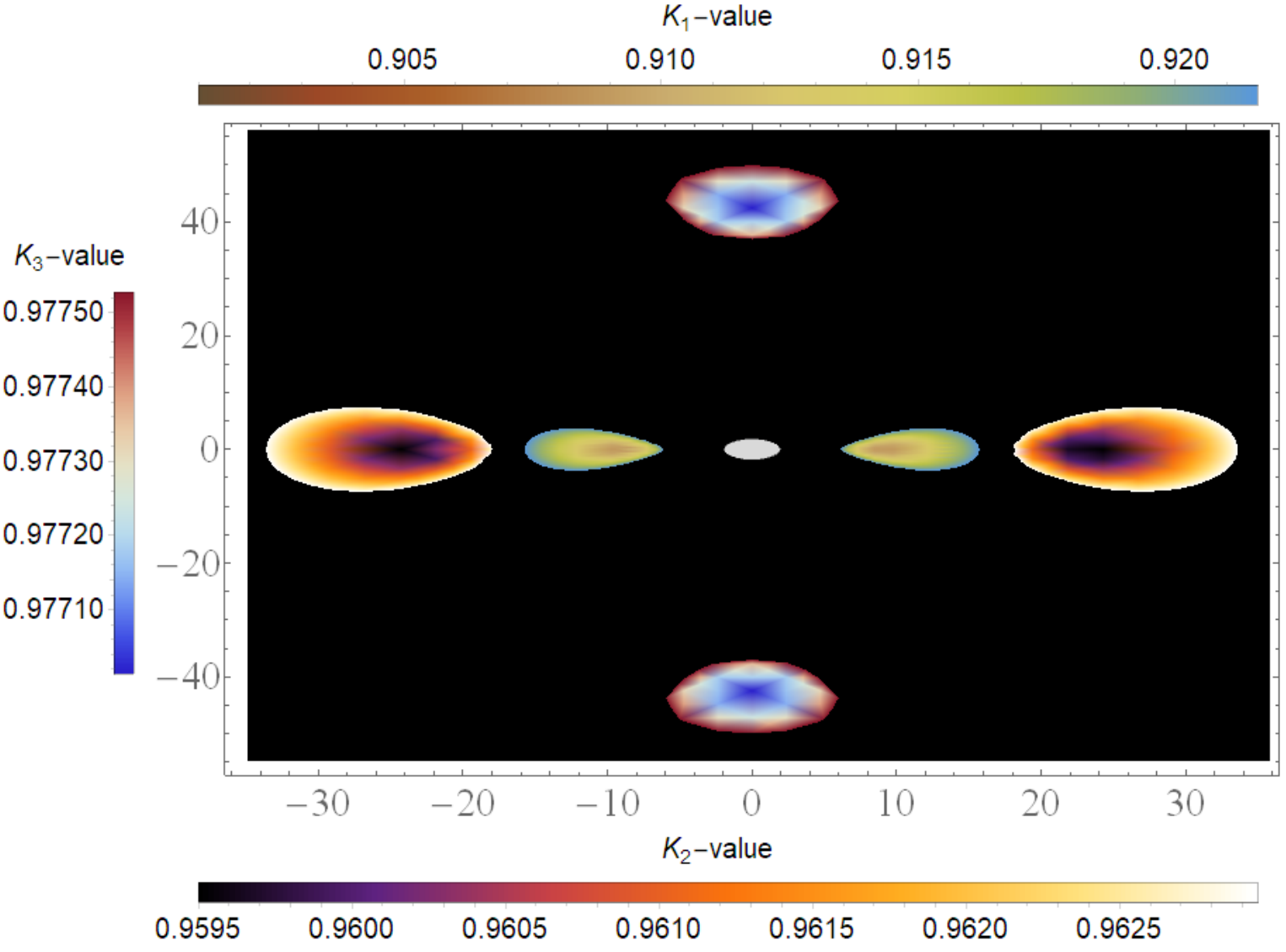}
\includegraphics[width=8cm]{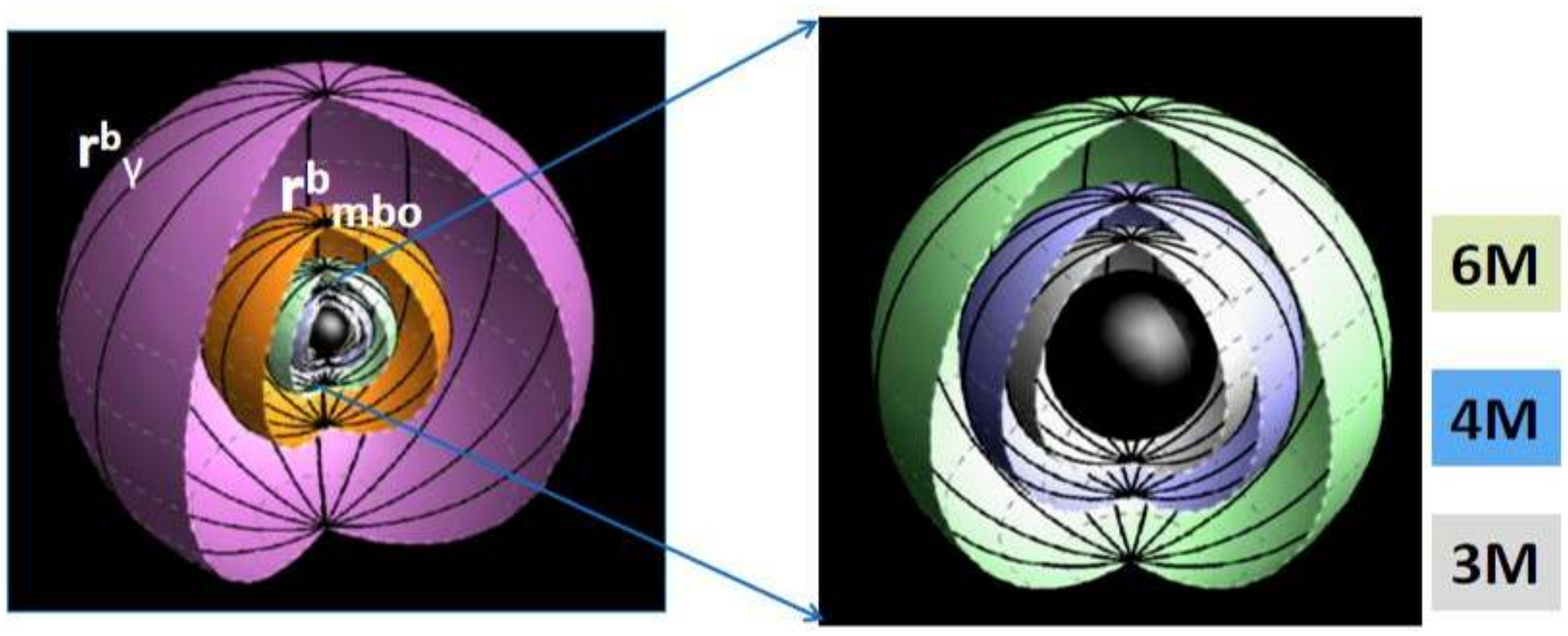}
  \caption{{Left panel: density plots of  tori orbiting the Schwarzschild central \textbf{BH} (gray surface). The tori parameter $K$ is shown in the plot lagend. The fluid specific angular momentum,  from the torus closest to \textbf{BH}  to   the most distant torus, is : $\ell_1=\sqrt{15}$, $\ell_2=\sqrt{28}$, $\ell_3=\sqrt{46.7}$. Correspondingly there are  parameters $K_1$, $K_2$ and $K_3$. }
Center and  right panel: Stability spheres for the Schwarzschild spacetime. Black sphere is the central $\textbf{BH}$. --See Eqs\il(\ref{Eq:swan1}) and  Figs\il\ref{Fig:KKNOK}.
Cusps of accreting tori are in $]r_{mbo},r_{mso}[=]4M,6M[$, the center in $[r_{mso},r_{mbo}^b[$.
Cusps of open cusped proto-jets configurations are in $]r_{\gamma},r_{mbo}[=]3M,4M[$, the center in $[r_{mbo}^b,r_{\gamma}^b[$. Configurations in $r>r_{\gamma}^b$ are quiescent.
\label{Fig:SIGNS} }
\end{figure}

Introduced in  \cite{ringed} as
\textbf{RAD} rotational law,  $\ell(r)$ (leading \textbf{RAD} function), in the frame of the \textbf{RAD} clusters, provides also the  misaligned toroids \textbf{RAD} distribution orbiting the central static attractor, in the ringed disks
\footnote{{Angular momentum (\ref{Eq:lqkp})  is a  well  known    function of the accretion  disks models, representing  an upper boundary condition on the disk rotation,with the respect to the  "Bondi case", distinguishing between the slow rotating disks, often referred to as "Bondi flows" spherically-symmetric (non-rotating, small accretion rates) accretion  \cite{Bondi}, and fast rotating disks. Considering therefore this last  case, it is assumed that an accretion  disk must have  an extended region where matter has a  large  centrifugal component, i.e., $\ell\geq\ell(r)$--\cite{abrafra}.}}.
The specific angular momentum $\ell$  also parameterizes each torus in the \textbf{RAD}, together with  the further $K$-parameter.  In general, each individual torus can be characterized by a radial profile $\ell(r)$ satisfying the condition $d\ell(r)/dr\geq0$ (in each torus); here we consider the simplest  limiting case $\ell=$constant corresponding to marginally stable tori. The toroidal equilibrium structures have the central radius  given by condition $\ell=const=\ell(r_c)$ fulfilled at the part of $\ell(r)$ curve corresponding to stable circular geodesics, at $r>r_{mso}/M=6$; crossing point  $\ell=\ell(r_\times)$ at $r_{mbo}<r<r_{mso}$ corresponds to the cusped toroidal structure ($r_{mbo}/M=4$)--Figs\il\ref{Fig:SIGNS}.
Figs\il(\ref{Fig:SIGNS})--left shows an example of  "orthogonal  tori" (i.e tori with relative misalignment  angle  $\vartheta=\pi/2$  where in an \textbf{eRAD}  all tori have $\vartheta=0$), where    we  also show  different $K$ values  in the torus,  from the maximum  (corresponding to  the outer region)  to the minimum  (corresponded to the inner region). All these possible configurations at different $K$s, as evidenced in Figs(\ref{Fig:Quadr}), are possible being  no  collision emergence, but eventually in the later phases which will be defined  by a variation of the model parameters values.  In Sec.\il(\ref{Sec:doc-ready}) we established  the morphological characteristics of these  tori and the  collision occurrence.
%
\begin{figure}
\includegraphics[width=8.69cm]{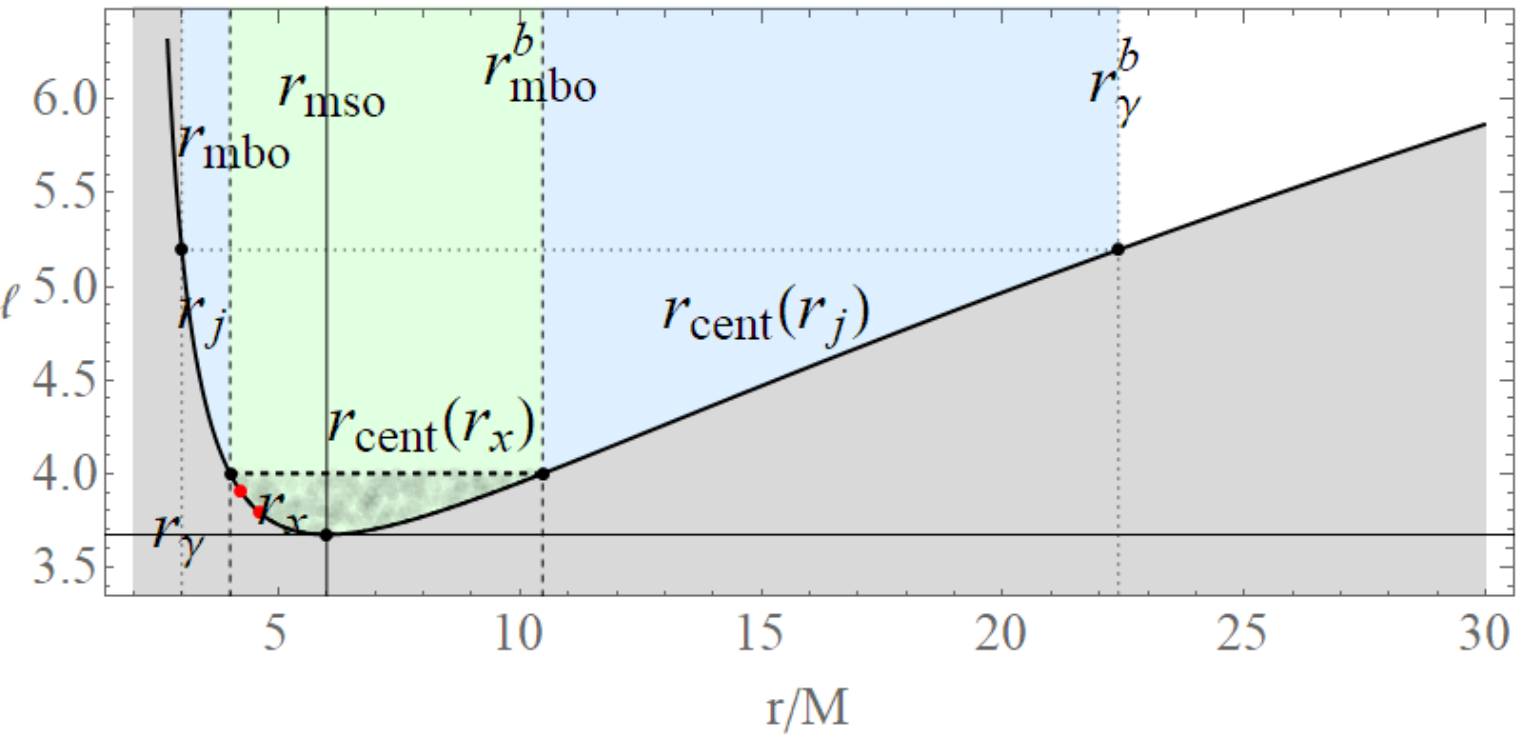}
\includegraphics[width=8.69cm]{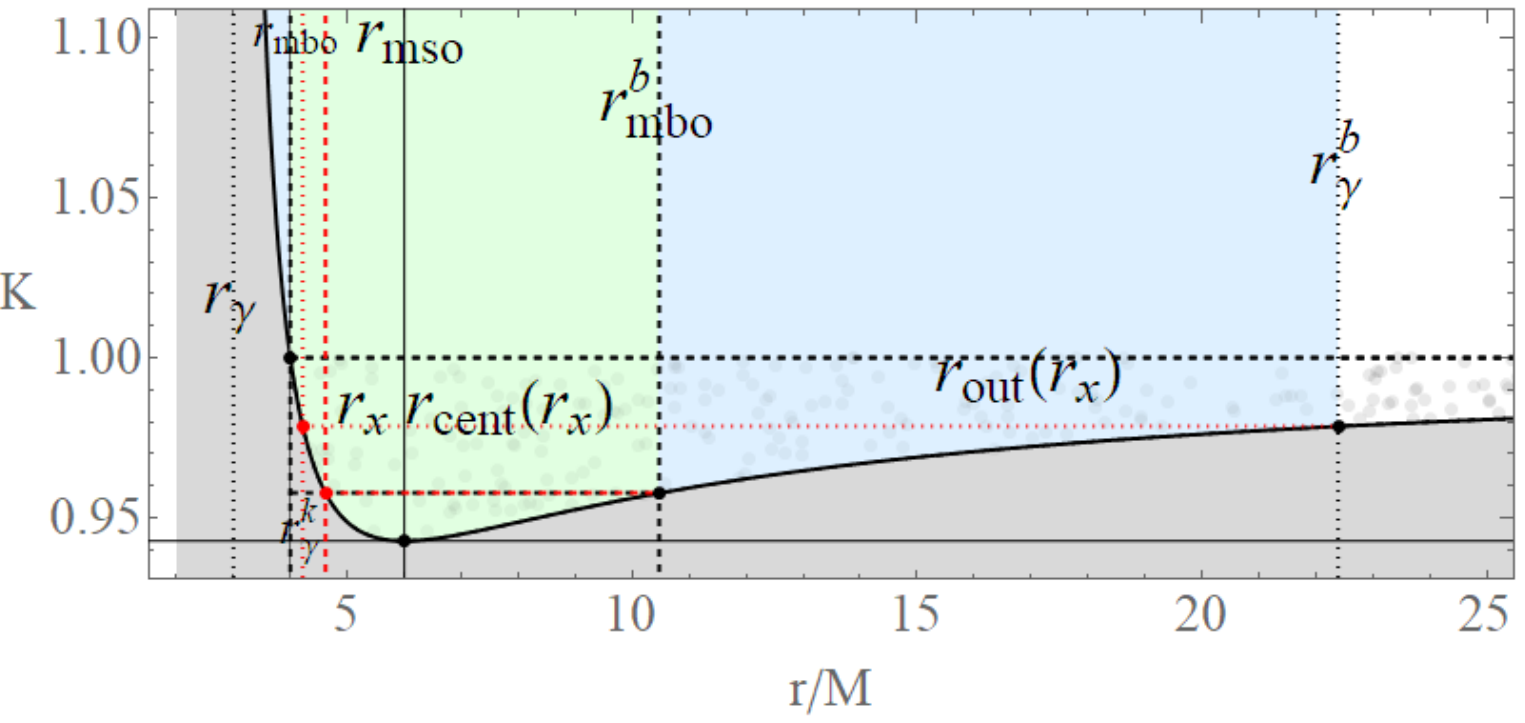}
\\
\includegraphics[width=3.69cm]{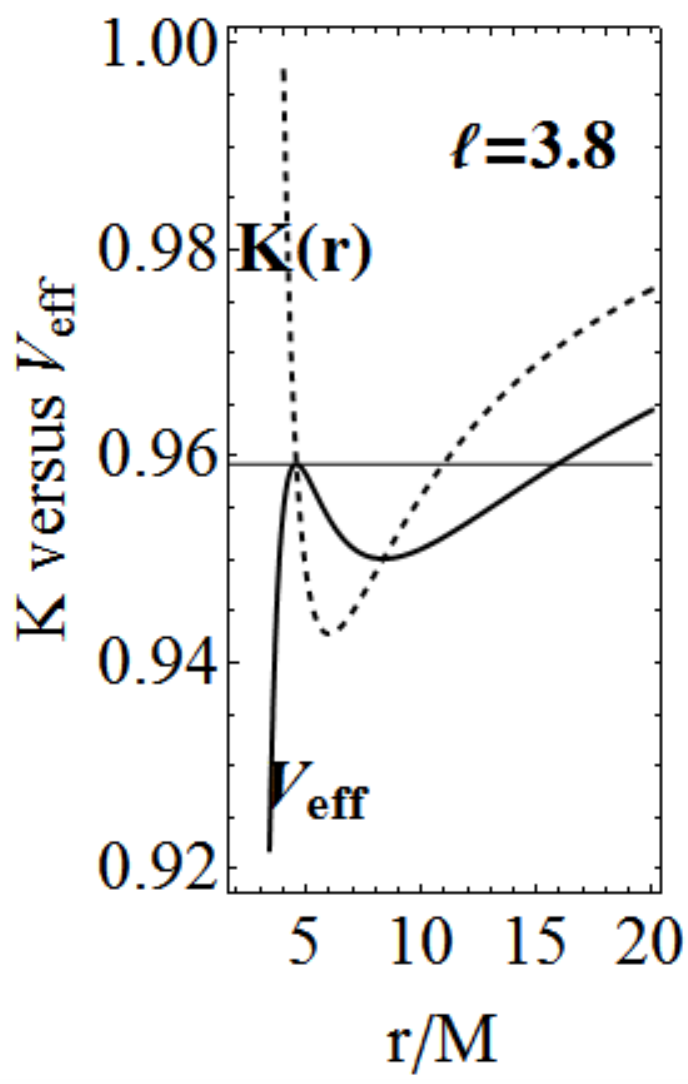}
\includegraphics[width=5.69cm]{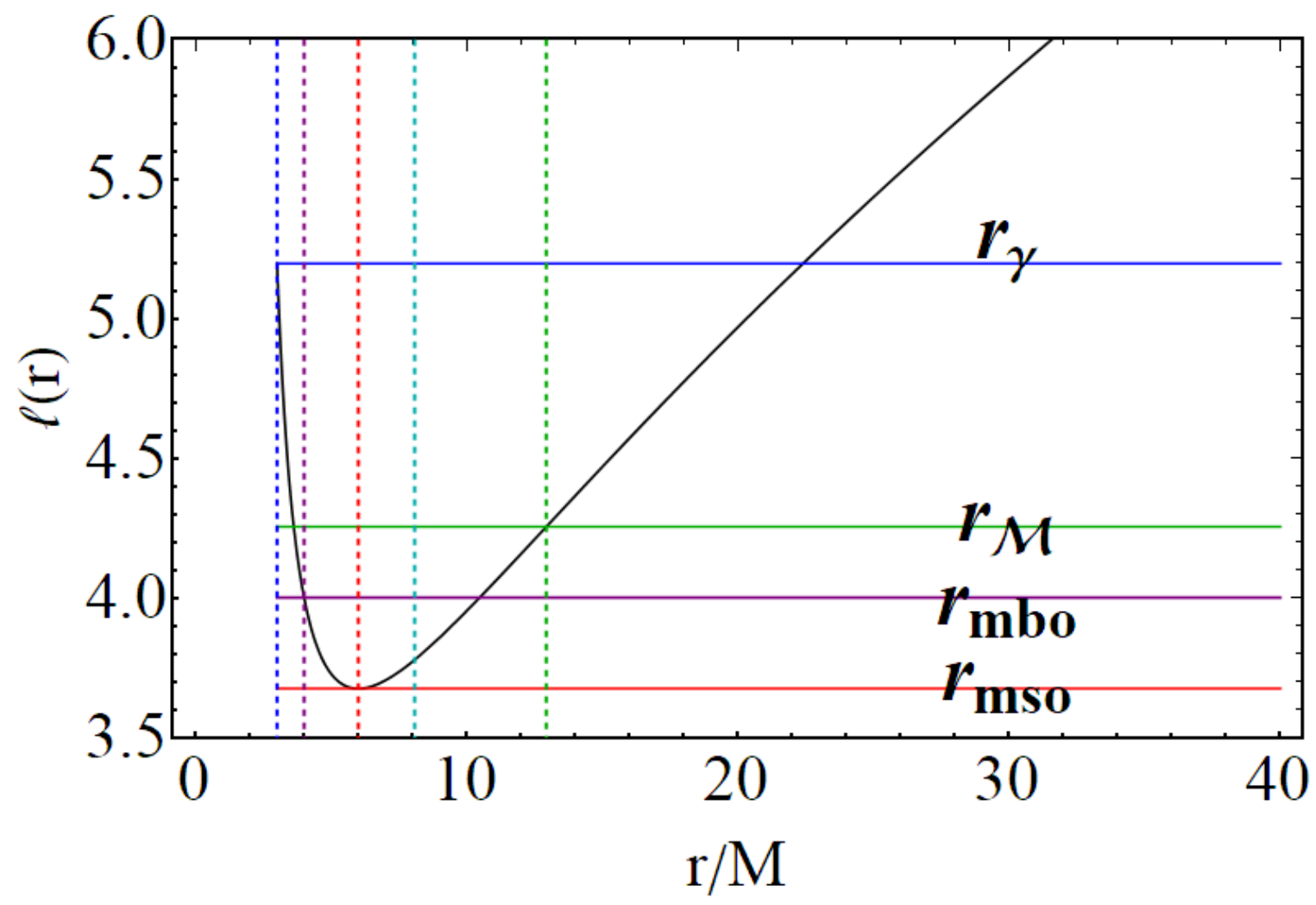}
\includegraphics[width=5.69cm]{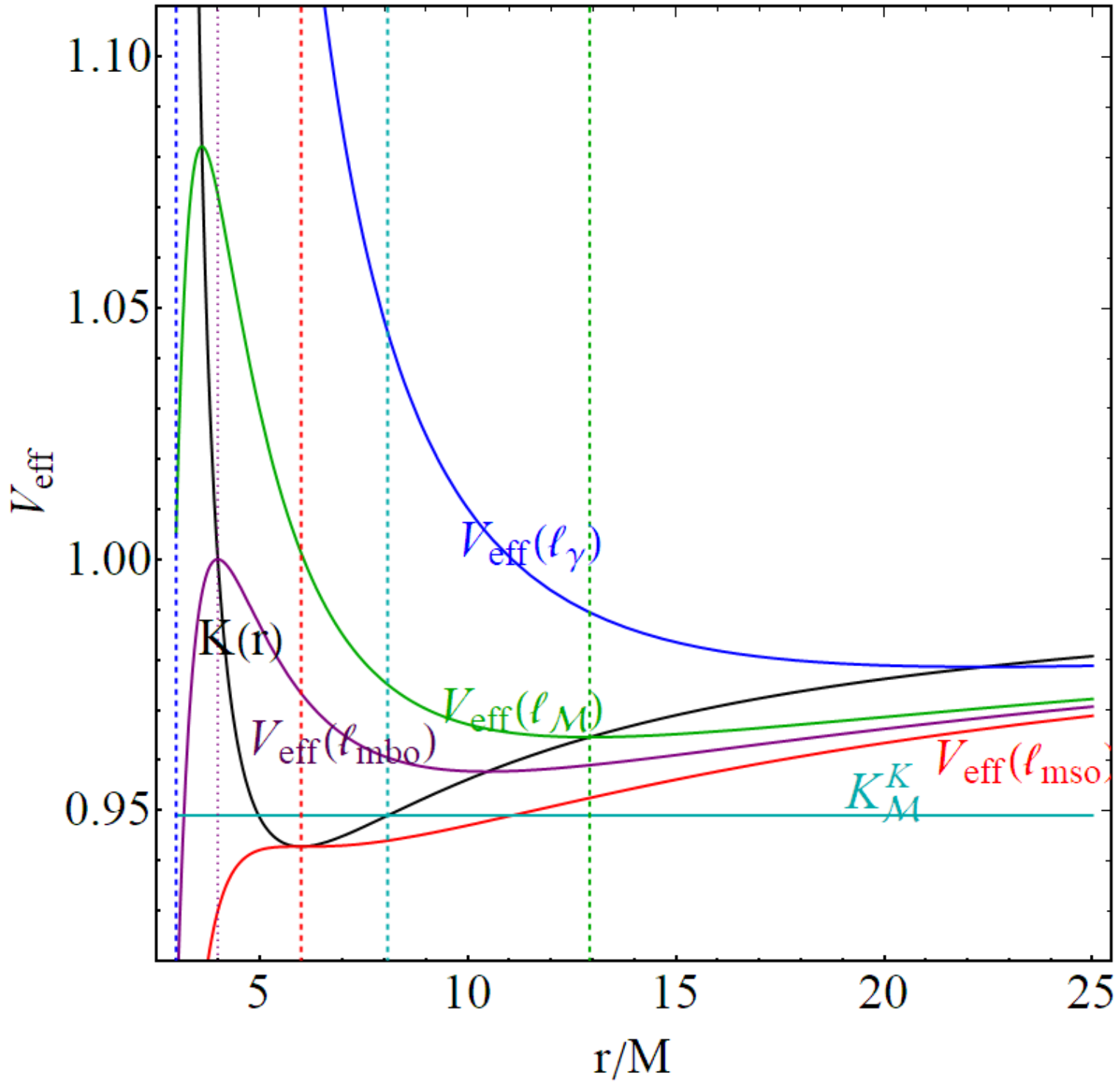}
\caption{Upper Left panel: \textbf{RAD} (and each torus component) specific angular momentum distribution   Eq.\il(\ref{Eq:lqkp}) as function of the radius $r/M$ on the  rotation plane of each torus.
Upper Right panel: $K-$function for the \textbf{RAD} (and each torus component) as function of $r/M$.
Radii $r_{\gamma}, r_{mso},r_{mbo}$ and $r^b_{\gamma}, r^b_{mbo}$, defined in Eq.\il(\ref{Eq:swan1}) are also shown--
see also Figs\il\ref{Fig:SIGNS} and Figs\il\ref{Fig:solidy}. Regions of existence for the center of different kind of  configurations and the respective critical points are emphasized with different colors--see Sec.\il(\ref{Sec:dig}). Below panels. Left: function $K(r)$ of Eq.\il(\ref{Eq:sincer-Spee}), the effective potential $V_{eff}$ as function of $r/M$ for the fluid specific angular momentum $\ell=3.8$.  Center: location of radii $r_{\mathcal{M}}$ in the specific angular momentum distribution $\ell(r)$. Right: effective potential $V_{eff}$ as function of $r/M$ evaluated  on different $\ell_{\bullet}\in\{\ell_{\gamma},\ell_{mbo},\ell_{mso},\ell_{\mathcal{M}}\}$--see Table\il(\ref{Table:pol-cy}). Black curve is the function  $K(r)$  of Eq.\il(\ref{Eq:sincer-Spee}).
\label{Fig:KKNOK} }
\end{figure}
Since   the spacetime is spherical symmetric, the toroidal  configurations are determined by quantities $\Qa\equiv\ell^2$ (which reflects the fact that the  $\ell$counterrotating tori, correspondent to condition $\ell_i\ell_j<0$ can be described {as $\ell$corotating tori of   the \textbf{eRAD},
in the sense that  their   characteristics are similar to the \textbf{eRAD} $\ell$corotating  tori, for example, for  their   stability   in \textbf{RAD} including cusped surfaces}.
Then we are able to characterize  the individual toroidal structures by the function $K(r)$
governing the extremal points of the effective potential (and the pressure $p$), as the extreme of the effective potential corresponds to the Keplerian orbits where $\ell=\ell(r)$--Figs\il\ref{Fig:KKNOK}.
The values of $K=$const govern, for fixed $\ell=$const, the concrete closed, equipotential surfaces.
Parameters $\ell=$constant, $K=$constant determine uniquely the tori in the Schwarzschild geometry.
For fixed torus center  $r_{cent}$ (maximum density points), the  function $K(r)$ of Eq.\il\ref{Eq:lqkp} provides  an independent tori parameter regulating the torus extension on \emph{its} equatorial plane, the torus density, the emergence of hydro-dynamical instability, the torus thickness and other morphological and dynamical characteristics of the torus at constant $\ell$ (we refer, for an  extensive discussion on the role of $\ell=$constant, to \cite{ringed,long,dsystem}).
 Torus evolution towards accretion  involves generally a decrease of the momentum magnitude $\ell$ and (but not always) an increase  of $K$-parameter--see discussion in  \cite{pugtot}.
In the case of equal  $\ell$, \textbf{RAD} tori  have equal maximum  pressure  (and  density)-points (eventually also equal minimum pressure points) but not in general  equal geometric center-- which  depends on  $K$ (we study this case in details in  Sec.\il(\ref{Sec:doc-ready})).
More generally,  many of the  tori essential  characteristics   can be obtained from the curves Eq.\il(\ref{Eq:lqkp}) constant.
 Equal  $\ell$ tori, not possible in the \textbf{eRAD} frame, might be possible configurations in the \textbf{RAD} because of the  different tori inclination angles. They  present a doubled collisional region, minimized in case of torus maximum inclination--$\vartheta_{ij}=\pm\pi/2$--coinciding with the toroidal section with $\min(K(\ell))$ and having equal maximum density points. From this point of view then  misaligned accretion tori around static attractors are {"advantaged" with respect to the  \textbf{eRAD} configurations, in the sense that they are subjected to constraints and instabilities  in many ways less complex than in the case of the Kerr spacetimes and, analogously,   formation and presence of misaligned structures around static attractors are likely to be expected.}

\subsubsection{Condition $\ell(r)$=constant}\label{Sec:dig}
\begin{figure}
\includegraphics[width=8.69cm]{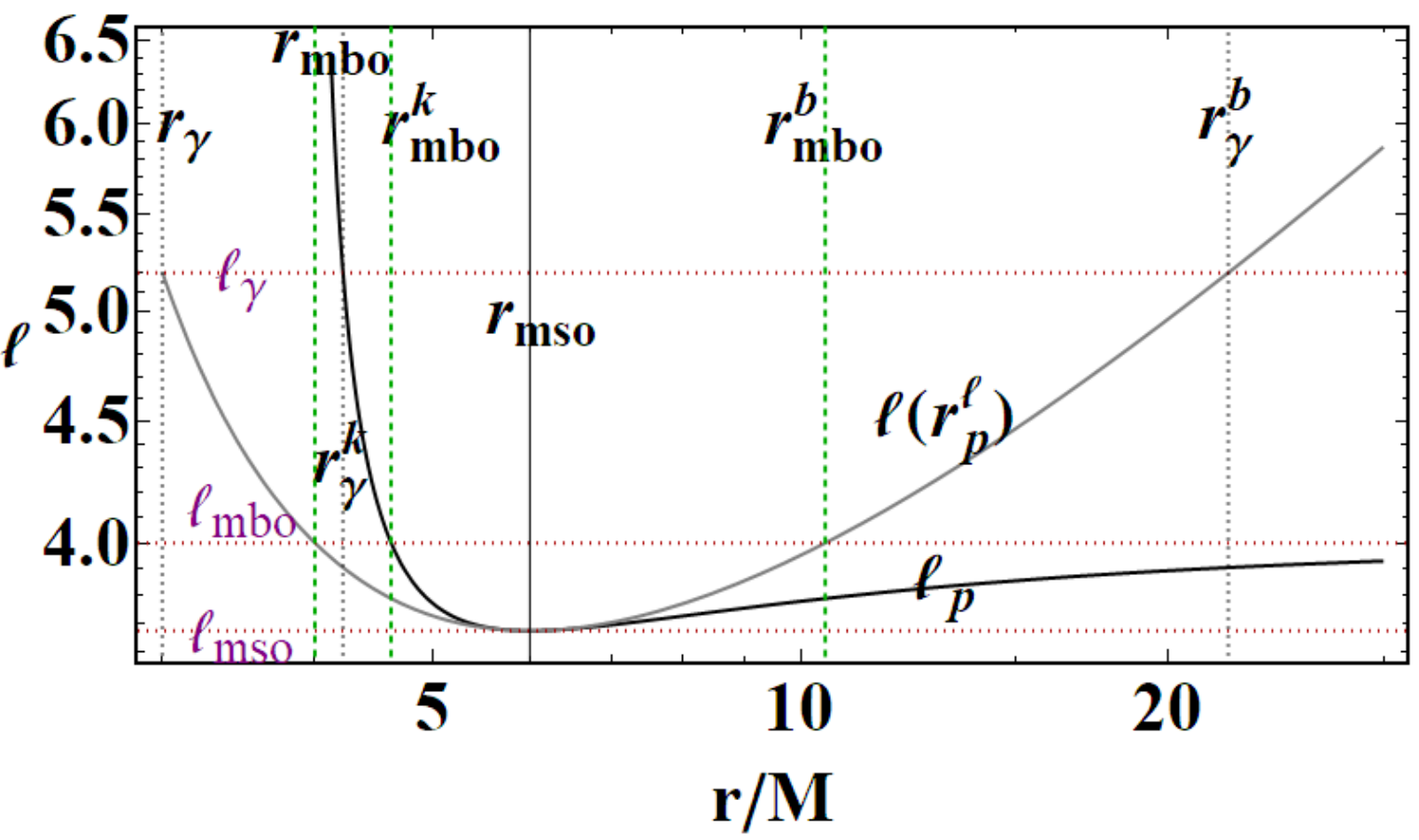}
\includegraphics[width=8.69cm]{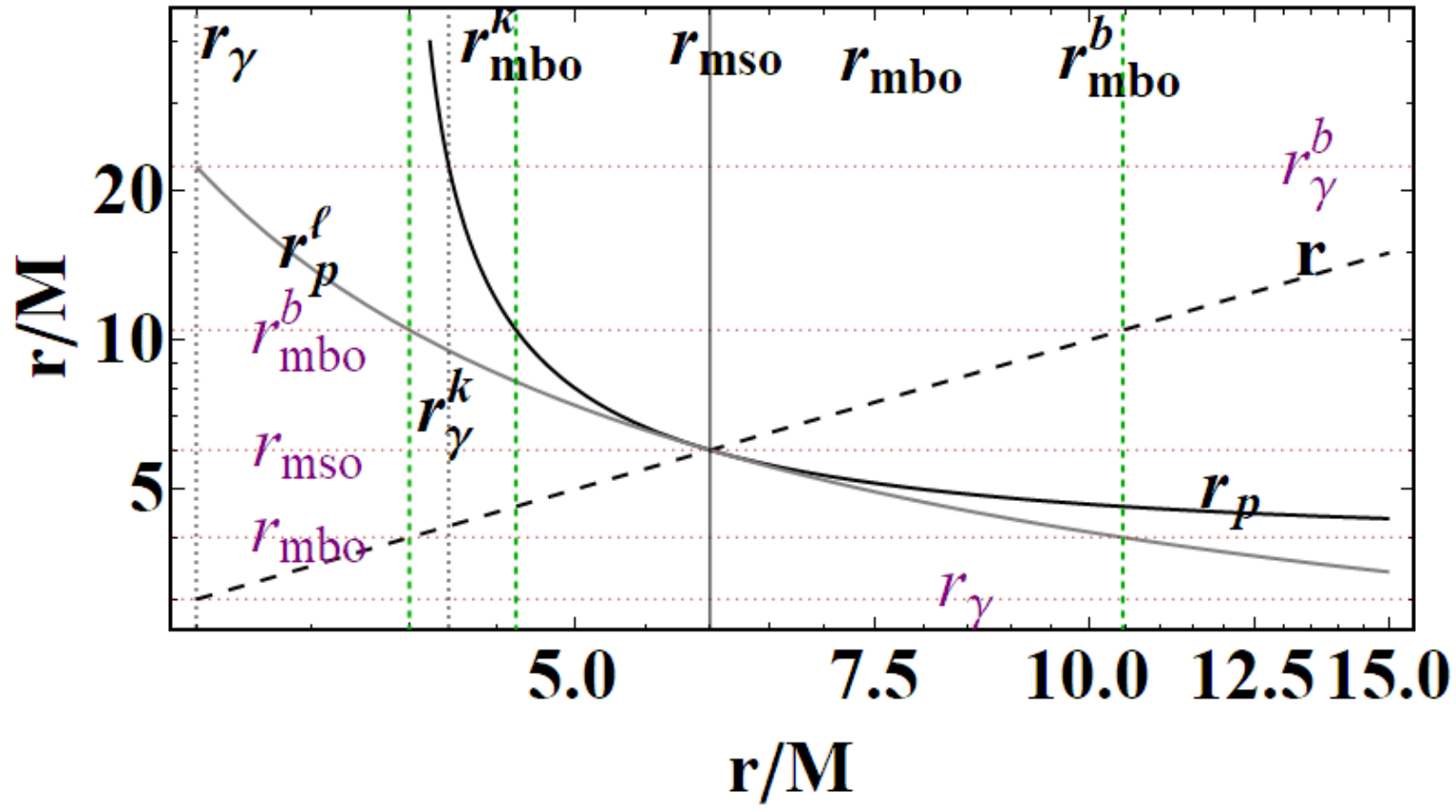}
\includegraphics[width=7.4cm]{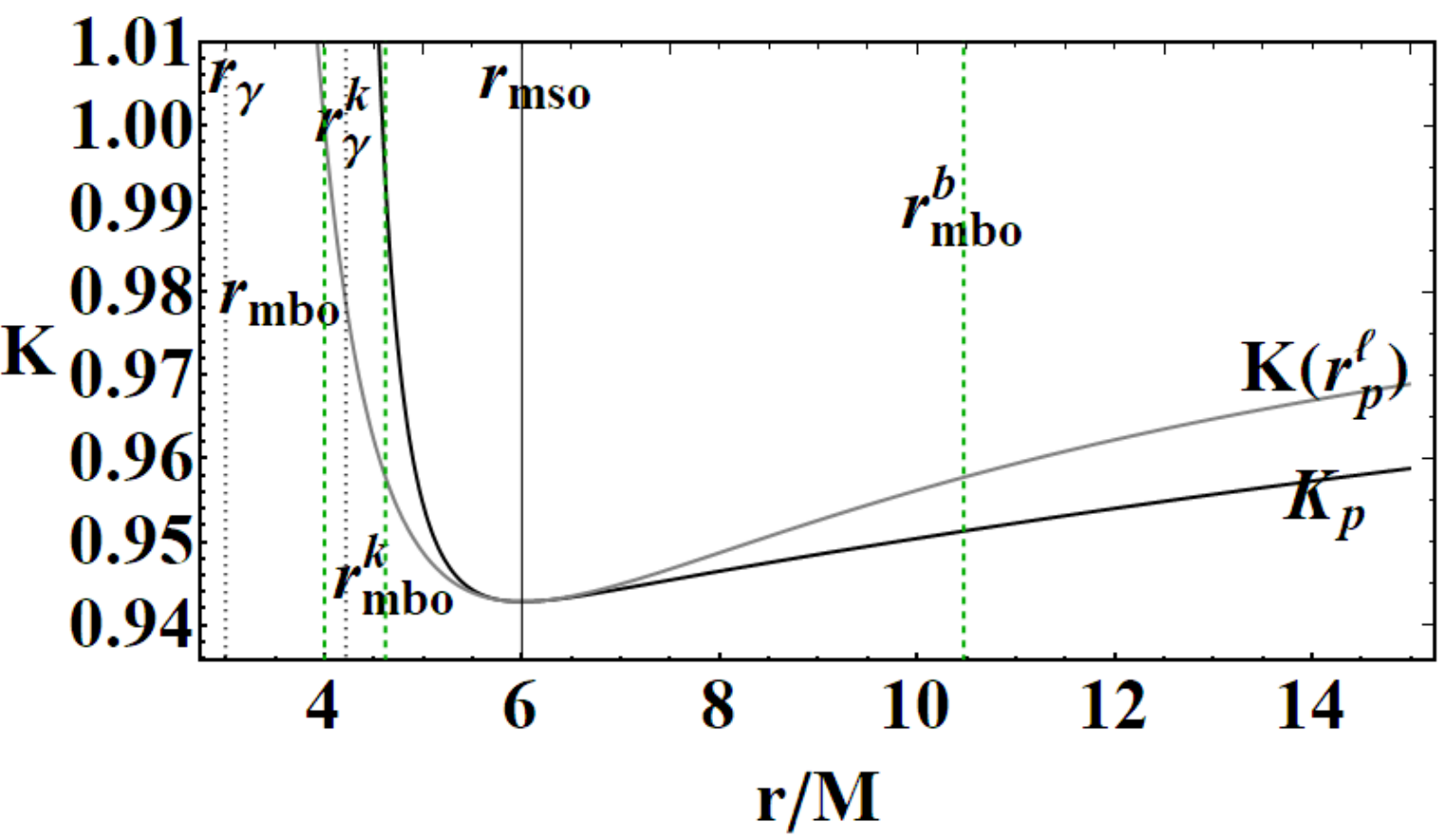}\\
\caption{Plots of quantities $\ell_p$, $r_p$   and $K(r_p)=K_p$ defined in Eq.\il(\ref{Eq:lqkp}) and Eq.\il(\ref{Eq:Krrp}) as function of $r/M$.
Radii $r_{\gamma}, r_{mso},r_{mbo}$ and $r^c_{\gamma}, r^c_{mbo}$, $r^K_{\gamma}, r^K_{mbo}$ are also shown--
see also Figs\il\ref{Fig:KKNOK}.
\label{Fig:solidy} }
\end{figure}

Curves $\ell(r)=$constant identify the center $r_{cent}$  (maximum pressure points) and, eventually, the  instability point $r_{\times}<r_{mso}<r_{cent}$ of an cusped  torus with cusp $r_{\times}$. {Therefore to obtain this solution we solve equation  $\ell(r)=\ell(r^{\ell}_p)$ for the radius  $r_{p}^{\ell}$ as function of the radial $r$ variable  finding:}
\bea\label{Eq:schaubl}&&
r_p^{\ell}=2 \left[\sqrt{\frac{r^2 (2 r-3)}{(r-2)^4}}+\frac{(r-1)r}{(r-2)^2}\right],\quad
 K(r_p^{\ell})=2 \sqrt{-\frac{(r-2)^2}{r \left[6 \left(\sqrt{2 r-3}-1\right)+r \left(r-4 \sqrt{2 r-3}+2\right)\right]}}
\eea
{where $r\in[2,r_{\gamma}^b]$ and $ K(r_p^{\ell})$ is the  corresponding energy curve from Eq.\il(\ref{Eq:sincer-Spee}) --Figs\il\ref{Fig:solidy}.  The  curves  $\ell=$constant on the  \textbf{RAD}  rotational law  in Eq.\il(\ref{Eq:lqkp}) identify a torus in the globular distribution, specifically one point  which is the  torus center (pressure and density maximum point). The other point has an essential role for the  torus  being  eventually a cusp point}. Radius $r_p^{\ell}(r)$ gives, with  equal  $\ell$, the radii  $r_{cent}$ or $r_{crit}$: for closed cusped tori, the critical point $r_{crit}$ is the  accretion point as function of the other radius of the couple. $K(r_p^{\ell})$ provides the $K$-parameter value of the maximum pressure $r_{cent}$ or minimum  ($r_{crit}$) (if this exists)--Figs\il\ref{Fig:solidy}. In this way we expressed the disk center as function of  the instability point and vice versa. This finally turns in a  \textbf{RAD} parametrization  in terms of disks pressure gradients, instead of  its  rational law. The radius $r_{\gamma}^b=M$  has an essential role in determining some particular tori, as we will see below.

The analysis of these conditions  provides the     ranges for the location of the accretion tori edges (cusps for inner edge  ``accreting tori'') and  the center of the open configurations.
Moreover we  discuss more deeply   the role of the radii of the spacetime structures, significant int he determination of the toroidal structures.
Therefore we obtain the   conditions for the existence of   the pressure and density critical points    of the instabilities and their location in the torus.  Considering also  \cite{open,long},  we can identify spheres $\mathcal{S}$ and spherical corona $\mathbf{R}_x$ with $x\in\{0,...,5\}$, centered on the attractor, relevant to this analysis and defined from the condition on the  radial distance  $r$ as follows:
\bea&&\label{Eq:swan}
\Sa_+: r\leq r_+;\mathbf{R_0}: r\in]r_+,r_{\gamma}[,\quad \mathbf{R_1}: r_j\in]r_{\gamma},r_{mbo}[
\eea
$ \mathbf{R_1}$ is the spherical annulus, as  defined by the radius range, where the cusp points for the open surfaces are located  $r_j$.
Notation is as follows: $r_+$ is the  \textbf{BH} horizon, $r_{\gamma}=3M$ is the marginally circular orbit (last photon orbit), $r_{mbo}=4M$ marginally bounded orbit, $r_{mso}=6M$ marginally stable orbit. $\mathbf{R_1}$ locates the  proto-jets cusps.
 Note that the radius of this spherical corona   $r_{\mathbf{R_1}}=M$ see \cite{proto-jet,long,open}  (the volume where such proto-jets  emissions are possible is $\mathcal{V}_{\mathbf{R_1}}={148 M^2\pi }/{3}\approx154.985M$).
(Note that  the regions  evaluated here, are those of the space in Figs\il(\ref{Fig:SIGNS}), regions of existence of the equatorial sections (on  torus symmetry each plane) of the  surfaces in  Figs\il(\ref{Fig:Quadr})).
We   therefore have
\bea\label{Eq:swan1}
&&\mathbf{R_2}: r_{\times} \in]r_{mbo},r_{mso}[,\quad
  \mathbf{R_3}: r_{cent} \in]r_{mso},r_{mbo}^b[, \quad r_{\gamma}^{b}=6 \left(\sqrt{3}+2\right)M=22.3923M
\eea
--Figs\il\ref{Fig:KKNOK},\ref{Fig:solidy},\ref{Fig:SIGNS}.
 Radius $r_{mbo}^{b}$  is solution  of\footnote{In general we adopt the following notation: for any quantity $\mathbf{Q}$ and radius $r_{\bullet}$ we adopt the notation $\mathbf{Q}_{\bullet}\equiv \mathbf{Q}(r_{\bullet})$,  for example
 there is $\ell_{\mso}\equiv\ell(r_{\mso})$.}  $\ell(r)=\ell_{mbo}$.
The spherical annuli   $\mathbf{R_2}$  and $\mathbf{R_3}$ locate   the  cusps, $r_{\times}$, and centers points, $r_{cent}$, of the  cusped tori $\cc_{\times}$.   $\mathbf{R_2}$, locating the inner edge $r_{\times}$ of an accreting tori,  is  a range whose linear  extension on the equatorial plane is   $r_{\mathbf{R_2}}=2M$ and with total  volume  $\mathcal{V}_{\mathbf{R_2}}={608 \pi M^3}/{3}\approx 636.696M^3$.
$\mathbf{R_3}$, locating  the center of the  cusped tori (tori with $\ell\in]\ell_{mso},\ell_{mbo}[$ where cusped closed configurations are possible). The  radius of this region is
$r_{\mathbf{R_3}}=4.47214M$ and total volume is  $\mathcal{V}_{\mathbf{R_3}}=3905.77M^3$.
The range $(\mathbf{R_2}\cup \mathbf{R_3})=]r_{mbo},r_{mbo}^b[$ is the maximum extension of the region in the torus between the  center and inner range  of an  accreting torus, the extension of this range is $r=6.47214M$ while its  volume is $\mathcal{V}=4542.46M^3$.
These configurations, especially if quiescent, can  achieve  considerable size, even  with  equatorial elongation $\lambda>200M $. Such discs therefore may be affected by several  factors influencing their formation and  stability as for example   their self-gravity.
We note then that the location of the point of maximum density in the disk in $r_{cent}\in \mathbf{R_3}$   remains rather close to the cusp $r_{\times}$, this point, however, does not coincide with the geometric maximum point in the disk, as we will study in detail in Sec.\il(\ref{Sec:doc-ready},\ref{Fig:orto}).
Then there is
\bea
   && \mathbf{R_4}:r \in]r_{mbo}^b,r_{\gamma}^{b}],\quad\mathbf{R_5}:r >r_{\gamma}^{b} \quad
\quad r_{mbo}^b=2 \left(\sqrt{5}+3\right)M\approx10.4721M.
\eea
Radius   $r_{\gamma}^{b}$ is solution of  $\ell(r)=\ell_{\gamma}$.
The annulus  $\mathbf{R_4}$ locates  the centers of the surfaces associated to the  open cusped solutions. Finally $\mathbf{R_5}$ indicates the  region, open from  above, where   the  centers  of the (equilibrium or {quiescent} i.e. with no cusp) tori  with moment $\ell>\ell_{\gamma}$ are located and consequently  there is $r_{cent}>r_{\gamma}^b$, in  Sec.\il(\ref{Sec:zeipel})  on the other hand we shall discuss the possibility of further limitations on this extended region.
(Then $\mathbf{R_4}$ locates  the center of  configurations with $\ell\in]\ell_{mbo},\ell_{\gamma}[$, where proto-jets (the limiting  cusped open configurations) are possible; the radius extension and volume are $r_{\mathbf{R_4}}=11.9202M$ and $\mathcal{V}_{\mathbf{R_4}}=42220.5M^3$ respectively).
 $\mathbf{R_5}$ locates
the center of the  configurations with $\ell>\ell_{\gamma}$. These quiescent  tori are located far away from the attractor  where  no minimal point of hydrostatic pressure can occur. (Note that the parameter ranges where cusped tori are possible is   smaller then the other characteristic parameter ranges.). Finally note  while in the Schwarzschild case the annulus are all concentric i.e. we can write  $\mathbf{R_0}<\mathbf{R_1}<\mathbf{R_2}<\mathbf{R_3}<\mathbf{R_4}<\mathbf{R_5}$ a condition that ensures many stability properties of this \textbf{RAD}, this relation is not a general properties of the Kerr geometries, but depends on the spin-to-mass ratio of the central \textbf{BH}, this condition differs for corotating fluids or counter-rotating fluids, furthermore in the \textbf{RAD} construction, in different spin-mass ranges of values it is necessary to relate together the annulus for corotating and counter-rotating fluids.
This analysis can be found in \cite{dsystem,Letter,Multy,long}.

 We also distinguish the
\emph{crossing spheres} as the  spherical regions  where  two tori, $T_1$ and $T_2$, crosses.
 In these spheres  the \emph{contact zones} are modeled with regions of one torus $T_1$ intersecting $T_2$, these are typically two symmetric elliptic  sections of the tori, whose areas  decrease as the tori inclination angle increases in  $\vartheta_{ij}\in [0, \pi/2]$. Tori with equal $\ell$ (irrespectively of $K$)  have  a crossing section which is  minimum   for a pair of  orthogonal tori  and maximum in the limiting case of $\vartheta_{ij}=0$.
All tori with equal $\ell$ have the same stability spheres (identified by location  of the centers)-- --Fig.\il\ref{Fig:SIGNS}.
\begin{figure}
\includegraphics[width=6cm]{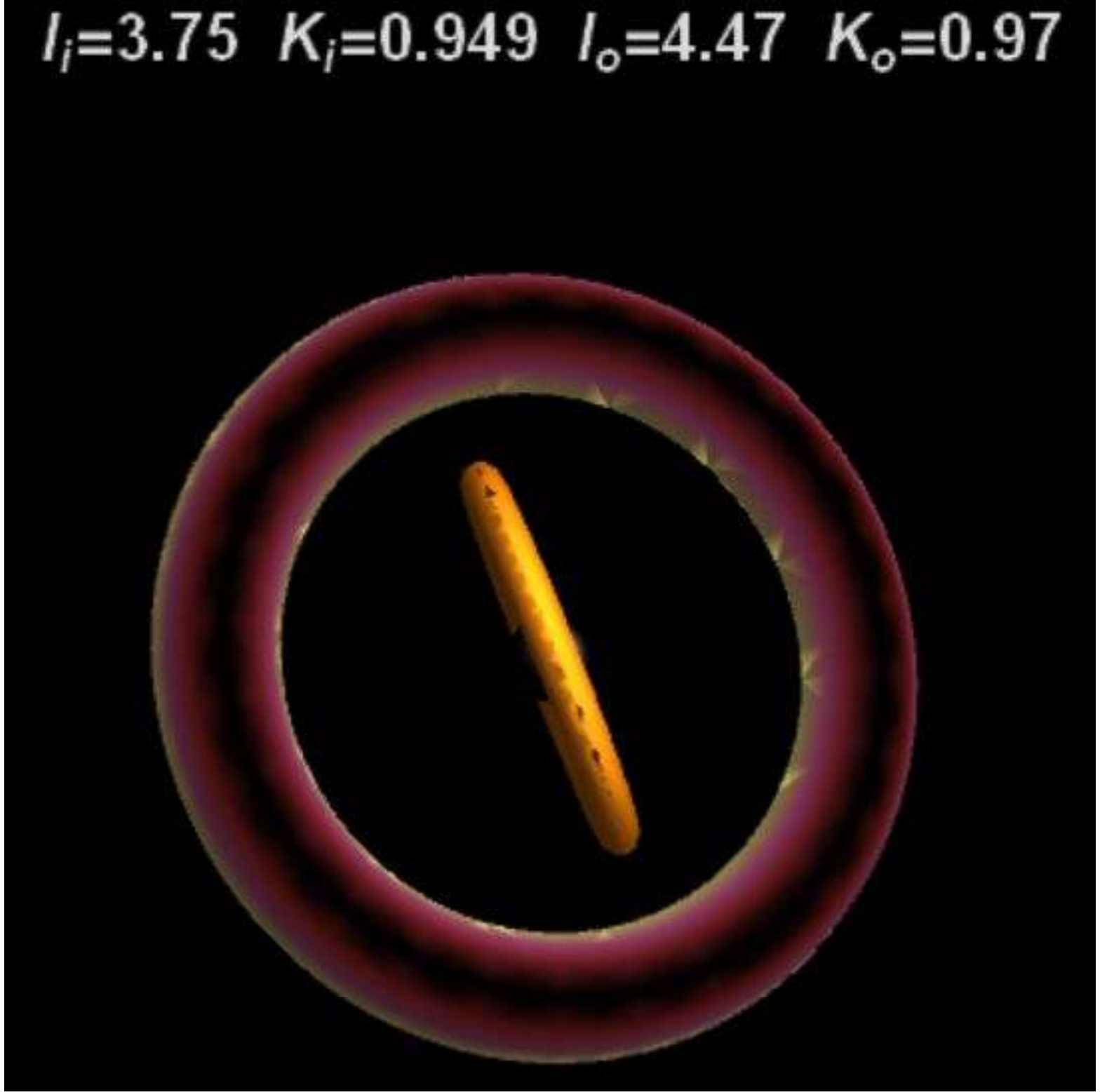}
\includegraphics[width=6cm]{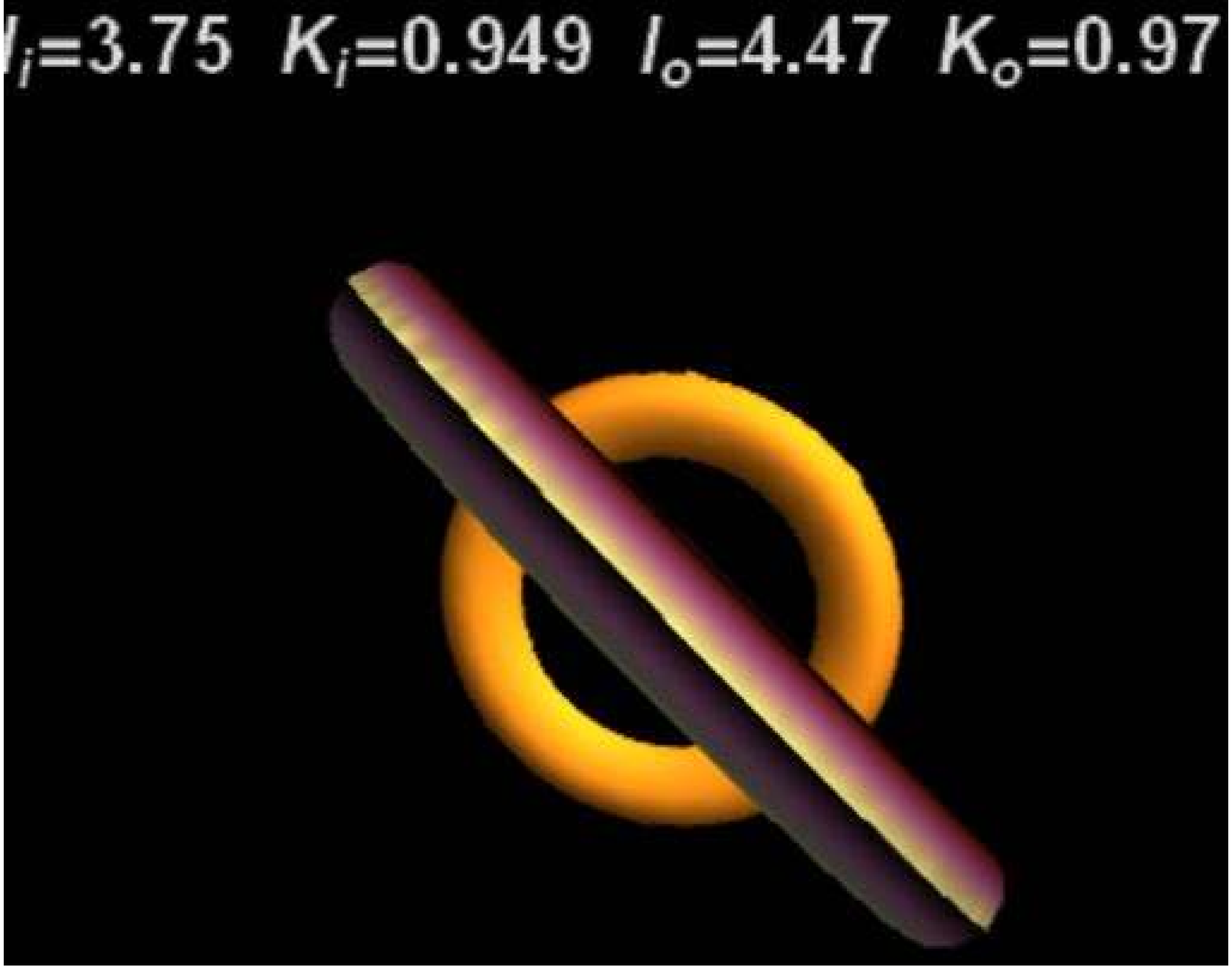}
\includegraphics[width=6cm]{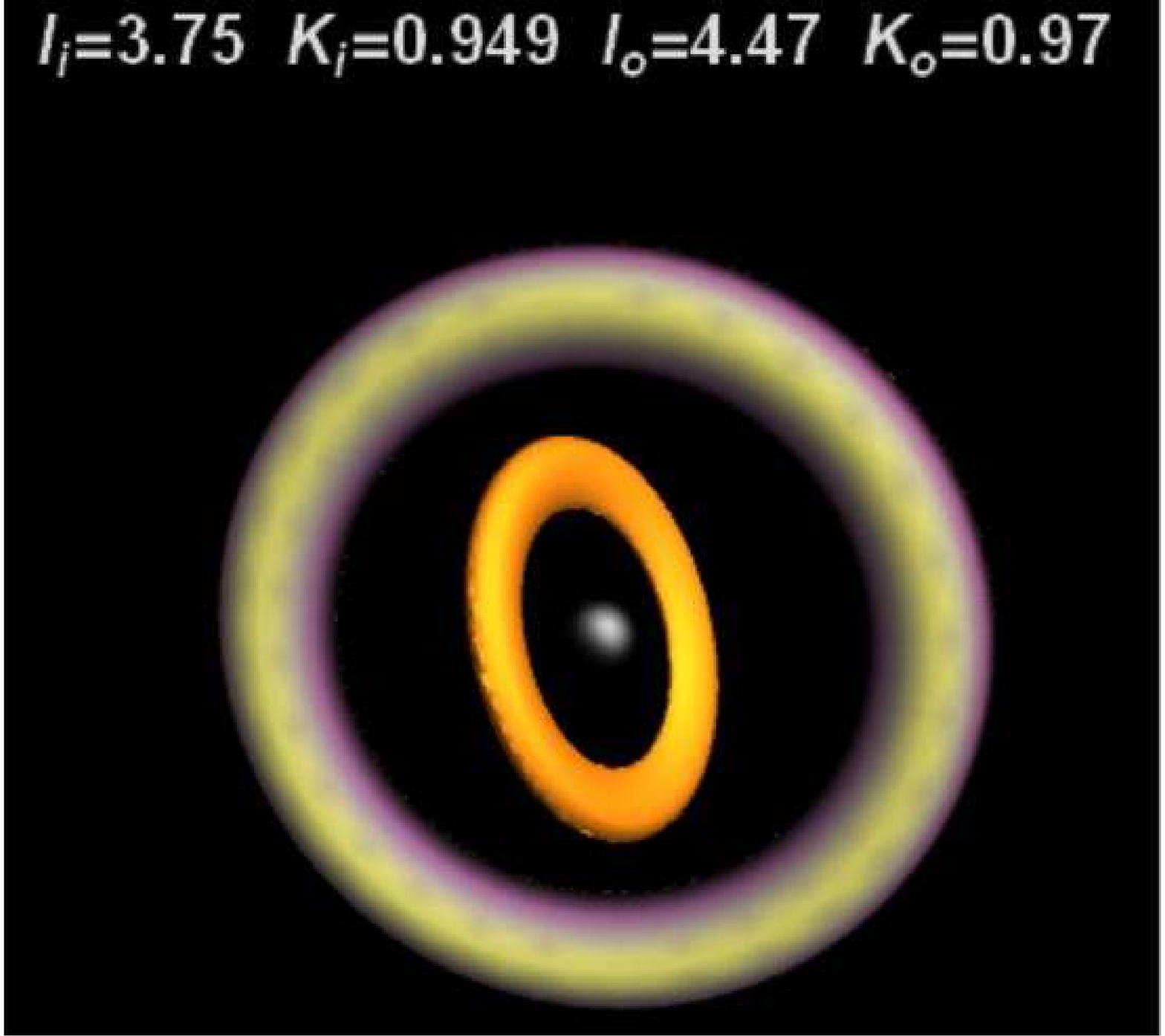}
\\
\includegraphics[width=5cm]{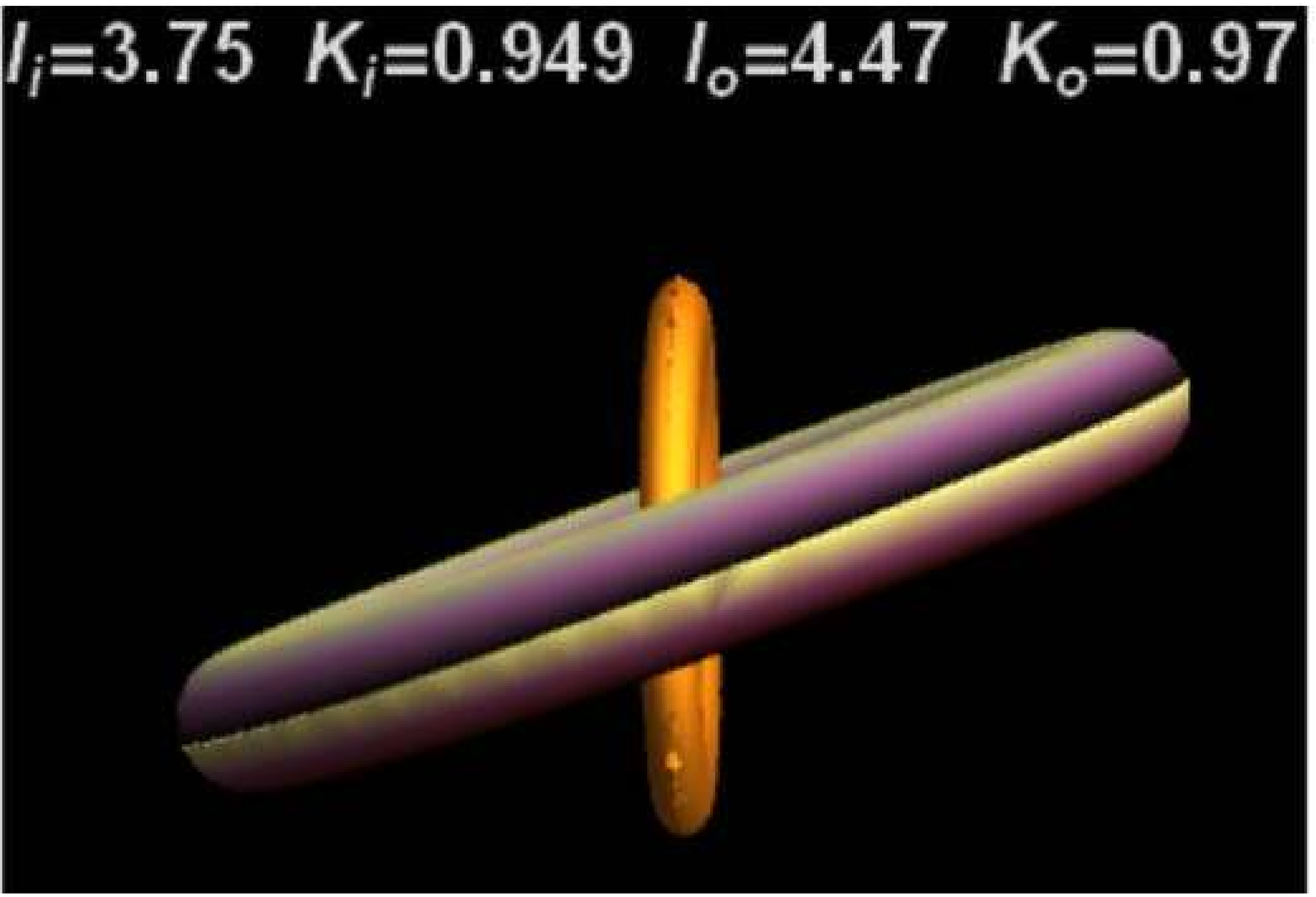}
\includegraphics[width=6cm]{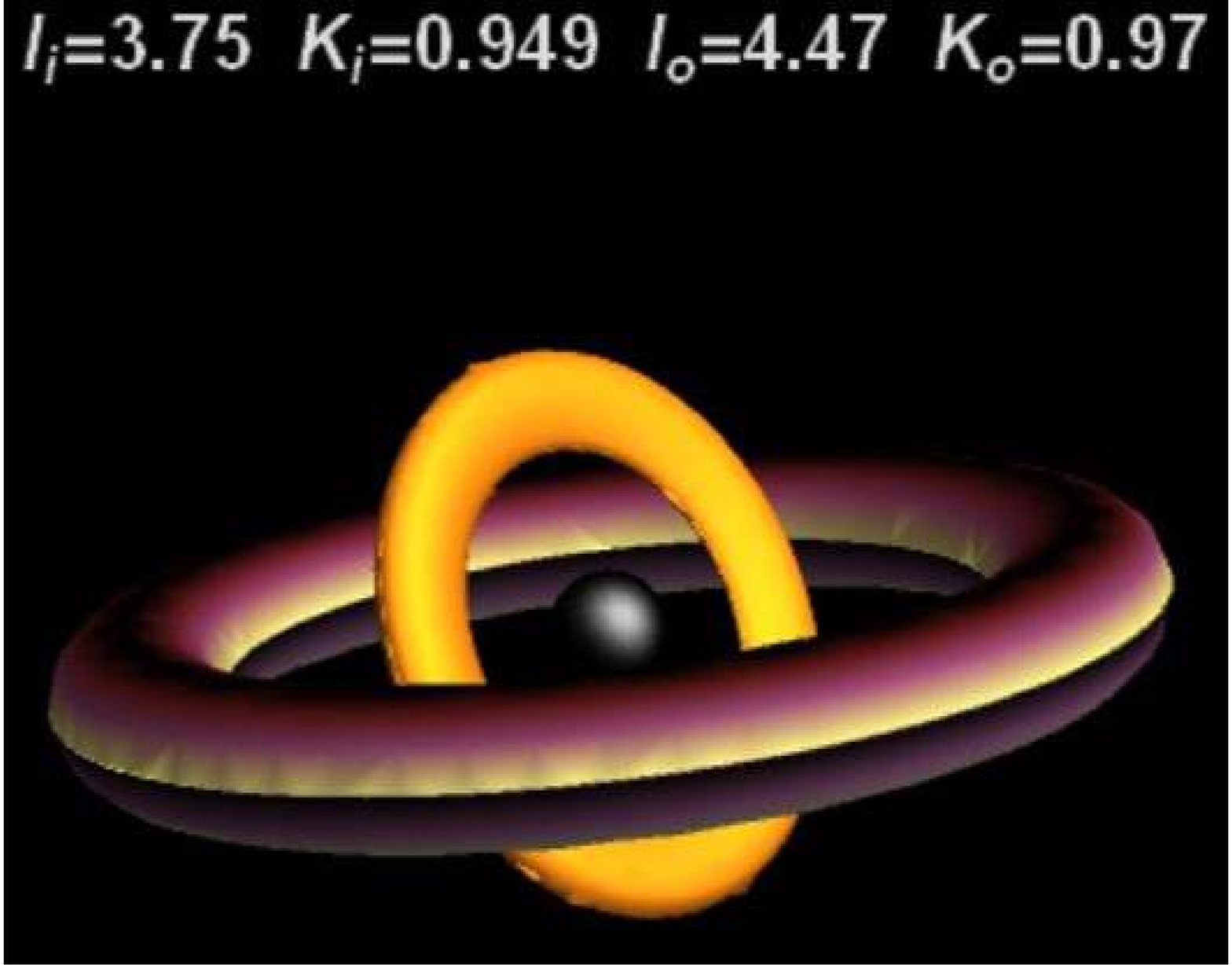}
\includegraphics[width=6cm]{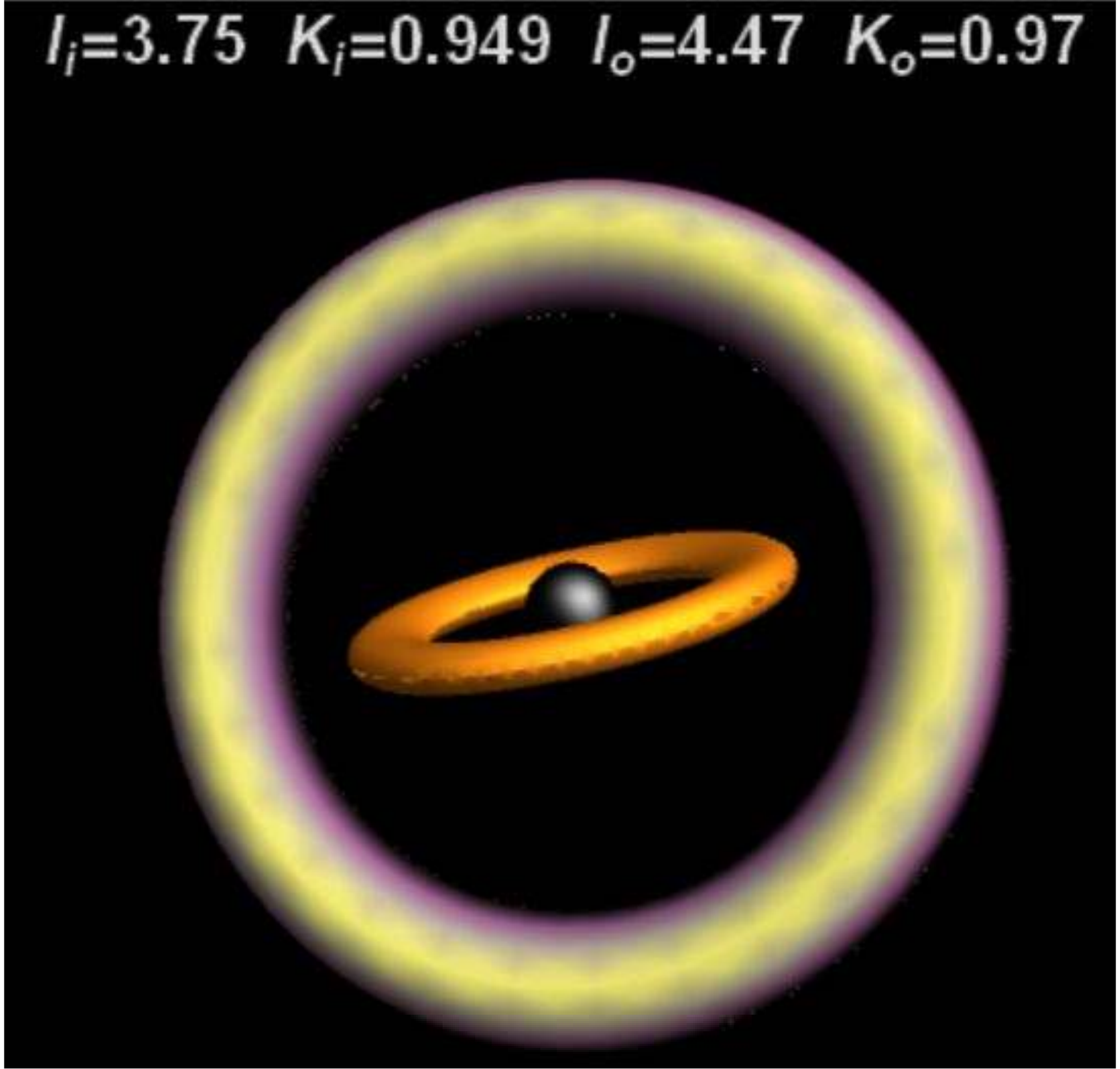}
  \caption{Density profile for \textbf{RAD} orbiting tori from 3D {HD} integration of the Euler equation (\ref{Eq:Eulerif0}).  Black region is the central Schwarzschild \textbf{BH}. Tori parameters $(\ell_i,\ell_o)$ (fluid specific angular momenta) and $(K_i,K_o)$ are  signed in the picture for the outer $(o)$ and inner torus $(i)$. The \textbf{RAD} is shown at different angle views. In Figs\il(\ref{Fig:Previosue}) the case of three toroidal components are shown.}\label{Fig:3Dinte}
\end{figure}
In Figs\il(\ref{Fig:3Dinte}) we show the profile of a \textbf{RAD} constituted by two tori at different   viewing angles, in Figs\il(\ref{Fig:Previosue}) we show the case of a globule composed of three inclined tori orbiting around the  central Schwarzschild \textbf{BH}.  These different visualizations of the same globule show how  some toroids  can be obscured. In  \cite{next} we provide a classification of the different possible views of the \textbf{RAD} with any number of components.

We close  this discussion  with further  considerations on the density of the   tori components in the \textbf{RAD} globular structure.
 Firstly,  radius   $r_{\mathcal{M}}=2 \left(2 \sqrt{3}+3\right)M=12.9282M\in\mathbf{ R_4}=]r_{mbo}^b,r_\gamma^b[$, solution of  $\partial_r^2 \ell(r)=0$,   represents the radius   of maximum density of tori (the radius being located in the  stability spheres region where centers of tori are located--see Eqs\il(\ref{Eq:swan1})). As discussed in \cite{dsystem} this radius would   play  a probable role in the formation of \textbf{RAD}.
On the other hand, there is  $\ell_{\mathcal{M}}=4.25362M\in]\ell_{mbo},\ell_{\gamma}[$,  indicating   quiescent  tori or   prot-jet  configurations in their instability phases.  We note however that with regard to the issue of tori density  in the \textbf{RAD} frame and the maximum density point,  the static case has similarities with
 the  \textbf{eRAD} counter-rotating tori  (rather than to the  corotating cases), in fact  for the  corresponding problem for  Kerr geometry, there is for counter-rotating tori  $-\ell_{\mathcal{M}}^+(a)\in]-\ell_{mbo}^{+},-\ell^+_{\gamma}(a)[$ for any value of the \textbf{BH} spin $a/M$, contrary to the corotating case where, for large \textbf{BH} spins, i.e. $a\approx 0.934M$, there is  $\ell_{\mathcal{M}}^-(a)>\ell_{\gamma}^-(a)$.
 This characteristic can be also  found in the profile of  the  $K$ function of Eq.\il(\ref{Eq:sincer-Spee}), where there is a saddle point in $r_\mathcal{M}^K=8.079M$, solution of  $\partial_r^2 K(r)=0$ with  $K_\mathcal{M}^K=0.948996M$, which is a maximum point of the gradient  $\partial_r K(r)$--
Figs\il\ref{Fig:KKNOK}.  A discussion on the significance of these systems in the perturbative analysis of disk parameters $\ell$ and $K$  can be found  in \cite{ringed}, we observe from the values of  $\ell_{\mathcal{M}}$,  tori maximum density points is the  outer and larger  annulus  in Figs\il(\ref{Fig:SIGNS}). This extreme point is related to the density within each torus of  the agglomerate which  depends on the torus location. This  property occurs  because the  fluid  density,  as clear from integration of Euler's equation (\ref{Eq:Eulerif0}), depends on   $K(r)$ function, and consequently the variation of this function, at the  torus center $K(r_{cent})$,  does not increase monotonically  with distance  $r$  from the central static \textbf{BH} attractor but there is a maximum at $r_\mathcal{M}^K$.
\subsubsection{Condition $K(r)=$constant}\label{Sec:conK-r-con}
The analysis of  tori with equal $K$ is shown in  Figs\il\ref{Fig:KKNOK}.
We consider the condition $K(r)=K(r_p)$, determining the radii $r_p\neq r$. These radii identify two different tori, $T_1$ and $T_2$, having  $K_{cent}(T_1)=K_{\times}(T_2)$ (and more in general $r_{crit}$ which possibly can be $r_{\times}$)
 with  radii $r_1>r_2$ satisfying the conditions $K(r)=K(r_p)$. This general property has been widely used in the analysis of  \cite{ringed,dsystem,long}. We note here that for a  cusped  torus,  condition  $K(r,\ell_{cent})=K(r_p)$ identifies the inner edge  $r_{\times}$ and the a  torus center, moreover also the  \emph{outer edge} $r_{out}^{\times}$ of the accreting tori when we fix $\ell$ for each solution. In the general  case it provide a boundary value (providing a rough assessment of the outer edge) of the location of the outer edge of the torus with cusp in $r_1$
 ({In the following when not otherwise specified or clear from the context, with notation $r_{crit}$ we mean critical points of the effective potential,  the enter  $r_{cent}$ and the cusp $r_{\times}$.})
There is then:
\bea&&\label{Eq:Krrp}
K(r)=K(r_p)\equiv K_p=\sqrt{\frac{(r-2) r^2}{r^3-\frac{16 (r-3)^3}{(r-4) (r-2)}}}: \quad r_p(r)\equiv\frac{4}{r-4}+4,
\quad
\ell_p\equiv\ell(r_p)=\sqrt{\frac{16 (r-3)^3}{(r-4) (r-2)^2}},
\eea
--Figs\il\ref{Fig:solidy}, with   $r_p\in[4M,6M]$ for the  cusped tori, and we evaluate in  $\ell_p$  the corresponding rotational law
$\ell_p\equiv\ell(r_p)$ which is essentially used here to locate  and characterize possible unstable states of the tori  in \textbf{RAD}, corresponding to the pair of $ Ks$  parameters  similarly to the case in Eq.\il(\ref{Eq:schaubl}). In general condition  $\mathbf{P}=$costant  for one of both the parameters $\mathbf{P}\equiv(\ell,K)$ plays an important role in the \textbf{RAD} characterization.
There are the radii
$r_{mbo}^k=\frac{1}{2} \left(\sqrt{5}+7\right)M\approx4.61803M$ such that $K(r_{mbo}^k)=K(r_{mbo}^b)$, and   $r_{\gamma}^k=\frac{6}{11} \left(\sqrt{3}+6\right)M\approx4.21748M$ such that $K(r_{\gamma}^k)=K(r_{\gamma}^b)$.
On different planes (different polar $\theta$ angles), tori at equal specific angular momentum but  with different  \emph{boundaries spheres}  having radii $r=r_{inner}$ and $r=r_{out}$, are in the equal center sphere
$r_{inner}^1<r_{inner}^2 <r_{cent}<r_{out}^2<r_{out}^1$ that is, they are concentric according to the limitations in \cite{dsystem}.
\begin{figure}
\includegraphics[width=4.6cm]{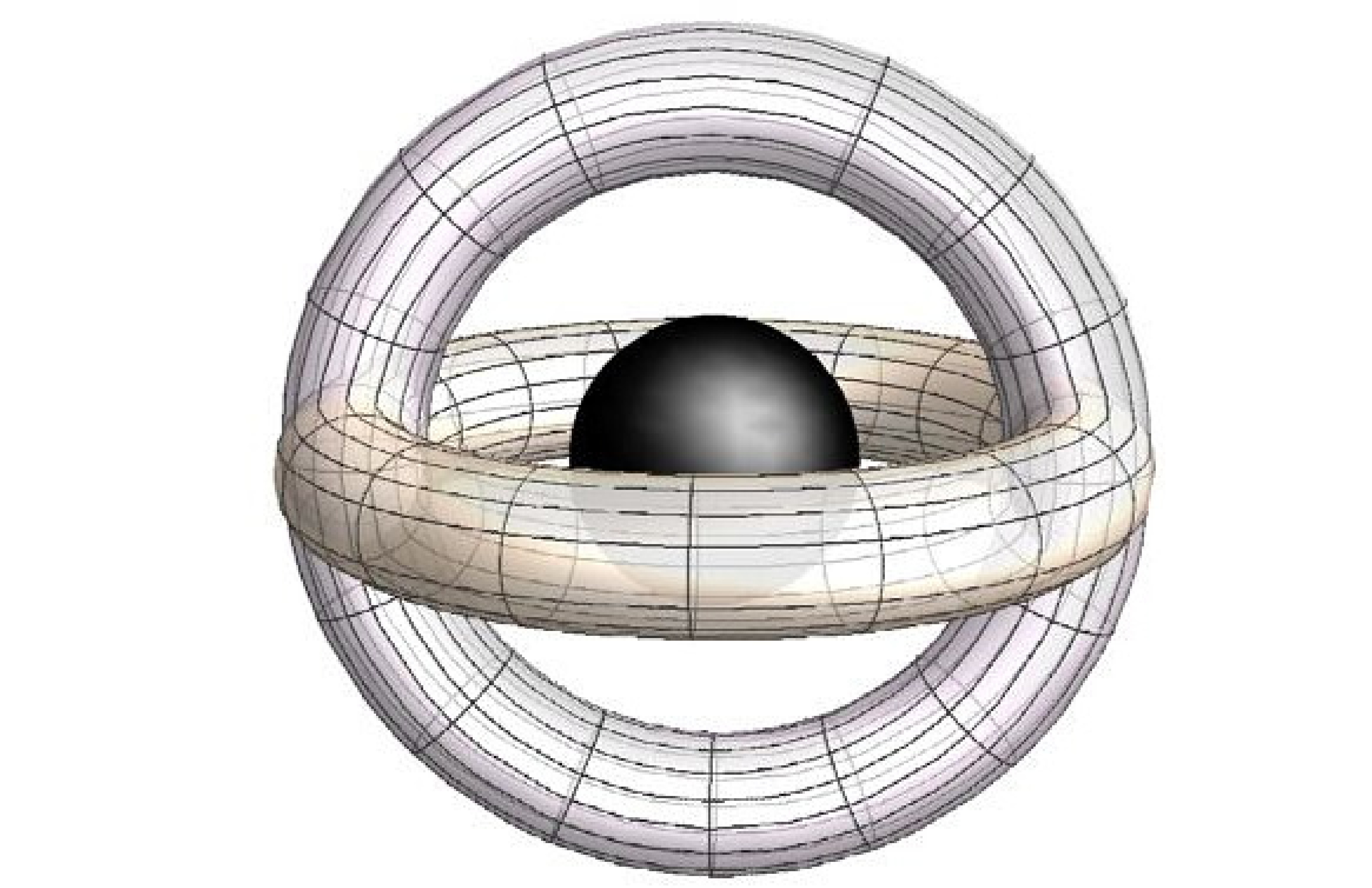}
\includegraphics[width=4.6cm]{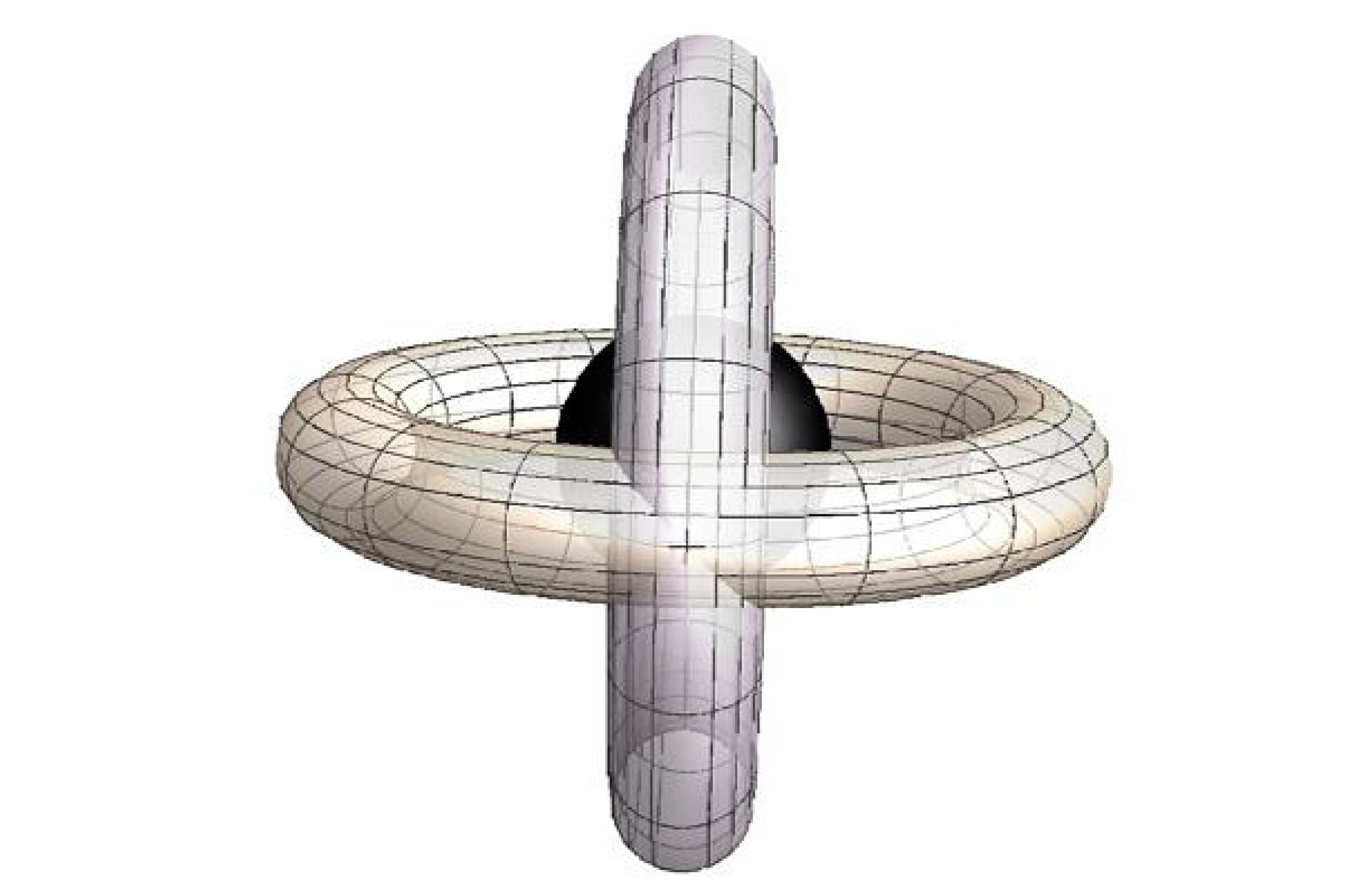}
\includegraphics[width=4.6cm]{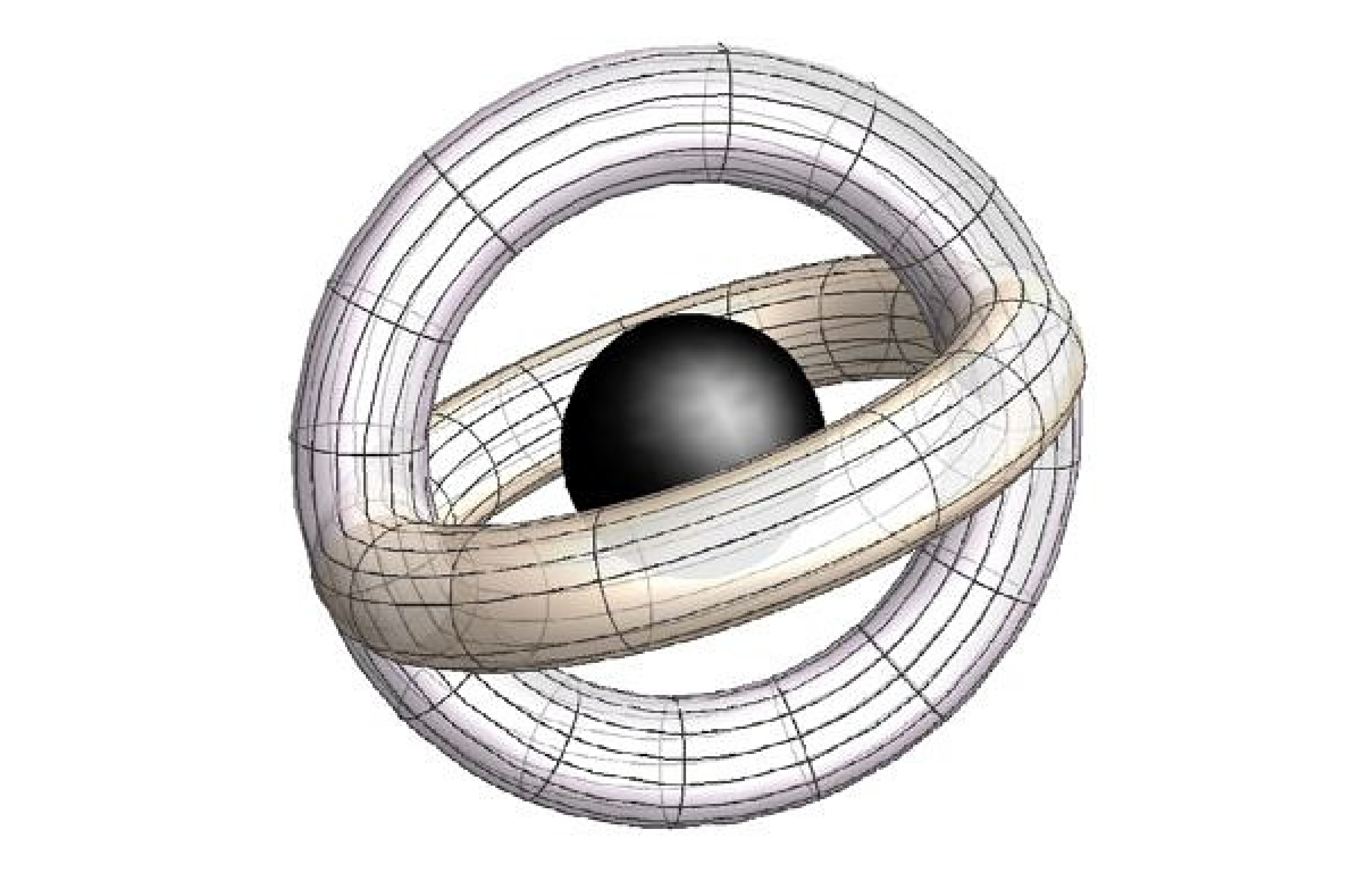}
\includegraphics[width=4.6cm]{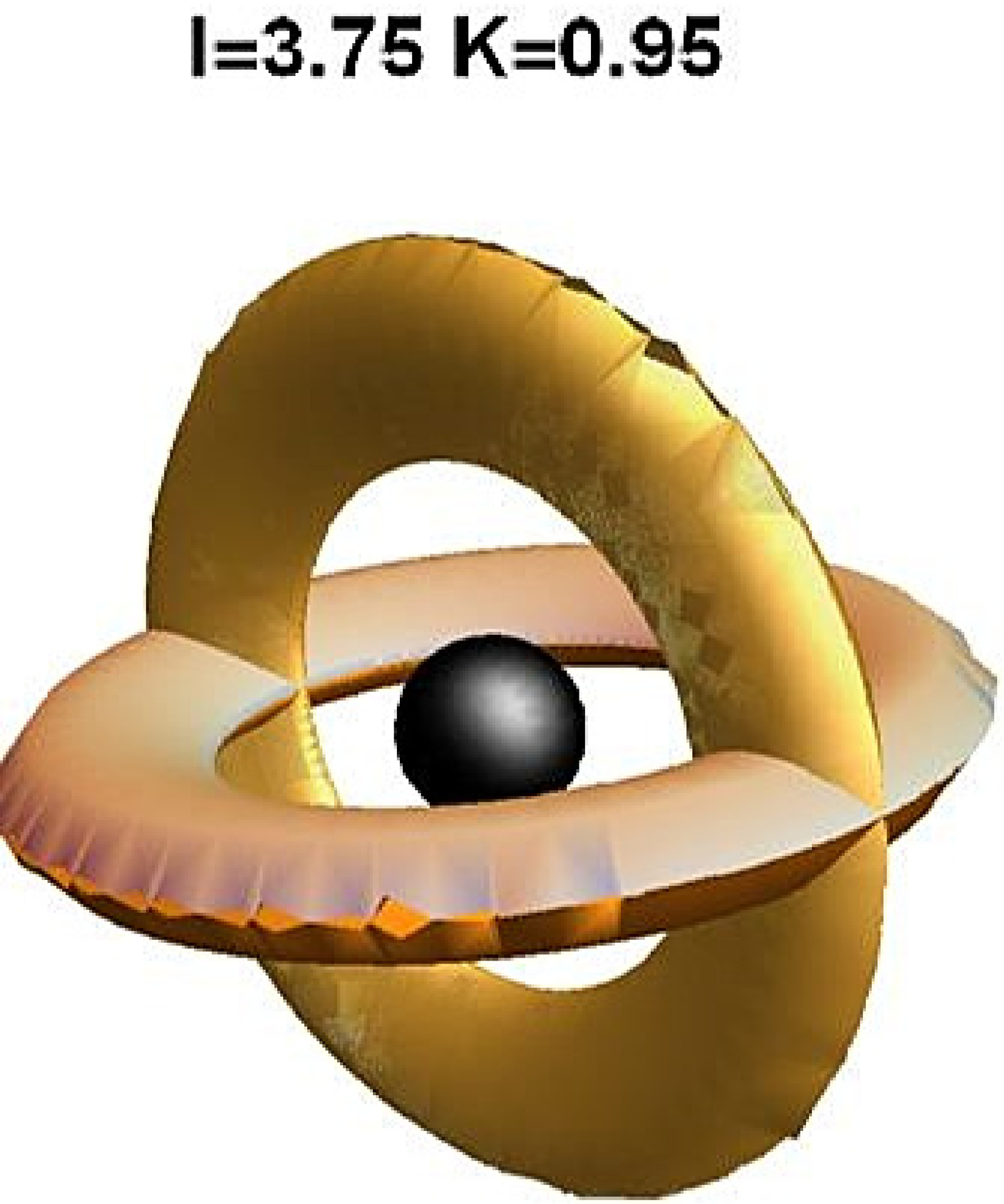}
\includegraphics[width=7.4cm]{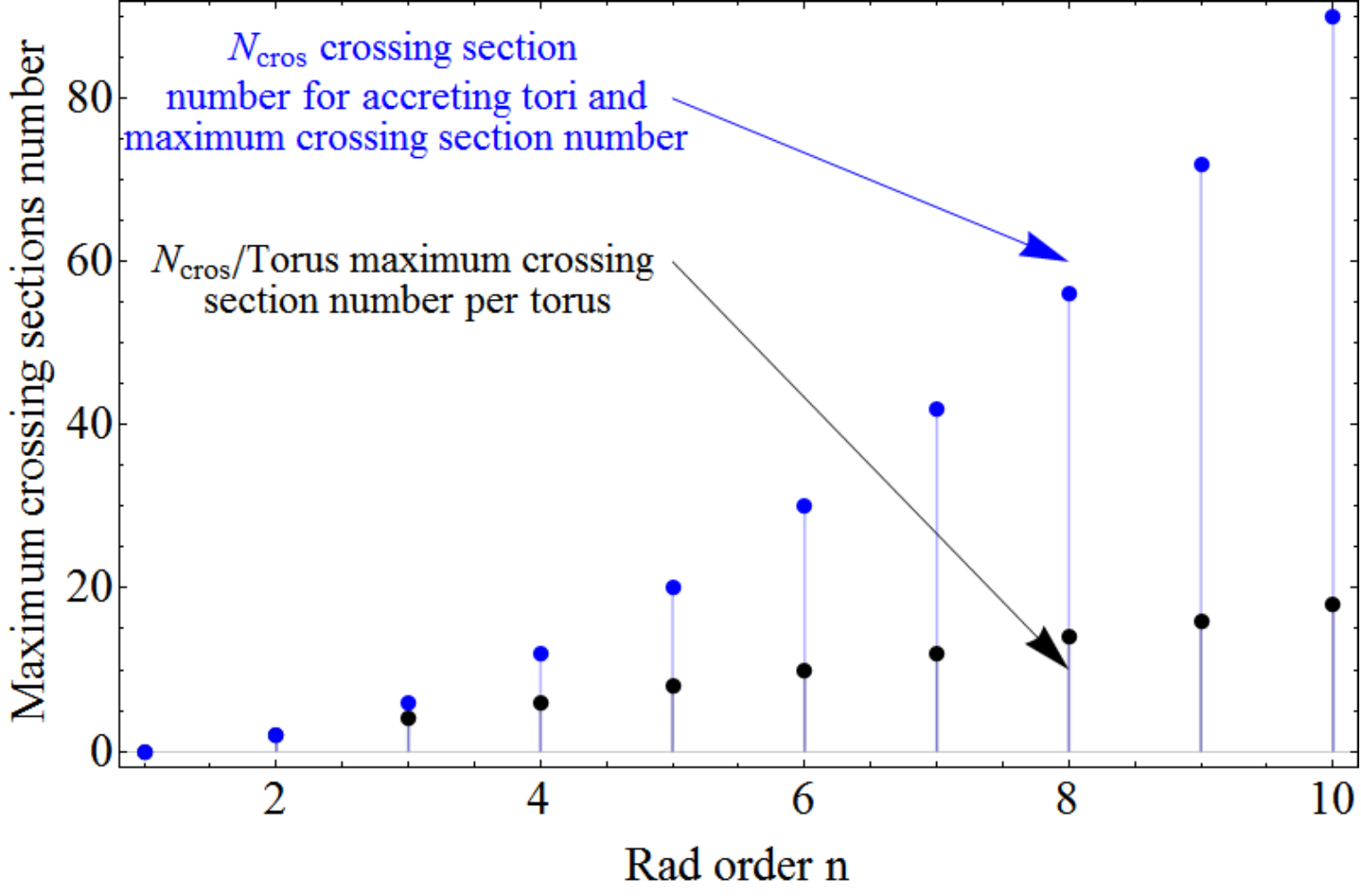}
\caption{Upper panels: Representation of the orthogonality condition between two misaligned \textbf{RAD} tori (tori inclination angle $\theta=\pi/2$). Below panels. \emph{Left}: Density profiles from (\ref{Eq:Eulerif0}) of two orthogonal limiting  misaligned tori $\ell$ is the fluid specific angular momentum and $K$ is a parameter regulating the torus elongation and density. Black region is the central Schwarzschild \textbf{BH}. \emph{Right}: Maximum number of crossing sections for tori relative inclination angle $\vartheta_{ij}\neq 0$ as function of \textbf{RAD} order (tori number) $n$.
\label{Fig:orto} }
\end{figure}
For $n$ \emph{accreting} tori (with different angles and sizes), the \emph{total number of }  crossing sections are,
$N_{cross}=2\sum_{k=1}^{n}(k-1)=n(n-1)$ (it is clear this results hold for any polar relative angle $\vartheta_{ij}\neq 0$ between two tori $i$ and $j$, at limiting  $\vartheta_{ij}= 0$  there is $n=1$). We note that the  cusped tori,  because of  the symmetries are all crossing,  then $N_{cross}$ is  the maximum number of crossing sections possible for $n$ generic (cusped or quiescent) tori  and the number of crossing section for cusped tori. Each torus   in a \textbf{RAD} of the order  $n$ has a maximum of  $2(n-1)$ crossing sections
--Figs\il(\ref{Fig:orto},\ref{Fig:Previosue}) and Figs\il(\ref{Fig:sepa},\ref{Fig:DopoDra}).
{In Appedix\il(\ref{Sec:zeipel})  we address  the question of whether there is an external limit to the formation of an most external torus in the  globular multipole, centered for example in $r>r_{\gamma}^b$--see Eq.\il(\ref{Eq:swan1})  which is  seemingly not existent  in Schwarzschild nor in Kerr \textbf{BH} spacetime  but  possibly  present in other \textbf{BH}  geometries  with,  for example, cosmological constant (and arguably present also in many naked singularity geometries. (To all appearances  the non-existence of such superior limit  is routed by considerations internal at the model  which includes the significant role of the  background geometry, but probably  also supported by external considerations that must be also considered, for example on the tori  microphysics (plasma models, role of magnetic field) or most likely the interaction with the embedding galactic material the \textbf{RAD} interacts with that would constraint the presence of such torus.) This relevant problem would be linked to the upper limit imposed on the mass, spin and  radius of the extended orbiting  object constituted by a ("self-gravitating")    \textbf{BH} nucleus embedded  in     a (multipole) shell  of gravitating orbiting  tori. (The definition of the radius is also given in Sec.\il(\ref{Sec:zeipel})). Finally, it is clear that the question of the eventual maximum distance of the outermost orbiting torus from the central attractor is all the more relevant for the eventual issue  of the formation, evolution  and stability of the \textbf{RAD} centered in al Kerr  \textbf{BH}, where the retrograde tori would be  probably  formed as exterior shells.

We conclude with some notes on the procedure we adopt in the next section.
In our analysis we  consider the quantities   $ (\mathbf{P, M, E})$.
\begin{description}\item{--}
$ \mathbf{P} $ in  the set of  model parameters introduced in Sec.\il(\ref{Sec:Misal}) and   in Figs\il(\ref{Fig:KKNOK},\ref{Fig:solidy}). There are then $2n$ parameters for $n$ \textbf{RAD} tori, constituted by the couples $(\ell,K)$ for each torus, points on the curves $\ell(r)$ and $K(r)$ of Eq.\il(\ref{Eq:lqkp}) and
Eq.\il(\ref{Eq:sincer-Spee}), respectively.
\item{--}
$ \mathbf{M} $ is the set of tori morphological characteristics, functions of the   $\mathbf{P}$ parameters   for example
$ \mathbf{M} =\{r_{inner},  r_{cent}, r_{out}, r_j, h, \lambda, \bar{\lambda},...\}$ see Table\il(\ref{Table:pol-cy}) and studied in Sec.\il(\ref{Sec:doc-ready}) as functions of $\mathbf{P}$.
\item{--}
Then there are the  quantities relating to the \textbf{BH} accretion disks energetics
 $ \mathbf{E} $ studied in Sec. (\ref{Sec:energ-RAD-poli}), quantities linked to the mass accretion rates. We explored $\mathbf{E}$ in terms of $\mathbf{P}$ and $\mathbf{M}$ combing the analysis in  Sec.\il(\ref{Sec:Misal}) and Sec.\il(\ref{Sec:doc-ready}).
 \end{description}
Considering also the  possible observational feedbacks from  the \textbf{RAD} structures  we in fact focus in our analysis  on
each quantities $(\mathbf{P,M,E})$ in terms of the one or two others, in this way constraining one characteristic of \textbf{RAD} toroidal component  to the others distinguishing classes of tori and narrowing these classes through the combinations of  the correlated analysis on the correspondent other quantities. For example in Sec.\il(\ref{Sec:energ-RAD-poli}) we relate  $\mathbf{E}$  to the inner edge of the disk, which is constrained  in Sec.\il(\ref{Sec:limiting}) to the possibility of tori collision in the  \textbf{RAD}, then in Sec.\il(\ref{Sec:morph}) we comprehensively   study  the   torus inner edge as function of $\textbf{P}$  and correlating to several others morphological characteristics basing on $\textbf{P}$ dependence.
These results  would  serve as a guideline for  possible observational identification of a \textbf{RAD}\footnote{
Observation of a quantity Y is therefore connected with an torus features  or model parameter X often not univocally and we correlate it with the possible information on further  Z characteristics,
through $\{ X,Y,Z\}$ in $\{\mathbf{P,M,E}\}$ we deduce the \textbf{RAD} structure. This is in fact the principle under the analysis for example of  \textbf{BH}  spin using energy extraction  observed for example in jet emission, in this case we also include a third ingredient which is the characteristics of \textbf{BH} parameter. We recall how \textbf{M} and \textbf{E} are used in all other \textbf{BH}-disk models a  to deduce  disk properties $\mathbf{\textbf{M}}$ as inner edge of accreting disk to deduce the \textbf{BH} spin or viceversa.}.
Here  we look for the greatest number of morphological characteristics \textbf{M} and of the most meaningful and numerous functional combinations in order to represent a complete \textbf{RAD} model setup,  correlating then the  different aspects. One or more of these characteristics  quantities \textbf{M},  \textbf{E} or even \textbf{P} may correspond to an entire  class of objects,  determining a  class of components rather then one specific \textbf{RAD}
Moreover we note  that our analysis is in scale of mass of the \textbf{BH} therefore this a free  parameter  of the model.
In Sec.\il(\ref{Sec:energ-RAD-poli})  we also introduce  a further independent parameter to narrow the classes of \textbf{RADs} considering the polytropic index and polytropic constant, distinguishing classes of polytropic for the \textbf{RAD} orbiting a central static \textbf{BH}--see also  \cite{mnras}.

\subsection{Morphological constraints on  disks stability}\label{Sec:mirpj}
At fixed $\ell$, the  torus    reaches its  maximum elongation   $\lambda_{\times}$ on the equatorial plane as cusped surface. The outer tori  have  larger magnitude of the  specific angular momentum   leading  in general  to  a larger elongation $\lambda$.
\begin{figure}
\includegraphics[width=16cm]{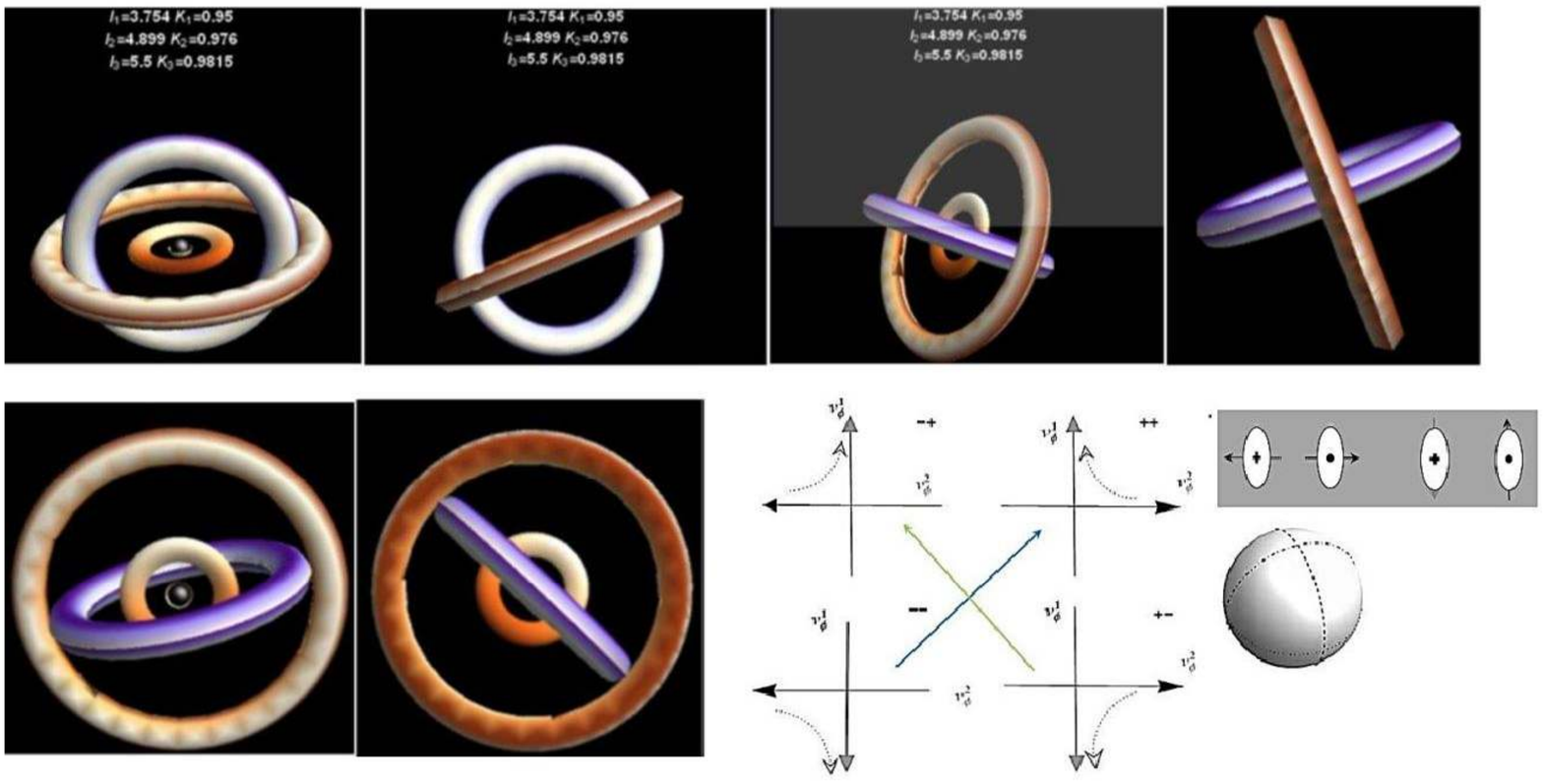}
\caption{Left: 3D HD integration of Euler equations (\ref{Eq:Eulerif0}) for density profiles of misaligned tori with parameter $K$ and fluid specific angular momentum $\ell$ as in figures. Black sphere is the central \textbf{BH}. Different angles of observation are shown. Tori parameters, specific angular momentum $\ell$ and $K$ parameter are as follows from the closes torus to the furthest to the central attractor, i.e.  on torus center, maximum density point, $r_{cent}^1<r_{cent}^2<r_{cent}^3$ there are: $(\ell_1=3.754, K_1=0.95)$,
$(\ell_2=4.899, K_2=0.976)$, $(\ell_3=5.5, K_3=0.9815)$. The case of a cluster composed by two tori  is in Figs\il(\ref{Fig:3Dinte}).  Right panel: schematic representation of fluid velocities for  two orthogonal \textbf{RAD} tori , velocities $v_{\phi}^1$ and $v_{\phi}^2$ of $T_1$ and $T_2$ shown on orthogonal direction. Four (symmetric) configurations are possible: $\{++,-+,--,--\}$ respectively according to the figures, (arbitrary) relative sign of the proper fluid angular velocity $\ell$. Black arrows indicate the  velocity directions and explain the  sign convention $\{++,-+,--,--\}$.
Colored diagonal arrow marks corresponding velocities  schemes for a torus  according to the symmetries. Inside gray right panel representation of ingoing  fluid (dot $\bullet$), and outgoing (plus $ \mathbf{+}$) from the figure for a torus.
\label{Fig:Previosue} }
\end{figure}
Figs\il(\ref{Fig:Previosue})  show  a  \textbf{RAD} with  three tori  from  different view angles.  Similarly to the case of two tori  in Figs\il(\ref{Fig:3Dinte}) it is clear  the obscuration effect on  an internal tori of the configuration  and  the similarity, depending on the  viewing  angle, with the  case of two tori. The diagrams  show the toroidal velocities  (in BL-frame adapted to each torus), considering in the orthogonal case (relative tilt  angle $\vartheta=\pi/2$),  for example in Figs\il(\ref{Fig:orto}),  having  two  degrees of freedom for  the fluid velocity, for two tori  having four  possible cases which are the equivalent of the  { $ \ell$corotating or $\ell$counterrotating tori} in the Kerr \textbf{BH eRAD}. Obviously this has a great relevance in the case of Schwarzschild  \textbf{BH}  in the occurrence  fluids collision.
On the other hand, for the $T_i<T_o$
quiescent (not cusped) tori  ($T_i$ is the closest to the central \textbf{BH}), there is  $\ell_i< \ell_o$ and $r_{out}^i<r_{inner}^o$--see notation in  Table\il(\ref{Table:pol-cy}).
Note that there can be    $\ell_i\neq \ell_o$  and  $K_i\neq K_o$ and $r_{out}^i=r_{inner}^o$.
For the cusped  configurations   conditions are simplified because   fixed  by   the   $\ell$ parameter only.
In general, for  $\ell_i < \ell_o$  there is
$r_{\times}^o < r_{\times}^i<r_{cent}^i<
 r_{cent}^o$ where it is  $r_{out}^i<r_{out}^o$. It follows in particular  that
the outer  cusped  torus incorporates (not just collides with) the inner cusped torus, analogously to the \textbf{eRAD} case. This aspect is in fact   independent of the tori inclination.  The orthogonal case, $\vartheta_{ij}=\pi/2$, determines the minimum cross  sections (the cross section regions are minimized to the minimum $\la_{\min}=\min \{\la_i\}_{i=1}^n$, $\la_i$ is the tori area cross section on   the $i$-tori equatorial plane).
Notably, in the \textbf{eRAD} the accreting outer  torus  leads to accretion
of the material towards the central \textbf{BH}, totally incorporating  the inner cusped or quiescent torus, viceversa, the  tori inclination reduced this phenomenon to a collisional effect, realizing the case of multi accretion  as a collision between emergent accreting tori which may occur in a transient phase of \textbf{BH} accretion disk life  where the tori collision   occurs, in the orthogonal case, in  the regions $2\la_{\min}$. This could in fact represent a complex but effective mechanism to increase the accretion rates of the central black hole, which acquires  also angular momentum of the in-falling matter from different  inclination angles, leading to a far more intriguing  phenomenon  of spin change and possibly runaway instability.
It remains however to consider in this scenario   the  time scales of the formations of these disks, and time scales on the  tori interaction processes. A dynamical analysis of the  systems from  the initial conditions developed here  may provide such evaluations.
\begin{figure}
\includegraphics[width=6.4cm]{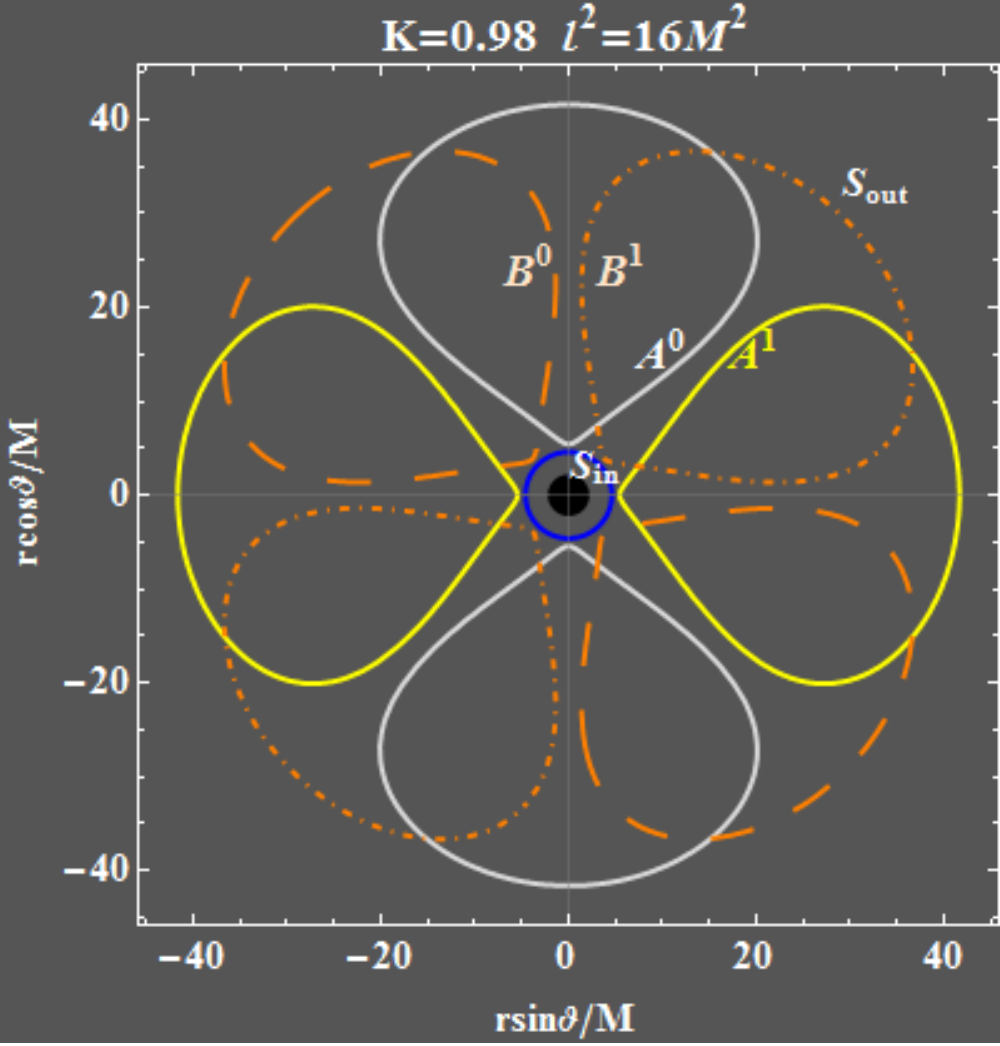}
\includegraphics[width=6.4cm]{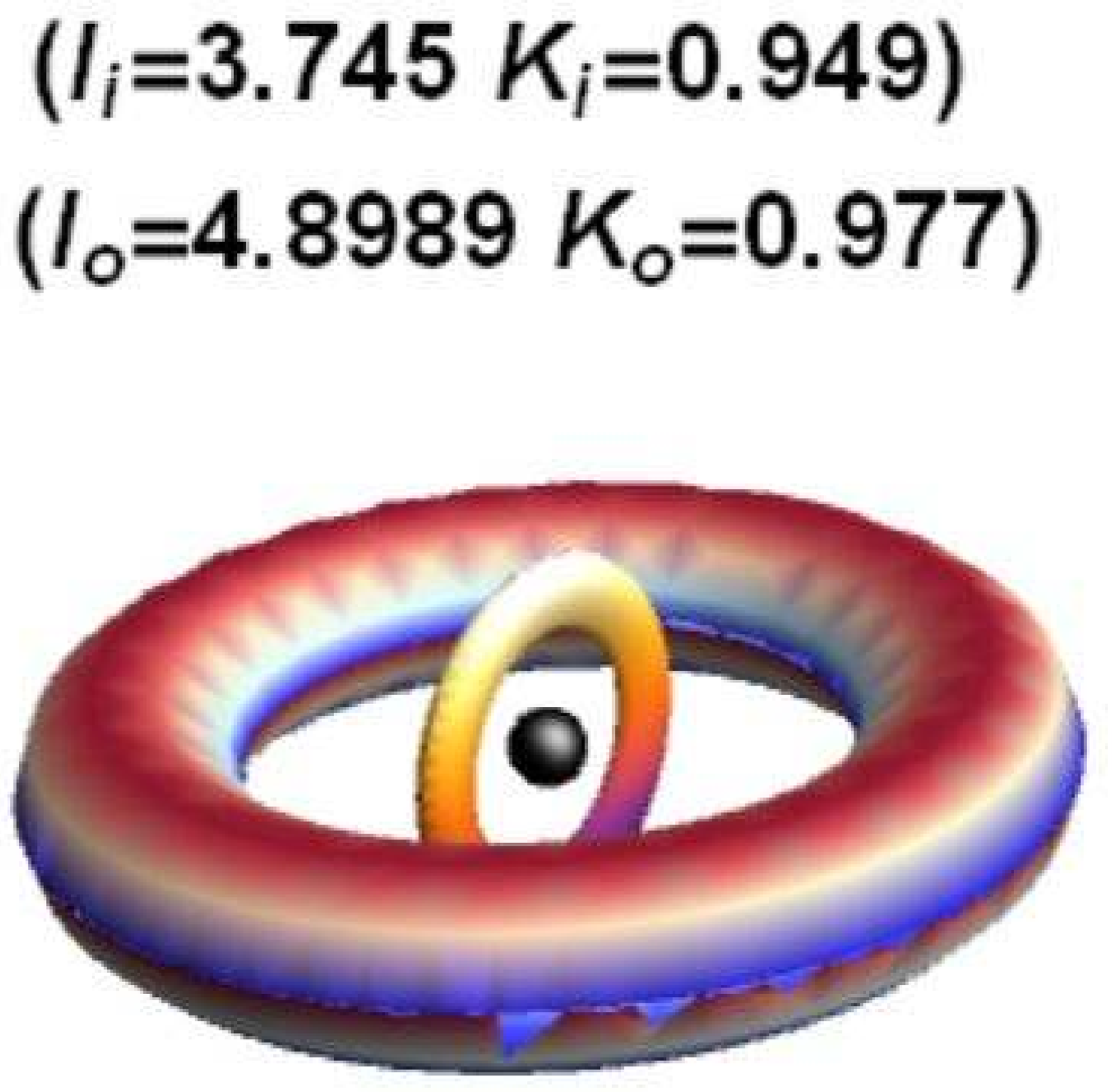}
\caption{Right: density profiles, 3D HD integration of the Euler equations (\ref{Eq:Eulerif0}) for two  orthogonal  \textbf{RAD} tori  defined by the specific fluid angular momentum $\ell$ and $K$ parameter as indicated in figure. Left: Equi-pressure surfaces. These are found from explicit equations  Eq.\il(\ref{Eq:condizioni}), for $\mathbf{A^0}$-model: $\mathcal{B}(x,y)=x$ and $\mathcal{Z}(x,y)=y$; $\mathbf{A^1}$-model: $\mathcal{B}(x,y)=y$ and $\mathcal{Z}(x,y)=x$; $\mathbf{B^0}$-model: $\mathcal{B}(x,y)= \frac{x-y}{\sqrt{2}}$ and  $\mathbf{B^1}$-model: $\mathcal{Z}(x,y)= \frac{x+y}{\sqrt{2}}$,
$\mathcal{B}(x,y)= \frac{x+y}{\sqrt{2}}$ and  $\mathcal{Z}(x,y)= \frac{-x+y}{\sqrt{2}}$ $(\theta=\pi/4)$. These are cross sections of the rigid Boyer surfaces. Boundary spheres $\Sa_{in}$ for the inner edge and $\Sa_{out}$ for the outer edge are also shown. Central black region is the Schwarzschild \textbf{BH}. According to $(\ell,K)$ as signed in figure.
\label{Fig:sepa} }
\end{figure}
A \textbf{RAD} and therefore to a greater extent a globulous  will be characterized by a region an internal vacuum characterized by different effects where accretion from the inner torus occurs and related phenomena as for example jet launch. Radius of this region  is clearly $]r_+,r_{inner}^i[$ that is from the \textbf{BH} horizon to the inner edge of the inner torus of the \textbf{RAD}. Other vacuum regions are present, corresponding to the spacing among tori, this discrete structure typical of \textbf{RAD} and \textbf{eRAD} systems. These $n-1$ regions, for an $n$components RAD, have radius
of radius $\bar{\lambda}\equiv r_{inner}^o-r_{out}^i $, in a given couple,  spacing between the inner edge of the outer torus and outer edge of the inner torus.
Finally in Sec.\il(\ref{Sec:fasc-eur}) we provide explicit expression  of the
 tori equatorial sections and discuss some limiting surfaces.
\section{Geometry of RAD accreting misaligned tori}\label{Sec:doc-ready}
 In this section we investigate the  \textbf{RAD} geometry and morphology considering the parameters $(\ell,K)$.
We can evaluate  \textbf{RAD} tori geometric features, as the thickness and  the elongation on the equatorial plane, depending on  the \textbf{RAD}   parameters as the $K$-parameter,  the specific angular momentum $\ell$, location $r_{\times}$ of the inner edge of cusped tori.
We consider particularly  the  torus  elongation  $\lambda(\ell,K)$, or the location of inner edge, the
location of the torus  center  $r_{cent}$,  i.e. point of maximum density and hydrostatic pressure, the location of the geometric  maximum $r_{\max}\equiv x_{\max}$  of the  \textbf{RAD} tori. We introduce also  the  torus thickness $S\equiv2h_{\max}/\lambda$, where $h_{\max}\equiv y_{\max}$ is the torus height, location of the outer edge torus $r_{out}$, in the particular case of cusped disk where  these quantities depend one parameter ($\ell_{crit}$ or $K_{crit}$ or $r_{crit}$) only. With notation $crit$ we mean  the notation for quantities calculated at the critical points of the torus potential, i.e. either the  centers or the cusps.
The evaluation of the geometrical thickness  has an essential role for example in the evaluation of the effects of disc-seismology as clarified in  \cite{next}.

\subsection{Morphological characteristics and stability: presence of a cusp
}\label{Sec:morph}
We list below the  main morphological properties of the orbiting tori in the cluster, considering particularly the cases when a cusp is present, indicating the emergence of accretion  conditions for the  toroidal configurations.
{Properties listed below are found in straightforward way  by considering the  toroidal surfaces solutions, and solving the related algebraic or maximization  problem.}
\begin{enumerate}
\item{\textbf{The torus elongation $\lambda(\ell,K)$.}}

The  toroidal elongation  $\lambda(\ell,K)$  of the (cusped or quiescent) tori on their  equatorial plane is given as
\bea&&\label{Eq:lam}
\lambda\equiv \frac{2 \tau  \cos \left(\frac{1}{6} \left[2 \cos ^{-1}(\alpha )+\pi \right]\right)}{\sqrt{3}},
\eea
--Figs\il(\ref{Fig:conicap}), where $(\alpha,\tau)$ are defined as follows
\bea
\alpha\equiv\left[\frac{8-9  \ell^2 (\mathbb{K}-1) \mathbb{K} (3 \mathbb{K}-1)}{\mathbb{K}^3 \tau ^3}\right],\quad\tau\equiv\sqrt{3} \sqrt{-\frac{\ell^2}{\mathbb{K}}+\ell^2+\frac{4}{3 \mathbb{K}^2}},\quad\mathbb{K}:\; K\equiv\sqrt{1-\mathbb{K}}.
\eea
\item\textbf{{Inner torus edge  $r_{inner}(\ell,K)$, outer torus edge $r_{out}(\ell,K)$, radius $r_{inner}^{BH}(\ell,K)$ of the innermost configuration}--Figs\il\ref{Fig:VClose},\ref{Fig:conicap}.}
We evaluate  $r_{inner}(\ell,K)$, $r_{out}(\ell,K)$ and $r_{inner}^{BH}(\ell,K)$  as follows:
\bea\label{Eq:mer-panto-ex-resul}
&&
r_{out}\equiv \frac{2 \left[\mathbb{K} \tau  \cos \left(\frac{1}{3} \cos ^{-1}(\alpha )\right)+1\right]}{3 \mathbb{K}},\quad r_{inner}\equiv\frac{2 \left[\frac{1}{\mathbb{K}}-\tau  \sin \left(\frac{1}{3} \sin ^{-1}(\alpha )\right)\right]}{3},
 \\\nonumber
 &&\mathbf{r_{inner}^{BH}}\equiv\frac{2\left[\frac{1}{\mathbb{K}}-\tau  \sin \left(\frac{1}{6} \left[2 \cos ^{-1}(\alpha )+\pi \right]\right)\right]}{3}.
\eea
The innermost configuration  is a closed solution of Euler equations (\ref{Eq:Eulerif0}), close to the horizon and coincident  with the inner ``Roche lobe'' of cusped torus--see Fig\il(\ref{Fig:Quadr})-left panel torus $(\mathbf{III})$--\cite{pugtot}. Therefore distance $\lambda_{in}^{BH}$
\bea
&&\lambda_{in}^{BH}\equiv r_{inner}-r_{inner}^{BH}=\frac{2}{3} \tau  \left[\sin \left(\frac{1}{6} \left[2 \cos ^{-1}(\alpha )+\pi \right]\right)-\sin \left[\frac{1}{3} \sin ^{-1}(\alpha )\right]\right],
\eea
vanishes for an cusped (accreting) torus.
\item\textbf{{The center $r_{cent}(\ell)$ of maximum density (and hydrostatic pressure) and  inner edge of accreting torus $r_{\times}(\ell)$ as function of the specific fluid angular momentum $\ell$}---Figs\il\ref{Fig:VClose},\ref{Fig:SMGerm}.}

The center $ r_{cent}(\ell)$ and the instability point $r_{\times}(\ell)$ are functions of the fluid specific  angular momentum $\ell$ only and can be expressed as
\bea&&\label{Eq:rcentro}
r_{cent}(\ell)\equiv\frac{1}{3} \left[\ell^2+2 L_{ \ell} \cos \left(\frac{1}{3} \cos ^{-1}(L_{\mathcal{ll}})\right)\right],
\quad r_{\times}(\ell)\equiv\frac{1}{3} \left[\ell^2-2 L_{ \ell} \cos \left(\frac{1}{3} \left[\cos ^{-1}(L_{\mathcal{ll}})+\pi \right]\right)\right],
\eea
where  $(L_{\mathcal{ll}}, L_{ \ell})$ are defined as follows:
\bea&&  L_{\mathcal{ll}}\equiv \frac{\ell^2 \left(\ell^4-18 \ell^2+54\right)}{L_{ \ell}^3},\quad L_{ \ell}\equiv\sqrt{\ell^2 \left(\ell^2-12\right)}.
\eea
\begin{figure}
  \includegraphics[width=8cm]{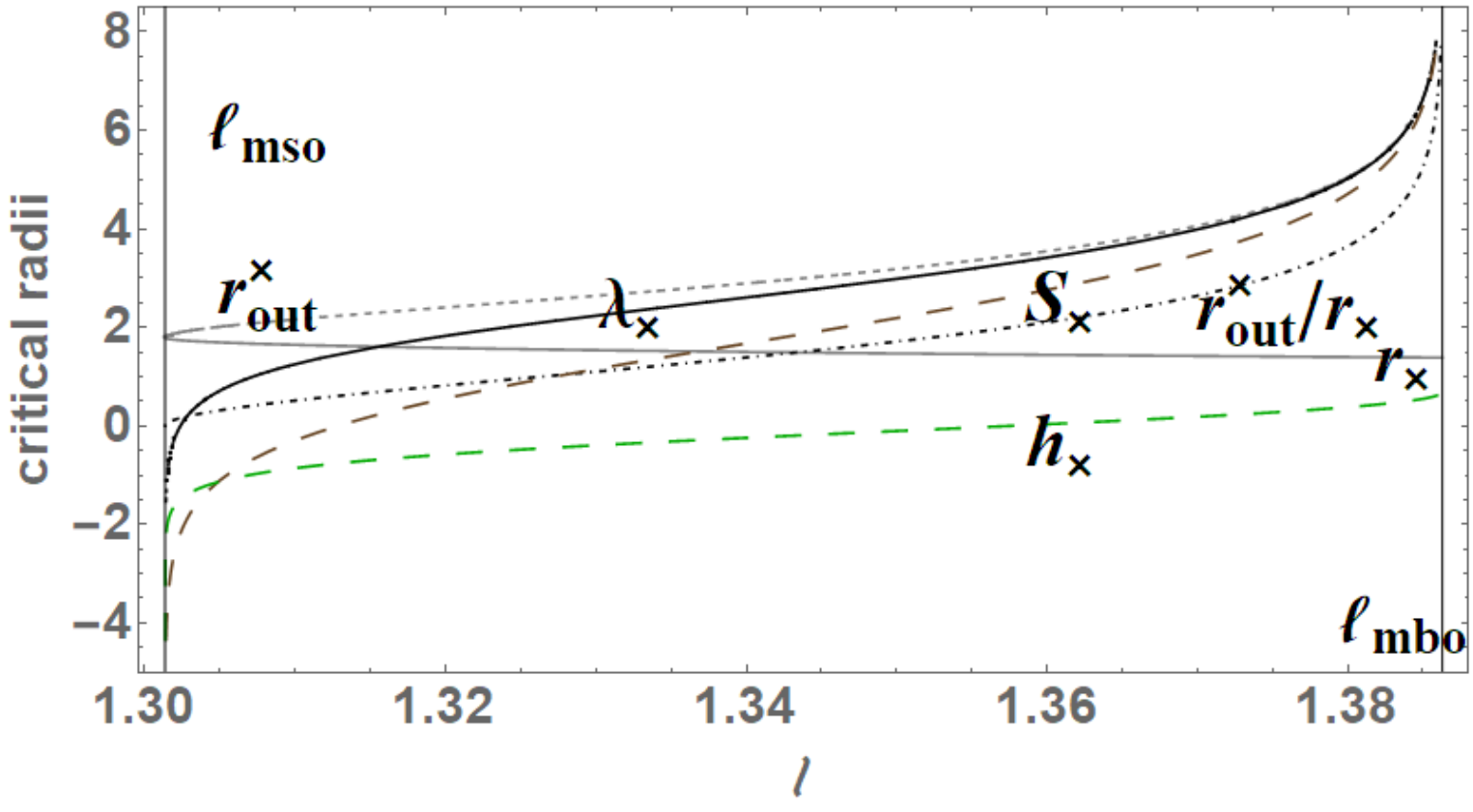}
  \includegraphics[width=8cm]{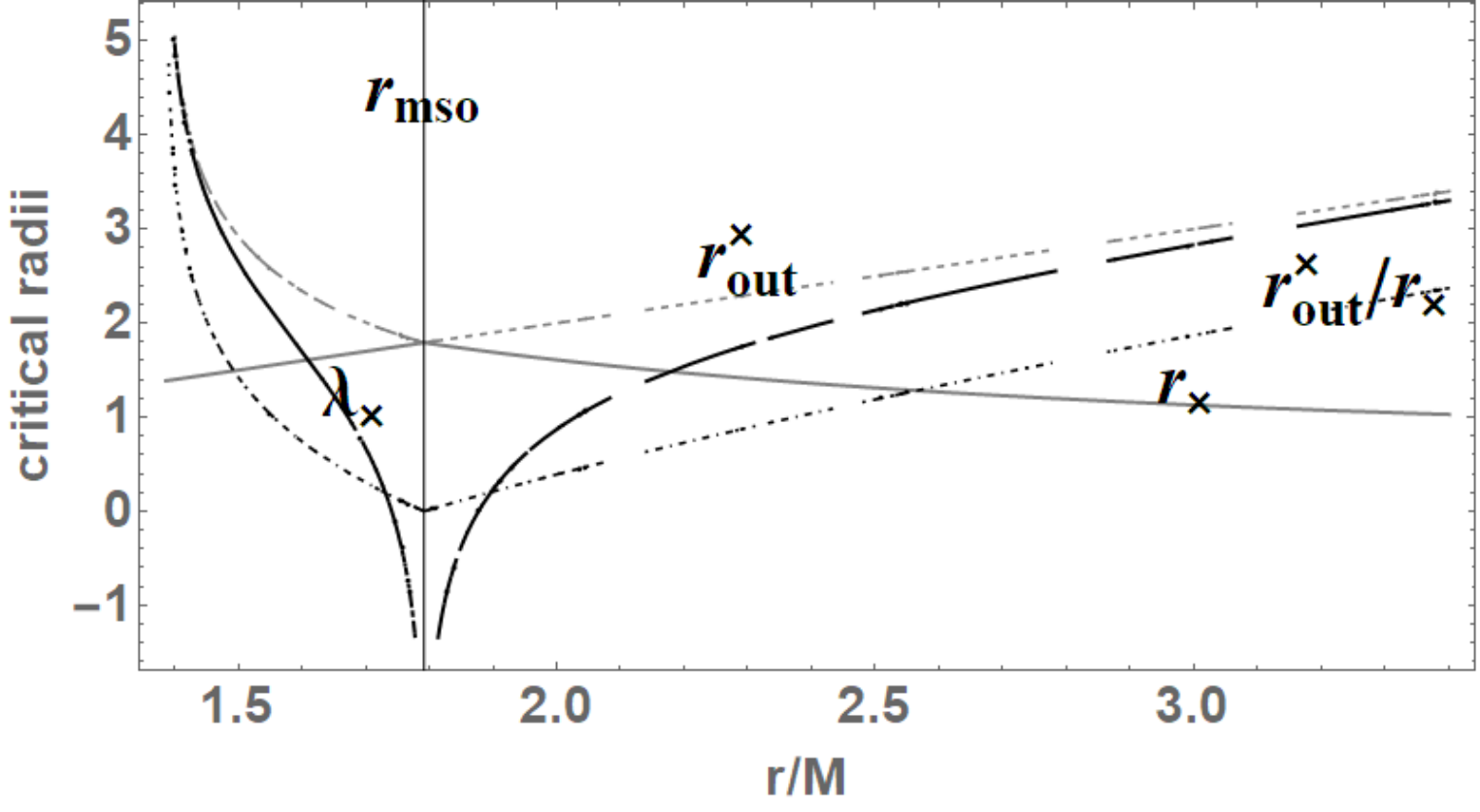}
  \caption{Cusped \textbf{RAD} tori: elongation $\lambda$, maximum height of the torus cross sections $h$, outer edge $r_{out}$ of accreting tori, tori geometrical thickness $\Sa$ as functions of the fluids specific angular momentum $\ell$ (left panel) and radius $r/M$ (right panel).
  Note the limiting values $\ell_{mso}$ and $\ell_{mbo}$ in the left panel, and $r_{mso}=6M$, $r_{mbo}=4M$, $r_{\gamma}=3M$ in right panel--see  Eqs\il(\ref{Eq:rcentro}, \ref{Eqs:rssrcitt}, \ref{Eq:over-top}, \ref{Eq:dan-aga-mich}).}\label{Fig:SMGerm}
\end{figure}
\item\textbf{{The $K$-parameter  $K_{cent}(\ell)$ at the  center of maximum density (and hydrostatic pressure), and at the  inner edge of accreting torus $K_{\times}(\ell)$ as functions of the fluid specific  angular momentum $\ell$}.}

We can evaluate the distribution of $K$-parameters in the \textbf{RAD} in terms of the leading function $\ell(r)$ in Eq.\il(\ref{Eq:lqkp}), in this way we express the correlation between $K$ and $\ell$ in the  misaligned tori:
\bea&&\label{Eq:grod-lock}
K_{cent}(\ell)\equiv\frac{\sqrt{\frac{\left[\ell^2+2 L_{ \ell} \cos \left\{\frac{1}{3} \cos ^{-1}(L_{\mathcal{ll}})\right\}-6\right] \left[ \ell^2+2 L_{ \ell} \cos \left(\frac{1}{3} \cos ^{-1}[L_{\mathcal{ll}}]\right)\right]^2}{ \ell^2 \left[3 \ell^4+2 \left(2  \ell^2-15\right) L_{ \ell} \cos \left[\frac{1}{3} \cos ^{-1}(L_{\mathcal{ll}})\right]-39  \ell^2+2 L_{ \ell}^2 \cos \left(\frac{2}{3} \cos ^{-1}[L_{\mathcal{ll}}]\right)+54\right]}}}{\sqrt{3}},
 \\\label{Eq:us-doi}
 &&K_{\times}(\ell)\equiv\frac{\sqrt{\frac{\left[ \ell^2-2 L_{ \ell} \sin \left(\frac{1}{3} \sin ^{-1}[L_{\mathcal{ll}}]\right)-6\right] \left[ \ell^2-2 L_{ \ell} \sin \left(\frac{1}{3} \sin ^{-1}[L_{\mathcal{ll}}]\right)\right]^2}{ \ell^2 \left[3 \ell^4+2 \left(15-2  \ell^2\right) L_{ \ell} \sin \left[\frac{1}{3} \sin ^{-1}(L_{\mathcal{ll}})\right]-39  \ell^2-2 L_{ \ell}^2 \cos \left(\frac{2}{3} \sin ^{-1}[L_{\mathcal{ll}}]\right)+54\right]}}}{\sqrt{3}}.
\eea
shown in Figs\il\ref{Fig:VClose},\ref{Fig:conicap}, see also Figs\il\ref{Fig:woersigna} where we also  study  these quantities assuming  $K=K_{crit}=K(r)$ or
$\ell=\ell_{crit}=\ell(r)$ as defined in  Eq.\il(\ref{Eq:lqkp}), in the planes $(\ell,r)$ and $(K,r)$, respectively. In doing so we parameterize  the cusped tori  in terms of one parameter, $\ell$ or $K$,  and one  the radii,  showing the role of limiting curves $\ell_{crit}$ and $K_{crit}$ setting the boundary values on the existence of misaligned tori. These curves determine the formations of \textbf{RAD} tori in a bounded region of the   $(\ell,r)$ or $(K,r)$ planes.  Consequently this analysis identifies also the    sets of  \textbf{RAD} tori governed by special relations appearing in these planes, typically there are sets of toroids with equal $R\in\{r_{out},r_{cent},r_{inner},\lambda\}$ or other characteristics as the torus thickness.
Specifically, we are interested here in determining the collisional emergence  among  the \textbf{RAD} tori.  This problem can be faced  by exploring     the condition  $r_{inner}^o=r_{out}^i$ ($()^o$ and $()^i$ refer to the outer  and inner torus respectively)  on the spherical radius $r$ considering the conditions on the $\ell$ and  $K$ parameters.
\begin{figure}
  \includegraphics[width=8cm]{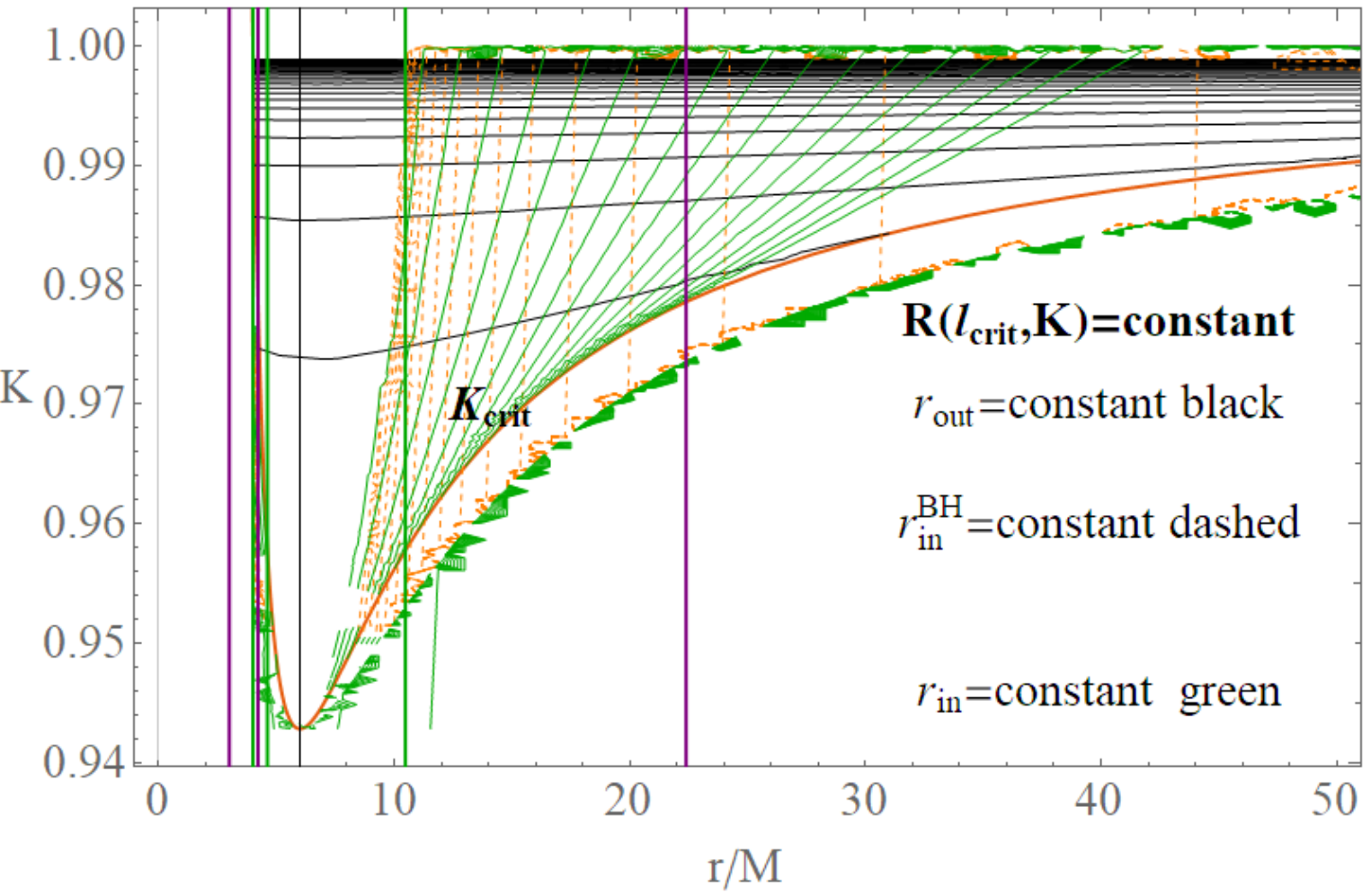}
  \includegraphics[width=8cm]{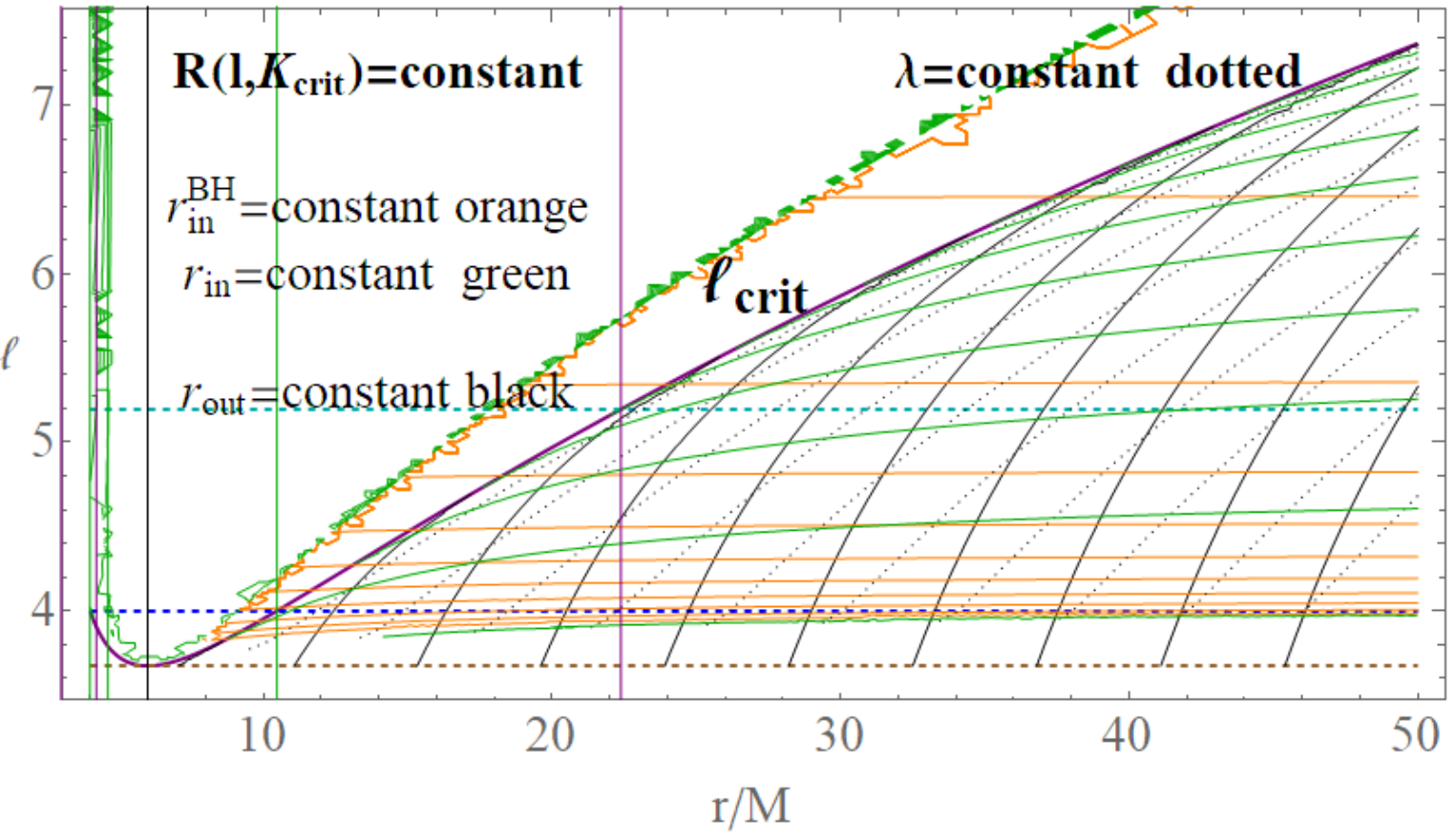}
  \includegraphics[width=8cm]{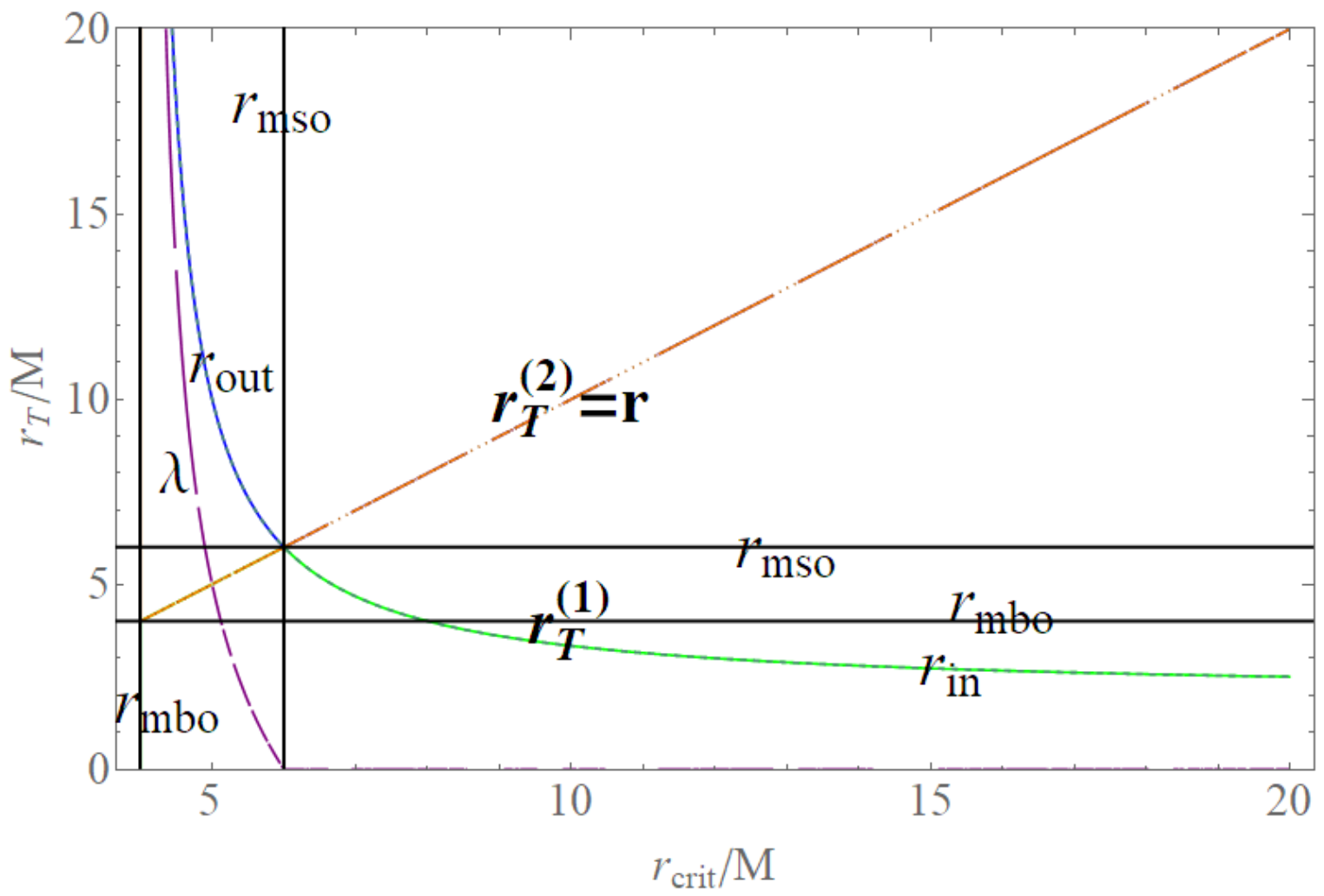}
  \includegraphics[width=8cm]{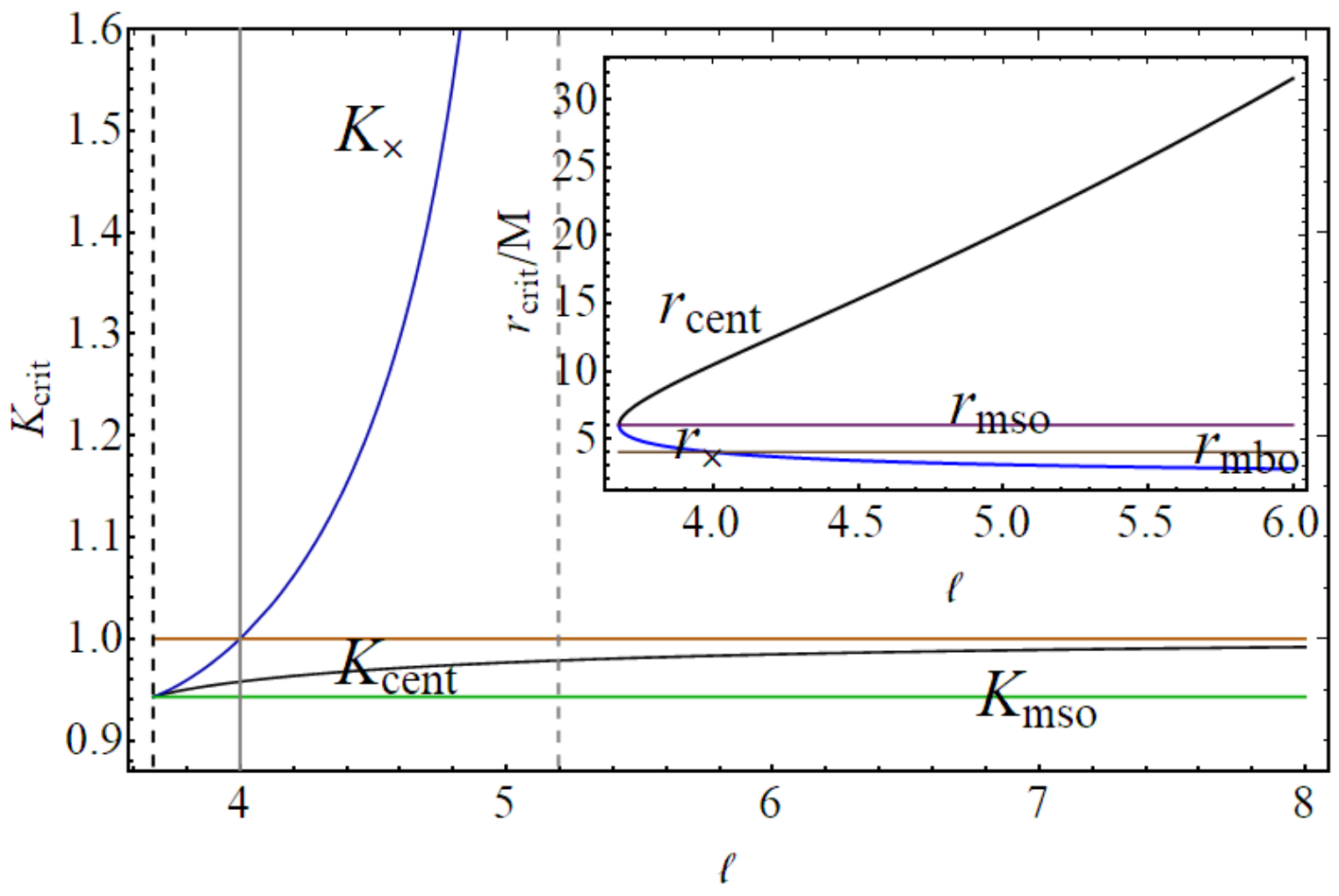}
  \caption{Upper panels: Plot of quantities  $R (\ell_{crit}, K)$=constant, where $R (\ell_{crit}, K)$ is the inner edge  $r_{inner}(\ell,K)$, outer edge $r_{out}(\ell,K)$, radius $r_{inner}^{BH}(\ell,K)$ of the innermost configuration in Eq.\il(\ref{Eq:mer-panto-ex-resul}),  in the plane $(K, r/M)$ considering $\ell=\ell_{crit}$ equal to the Keplerian specific angular momentum of the fluids--
we note the limit  $r=3 M$ and $r= 6 M$ and the limiting curve $K_{crit}=K(r)$ in Eq.\il(\ref{Eq:lqkp}). Right panel:  quantities  $R (\ell, K_{crit})$=constant
in the plane $(\ell,r)$ where $K=K_{crit}=K(r)$ of  Eq.\il(\ref{Eq:lqkp})--note that the limiting curve is $\ell_{crit}=\ell(r)$ see also Eq.\il(\ref{Eq:lam}) for the  torus elongation $\lambda$.  Below panels. Left: elongation $\lambda$  as function of $r_{crit}$, note the limit $r_{crit}>0$--see also  Figs\il\ref{Fig:woersigna}. We considered the conditions  $\mathbf{[\Im]}:\il r_T(r_{crit})=r_T(K_{crit},\ell_{crit})$ providing  the  two solutions: $r_T^{(1)}\equiv ({2M r})/({r-4M})$  and $r_T^{(2)}\equiv r$, for $r>4M$.
relating radius $r$ to the critical radius $r_{crit}$. Right panel: $K_{crit}$, for the accreting cusps (K<1) and proto-jet cusps $K\geq1$ and $K_{cent}\in[K_{mso},1[$ as functions of $\ell$
in Eq.\il(\ref{Eq:grod-lock}). Inside panel
$r_{crit}$ as function of $\ell$
Eq.\il(\ref{Eq:rcentro}). }\label{Fig:VClose}
\end{figure}
This issue can be addressed by   studying  the
condition $\mathbf{[\Im]}:\il r_\mathrm{T}(r_{crit})=r_\mathrm{T}(K_{crit},\ell_{crit})$, for any  $r_\mathrm{T}\in\{r_{inner},r_{out},r_{inner}^{BH}\}$,   providing the following two solutions: $r_T^{(1)}\equiv ({2M r})/({r-4M})$  and $r_\mathrm{T}^{(2)}\equiv r$, for $r>4M$,
relating radius $r$ to the critical radius $r_{crit}$, as in Figs\il\ref{Fig:VClose} where we also show the relations with inner and outer edges of the toroids.
For the evaluation of the quantities of Figs\il\ref{Fig:conicap}, we used the class of  accreting models (\textbf{\texttt{AC}}) defined   with $\ell=
 \ell_{mso} + \ell_{d}^{\epsilon}$, where $\ell_{d}^{\epsilon}\equiv(\ell_{mbo}-\ell_{mso})/\epsilon$  with  $K =K_{crit}$.
\begin{figure}
  \includegraphics[width=8.5cm]{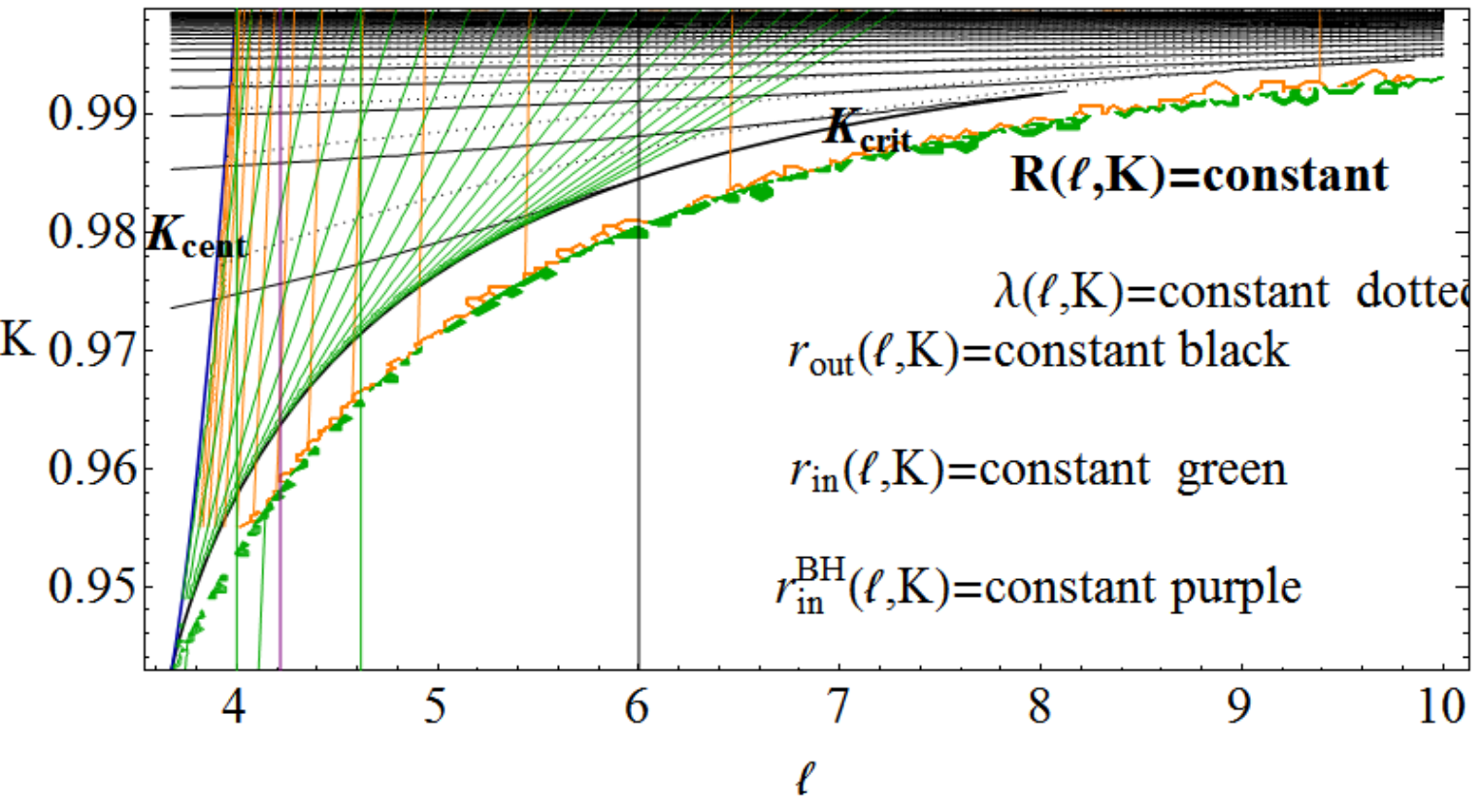}
  \includegraphics[width=8.5cm]{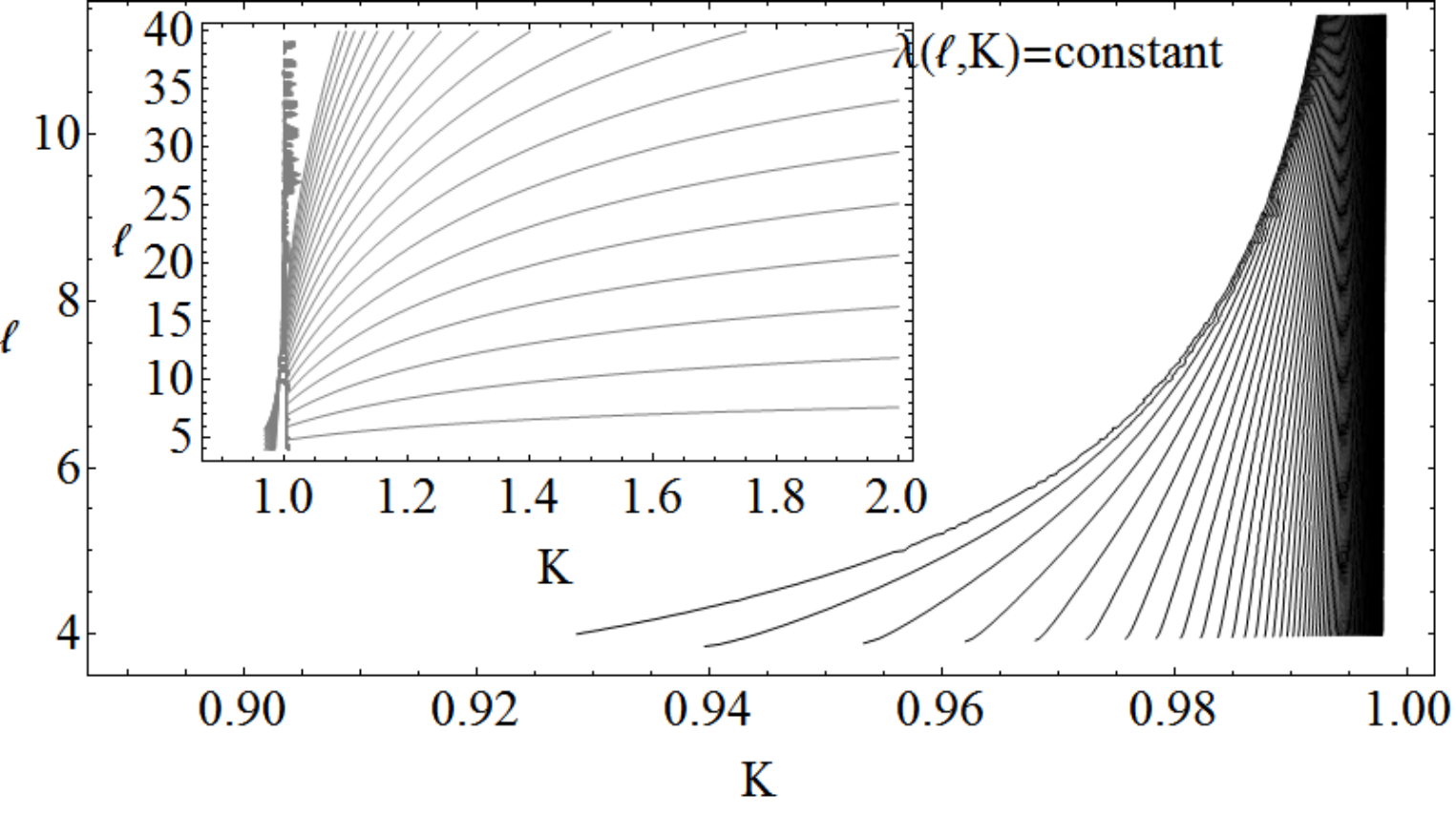}\\
   \includegraphics[width=8.5cm]{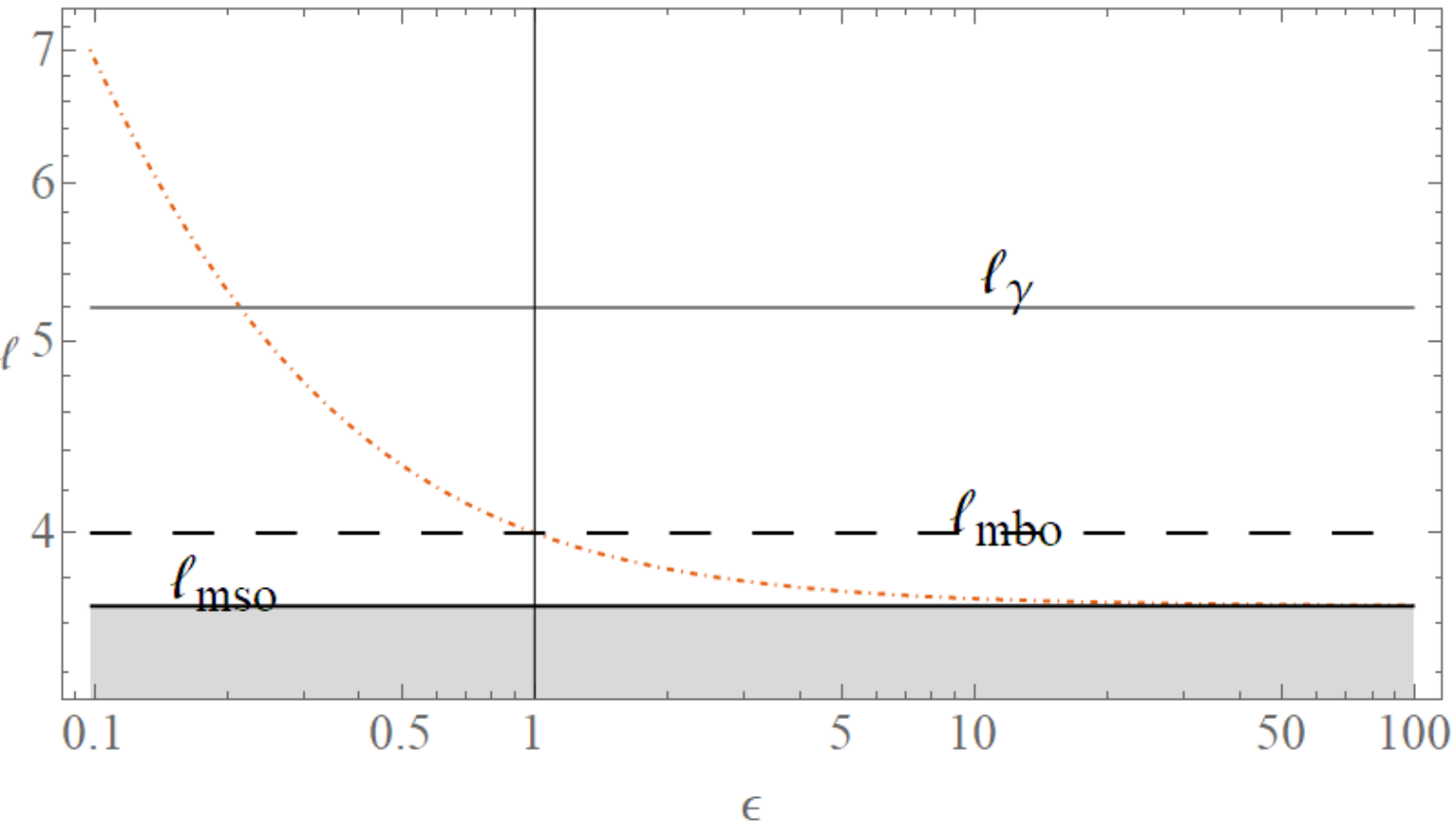}
  \includegraphics[width=8.5cm]{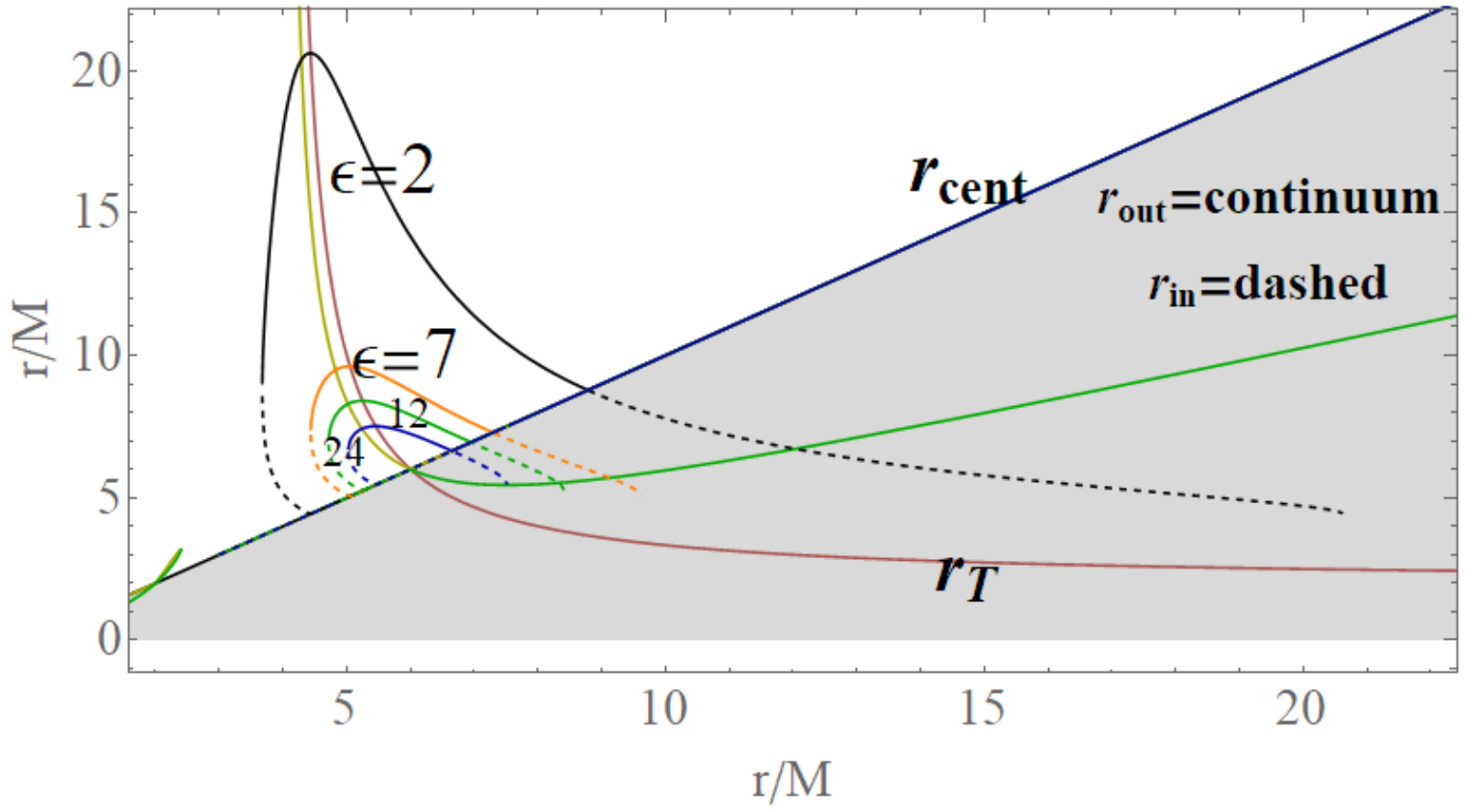}\\
  \includegraphics[width=8.5cm]{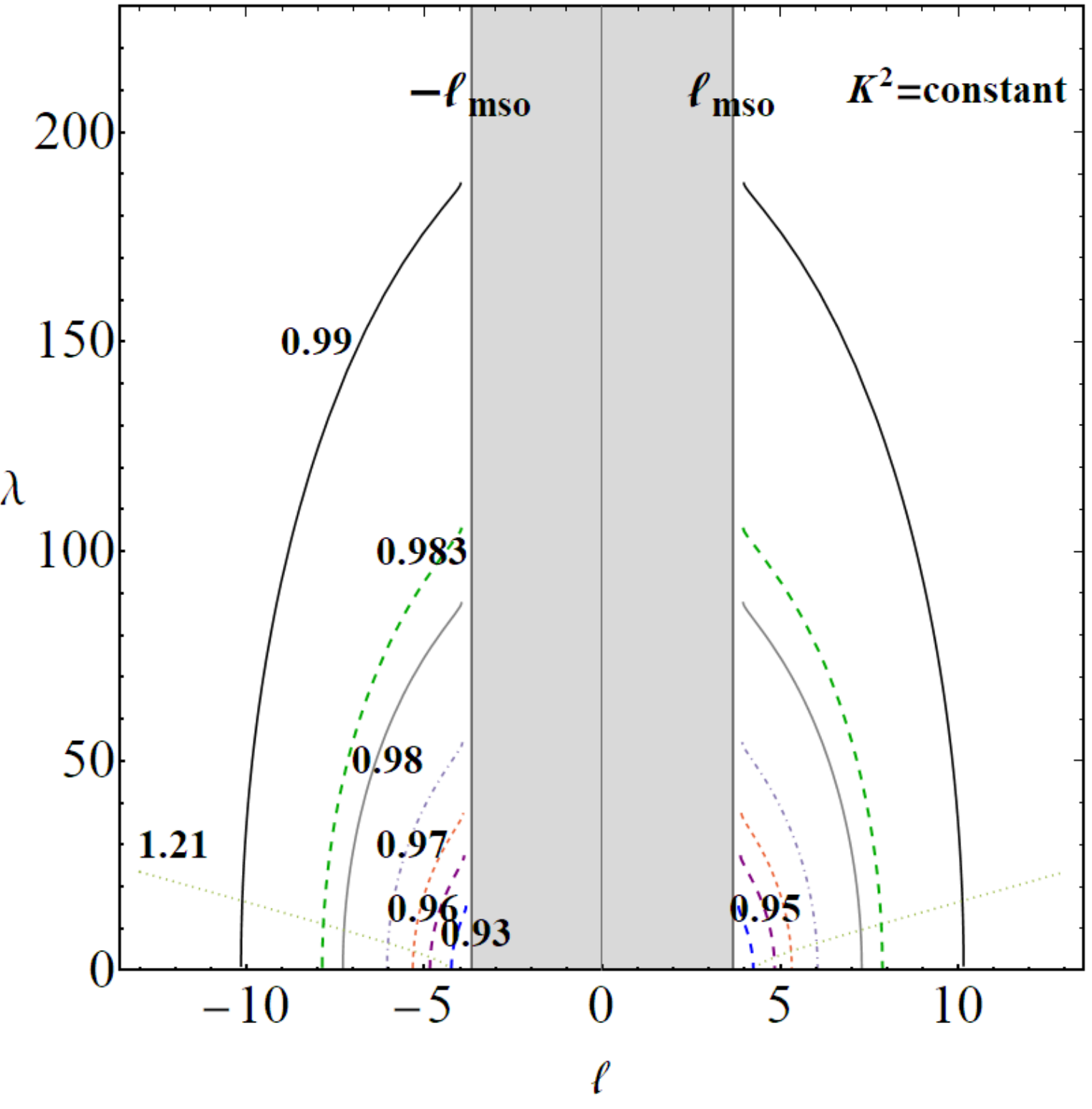}
  \caption{Upper panels. Left: Quantities  $R (\ell_{crit}, K)$=constant, as listed in figure, where  $r_{inner}(\ell,K)$ is the inner edge, $r_{out}(\ell,K)$  is the torus outer edge,  $r_{inner}^{BH}(\ell,K)$ is the radius of the innermost configuration in Eq.\il(\ref{Eq:mer-panto-ex-resul}),  and   $\lambda$  is the torus elongation. Curves are sets of toroids with equal  $R (\ell_{crit}, K)$, in the plane $(\ell, K)$, configurations are bounded by the curves $K_{cent}, K_{crit}$ of Eq.\il(\ref{Eq:grod-lock}). Right panel: \textbf{RAD} tori with equal   elongations $\lambda$=constant
as function of $\ell$ and $K$; for fixed $\ell$ curves, fixed elongation states for different $K$.
For fixed  $K$ and  $\lambda$,  we identify classes of tori  with different specific momentum but equal elongations and $K$ parameter (regulating several tori characteristics as the mass accretion rates).
Center line. Left panel: plot of $\ell$ as function of  $\epsilon$ in the accreting models (\textbf{\texttt{AC}}). Right panel: inner and outer tori edge of Eq.\il(\ref{Eq:mer-panto-ex-resul}) in different (\textbf{\texttt{AC}}) models, as functions of $r$, note the smaller is $\epsilon$ (model parameters) the larger is the torus, $r_T$ is in Figs\il\ref{Fig:VClose}. Bottom panel: curves of   elongations $\lambda$  of the tori with equal $K$ as function of the fluid specific angular momentum, constructed according to  Eq.\il(\ref{Eq:lam}).} \label{Fig:conicap}
\end{figure}
The geometric toroids thickness  is a significant feature of  the ringed structure which actually governs many properties of the disk.  In general \textbf{RAD} tori may be also very large at  large distance from the attractor, having large specific angular momentum magnitude (in these cases tori self-gravity should be considered  relevant). As a consequence of this analysis we also give an assessment of the toroids in the \textbf{RAD} in terms of their thickness.
 We start  considering the point of maximum thickness in the toroid which  generally does not coincide with the point of  of maximum pressure and density.
 \item\textbf{Geometric maximum  radius of the torus surface $r_{\max}^o(K,\ell)$ and the innermost surface $r_{\max}^i(K,\ell)$, the maximum $h_{\max}^o(K,\ell)$ of the torus surface as functions of $K$ and $\ell$. }

Misaligned tori  orbiting on their equatorial plane around a static attractor  are essentially axially symmetric, the maximum (geometric maximum) of their associated surface,   is located at distance  $r_{\max}^o$ from the central attractor
\bea\label{Eq:mat-ye}
&&
r_{\max}^o\equiv\sqrt{\frac{K^2 \ell^2}{K^2-1}+4 \sqrt{\frac{2}{3}} \psi  \cos \left[\frac{1}{3} \cos ^{-1}( \psi_\pi)\right]},\quad r_{\max}^i\equiv\sqrt{\frac{K^2 \ell^2}{K^2-1}-4 \sqrt{\frac{2}{3}} \psi  \sin \left[\frac{1}{3} \sin ^{-1}(\psi_\pi)\right]},
\eea
while the maximum height reads
\bea
\label{Eq:xit-graci}
&&
h_{\max}^o\equiv\sqrt{-\frac{K^2 \ell^2}{K^2-1}+\frac{\left[3 K^4 \ell^2 \sec \left(\frac{1}{3} \cos ^{-1}[ \psi_\pi]\right)+4 \sqrt{6} \left(K^2-1\right) \psi \right]^2}{24 \left(K^2-1\right)^4 \psi ^2}-4 \sqrt{\frac{2}{3}} \psi  \cos \left[\frac{1}{3} \cos ^{-1}(\psi_\pi )\right]},
\\
&&\psi_\pi \equiv-\frac{3}{4}\sqrt{\frac{3}{2}} \left(K^2-1\right)^2 \psi, \quad\psi \equiv\sqrt{-\frac{K^4 \ell^2}{\left(K^2-1\right)^3}},
\eea
$r_{\max}^i$ is the location of maximum point for the inner Roche lobe close to the central \textbf{BH},
$h_{\max}^o$ is the semi height of the disk.
Note that there is  $r_{\max}^o\neq r_{cent}$--the point of maximum density and (hydrostatic) pressure in the torus is not coincident with the position of the geometric maximum of the torus surface.
Moreover, these functions, as well as most of the quantities characterizing the disk, depend on the even powers of $K$ and $\ell$ and show an explicit dependence on the limiting cut-off value $(K^2-1)=0$.
These quantities are represented in Figs\il\ref{Fig:lom-b-emi}, where we also consider them  depending on the fluid specific angular momentum $\ell$ and $K$. We also show  condition for $r_{\max}^o= r_{cent}$, and  the value of the torus  surface at its geometric center  $h(r_{\max}^o)$ and at its maximum density center $h(r_{cent})$.
\begin{figure}
   \includegraphics[width=8.5cm]{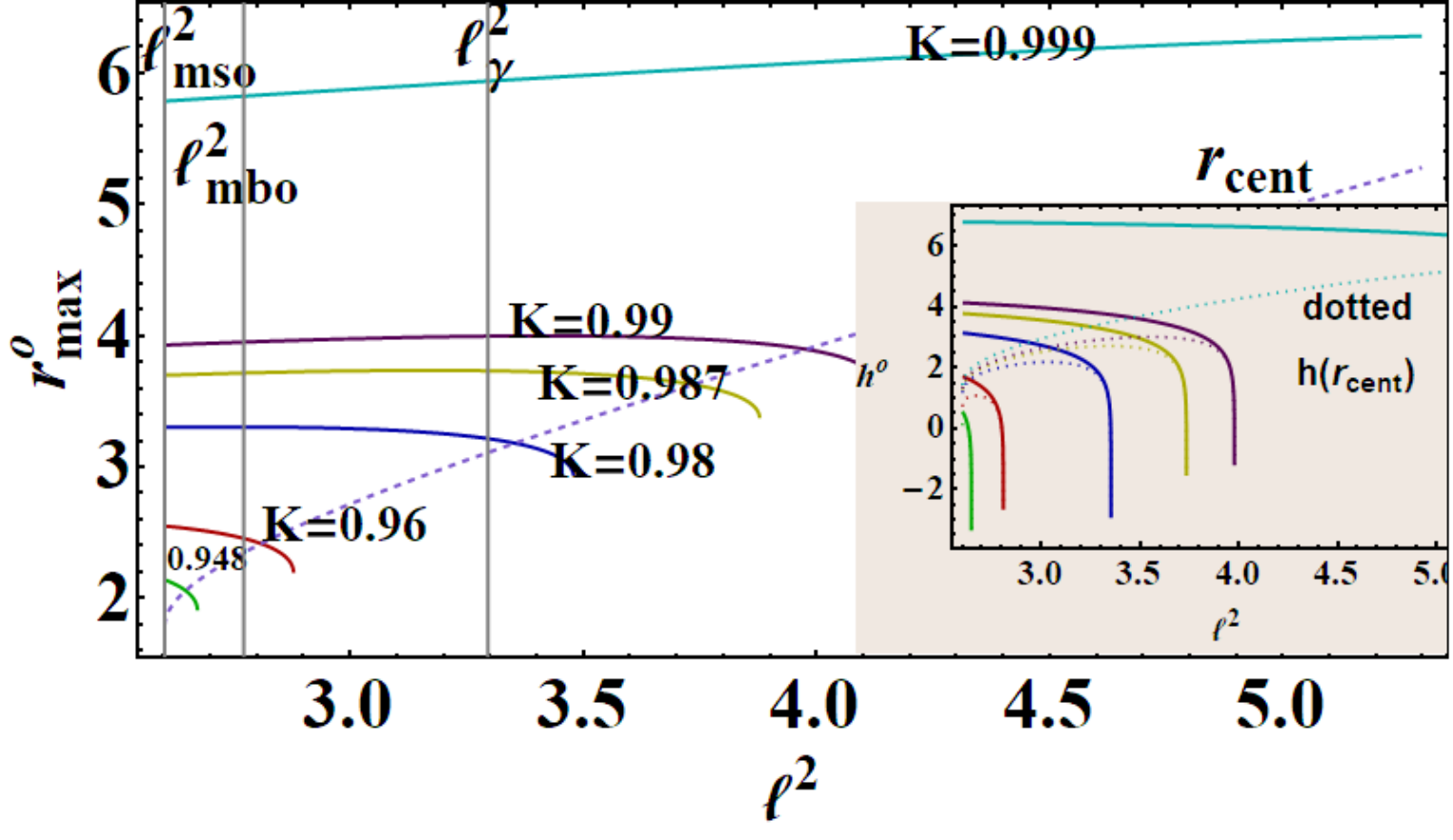}
  \includegraphics[width=8.5cm]{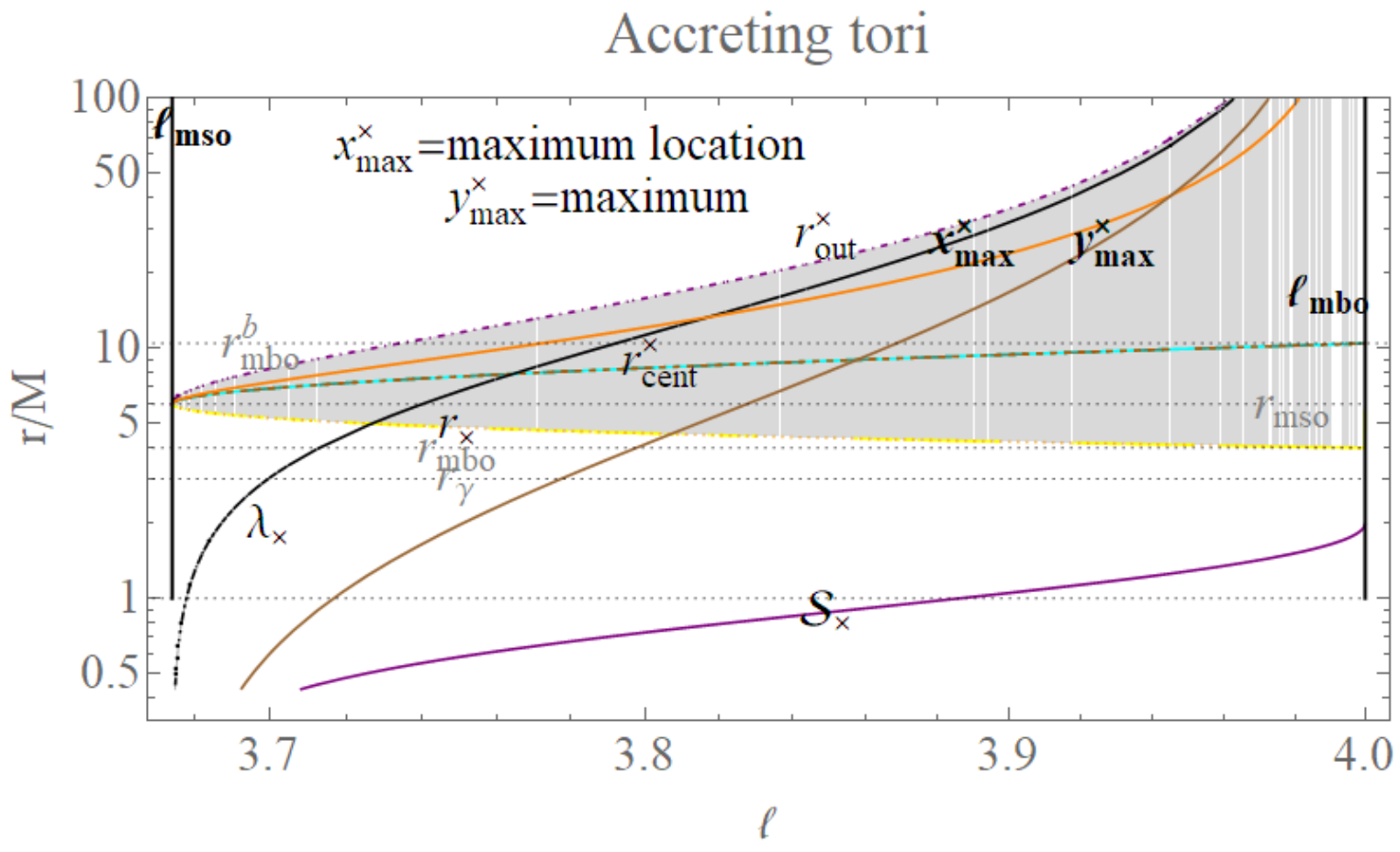}
  \includegraphics[width=8.5cm]{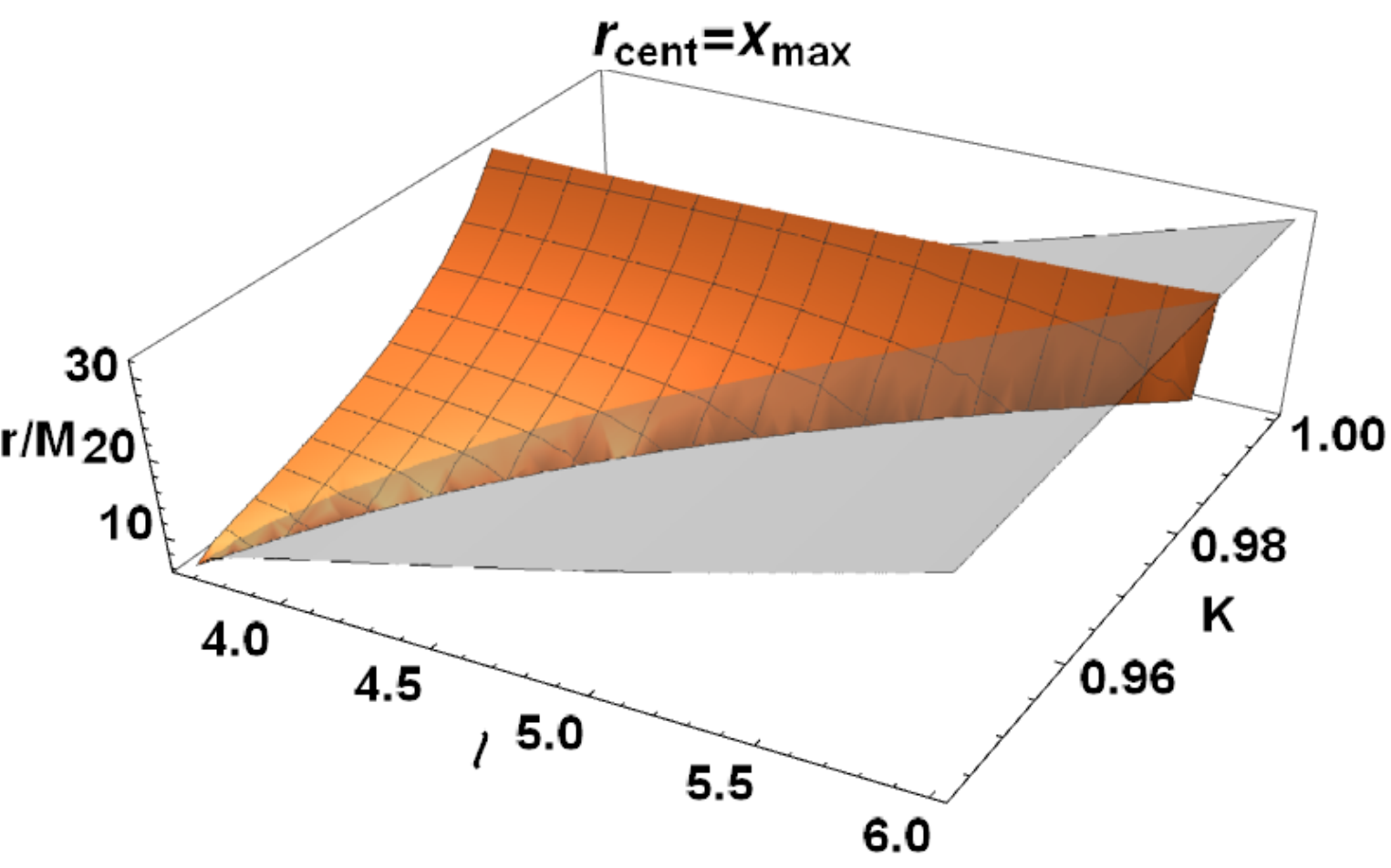}
    \includegraphics[width=8.5cm]{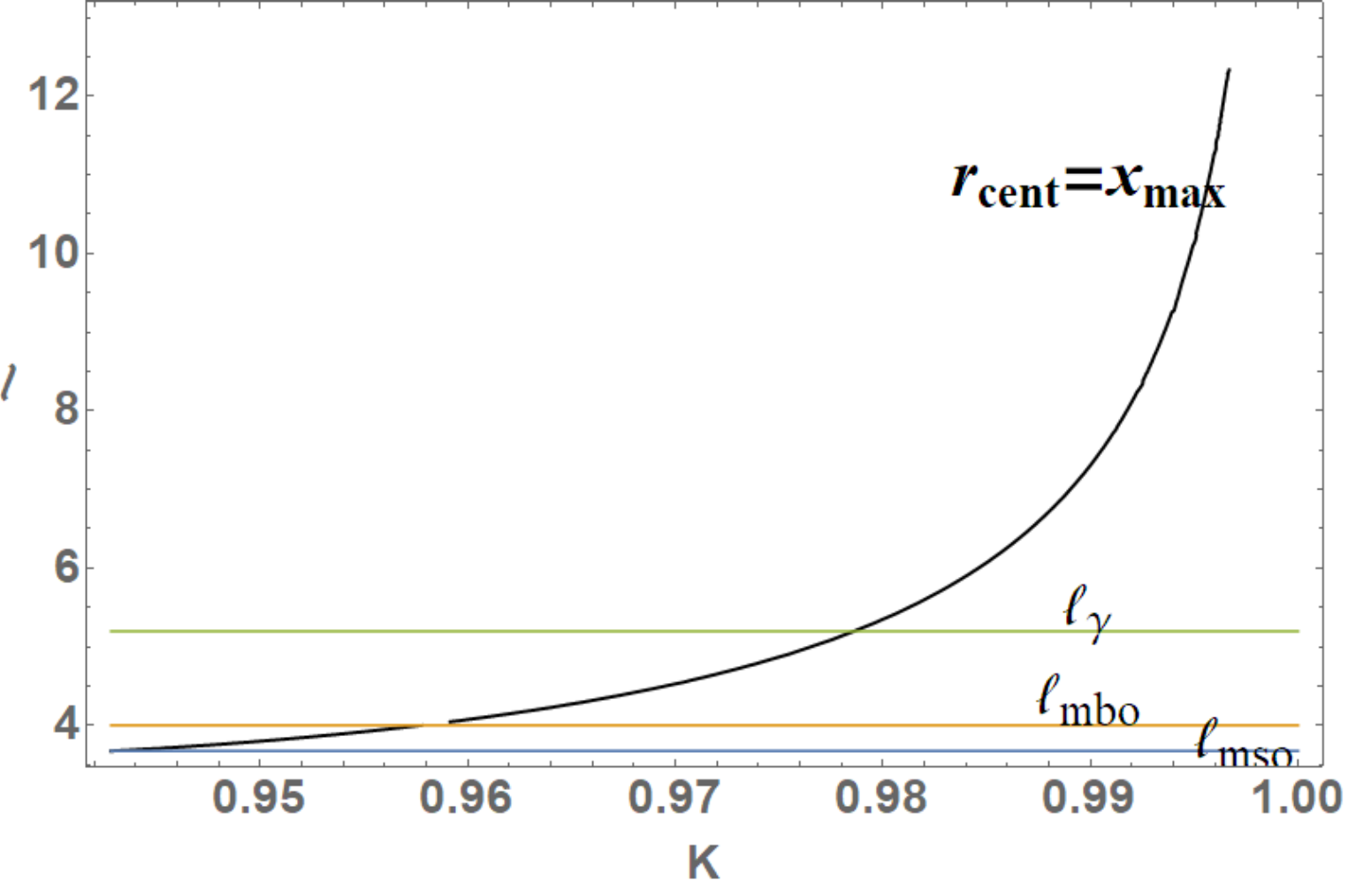}
  \caption{Upper panels. Left: Geometric maximum  radius of the torus surface $r_{\max}(K,\ell)$, the maximum $h_{\max}(K,\ell)=y^{\times}_{max}$ of the torus surface as functions of $K$ and $\ell$ as defined in Eq.\il(\ref{Eq:xit-graci}). $r_{cent}$ is the location of maximum density  point in the disk, in general there is  $r_{cent}\neq r_{\max}(K,\ell)$. Inside panel:  $h(K,\ell)$ and $h(r_{cent})$ (dotted) torus surface correspond at its center $r_{cent}$ to different values of $K$. Right: The inner edge  $r_{\times}(\ell)$,   the outer edge,
 $r_{out}^{\times}(\ell)$, the center $r_{cent}^{\times}(\ell)$, the elongation $\lambda_{\times}(\ell)$ of the accreting tori, thickness $\Sa_{\times}\equiv 2y^{\times}_{max}/\lambda_{\times}$,
 $y^{\times}_{max}(\equiv h(K,\ell))$ maximum  and location of geometrical maximum $x^{\times}_{max}(\equiv r_{\max}(K,\ell))$   as functions of  the specific fluid angular momentum $\ell$. Note,  for accreting tori  the thickness is $\Sa_{\times}\geq1$ only for $\ell\gtrsim 3.87$. Bottom panels-coincidence $r_{cent}=x_{max}$ (centrum of maximum density/ geometrical centrum). Left panel: surfaces $r_{cent}$ (gray), $x_{max}$ (orange) as functions of $(\ell,K)$. Right panel: classes of tori with $r_{cent}=x_{max}$.}\label{Fig:lom-b-emi}
\end{figure}
So there is a special class of toroids where  $r_{cent}=r_{\max}^o$.
\textbf{RADs} with  cusped misaligned  configurations   constitute a particularly significant case, therefore we  focus on the morphological quantities of the toroidal components when the inner \textbf{RAD} torus is cusped  (in accreting phase).  In this case, $\ell$ is the only independent model parameter.
\item{\textbf{Torus height $h_{\max}^o(r_{\times})$ as function of the  cusp location.}}

Thus, considering Eqs\il(\ref{Eq:xit-graci}) and Eqs\il(\ref{Eq:lqkp}),  we evaluated the   torus height $h_{\max}^o(r_{\times})$ as function of the  cusp location $r=r_{\times}\in]r_{mbo},r_{mso}[$ and similarly the  locations  $r_{\max}^o(r_{crit})$ and $r_{\max}^i(r_{crit})$, as function of $r_{crit}$
{\small
\bea\nonumber
&&
h_{\max}^o(r_{\times})=\sqrt{-2 \sqrt{6} \sqrt{\frac{(r_{\times}-3) (r_{\times}-2)^2 r_{\times}^4}{(r_{\times}-4)^3}} \sec \left[\frac{1}{3} \cos ^{-1}(\psi_\rho )\right]+\frac{9 (r_{\times}-2)^2 r_{\times}^2 \sec ^2\left[\frac{1}{3} \cos ^{-1}(\psi_\rho )\right]}{8 (r_{\times}-4) (r_{\times}-3)}+\frac{(r_{\times}-2) (5 r_{\times}-18) r_{\times}^2}{(r_{\times}-4)^2}},
\\\label{Eqs:rssrcitt}
&&
r_{\max}^o(r_{crit})=\sqrt{4 \sqrt{\frac{2}{3}} \psi_{\lambda } \cos \left[\frac{1}{3} \cos ^{-1}\left(-\frac{3}{4} \sqrt{\frac{3}{2}} \psi_{\lambda } \psi _{\sigma }^2\right)\right]+\frac{r^2}{(r-3) \psi_{\sigma }}},
 \\\label{Eq:pos-spe}
 &&r_{\max}^i(r_{crit})=\sqrt{\frac{r^2}{(r-3) \psi _{\sigma }}-4 \sqrt{\frac{2}{3}} \psi _{\lambda } \cos \left[\frac{1}{3} \left(\cos ^{-1}\left[-\frac{3}{4}\sqrt{\frac{3}{2}} \psi _{\lambda } \psi _{\sigma }^2\right]+\pi \right)\right]},
\eea
where
\bea
\psi _{\sigma }\equiv\frac{4-r}{(r-3) r},\quad\psi_{\lambda }\equiv\sqrt{-\frac{(r-2)^2 r}{(r-3)^2 \psi _{\sigma }^3}}, \quad\psi_\rho\equiv-\frac{3 \sqrt{\frac{3}{2}} (r_{\times}-4)^2 \sqrt{\frac{(r_{\times}-3) (r_{\times}-2)^2 r_{\times}^4}{(r_{\times}-4)^3}}}{4 (r_{\times}-3)^2 r_{\times}^2},
\eea}
 we used  $K=K_{crit}$ and $\ell=\ell_{crit}$ in Eq.\il(\ref{Eq:xit-graci}),
 parameterizing these quantities  for the cusp location.
 These quantities are represented in Figs\il\ref{Fig:woersigna} and \ref{Fig:SMGerm}.
\begin{figure}
  \includegraphics[width=9cm]{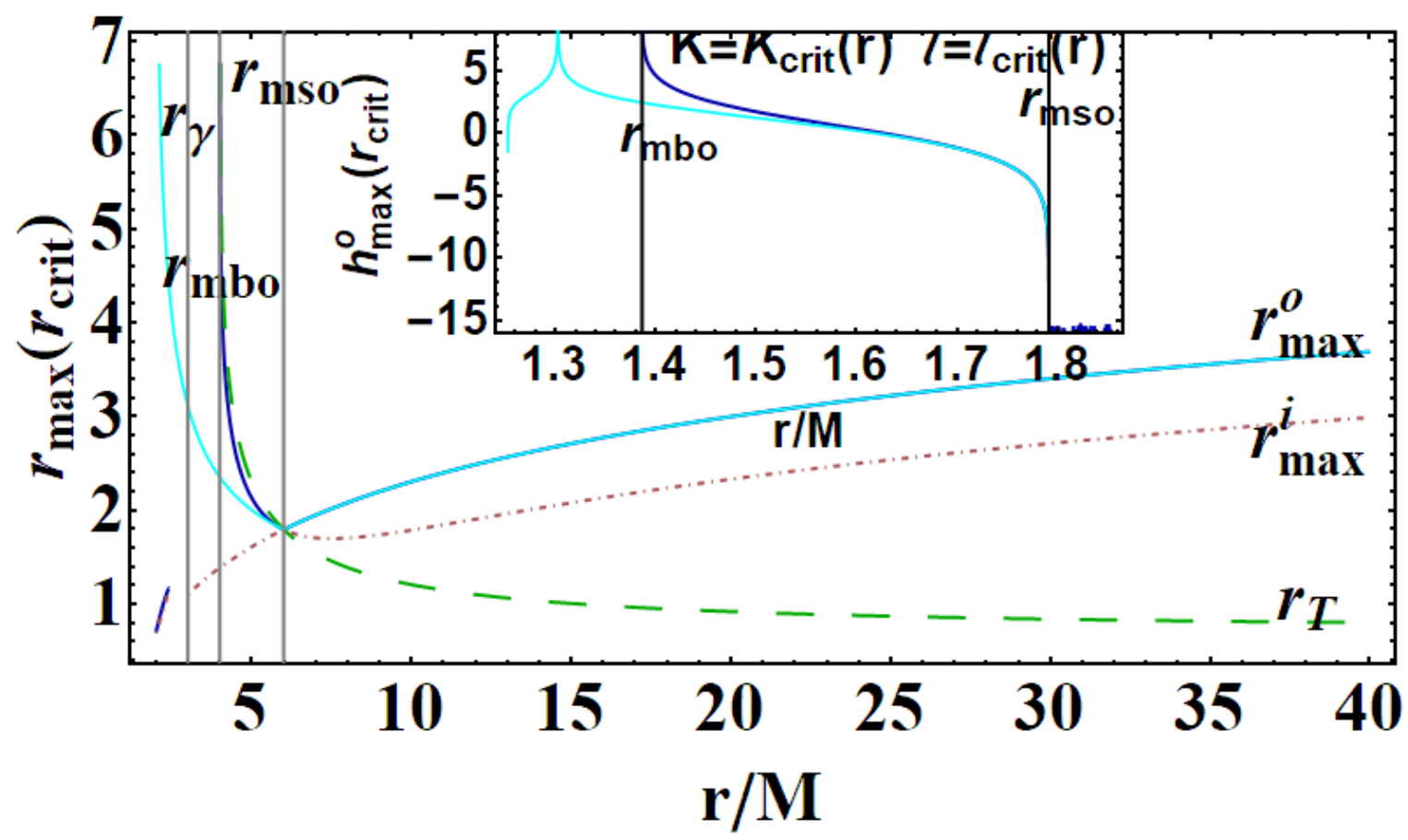}
  \includegraphics[width=8cm]{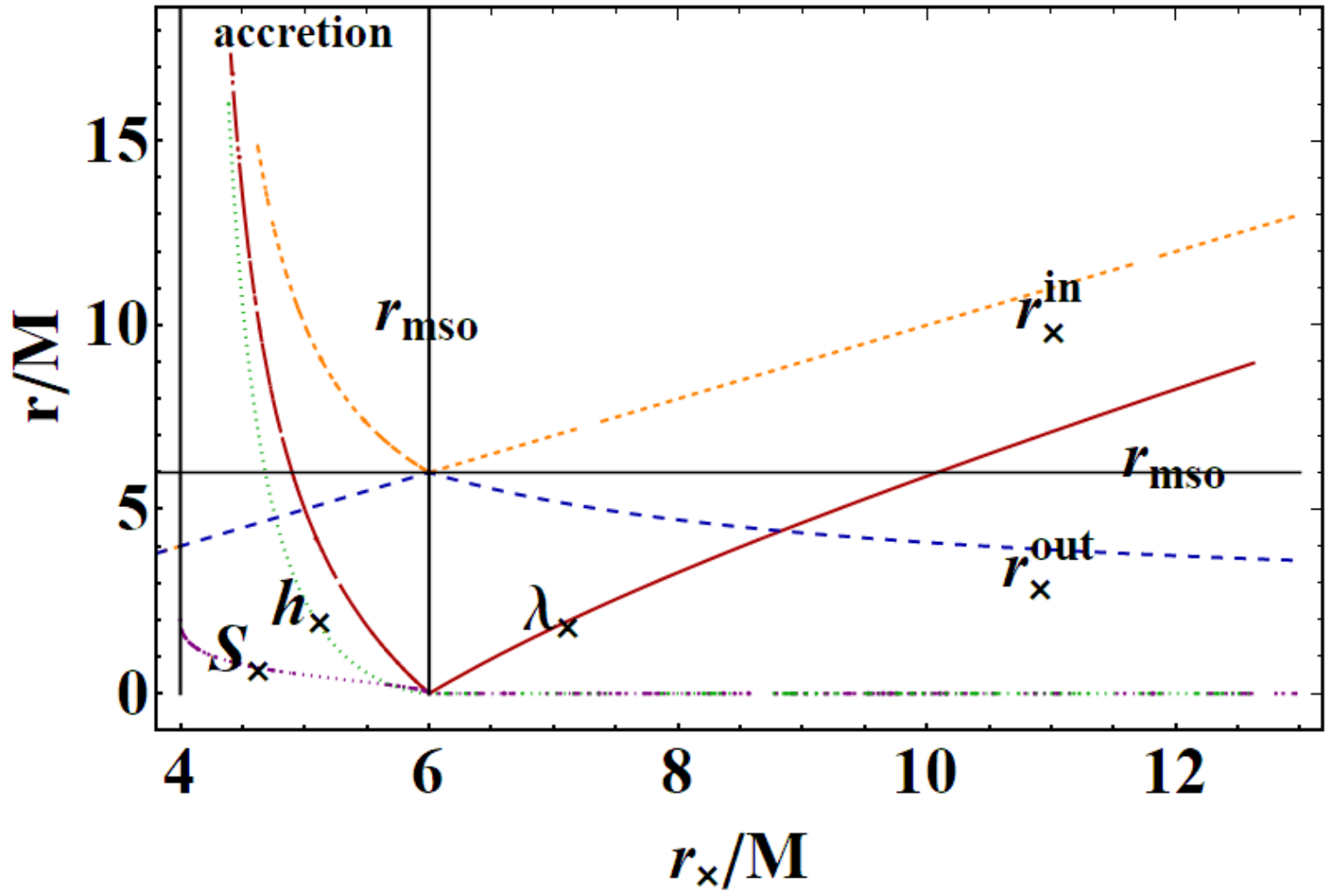}
  \caption{Left panel: The torus height $h^o(r_{\times})$ and the locations  $r_{\max}^o(r_{crit})$ and $r_{\max}^i(r_{crit})$ as functions of  accreting radius (cusped location) $r=r_{\times}\in]r_{mbo},r_{mso}[$ in Eq.\il(\ref{Eqs:rssrcitt}). Right panel: elongation $\lambda$, maximum highness $h_{\times}$ and thickness $\Sa_{\times}$ as functions of the cusp $r_{\times}$ radius in Eq.\il(\ref{Eq:over-top}). }\label{Fig:woersigna}
\end{figure}
It is clear that we can similarly  express the torus  elongation $\lambda$  and thickness in Eq.\il(\ref{Eq:lam}) and   Figs\il\ref{Fig:VClose},\ref{Fig:conicap} as functions of $r_{crit}$.
\item{\textbf{Outer and inner edges  as function of the cusps  $(r_{out}^{\times}(r_{\times}),r_{inner}^{\times}(r_{\times}))$.}}

Thus, the outer and inner edges of an accreting torus as function of $r_{\times}$ are
\bea&&\label{Eq:over-top}
r_{out}^{\times}(r_{\times})=\frac{2}{3} \left[\sqrt{\frac{(r_{\times}-6)^2 r_{\times}^2}{(r_{\times}-4)^2}} \cos \left[\frac{1}{3} \cos ^{-1}\left(-\frac{(r_{\times}-6) r}{(r_{\times}-4) \sqrt{\frac{(r_{\times}-6)^2 r_{\times}^2}{(r_{\times}-4)^2}}}\right)\right]+\frac{r_{\times}}{r_{\times}-4}+r\right],
\\
&&
r_{inner}^{\times}(r_{\times})=\frac{1}{3} \left[\frac{r_{\times}^3}{(r_{\times}-2)^2}-2 \sqrt{\frac{r_{\times}^3 \left[\frac{r_{\times}^3}{(r_{\times}-2)^2}-12\right]}{(r_{\times}-2)^2}} \cos \left[\frac{1}{3} \left(\cos ^{-1}\left[\frac{r_{\times}^3 \left(\frac{r_{\times}^6}{(r_{\times}-2)^4}-\frac{18 r_{\times}^3}{(r_{\times}-2)^2}+54\right)}{(r_{\times}-2)^2 \left(\frac{r_{\times}^3 \left(\frac{r_{\times}^3}{(r_{\times}-2)^2}-12\right)}{(r_{\times}-2)^2}\right)^{3/2}}\right]+\pi \right)\right]\right],
\eea
from which we derive critical elongation $\lambda_{\times}$, and thickness $\Sa_{\times}=2 h_{\times}/(\lambda_{\times})$ of the cusped tori--Figs\il\ref{Fig:woersigna}.
Note that  $(r_{inner}^{\times}(r_{\times}),r_{out}^{\times}(r_{\times}))$  combine solutions
$r=r_{\times}$ an $r=r_{out}$ as clear from Figs\il\ref{Fig:woersigna}.
\end{enumerate}
It is convenient to express  these  quantities eliminating the radial dependence (distance $r$ from the central $\mathbf{BH}$ and dependence from the spheres radius $r$) by writing  them in terms of the pair $(\ell, K)$.
Specifically, there is
\begin{enumerate}
\item \textbf{Fluid specific angular momentum $\ell$ as function of $K$-parameter.}
 \bea\label{Eq:crititLdeval}
 &&
 \ell_{crit}^o(K)\equiv\frac{\sqrt{-\frac{-27 K^4+K \left(9 K^2-8\right)^{3/2}+36 K^2-8}{K^2 \left(K^2-1\right)}}}{\sqrt{2}},\quad \ell_{crit}^i(K)\equiv\frac{\sqrt{\frac{27 K^4+K \left(9 K^2-8\right)^{3/2}-36 K^2+8}{K^2 \left(K^2-1\right)}}}{\sqrt{2}},
 \eea
 where
  $\ell_{crit}^o(K)> \ell_{crit}^i(K)>\ell_{mso}$ {and} $\ell_{crit}^i(K)\in[\ell_{mso},\ell_{\gamma}[$.
These momenta, showed in Figs\il\ref{Fig:leaduniUR} relate the specific angular momentum $\ell$ of the torus fluid to its $K$-parameter of Eqs\il(\ref{Eq:lqkp}).  Here $K$ is a free parameter for a quiescent torus ranging in $[K_{\min},K_{\max}]$, where  $K_{mso}<K_{\min}<K<\mathfrak{K}$,  and  $\mathfrak{K}= K_{\max}$ for torus in the accreting range of specific angular momentum values or
 $\mathfrak{K}=1$ otherwise. The couple $(K_{\min},K_{\max})$  is defined by  the $K(r)$ function of Eq.\il(\ref{Eq:lqkp})  at maximum,  $r_{\max}$, or minimum, $r_{\min}$, of the hydrostatic pressure, respectively. The function  $\ell_{crit}^o(K)$, evaluated for $K=K_{\max}$, provides the specific angular momentum of the torus whose instability accreting phase is associated  to the occurrence of the  value $K=K_{\max}$, viceversa $\ell_{crit}^i(K)$, evaluated for $K=K_{cent}$, provides the specific angular momentum of the torus whose center of maximum density  has  value $K=K_{cent}$. Lines $\ell=$constant, figuring a single torus, provide the couple $(K_{cent},K_{\max})$ when $\ell\in[\ell_{mso},\ell_{mbo}]$).
\begin{figure}
  \includegraphics[width=8cm]{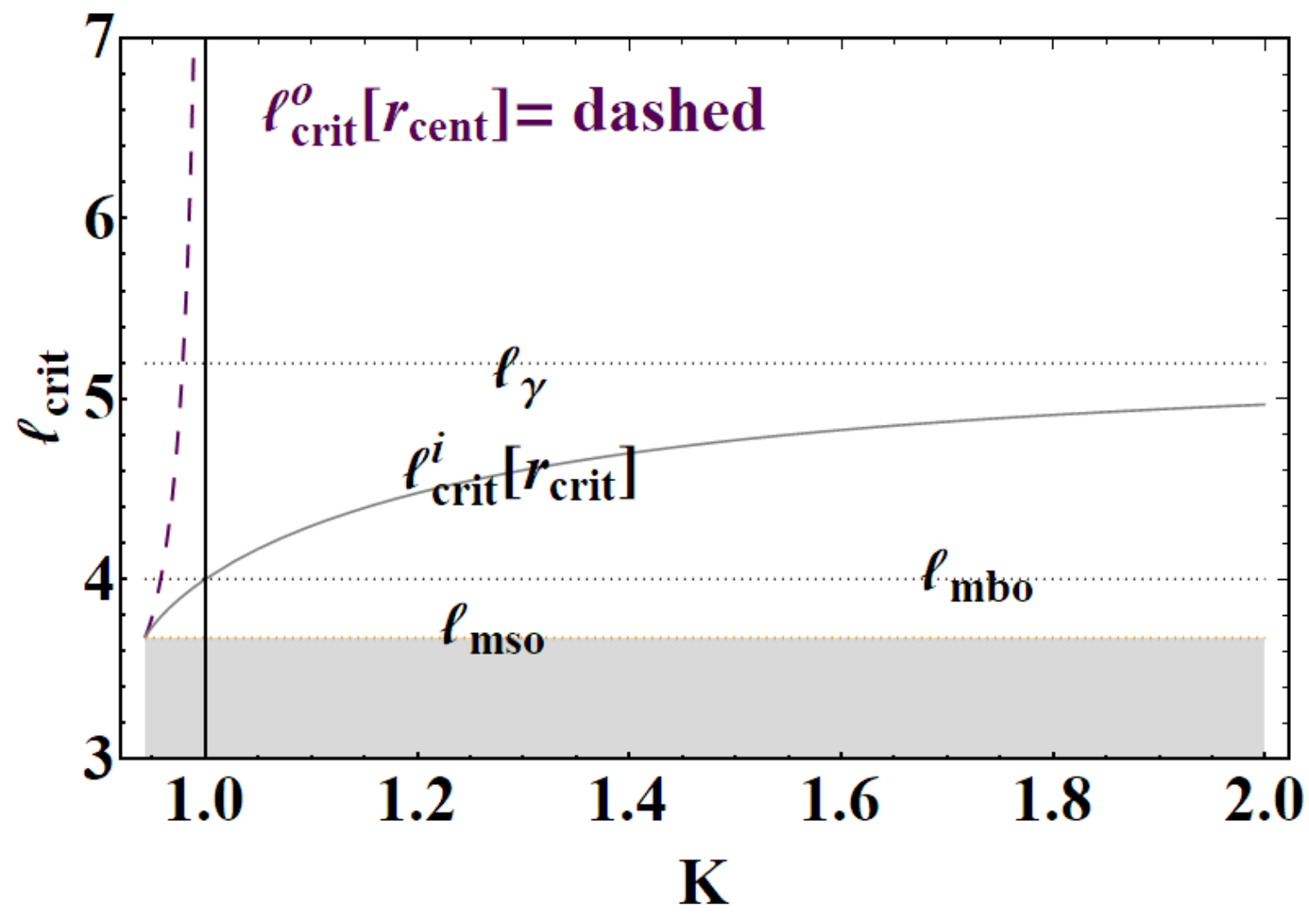}
 \includegraphics[width=8cm]{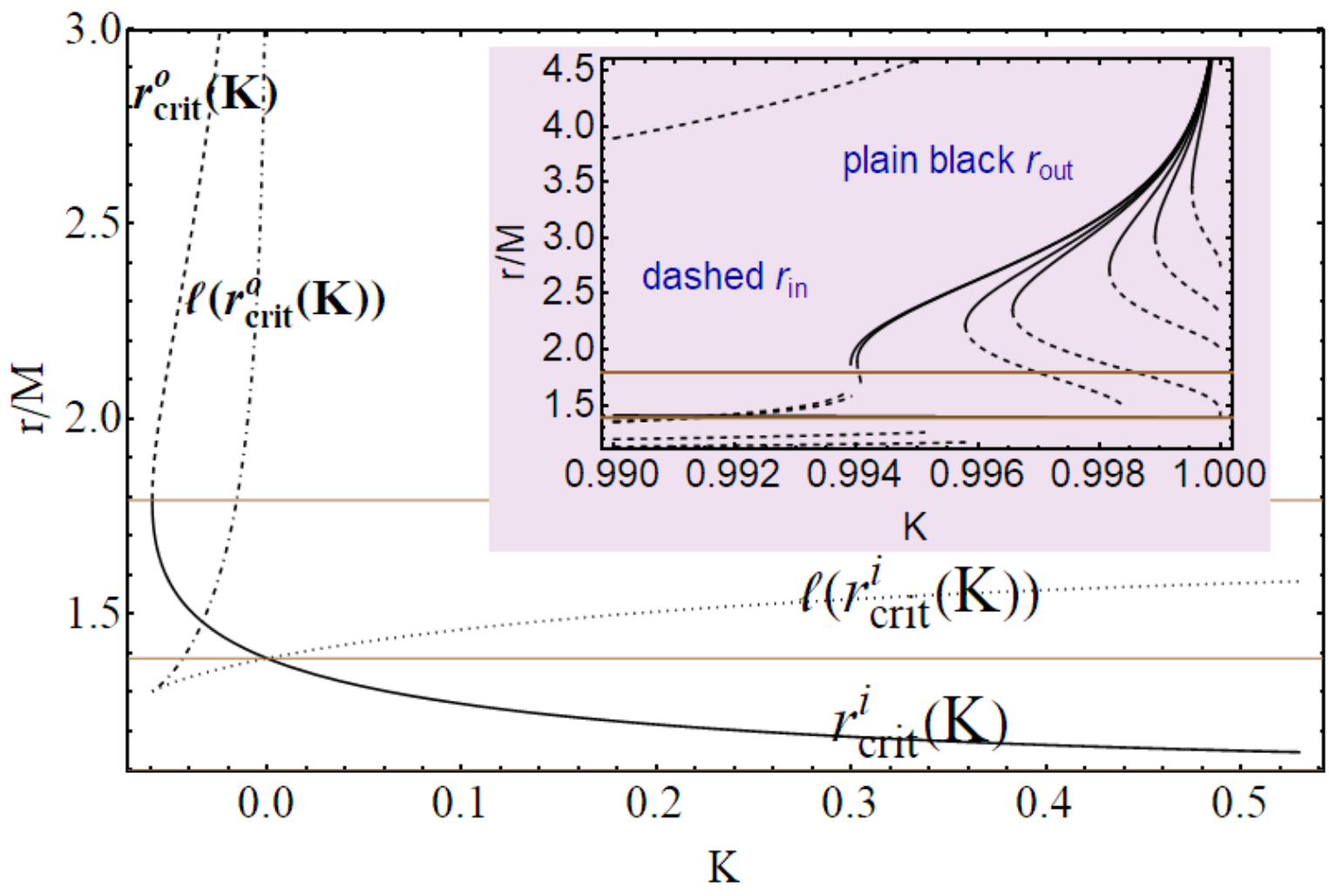}
  \caption{Left panel: Functions $\ell_{crit}^o(K)$ and $\ell_{crit}^i(K)$ of Eqs\il(\ref{Eq:crititLdeval})  as functions of $K$. Right panel:
$r_{crit}^o(K)$ and $r_{crit}^i(K)$  as functions of $K$ defined in Eq.\il(\ref{Eq:nicergerplto}), inside panel, curves $r_{inner}(K)$ and $r_{out}(K)$ in Eqs\il(\ref{Eq:mer-panto-ex-resul}) for different values of $\ell$.}\label{Fig:leaduniUR}
\end{figure}
 \item{\textbf{Tori critical radii   $r_{crit}(K)$   as a function of $K_{crit}$.}}

 Analogously, by using the relation $\ell(r)=\ell_{crit}^o(K)$ and
 $\ell(r)=\ell_{crit}^i(K)$, we find an expression for the critical radii   $r_{crit}(K)$ of the tori as a function of $K_{crit}$  in the form
 \bea\label{Eq:nicergerplto}
r_{crit}^o(K)\equiv -\frac{8}{K \left(\sqrt{9 K^2-8}+3 K\right)-4},\quad r_{crit}^i(K)\equiv\frac{8}{K \left(\sqrt{9 K^2-8}-3 K\right)+4},
 \eea
--Figs\il\ref{Fig:leaduniUR};
respectively,  there is $r_{crit}^i(K_{\times})=r_{inner}^{\times}$ (inner edge for accreting torus), and   $r_{crit}^o(K_{cent})=r_{cent}^{\times}$ (center of cusped configurations). We used  $K$ at the center of torus maximum density  providing therefore the  center of the correspondent cusped torus. (We note the other critical radius,  and respectively   momentum,   are related to the double corresponding toroidal configurations considered in the Sec.\il(\ref{Sec:doc-ready}), obtained as   solutions of the function $K(r)=$constant--see Fig\il(\ref{FIG:nicergerplto})).
\begin{figure}
  \includegraphics[width=7cm]{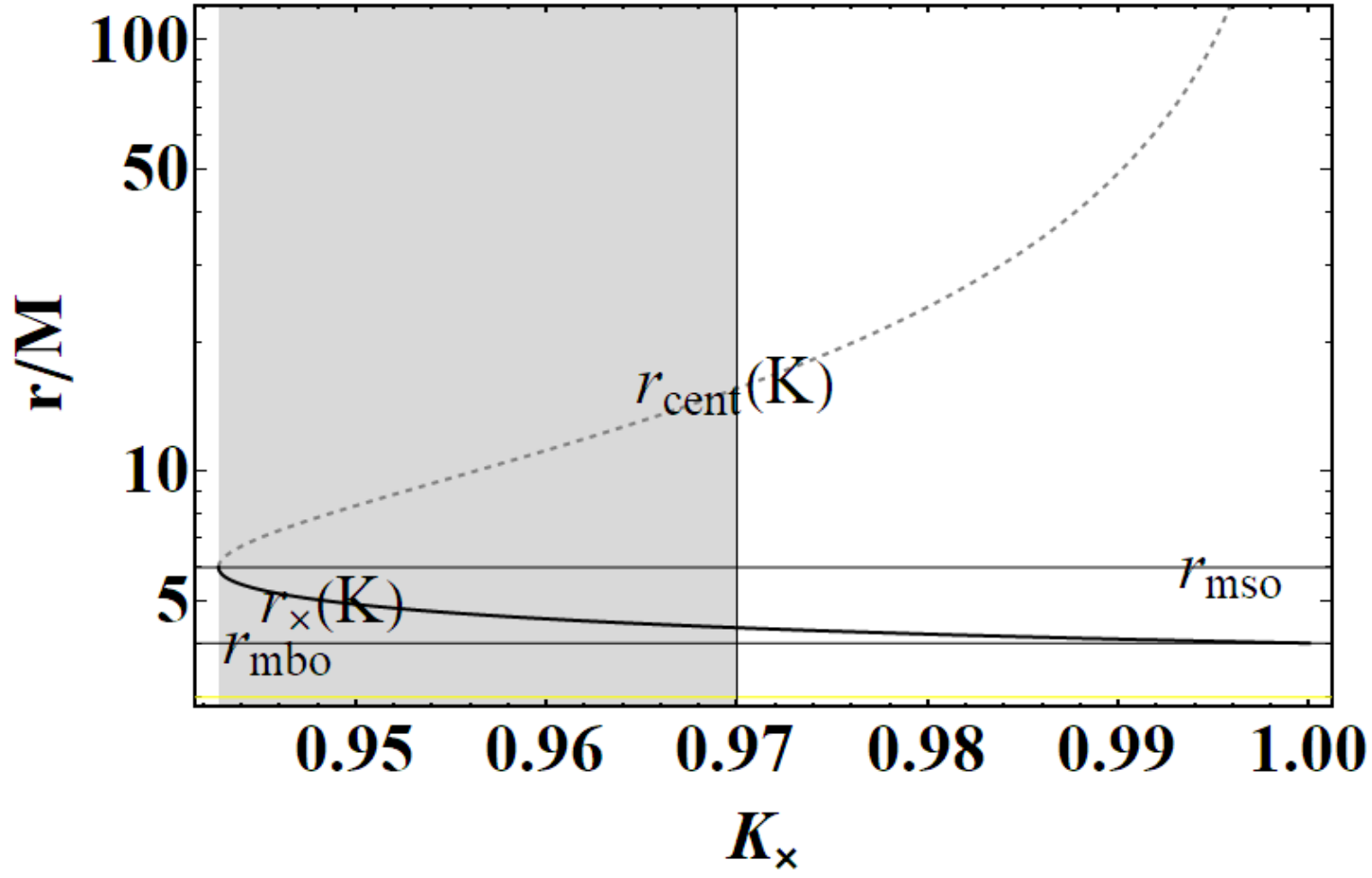}
  \caption{ $r_{cent}$ (maximum density point) and $r_{\times}$ critical points as functions of $K\in[K_{mso},1]$ defined in Eq.\il(\ref{Eq:nicergerplto}).}\label{FIG:nicergerplto}
\end{figure}
 Particularly for the cusped tori we expressed the inner edge,  $r_{\times}(\ell)$,   the outer edge,
 $r_{out}^{\times}(\ell)$, the center, $r_{cent}^{\times}(\ell)$, the elongation, $\lambda_{\times}(\ell)$, the thickness, $\Sa_{\times}\equiv 2h^{\times}_{max}/\lambda_{\times}$,
   the torus maximum  $h^{\times}_{max}$  and location of geometrical maximum,  {$x^{\times}_{max}=r^{o}_{max}$}, as function on the only free parameter $\ell$ in Fig.\il\ref{Fig:lom-b-emi}.
   \item
\textbf{{Outer edge of accreting torus  as a function of the specific angular momentum $\ell$.}}

To complete our analysis we introduce the following  explicit form for the outer margin of the cusped disk as a function of the specific angular momentum:
\bea&&\label{Eq:dan-aga-mich}
r_{out}^{\times}\equiv\frac{2 \ell^2 \hat{\mathbf{\psi}}_2}{3 \ell^2 \hat{\mathbf{\psi}}_2-\hat{\mathbf{\psi}}_0^3+6 \hat{\mathbf{\psi}}_0^2}+\frac{2 \hat{\mathbf{\psi}}_4 \sqrt{\frac{\ell^2 \left(-3 \ell^2 \hat{\mathbf{\psi}}_0^3 \hat{\mathbf{\psi}}_2+18 \ell^2 \hat{\mathbf{\psi}}_0^2 \hat{\mathbf{\psi}}_2+12 \ell^2 \hat{\mathbf{\psi}}_2^2+\hat{\mathbf{\psi}}_0^6-12 \hat{\mathbf{\psi}}_0^5+36 \hat{\mathbf{\psi}}_0^4\right)}{\left(-3 \ell^2 \hat{\mathbf{\psi}}_2+\hat{\mathbf{\psi}}_0^3-6 \hat{\mathbf{\psi}}_0^2\right)^2}}}{\sqrt{3}}
\eea
where
\bea\nonumber
&&
\hat{\mathbf{\psi}}_0\equiv\ell^2-2 \sqrt{\ell^2 \left( \ell^2-12\right)},\quad\psi\equiv\cos \left[\frac{1}{3} \left(\cos ^{-1}\left[\frac{\ell^2 \left(\ell^4-18 \ell^2+54\right)}{\left(\ell^2 \left[\ell^2-12\right]\right)^{3/2}}\right]+\pi \right)\right],
 \\\nonumber
 &&\hat{\mathbf{\psi}}_2\equiv-2 \left(2 \ell^2-15\right) \sqrt{\ell^2 \left(\ell^2-12\right)} y-2 \left(\ell^2-12\right) \ell^2 \hat{\mathbf{\psi}}_1+3 \left(\ell^4-13 \ell^2+18\right),
 \\\nonumber
&&\hat{\mathbf{\psi}}_1
\equiv\sin \left[\frac{1}{6} \left(4 \cos ^{-1}\left[\frac{\ell^2 \left(\ell^4-18 \ell^2+54\right)}{\left[\ell^2 \left(\ell^2-12\right)\right]^{3/2}}\right]+\pi \right)\right],
\\\nonumber
&&
\hat{\mathbf{\psi}}_3\equiv\left(1-\frac{(\hat{\mathbf{\psi}}_0-6) \hat{\mathbf{\psi}}_0^2}{3 \ell^2 \hat{\mathbf{\psi}}_2}\right) \sqrt{\frac{\frac{(\hat{\mathbf{\psi}}_0-6) \hat{\mathbf{\psi}}_0^2 \left(-3 \ell^2 \hat{\mathbf{\psi}}_2+\hat{\mathbf{\psi}}_0^3-6 \hat{\mathbf{\psi}}_0^2\right)}{3 \ell^2 \hat{\mathbf{\psi}}_2^2}+4}{\left(1-\frac{(\hat{\mathbf{\psi}}_0-6) \hat{\mathbf{\psi}}_0^2}{3 \ell^2 \hat{\mathbf{\psi}}_2}\right)^2}},
\\\nonumber
&&\hat{\mathbf{\psi}}_4\equiv\cos \left[\frac{1}{3} \cos ^{-1}\left(\frac{\frac{(\hat{\mathbf{\psi}}_0-6) \hat{\mathbf{\psi}}_0^2 \left[-2 \ell^2 \hat{\mathbf{\psi}}_2+\hat{\mathbf{\psi}}_0^3-6 \hat{\mathbf{\psi}}_0^2\right] \left(-3 \ell^2 \hat{\mathbf{\psi}}_2+\hat{\mathbf{\psi}}_0^3-6 \hat{\mathbf{\psi}}_0^2\right)}{\ell^4 \hat{\mathbf{\psi}}_2^3}+8}{\hat{\mathbf{\psi}}_3^3}\right)\right]
\eea
represented by Figs\il\ref{Fig:lom-b-emi},\ref{Fig:tookP},\ref{Fig:SMGerm}.
\end{enumerate}

\subsection{Limiting surfaces in the \textbf{RAD}}\label{Sec:limiting}
{Even in the simplest case of static background, the \textbf{RAD}  toroidal components are characterized by  boundary conditions depending  on  the distance, in the clusters, from the central attractor. In Sec.\il(\ref{Sec:sfer.J}) we consider more closely the geometric thickness of the misaligned tori, firstly  providing  definition of torus  thickness  and  then  exploring  the conditions   for  \textbf{RAD} globule host  differently   thick toroids.  Nevertheless, the \textbf{RAD}  tori are characterized by momentum distribution    $\ell(r)$ of Eq.\il(\ref{Eq:lqkp}), where this function represents  the upper boundary for the  so called spherical accretion.
These conditions are very significant  for the single component as a predominant feature that regulates many thermodynamic and oscillation  properties of the disk.
From the point of view of \textbf{RAD}, this analysis  predicts the  observational characteristics of the different components in which we can disentangle the globuli, as thickness and the distance from the central attractor.
We conclude in Sec.\il(\ref{Sec:ount}) with some  considerations on the outer toroidal surfaces of  the agglomerate
and tori collision conditions.}
\subsubsection{Asymptotic limits and conditions on the quasi-sphericity of the torus}\label{Sec:sfer.J}
 Disk geometric thickness underlies different aspects of the physics of the accretion disks, including  the disks oscillation  modes and  accretion rates.
Here we provide an evaluation of the geometric thickness of the misaligned tori considered here in the \textbf{RAD} frame, our goal is to establish  conditions under  which disks are geometrically thick considering the parameters of the model and the limit value  $\Sa=1$. Moreover we give the  tori distribution in the  \textbf{RAD} considering the characteristic of the geometric thickness.
For cusped tori, where there is  $\ell\in]\ell_{mso},\ell_{mbo}[$, the torus elongation $\lambda$ on its equatorial plane can reach very large values as $\ell \lessapprox\ell_{mbo}$ (simultaneously the center approaches  $r_{mbo}^b$), as clear from the  representation of the torus elongation in the Figs\il\ref{Fig:tookP}-gray region.
The torus thickness is
$\Sa=1$ for  $K=
0.975$ and $\ell=
3.887\in]\ell_{mso},\ell_{mbo}[$. In Figs\il\ref{Fig:torithick} and \ref{Fig:SMGermMs} we represented the classes of tori with equal geometrical thickness $\Sa$, in dependence of the parameters $\ell$ and $K$, splitting in the two large classes of geometrically thin $\Sa<1$ and geometrically thick $\Sa>1$ \textbf{RAD} tori, for quiescent and cusped configurations. We note that models $\mathbf{A}^{\pm}$ and $\mathbf{B}^{\pm}$, characterized by  $K=$constant, but different fluids specific angular momentum $\ell$ refer to analysis of equal $K$ configurations as in Eq.\il(\ref{Eq:Krrp}). From which it is clear that generally toroids are thicker as  larger is  $K$  and with large fluid specific angular momentum $\ell$ (generally associated with toroids located far away from the central \textbf{BH}).
Models $(\mathbf{A}^{+}, \mathbf{A}^-)$ and $(\mathbf{B}^{+}, \mathbf{B}^-)$ correspond to tori with equal geometrical thickness and $K$ parameters. Figure\il(\ref{Fig:torithick}) show then also the relative role of the fluid specific momentum $\ell$ and $K$ in the tori.
Configuration $\mathbf{D}$ is therefore a limiting toroid with minimum $K$ and $\ell$ possible for $\Sa_\mathbf{D}=1$.
It is clear that for  accreting configurations the  farthest  is the accreting torus  from the central attractor and the largest it is. It is also evident particularly from the Figs\il\ref{Fig:torithick} that for low magnitude of the specific angular momentum the thickness is essentially determined by $K$ parameter (i.e. fluid density) particularly for the accreting tori, viceversa the main governing parameter for quiescent tori at $\ell>\ell_{mbo}$ is the specific angular momentum.
\begin{figure}
   \includegraphics[width=7cm]{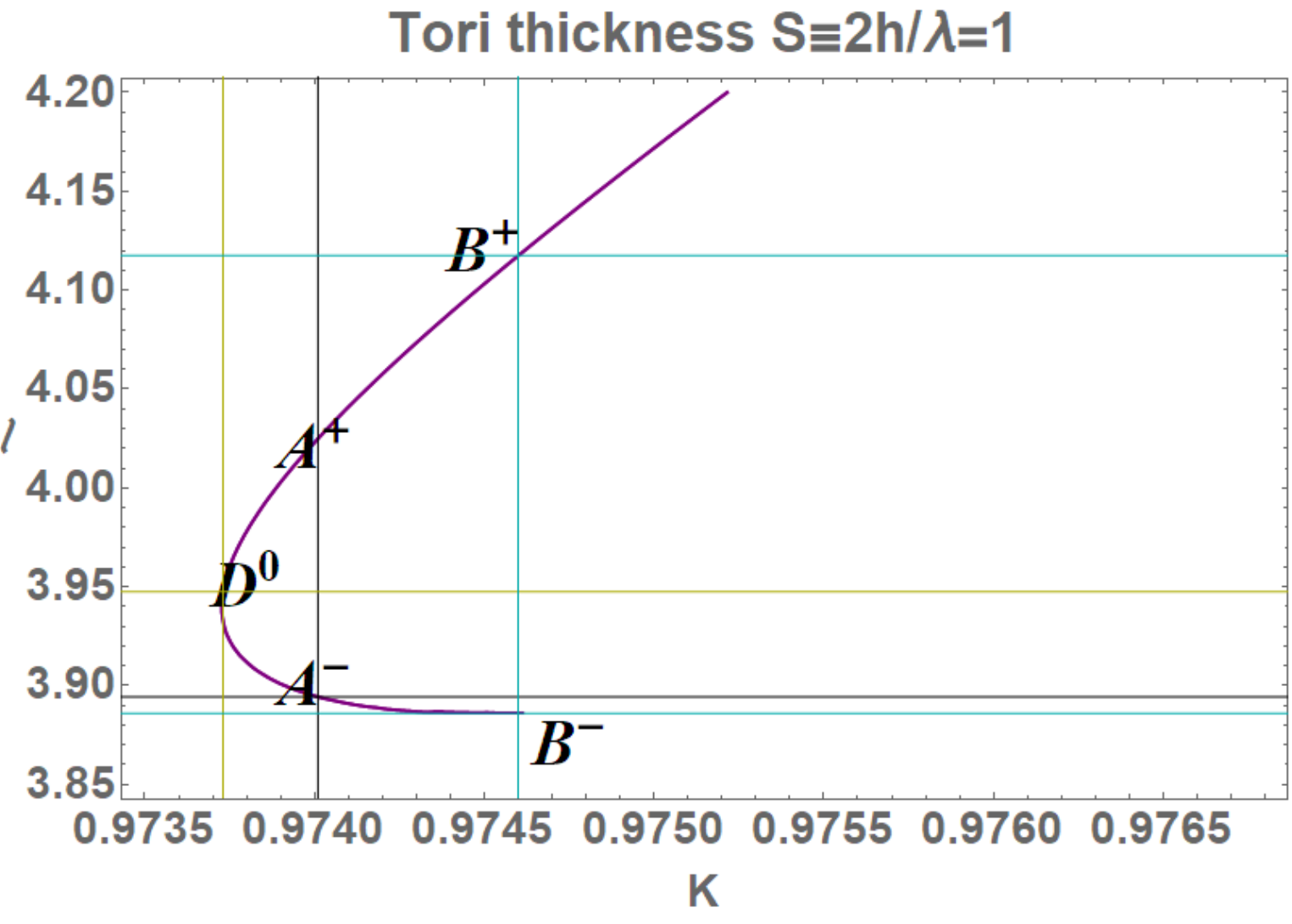}
  \includegraphics[width=7cm]{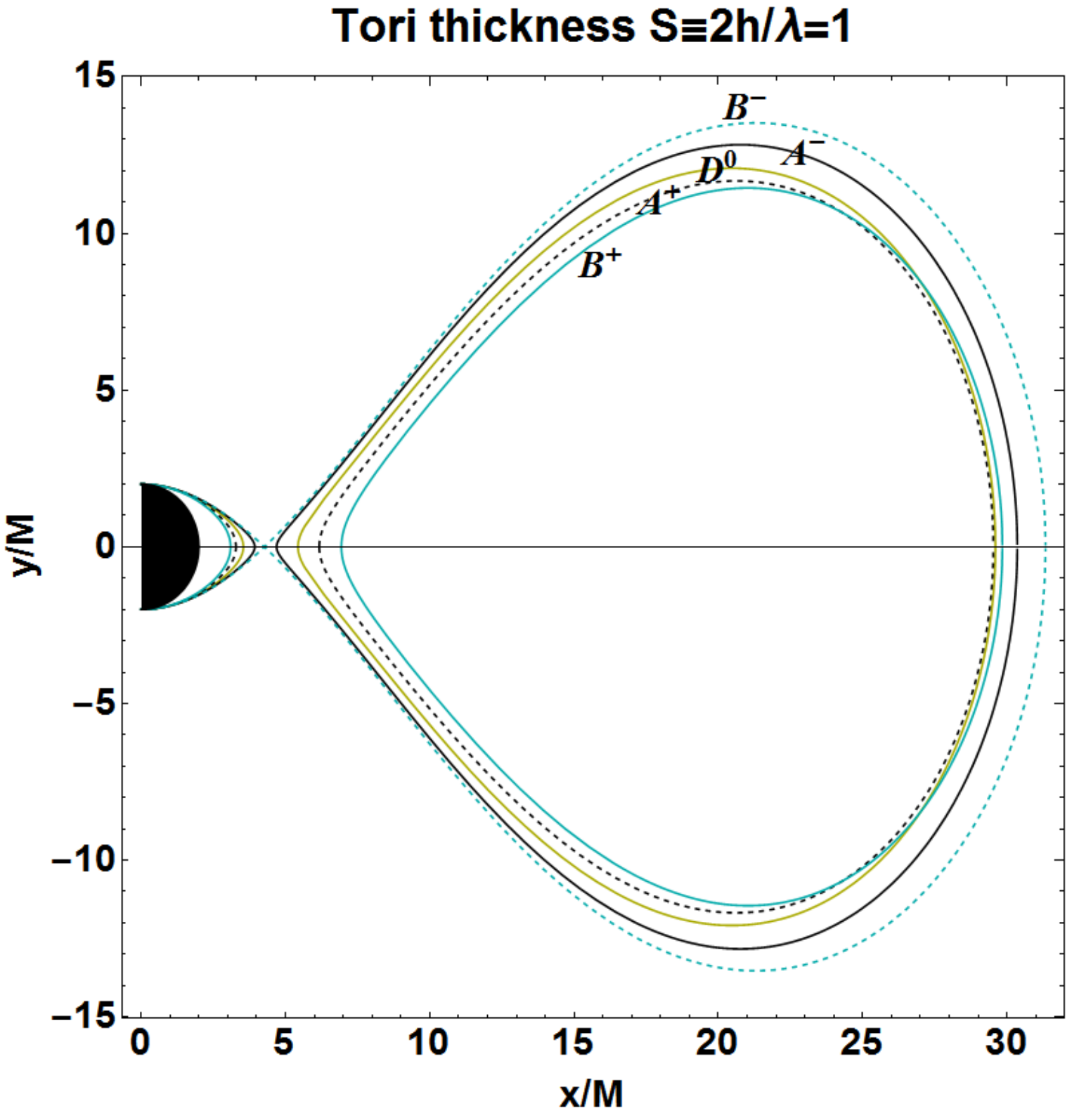}
  \includegraphics[width=7cm]{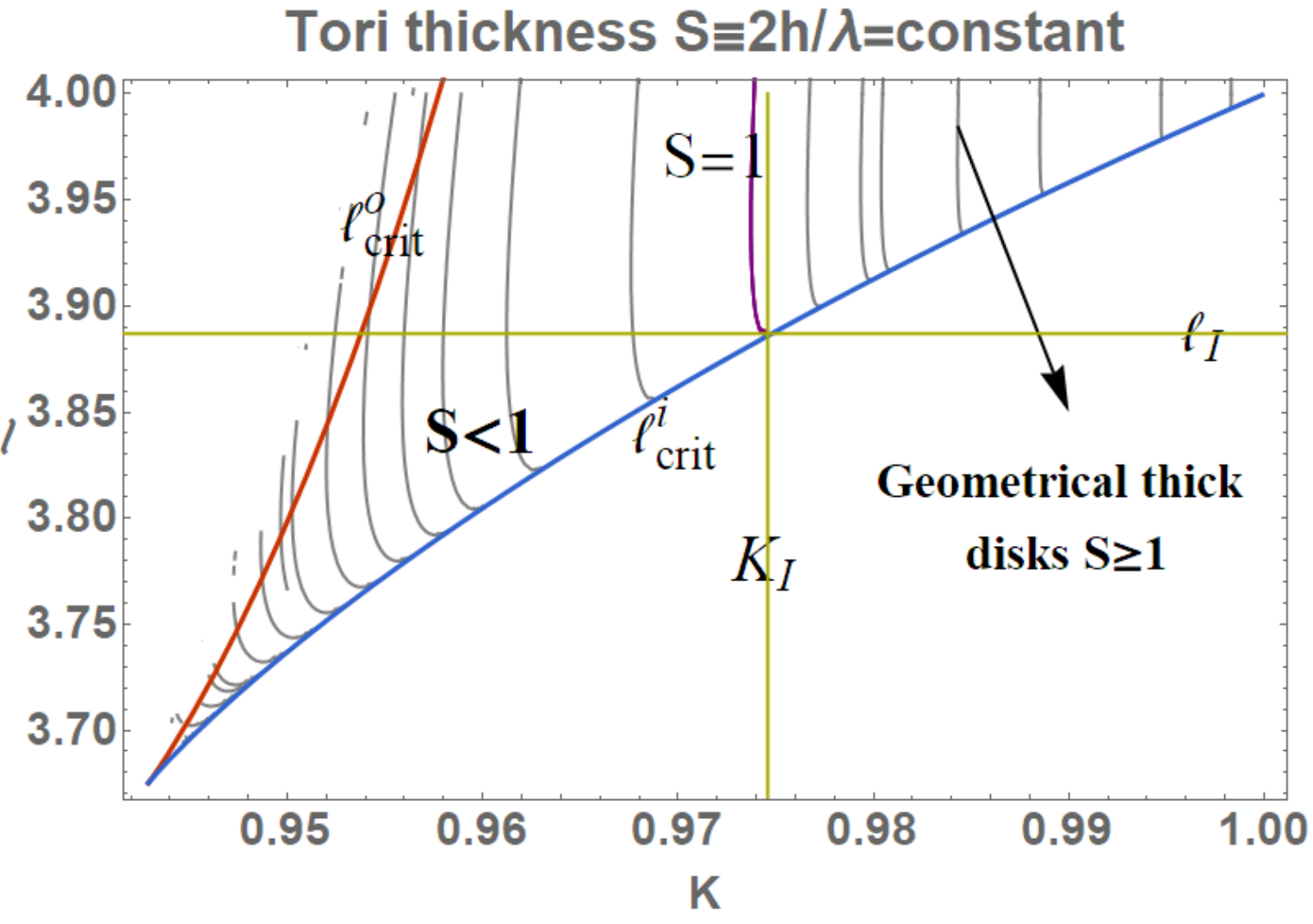}
  \includegraphics[width=7cm]{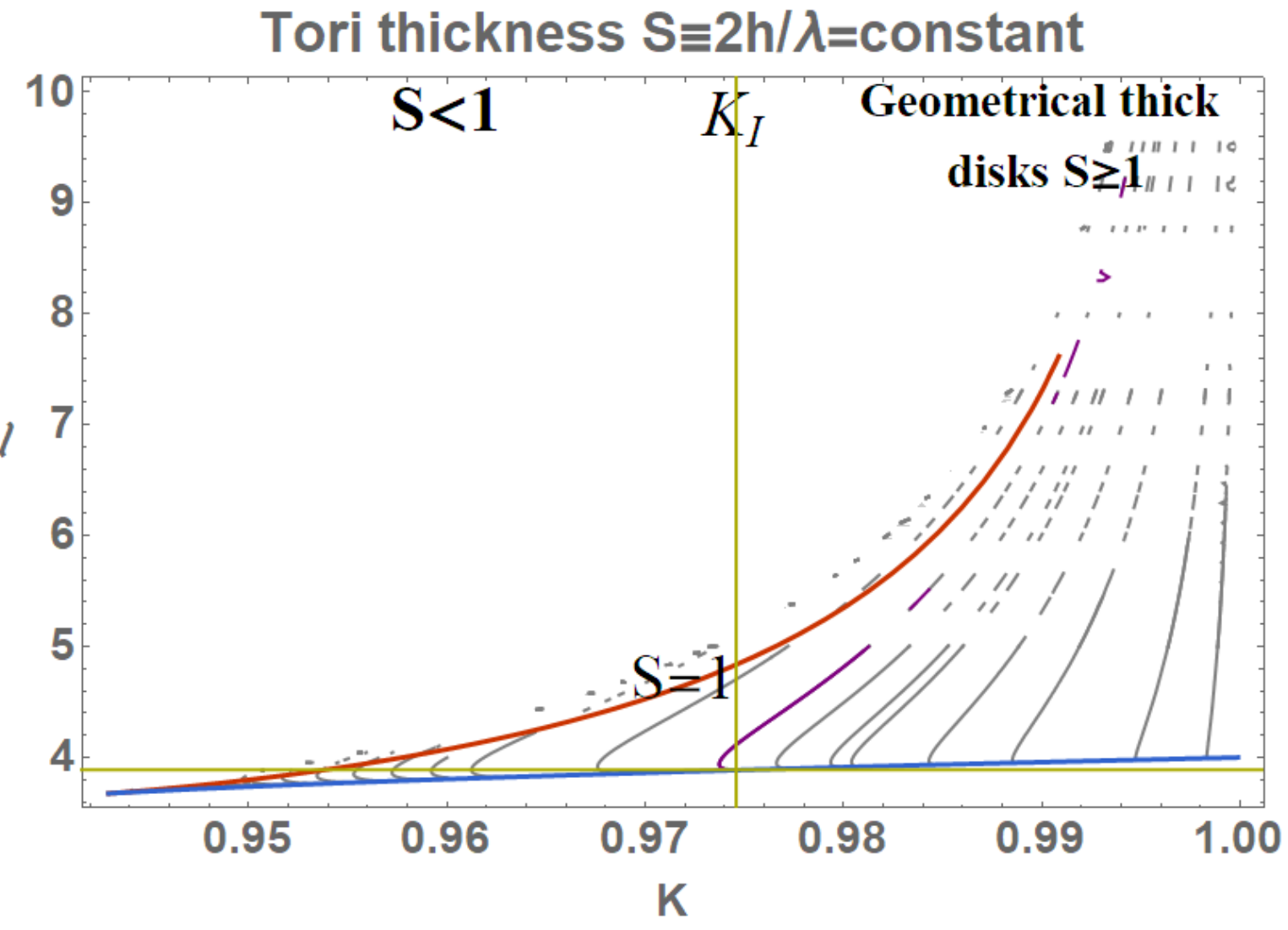}
  \caption{Upper left panel: Classes of tori  with thickness  $\Sa=1$, in the plane of parameters $\ell$ and $K$. ($\mathbf{A^{\pm}, B^{\pm}, D^{0}}$) are special tori models whose cross sections on their equatorial planes (which can be also different) are represented in right panel. Bottom panel: classes of tori with equal thickness $\Sa=$constant   in the plane of parameters $\ell$ and $K$. Limiting value $\Sa=1$ is shown. Left panel is the range $\ell\in[\ell_{mso},\ell_{mbo}]$, right panel explores the range $\ell>\ell_{mbo}$. $\ell_I$ and $K_{I}$ define the $\Sa=1$ case. See also Figs\il\ref{Fig:SMGermMs}. }\label{Fig:torithick}
\end{figure}
{Figs\il\ref{Fig:torithick} and  Figs\il\ref{Fig:SMGermMs}  show an example of  tori  having  equal thickness, or equal parameter $\mathbf{P}$ but different thickness.}
\begin{figure}
  \includegraphics[width=4cm,angle=90]{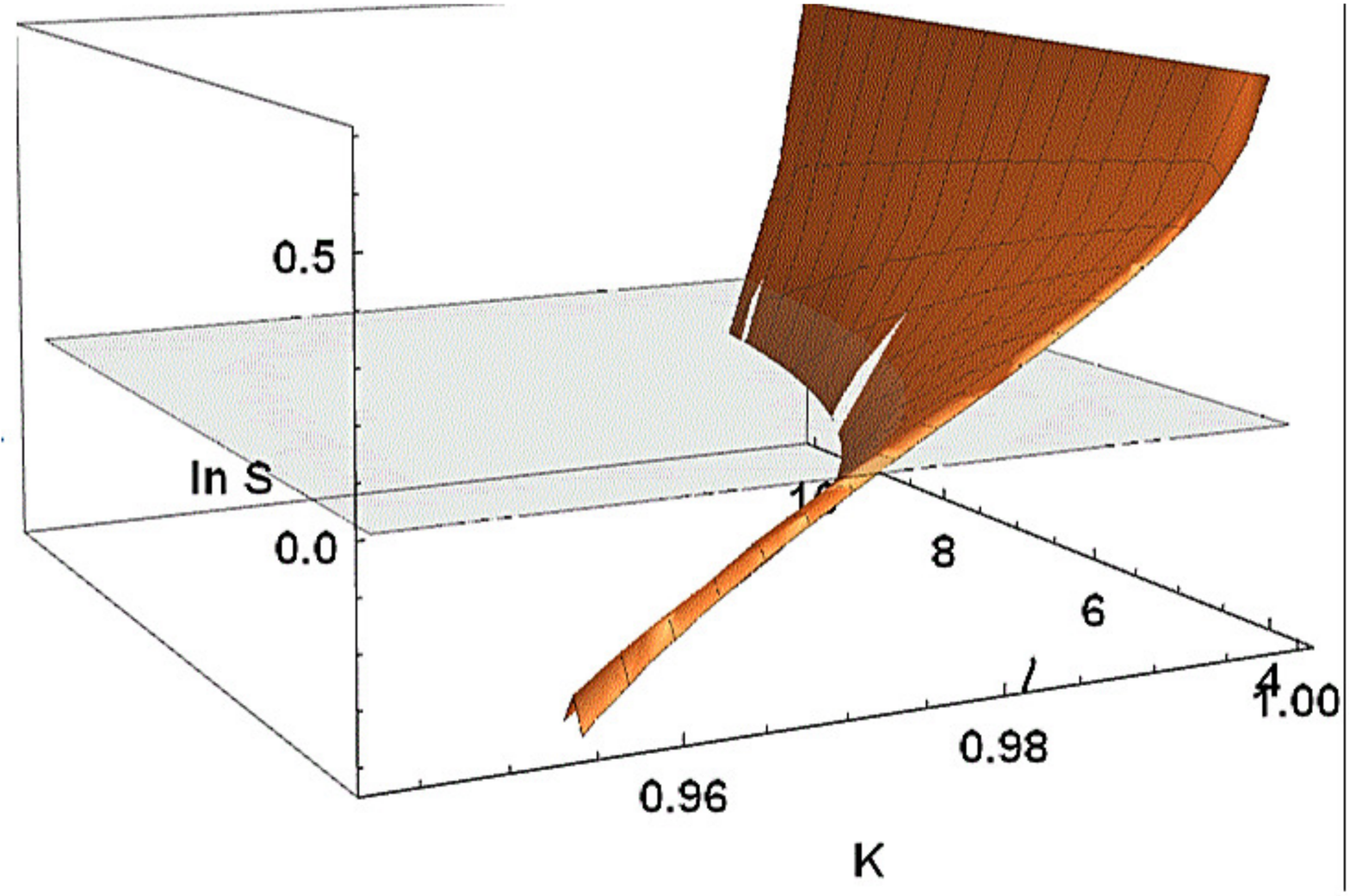}
   \includegraphics[width=6cm]{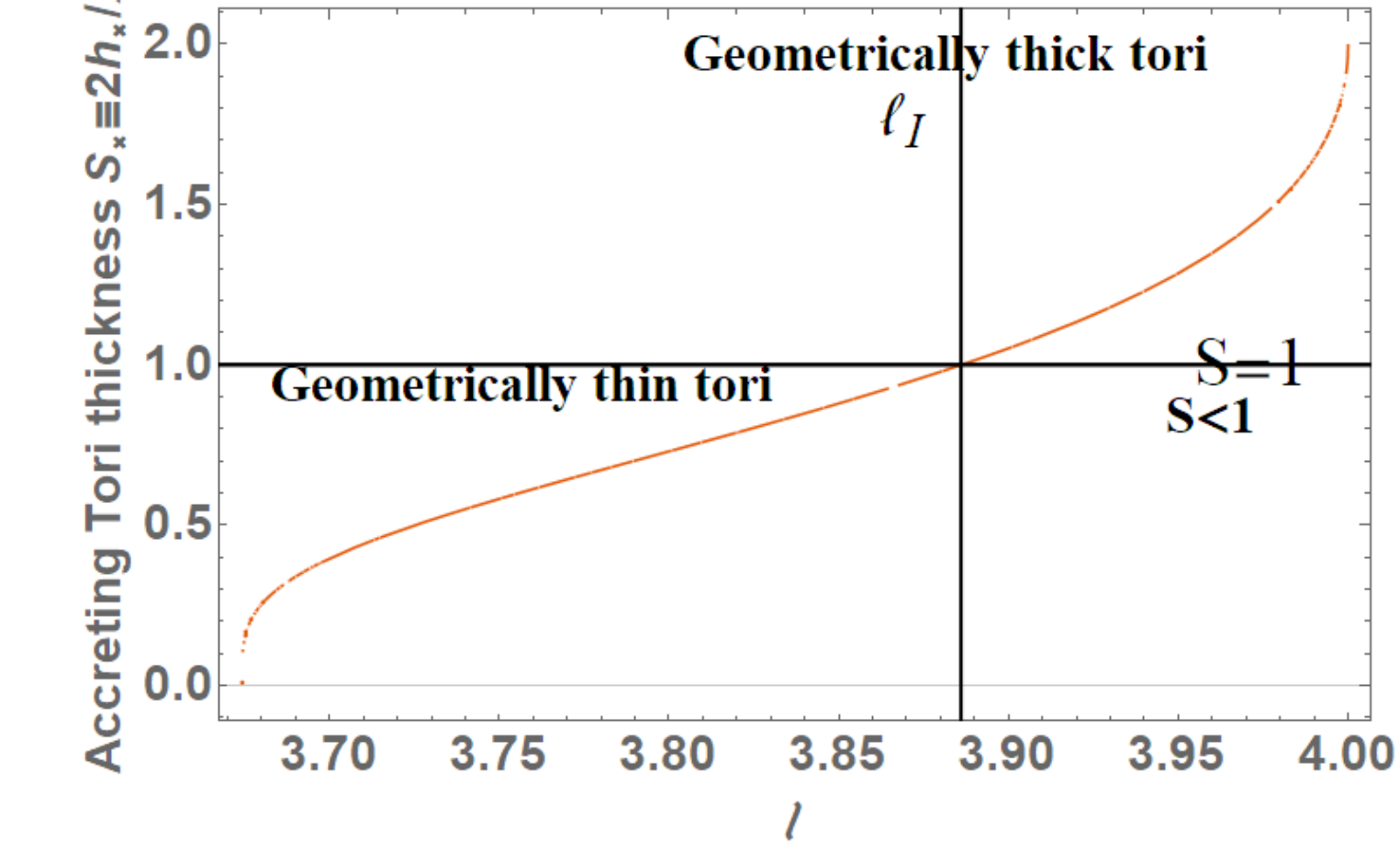}
  \includegraphics[width=6cm]{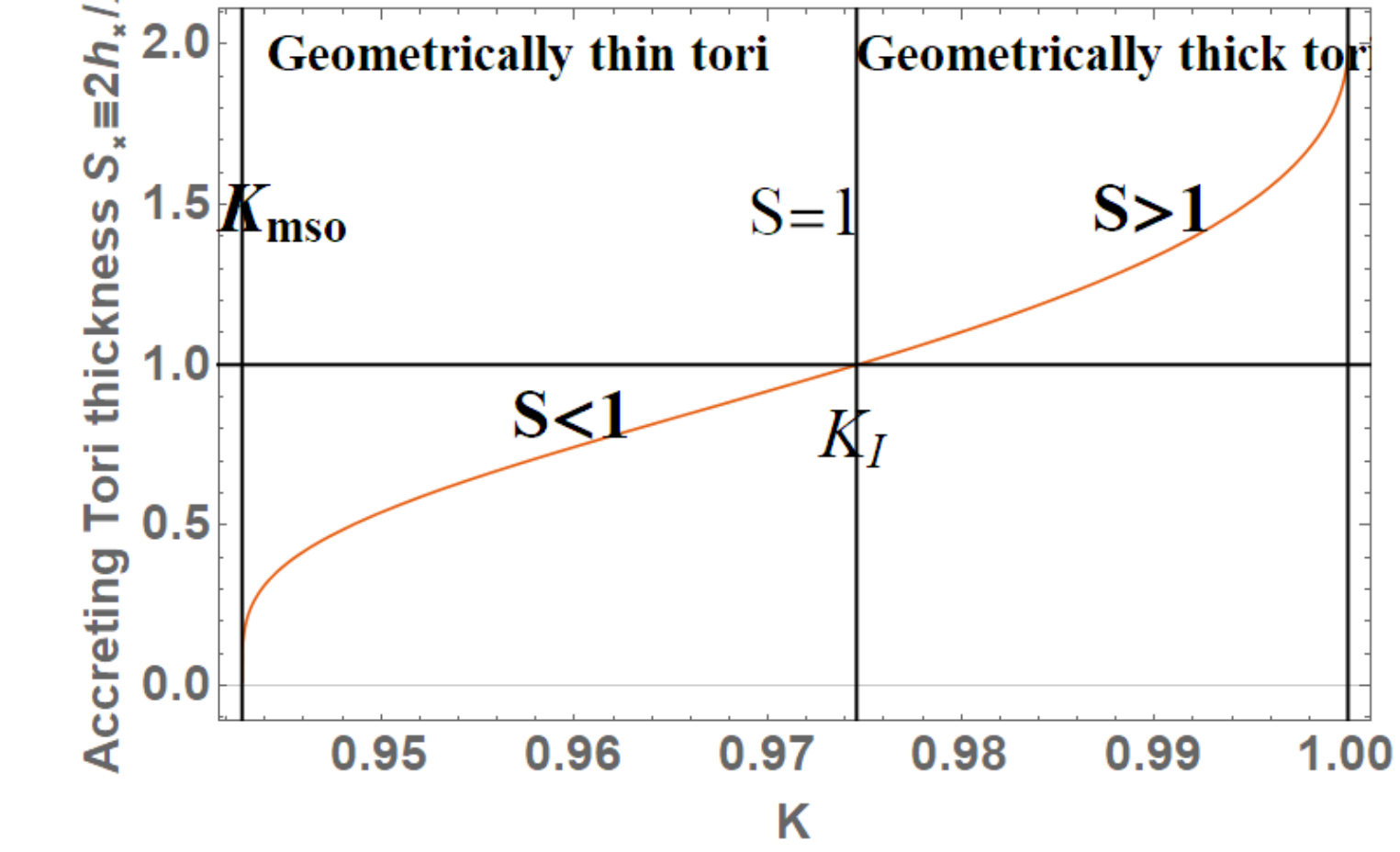}
  \caption{Left panel: $3D$ plot of the tori thickness $\Sa$ as function of tori parameters $\ell$ (fluid specific angular mometum) and $K$. The plane correspondent to the limiting value $\Sa=1$ is shown. Function define classes of tori with equal geometrical thickness. Center and left panels: thickness of cusped (accreting) tori as function of fluid specific angular momentum $\ell\in[\ell_{mso},\ell_{mbo}]$ (center panel) and $K\in[K_{mso},K_{mbo}]$ parameter (right panel). See also Figs\il\ref{Fig:torithick}. }\label{Fig:SMGermMs}
\end{figure}
Analysis in Figs\il(\ref{Fig:torithick}) shows that thickness $\Sa$ is essentially regulated by the $K$ parameters
 and therefore the energy function $K(r)$. This analysis also sets the condition for the formation of \textbf{RAD} as a globulus, shows moreover classes of tori at equal of similar thickness therefore having similar properties which are essentially determined by this parameter as some aspects of the oscillations, it is clear also the role of critical limiting
curves $\ell_{crit}(K)$.
Condition
$r_{cent}(\ell)<x_{\max}(\ell,K)$, showed in Figs\il\ref{Fig:lom-b-emi} is a peculiar aspect of thick disks
the limiting value $r_{cent}=x_{\max}(\ell)$ (for accreting tori) is for $\ell\approx
3.67423$, while the most general relation is shown in  Figs\il\ref{Fig:lom-b-emi}
from which we see that in general for large  $\ell$ and small  $K$ there is $r_{cent}(\ell)=x_{\max}(\ell,K)$,
 Figs\il\ref{Fig:lom-b-emi} also show that the only a limited portion of discs in the conditions to be cusped satisfy the  condition $r_{cent}=x_{\max}$ in general for small values of  $K$.

\subsubsection{Formation of the outer torus  in the  marginally collision sphere: $r_{out}^i\lessapprox r_{inner}^o$}
\label{Sec:ount}

The conditions for the emergence of the tori collision  is important to  establish the stability conditions of the \textbf{RAD}  including misaligned tori and also possibly the time scale of the involved processes that are potentially observable\footnote{
The analysis of collisional spheres is also relevant for setting constraints for the observation and recognition of a black hole in the conditions considered here and also clarifies the perspective  of the accretion ball, maximum  \textbf{BH} coverage i.e. that condition for which \textbf{BH} is in suitable conditions and a stage in its life surrounded by accreting tori considered here. This limiting condition fixes  in comparison with other embedded \textbf{BH} models  as, for example, self gravitating shells.
  }
We consider now the condition $r_{inner}^o\geq r_{out}^{\times}$,  featuring the existence of  an inner torus in accretion and a quiescent outer torus   on a  collision sphere (a sphere centered in the attractor and located at radius $r_{inner}^o$).
To this purpose we introduce the useful limiting radii
\bea\label{Eq:polaplot}
r_K^{in}\equiv r_K^-\quad\mbox{and}\quad r_K^{out}\equiv r_K^+,\quad\mbox{where}\quad
r_K^{\pm}\equiv\frac{1}{4} \left(\ell^2\pm\ell \sqrt{\ell^2-16}\right),
\eea
representing  the conditions $K=1$ with the potential and therefore set limits for location of the inner and outer edges of quiescent tori with $\ell>\ell_{mbo}=4$. Figure\il(\ref{Fig:tookP}) shows the limit for the existence of these specific radii.
\begin{figure}
  \includegraphics[width=10cm]{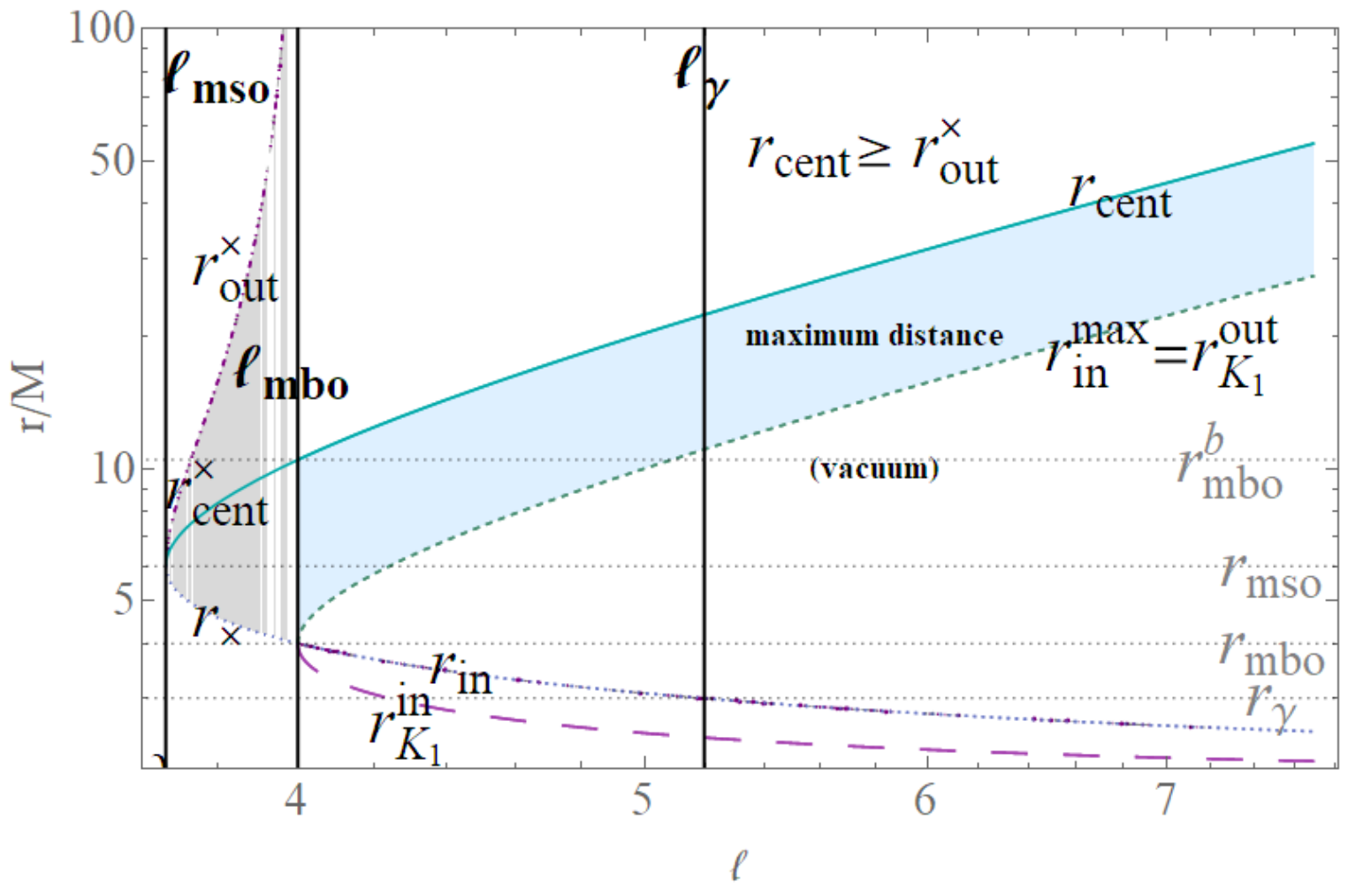}
  \includegraphics[width=8cm]{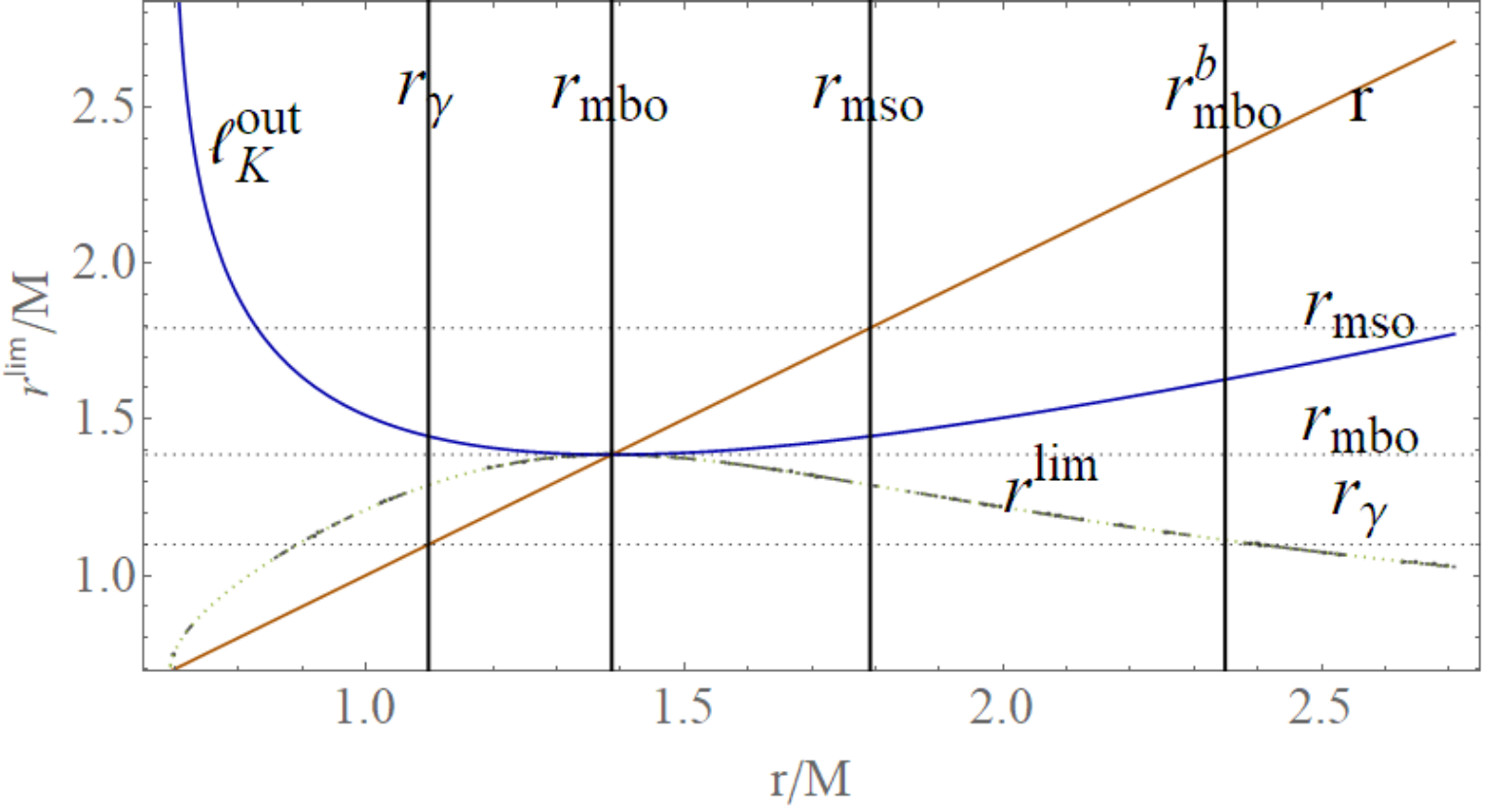}
  \includegraphics[width=8cm]{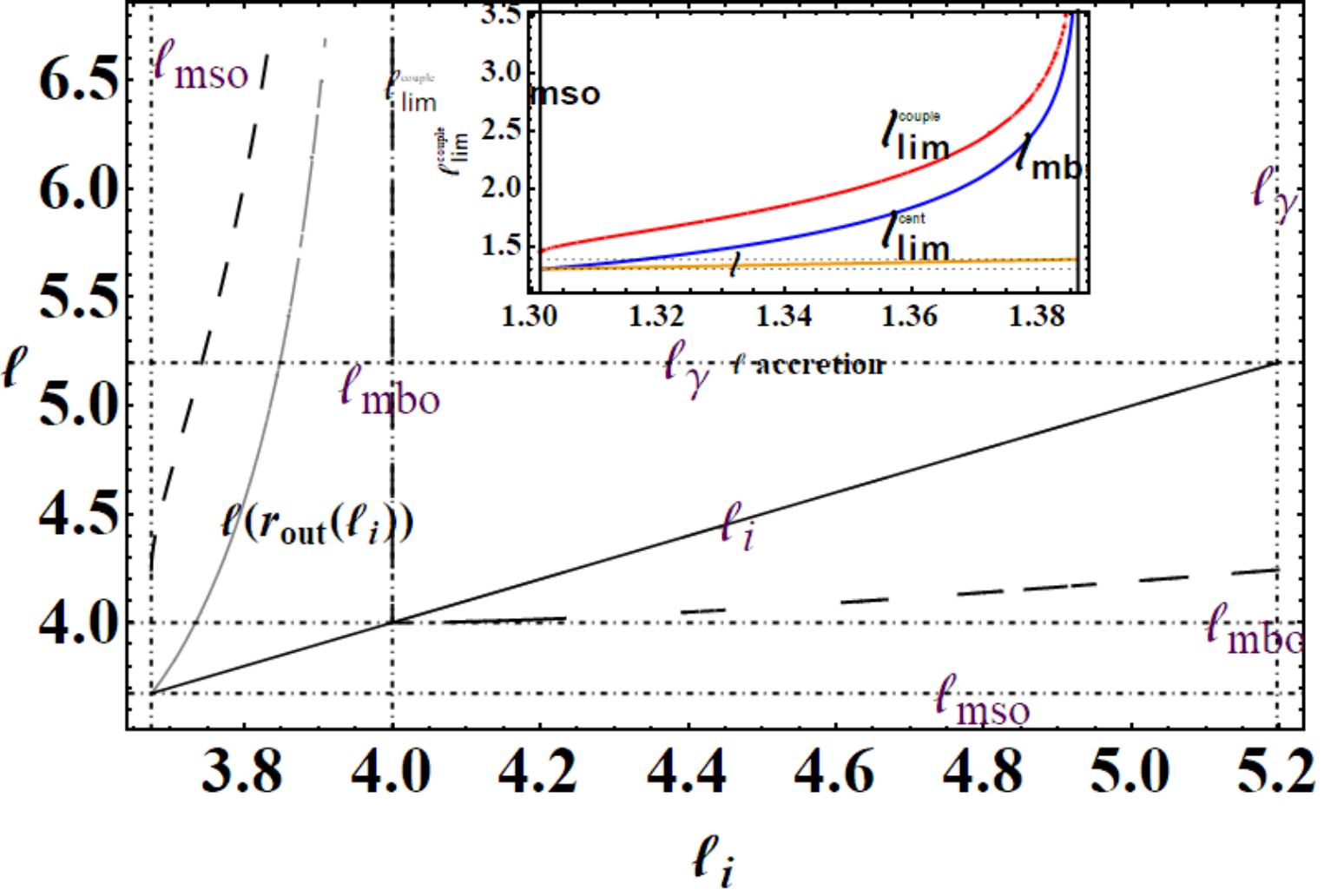}
  \caption{Existence of  \textbf{RAD} tori with inner torus in accretion  (extension is the gray region where inner and outer edge and center of accreting tori is represented.) Blue region represents the maximum extension of the inner edge of the quiescent torus at $\ell>4$, where $r_K^{in}$ and $r_K^{out}$ are in Eq.\il(\ref{Eq:polaplot}). There can be a vacuum region between the inner  and the outer torus. Quantities $r^{\lim}\in\{r_{inner}^{\lim},r_{out}^{\lim}\}$ are introduced in Eq.\il(\ref{Eq:this-leve}) and in  $\ell^{out}$ Eq.\il(\ref{Eq:vast-maj}).}\label{Fig:tookP}
\end{figure}
Figs\il(\ref{Fig:tookP}) show the analysis if tori spacing $\bar{\lambda}\equiv r_{inner}^o-r_{out}^{i}$ for a couple of tori, where $r_{inner}^o$ is the inner edge of the outer torus of the couple and  $r_{out}^{i}$ is the outer edge of the inner torus of the couple with respect to the central \textbf{BH}, therefore this is substantially an analysis of the emergence of tori collision according to the state, cusped or quiescent tori and others morphological features.
We can precise this relation considering the limits on the fluids angular momentum.
From Eq.\il(\ref{Eq:polaplot}) we obtain the limiting specific angular momenta
\bea\label{Eq:vast-maj}
\ell^{out}\equiv\frac{\sqrt{2}r_{out}}{\sqrt{r_{out}-2}},\quad \ell_{\lim}^{\texttt{\textbf{couple}}}\equiv\ell^{out}(r_{out}=r_{out}(\ell))=\frac{2 (\hat{\mathbf{\psi}}_3 \hat{\mathbf{\psi}}_4+1)}{3 \sqrt{\frac{r_{out}^{\times}}{2}-1} \left(1-\frac{(\hat{\mathbf{\psi}}_0-6) \hat{\mathbf{\psi}}_0^2}{3 \ell^2 \hat{\mathbf{\psi}}_2}\right)},
\eea
where $\ell_{\lim}^{\texttt{\textbf{couple}}}$  has been obtained  using Eq.\il(\ref{Eq:dan-aga-mich}) in
$\ell^{out}$, represented in  Figs\il\ref{Fig:tookP} and \ref{Fig:lom-b-emi}.
The function $\ell_{\lim}^{\texttt{\textbf{couple}}}(\ell)$ provides the specific angular momentum of the outer torus  to have $r_{inner}(\ell_{\lim}^{\texttt{\textbf{couple}}})=r_{out}^{\times}(\ell)$,   where
$r_{out}^{\times}(\ell)$ is the outer edge of the cusped torus (therefore  $\ell\equiv \ell_{\times}\in]\ell_{mso},\ell_{mbo}[$), defined in Eq.\il(\ref{Eq:dan-aga-mich}) and pictured in Figs\il\ref{Fig:tookP} and \ref{Fig:lom-b-emi}.
  The inner edge $r_{inner}(\ell_{\lim}^{\texttt{\textbf{couple}}})=r_K^+$ is the limiting value for the inner edge location at $K^{out}=1$, which is  the $K$-parameter of the outer torus being the superior limiting value of Eq.\il(\ref{Eq:vast-maj})    and showed in  Figs\il\ref{Fig:tookP} and \ref{Fig:lom-b-emi}, relating  the angular momentum of the inner and outer torus.
Note that condition   $K\lessapprox1$  implies that there is only one value $\ell$ for the outer torus with  inner  edge equal to the  outer edge of the inner accreting torus.
In Figs\il\ref{Fig:lom-b-emi} $r^{\lim}(r)=r_{out}^{\times}(\ell^{out})=r_{inner}^{\times}(\ell^{out})$, or at the limiting conditions on the  outer and inner  edge of the cusped torus  $r_{inner}^{\times}$  and $r_{out}^{\times}$ for  $\ell=\ell^{out}$. $\ell_{\lim}^{couple}$ is $\ell^{out}(r_{out}^{\times})$ or the limiting value of the specific angular momentum $\ell_{\lim}^{couple}$ (for  $K=1$) at the outer edge of the cusped torus.

In general, the inner edge of the outer torus $r_{inner}^{o}$ has to be
$r^{o}_{in}(\ell_{o})\in]r_{K}^+(\ell_{o}),r_{cent}(\ell_{o}),[\equiv \mathbb{S}_{o}^{\textbf{\texttt{couple}}}$, represented in figure Fig\il(\ref{Fig:tookP})  (light-blue band)  and we consider  a restriction of this region due to the condition $r_{K}^+(\ell_{out})<r_{out}^{in}(\ell_{in})$ (corresponding to  condition $K\in]K_{mso},1[$).
The limiting case  $r_{K}^+(\ell_{out})=r_{out}^{in}(\ell_{in})$ occurs for
$\ell_{\lim}^{\texttt{\textbf{couple}}}$ of Eq.\il(\ref{Eq:vast-maj})  with $\ell_{in}\in ]\ell_{mso},\ell_{mbo}[$,
(or $\ell_{\lim}^{\texttt{\textbf{couple}}}$ such that $K\lessapprox1$). It is clear that the larger is
$\ell_{out}=\ell_{\lim}^{\texttt{\textbf{couple}}}$, the greater is the  dimension of the outer torus.
The other necessary condition is $r_{cent}^{o}(\ell_{o})\geq r^{i}_{out}(\ell_{i}) $;
 if the inner torus is cusped, then we can eliminate the  dependence from the density parameter $K$.
Therefore, we consider  the following two conditions in this special case:
\textbf{(a)} The torus is cusped: $r_{cent}^{o}(\ell_{o})\geq r^{i}_{out}(\ell_{i}) $, obtaining the  relation $\ell_{o}(\ell_{i})$;
\textbf{(b)} The inner torus is \emph{not} cusped (quiescent phase): there is then $r_{cent}^{o}(\ell_{o})\geq r^{i}_{out}(\ell_{i},K_i) $, obtaining the  relation $\ell_{o}(\ell_{i},K_{i})$.
Note that  eventually,  in the \textbf{(a)} and  \textbf{(b)} cases,  we  could  search   for a relation  in terms of  $K$-parameters as a relation $K_{o}(K_{i})$.
We also obtain the following limits on the inner and outer radius:

\bea&&\nonumber
r_{inner}^{\lim}\equiv r_{inner}(\ell=\ell^{out})=\frac{2}{3} \left[\frac{r^2}{r-2}-2 \upsilon _i \sin \left[\frac{1}{3} \sin ^{-1}\left(\upsilon _{\text{ii}}\right)\right]\right]
\eea
where
\bea
&&\upsilon _i\equiv\sqrt{\frac{r^2 [(r-6) r+12]}{(r-2)^2}};\quad\upsilon _{\text{ii}}\equiv\frac{r^2 [r (r [2 (r-9) r+63]-108)+108]}{2 (r-2)^3 \upsilon _i^3},
\\
&&\nonumber \mbox{and}
\\&&\nonumber
 {\mbox{\small$
r_{out}^{\lim}\equiv r_{out}(\ell^{out})=\frac{2}{3} \left[\sqrt{2} \upsilon _{\text{vi}} \cos \left(\frac{1}{3} \cos ^{-1}\left[\frac{\frac{81 (r_{inner}^{\lim}-2) \left(r_{inner}^{\lim}\right)^2 \left[9 (r_{inner}^{\lim}-2) \left(r_{inner}^{\lim}\right)^2 (r-2)-2 r^2 \upsilon _v\right]\left[27 (r_{inner}^{\lim}-2) \left(r_{inner}^{\lim}\right)^2 (r-2)-4 r^2 \upsilon _v\right]}{4 r^4 \upsilon _v^3}+8}{\frac{2 \sqrt{2} \upsilon _{\text{vi}}^3}{\upsilon _{\text{vii}}^3}}\right]\right)\right.+$}}
\\&&\label{Eq:this-leve}
\left.+\upsilon _{\text{vii}}\right],
\\&&
\upsilon _{\text{iii}}\equiv\frac{8 r^2 ((r-6) r+12) \cos \left[\frac{2}{3} \sin ^{-1}\left(\upsilon _{\text{ii}}\right)\right]}{(r-2)^2},\quad
\upsilon _{\text{iv}}\equiv\frac{2 (r (2 r-13)+26) r^2}{(r-2)^2}+18;
\\
&&\upsilon _v\equiv-4 \left(\frac{4 r^2}{r-2}-15\right) \upsilon _i \sin \left[\frac{1}{3} \sin ^{-1}\left(\upsilon _{\text{ii}}\right)\right]-\upsilon _{\text{iii}}+3 \upsilon _{\text{iv}},
 \\
 &&
 \upsilon _{\text{vi}}\equiv\sqrt{\frac{243 (r_{inner}^{\lim}-2)^2 \left(r_{inner}^{\lim}\right)^4 (r-2) r^2-54 (r_{inner}^{\lim}-2) \left(r_{inner}^{\lim}\right)^2 r^4 \upsilon _v+8 r^4 \upsilon _v^2}{\left[9 (r_{inner}^{\lim}-2) \left(r_{inner}^{\lim}\right)^2 (r-2)-2 r^2 \upsilon _v\right]{}^2}}
,
\\
&&
 \upsilon _{\text{vii}}\equiv\frac{1}{1-\frac{9 (r_{inner}^{\lim}-2) \left(r_{inner}^{\lim}\right)^2 (r-2)}{2 r^2 \upsilon _v}}.
 \eea
Notice,  $r_{out}(K_{\times},\ell)$  and $r_{inner}(K_{\times},\ell)$ are the  outer and the inner edges of the  inner  cusped torus, and
 ($r_{out}(K_{cent},\ell)$, $r_{inner}(K_{cent},\ell)$) are both the $r_{cent}(\ell)$ of  the cusped torus, where  $r_{out}(K,\ell)$  and $r_{inner}(K,\ell)$ are in Eqs\il(\ref{Eq:mer-panto-ex-resul}), while
$(K_{\times}, K_{cent})$ are defined in Eq.\il(\ref{Eq:grod-lock}), and  $r_{cent}$ is in Eq.\il(\ref{Eq:rcentro}).
We introduce  the parameter $K_{crit}$ as
\bea&&\label{Eq:rac.tes.tec}
K_{r_{\wp}}(r_{\wp})\equiv\frac{r_{\wp}-2}{\sqrt{r_{\wp}(r_{\wp}-3)}}.\in\{K_{cent},{K_{\times}}\},\quad
r_{\wp}\in\{r_{out}^{\times},r_{cent}\},
\eea
which is  the value of  the $K$-parameter at the center of torus, $K_{cent}$,  or the value ${K_{\times}}$ associated to the  cusped  tori, expressed   as the function of   $r_{\wp}$  intended as center or the inner edge of the cusped torus.
The existence of a saddle point for  $K_{r_{\wp}}(r_{\wp})$ at $r_{\wp}=\frac{3}{8} \left(\sqrt{73}+13\right)=8.079M\equiv r_{\mathcal{M}}^K\in]r_{mso},r_{mbo}^b[$  has implications on the density of the \textbf{RAD} tori structure (see analogue argument for $\ell$ in \cite{ringed,long}).
Thus there is also
\bea
\ell_{cent}^{\lim}(\ell_{in})\equiv
\frac{2 \sqrt{\frac{2}{3}} \left[1-\frac{(\hat{\mathbf{\psi}}_0-6) \hat{\mathbf{\psi}}_0^2}{3 \ell^2 \hat{\mathbf{\psi}}_2}\right]\sqrt{\frac{(\hat{\mathbf{\psi}}_3 \hat{\mathbf{\psi}}_4+1)^5 \left[\frac{2 (\hat{\mathbf{\psi}}_3 \hat{\mathbf{\psi}}_4+1)}{3 \left[1-\frac{(\hat{\mathbf{\psi}}_0-6) \hat{\mathbf{\psi}}_0^2}{3 l^2 \hat{\mathbf{\psi}}_2}\right]}-2\right)^2}{\left[1-\frac{(\hat{\mathbf{\psi}}_0-6) \hat{\mathbf{\psi}}_0^2}{3 l^2 \hat{\mathbf{\psi}}_2}\right]^5}}}{3 (\hat{\mathbf{\psi}}_3 \hat{\mathbf{\psi}}_4+1) \left(\frac{2 (\hat{\mathbf{\psi}}_3 \hat{\mathbf{\psi}}_4+1)}{3 \left[1-\frac{(\hat{\mathbf{\psi}}_0-6) \hat{\mathbf{\psi}}_0^2}{3 \ell^2 \hat{\mathbf{\psi}}_2}\right]}-2\right)^2},
\eea
considered as
function of $\ell=\ell_{\in}\in]\ell_{mso},\ell_{mbo}[$,
for the \emph{center} $r_{cent}^o$ to have  the outer torus  located on the outer margin of the inner accreting one, i.e. $r_{cent}^o=r_{out}^i$ (this can be obtained from Eqs\il(\ref{Eq:rcentro}) and (\ref{Eq:dan-aga-mich})  assuming $r^{\times}_{out}(\ell_i)=r_{cent}(\ell_o)$). Therefore, there  has to be $\ell_{o}>\ell_{cent}^{i}=\ell_{cent}^{\lim}(\ell_{i})$.
The center of cusped  torus is located in  $r_{cent}^{\times}\in[r_{mso},r_{mbo}^b[=[6M,10.47M]$, whereas the outer edge can extend  to very large  distance from the attractor.
 For the analysis of this case, we return to the  (\textbf{\texttt{AC}})-model  of Fig\il(\ref{Fig:conicap}) modifying some definitions. Specifically,
we introduce the specific angular momentum for case {\small\textbf{(1)}} $\ell(\epsilon)=({\ell_{mbo}-\ell_{mso}})/\epsilon+\ell_{mso}$, such that  for $\epsilon=1$ there is  $\ell=\ell_{mbo}$ and,
asymptotically, for $\epsilon\rightarrow\infty$, there is  $\ell\rightarrow\ell_{mso}$. These new  definitions restrict the analysis to cusped tori or quiescent tori in the angular momentum range $]\ell_{mso},\ell_{mbo}[$.
In Figs\il\ref{FIG:multiplot} we considered  case {\small\textbf{(2)}}--
the momenta  $\ell_b(\epsilon)=({\ell_{\gamma}-\ell_{mbo}})/{\epsilon}+\ell_{mbo}$ such that,  for $\epsilon=1$, there is $\ell=\ell_{\gamma}$ and,
asymptotically, for $\epsilon\rightarrow\infty$, there is  $\ell\rightarrow\ell_{mbo}$, finally   in case {\small\textbf{(3)}}, there is the specific angular momentum $\ell_c(\epsilon)=({\ell_{\gamma}})/({\epsilon})+\ell_{\gamma}$, being
$\ell_c(\epsilon)=\ell_{\gamma}$ for $\epsilon\rightarrow\infty$, and $\ell_c(\epsilon)\rightarrow\infty$ for  $\epsilon=0$.
Therefore we studied
the momentum  $\ell^o$ such that  $r_{inner}^o=r_{out}^i$, the location of the torus center  $r_{cent}^o$.
\begin{figure}
  \includegraphics[width=7cm]{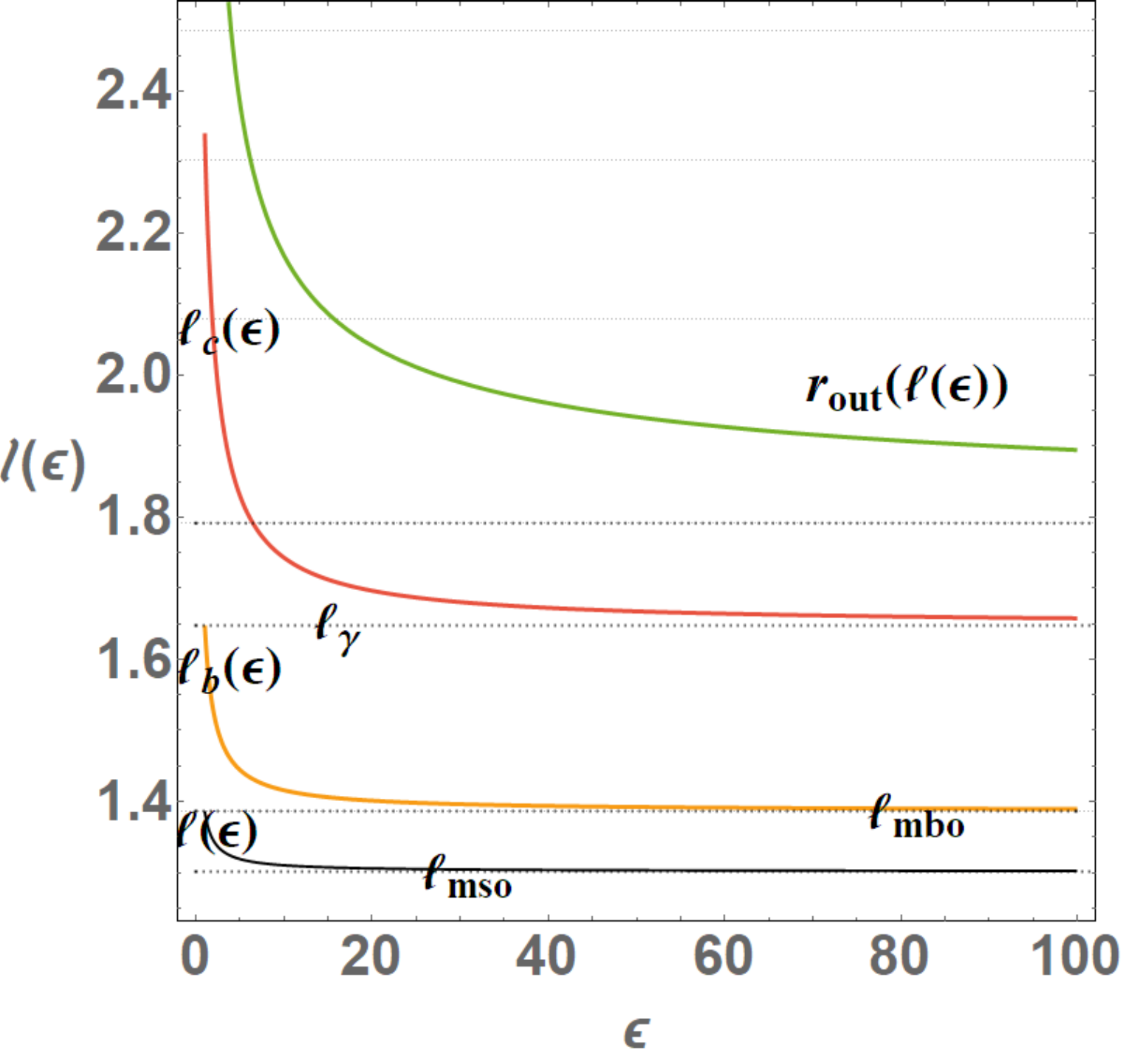}
  \includegraphics[width=9cm]{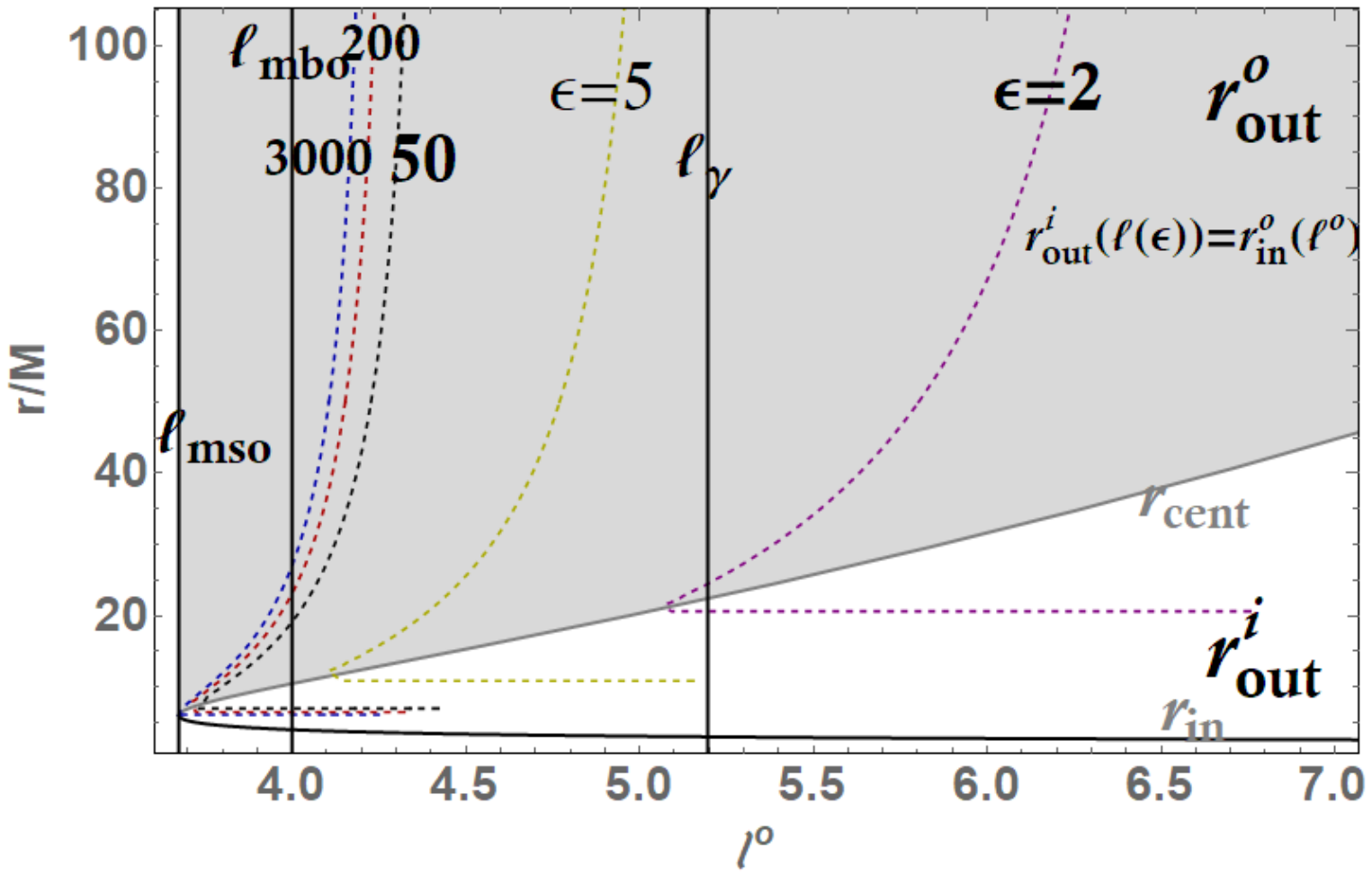}
  \caption{Left panel: specific angular momenta   $\ell(\epsilon)$, $\ell_b(\epsilon)$ and $\ell_c(\epsilon)$ as functions of $\epsilon$ such that for  $\epsilon=1$ there is $\ell_b=\ell_{mbo}$, $\ell_c=\ell_{\gamma}$ and
 for $\epsilon=\infty$ there is  $\ell=\ell_{mso}$, $\ell_b=\ell_{mbo}$   and $\ell_c=\ell_{\gamma}$ (accreting tori have angular momentum $\ell\in\mathbf{L1}$)  and $\ell_c=\infty$ for  $\epsilon=0$. The outer edge $r_{out}^{\times}$ of the accreting torus defined in Eq.\il(\ref{Eq:dan-aga-mich}) is at $\ell=\ell(\epsilon)$ is also shown as function of $\epsilon$. Right panel: Analysis of the marginally collision sphere. Inner $r_{inner}^o$ and outer edge $r_{out}^o$ the center $r_{cent}^o$ (location of maximum density and hydrostatic pressure) of the the outer torus having $r_{inner}^o=r_{out}^i$ i.e. the inner margin coincident with the outer edge of the inner accreting torus as function of the angular momentum of the outer torus $\ell^o$, for different values of $\epsilon$ where the specific angular momentum of the inner accreting torus is $\ell(\epsilon)$.}\label{FIG:multiplot}
\end{figure}
\begin{figure}
  \includegraphics[width=8cm]{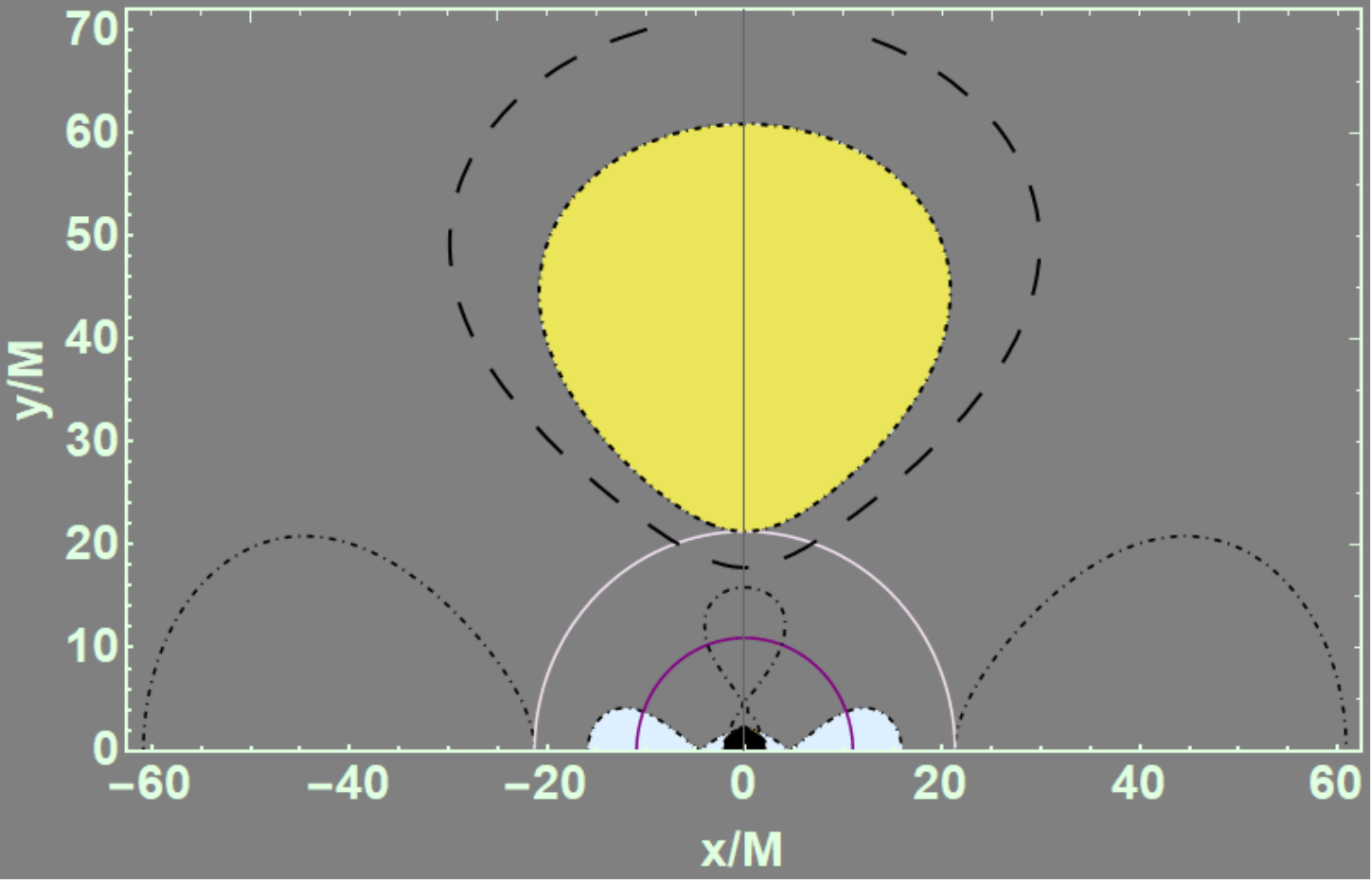}
  \includegraphics[width=7cm]{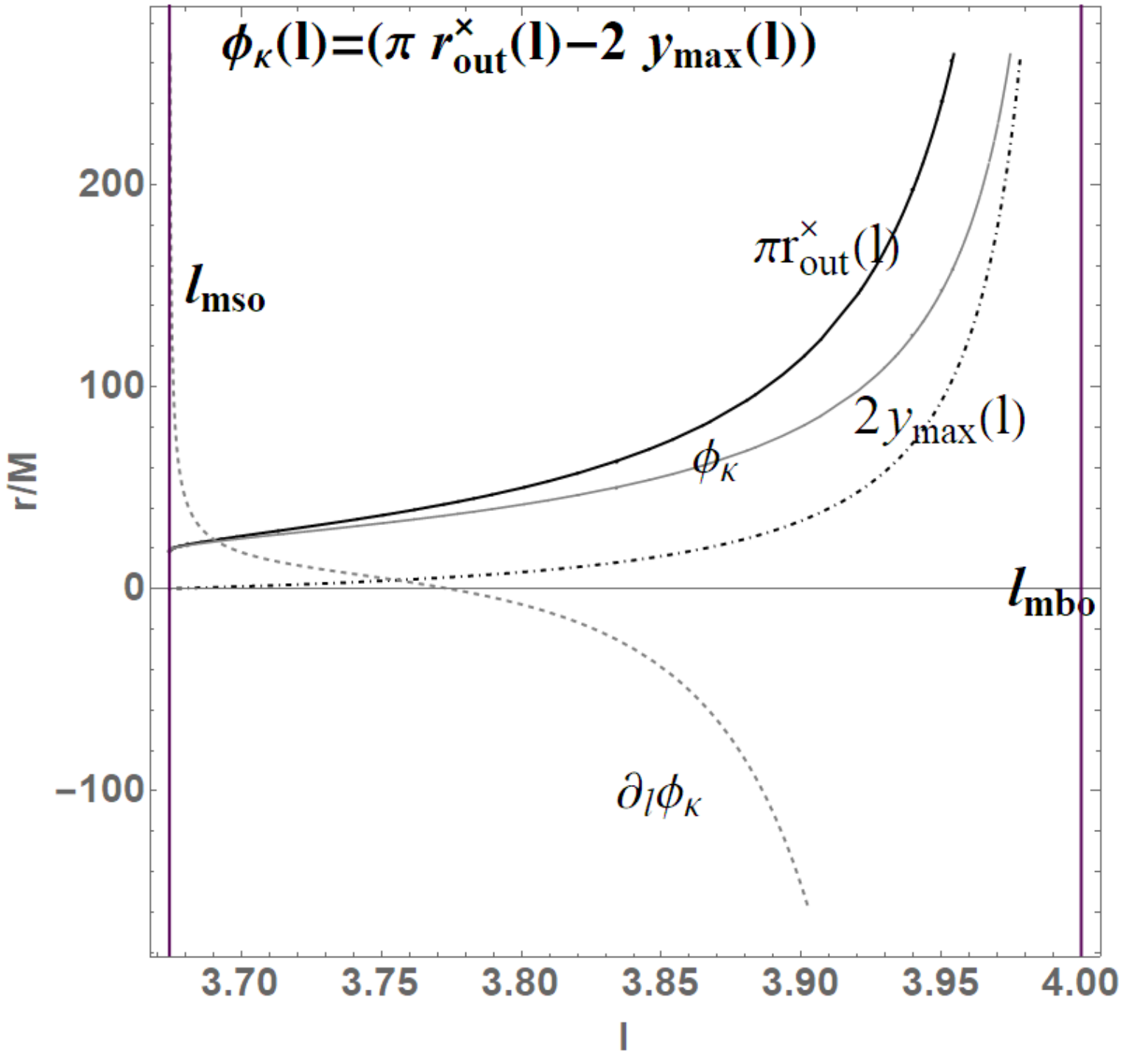}
  \caption{Collisional spheres and maximum \textbf{BH} converge: limiting toroidal surfaces in the orthogonal configurations.}\label{Fig:Forse1ce}
\end{figure}
In the following section we will reconsider models associated with these for the evaluation of the quantity of each torus  that are related to the energetic and depend on the thickness of the disks flow.
\section{RAD energetics and limiting open surfaces for misaligned tori}\label{Sec:er}
A \textbf{BH} and especially a \textbf{SMBH} can be  characterized   by unusual or extraordinary periods of activity of the \textbf{BH}, in terms of enhanced accretion rates and interrupted accretion periods and jet emissions, detectable in the alteration of the mass accretion rates, or  recognizable as mechanism at base for  high masses considered in the \textbf{SMBH}, with a contribution of matter in accretion that one can think maximized by the number of accreting orbiting tori .
Accretion, in the case of a globular model  would  prompt a   relatively fast  collapse  of the structure into the central \textbf{BH}  with a huge  mass spin contribution, with a great release of energy  and matter outburst.
The \textbf{RAD} energetic quantities $\mathbf{E}$ show two kinds of fundamental properties typical of geometrically thick disks. There is an explicit  dependence of  $\mathbf{E}$ on the polytropic index $\gamma$  and polytropic constant but as shown in Table\il(\ref{Table:Q-POs})   with an   factors allowing to define independent quantities and therefore study the quantities \textbf{E} varying \textbf{P}, and therefore with respect to the  morphological characteristics $\mathbf{M}$ {(quantities $(\mathcal{R},\mathcal{N})$ in  Table\il(\ref{Table:Q-POs})   are re-parameterizations of  $(\mathcal{O},\mathcal{P})$ respectively)}. This has an important consequence from the point of view of observation and recognition of such systems.
Many characteristic studied in Sec.\il(\ref{Sec:Misal}) and Sec.\il(\ref{Sec:doc-ready})  are thus independent from details on the specific polytropic but rather consequences of the geometry and pressure gradients.
These tori are therefore  determined by the title  angles-Figs\il(\ref{Fig:DopoDra}).

As discussed in Sec.\il(\ref{Sec:Misal}), the model adopted  for the single toroid of the globulus,  described as solutions of Eq.\il(\ref{Eq:sha-conf}) with a barotropic equation of state, provides open-cusped surfaces whose meaning is yet to be  fully  understood. These special surfaces  are associated with instabilities (in this case referred to as a topological features of the surfaces) having  a broad centrifugal component with  $\ell\in [\ell_{mbo},\ell_{\gamma}]$, and the  cusp  $r_{crit}=r_j\in[r_{mso},r_{mbo}]$ is located in a specific \emph{annulus}  externally  with respect to the accretion cusp region--Figs\il(\ref{Fig:SIGNS}). Therefore these structures, variously related to jet or proto-jets configurations are not directly correlated to  accretion. It is clear then that the   centrifugal component in competitions  with the attractive gravitational component of the force balance equations, prevails although not sufficiently hight with respect to a strong gravitational component  $ r_j\in[r_{mso},r_{mbo}]$ to stabilize the disk (absence of a cusp) as it occurs for the tori with specific momentum in $\ell>\ell_{\gamma}$.  
In Sec.\il(\ref{Sec:energ-RAD-poli}) we consider more specifically the polytropic fluids and the globuli characteristics  directly dependent.on the polytropic parameters typical and in particular from the characteristics of the \textbf{RAD} energetics
as the  accretion rates.
\subsection{Polytropic fluids and RAD energetics}\label{Sec:energ-RAD-poli}
These \textbf{RAD} structures pose the interesting question of the internal activity of the cluster and particularly the internal exchanges of energy and matter between the tori and the tori and central \textbf{BH}. As considered for the   \textbf{eRAD } system, \textbf{RAD} and globuli are characterized by a vivid internal activity made up of collision among tori, inner  accretion, internal jet launch, whether constrained  by open-proto-jet configurations considered here in Sec.\il(\ref{Sec:energ-RAD-poli}) or rather by jet emission from accretion jet correlation. This situation in case misaligned tori  considered here is even more clear, the internal activity can eventually even undermine the structure and its  stability  especially in the case of Kerr attractor.  The existence of these structures  shows the  possibility that a \textbf{BH}-system  can  be "dormant", quiescent, i.e. undergoing a period of low activity  intended interactions with clouds, stars of \textbf{BHs} companions  or more generally with the galactic environment. We  could picture this situation defined  cold or warm globulus  otherwise  the last having  a vivid internal life and exchanges of energy and matter. One essential difference with respect to other similar situations foreseen  embedded  \textbf{BH} considered  in the literature is the fact that we consider an  orbiting non-self gravitating \textbf{RAD} composed by objects with very different characteristic starting from an intricate multipoles structure.
 In this section we  provide   evaluations  of  quantities  related  to tori energetics such as the  mass-flux,  the  enthalpy-flux (evaluating also the temperature parameter),
and  the flux thickness--see \cite{abrafra,Japan}--Table\il(\ref{Table:Q-POs}).
 Considering    polytropic fluids with pressure $p=\kappa \varrho^{1+1/n}$, we listed these quantities in Table\il(\ref{Table:Q-POs}).
These quantities have been  obtained by considering  the flow thickness expressed through the density profiles.
The relativistic frequency $\Omega$  reduces  to Keplerian  values $\Omega_K$ at the edges of the accretion torus, where  the pressure forces   are vanishing
\footnote{It has been shown that for the Schwarzschild geometry there is a specific classification of eligible polytropics (see  \cite{Raine}), and a specific class of polytropics is characterized by a discrete  range of values for the index $\gamma$ \cite{mnras}.}
\begin{table*}
\caption{Quantities $\mathcal{O}$ and $\mathcal{P}$. There is  $\varpi=n+1$, with  $\gamma=1/n+1$  being the polytropic index. $\Omega_K$ is  the Keplerian angular velocity.   $W=\ln V_{eff}$,
$W_s\geq W_{\times}$ is the value of the equipotential surface, which is taken with respect to the asymptotic value, $ W_{\times}=\ln K_{\max}$ is $W$ at  the inner edge of accreting torus. $\mathcal{L}_{\times}/\mathcal{L}$ is the  fraction of energy produced inside the flow and not radiated through the surface but swallowed by central \textbf{BH}. $(W(r_{mbo})-W(r_{mso}))$  is the  maximum difference  of quantities in  $\mathcal{O}$ and $\mathcal{R}$.
  $\mathcal{L}$ representing the total luminosity, $\dot{M}$ the total accretion rate where, for a stationary flow, $\dot{M}=\dot{M}_{\times}$,
$\eta\equiv \mathcal{L}/\dot{M}c^2$ the efficiency, $\mathcal{D}(n,\kappa), \mathcal{C}(n,\kappa), \mathcal{A}(n,\kappa), \mathcal{B}(n,\kappa)$ are functions of the polytropic index and the polytropic constant.
}
\label{Table:Q-POs}
\centering
\begin{tabular}{|l|l|}
 \hline\hline
 \mbox{\textbf{Quantities}}$\quad  \mathcal{O}(r_\times,r_s,n)\equiv q(n,\kappa)(W_s-W_{\times})^{d(n)}$ &   $\mbox{\textbf{Quantities}}\quad  \mathcal{P}\equiv \frac{\mathcal{O}(r_{\times},r_s,n) r_{\times}}{\Omega_K(r_{\times})}$\\\hline\hline
\text{\textbf{[$\mathcal{R}$-quantities]}:}  $\mathcal{R}_{*}\equiv(W(r_{s})-W_{*})^\varpi
$&\text{\textbf{[$\mathcal{N}$-quantities]}:  } $    \mathcal{N}_{*}\equiv\frac{{r_*} (W(r_{s})-W_{*})^\varpi}{\Omega_K(r_*)}
$
 \\
\hline\hline
$\mathrm{\mathbf{Enthalpy-flux}}=\mathcal{D}(n,\kappa) (W_s-W)^{n+3/2},$&  $\mathbf{torus-accretion-rate}\quad  \dot{m}= \frac{\dot{M}}{\dot{M}_{Edd}}$  \\
 $\mathrm{\mathbf{Mass-Flux}}= \mathcal{C}(n,\kappa) (W_s-W)^{n+1/2}$& $\textbf{Mass-accretion-rates }\quad
\dot{M}_{\times}=\mathcal{A}(n,\kappa) r_{\times} \frac{(W_s-W_{\times})^{n+1}}{\Omega_K(r_{\times})}$
 \\
 $\frac{\mathcal{L}_{\times}}{\mathcal{L}}= \frac{\mathcal{B}(n,\kappa)}{\mathcal{A}(n,\kappa)} \frac{W_s-W_{\times}}{\eta c^2}$&     $\textbf{Cusp-luminosity}\quad  \mathcal{L}_{\times}=\mathcal{B}(n,\kappa) r_{\times} \frac{(W_s-W_{\times})^{n+2}}{{\Omega_K(r_{\times})}}$
\\\\
\hline
\end{tabular}
\end{table*}
To simplify this analysis
we consider the general  form
 $\mathcal{O}$ or    $\mathcal{P}$ form, as in  Table\il(\ref{Table:Q-POs}), where $q(n,\kappa)$ and $d(n)$ are general   functions of the polytropic index  $\gamma=1+1/n$ and polytropic constant $\kappa$. These quantities express   the mass flow rate through the cusp (mass loss, accretion rates)  $\dot{M}_{\times}$,  and the cusp luminosity $\mathcal{L}_{\times}$ (and the accretion efficiency $\eta$),
 measuring the
rate of the thermal-energy    carried at the  cusp--see\cite{abrafra,Japan,Blaschke:2016uyo,2017PhRvD..96j4050S} also \cite{long,Multy,Letter}.    The   $\mathcal{O}(r_\times,r_s,n)$ depend on the accretion sphere (inner edge of the cusped torus) and   the radius   $r_s$ which is  related to the  flow thickness, $\overleftarrow{\mathbf{h}}_{s}$, of the matter flow.
These quantities regulating  the tori energetics are shown in Fig.\il\ref{Fig:RegofreurOa}
as  functions of the cusp locations where the radius $r_s$ in the definitions of Table\il(\ref{Table:Q-POs}) has been fixed  arbitrarily,  preserving the geometric sense of this cut-off, and the  dependence of the polytropic index through  $\kappa=n+1$ (polytropic index $\gamma=\frac{1}{n}+1$)--Figs\il\ref{Fig:RegofreurO}:
\bea
&&\label{Eq:arbi-tra}
r_{\times}=\left(6-\frac{2}{\epsilon}\right)M,\quad
\frac{r_s}{M}=6-\frac{2}{\epsilon }-\epsilon_s=\frac{2(3d \epsilon - d-2 \epsilon +1)}{d \epsilon} ,
\eea
where
\bea\epsilon\geq1\quad \epsilon_s\equiv \frac{\epsilon_{ss}}{d}\in[0,\epsilon_{ss}],\quad(d\geq1)\quad\epsilon_{ss}\equiv\frac{4 \epsilon -2}{\epsilon},
\eea
and there is
\bea
 \lim_{\epsilon\rightarrow \infty}r_{\times}(\epsilon)=r_{mso},\quad r_{\times}(\epsilon=1)=r_{mbo},\quad
 \lim_{d\rightarrow \infty}r_{s}(d,\epsilon)=r_{\times}(\epsilon),\quad r_{s}(d=1,\epsilon)= r_+=2M.
\eea
(Note, models defined in Eq.\il(\ref{Eq:arbi-tra}), similar ti  \texttt{\textbf{(AC)}} of Sec.\il(\ref{Sec:doc-ready}) but defined through the evaluation of points  $r_s$ setting  $K_s$ and $r_{\times}$, setting  $\ell$ and $K_{\times}$, as in Figs\il\ref{Fig:RegofreurO}.)
\begin{figure}
%
  \includegraphics[width=5cm]{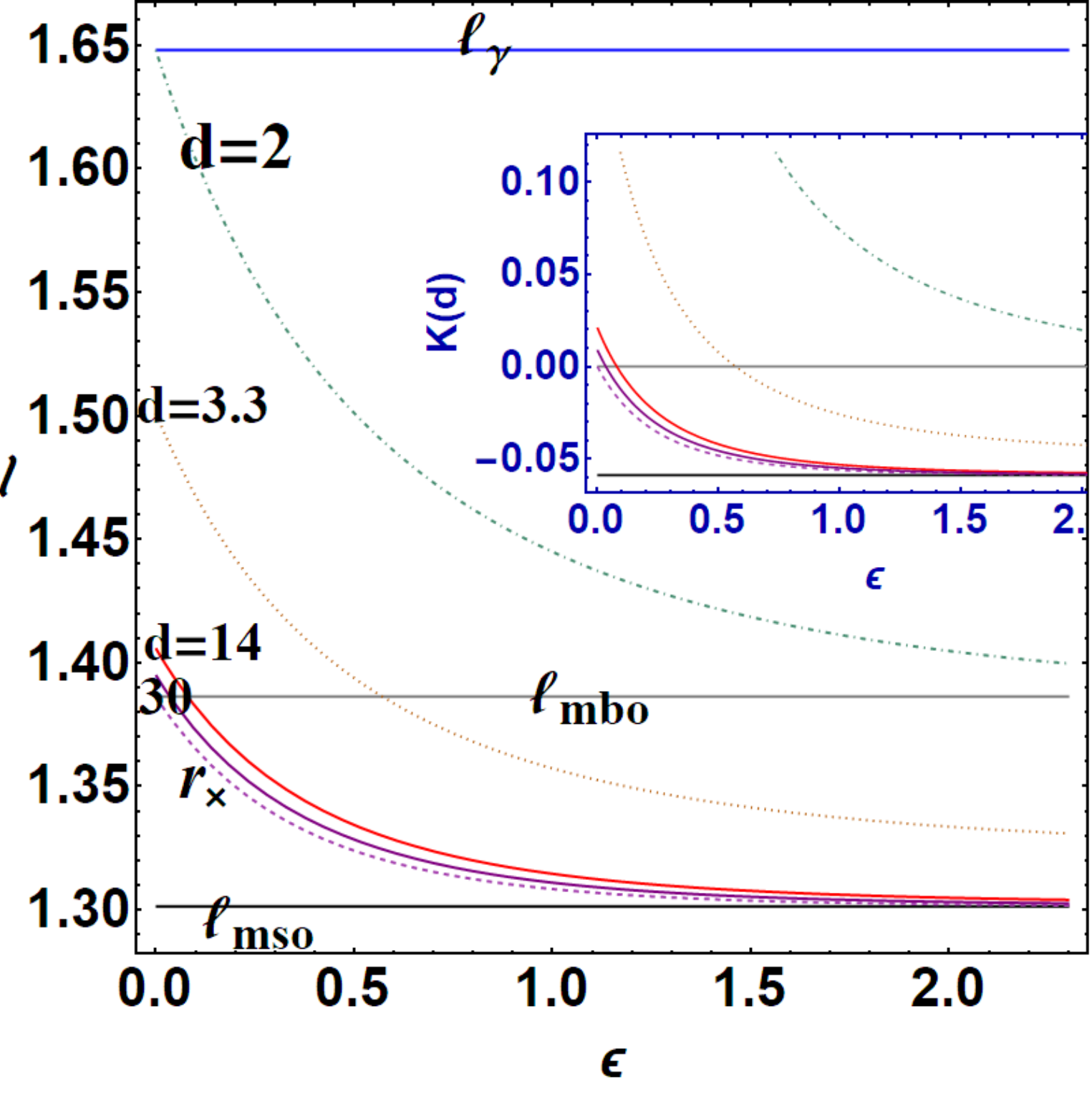}
   \includegraphics[width=7cm]{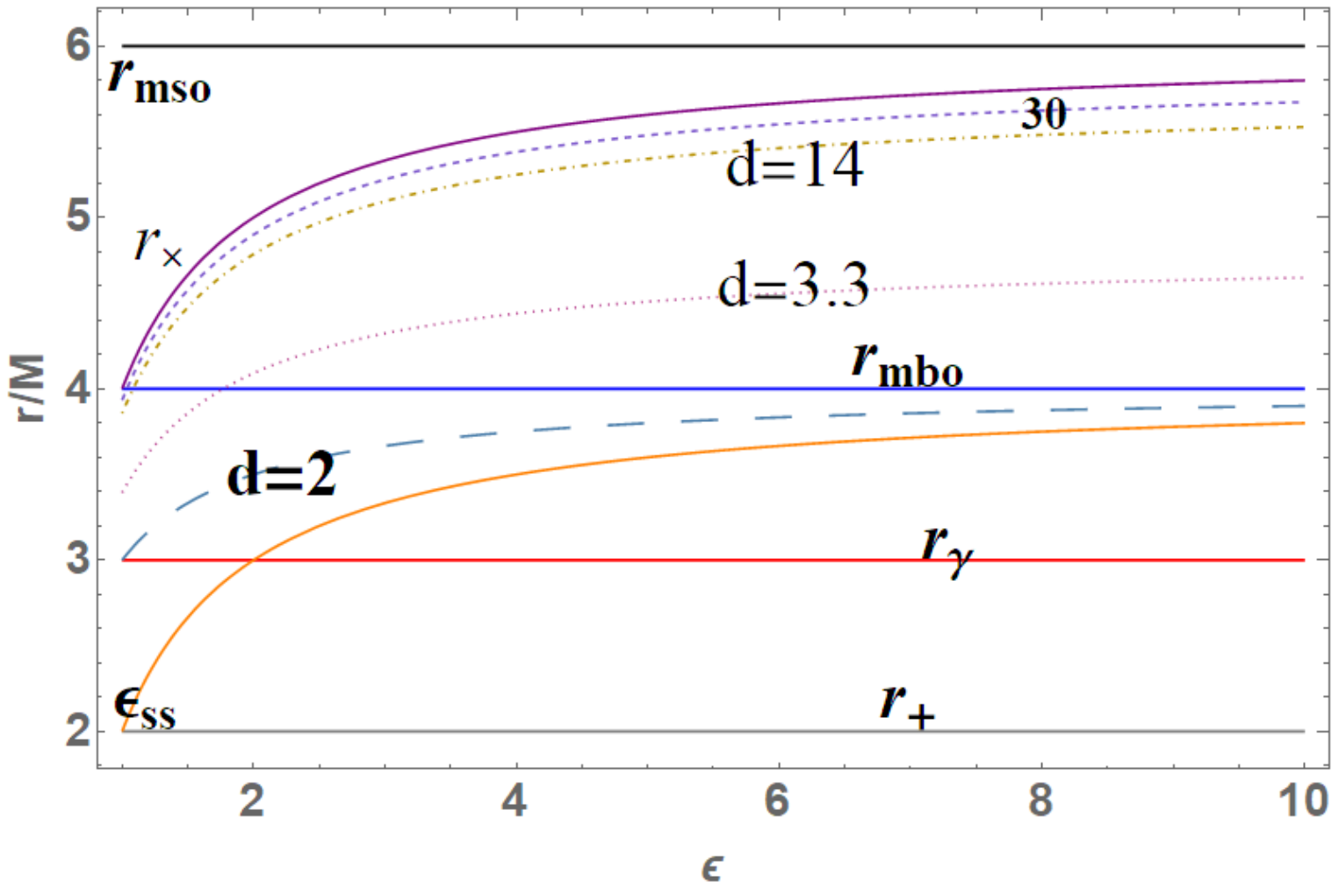}
    \includegraphics[width=3.4cm]{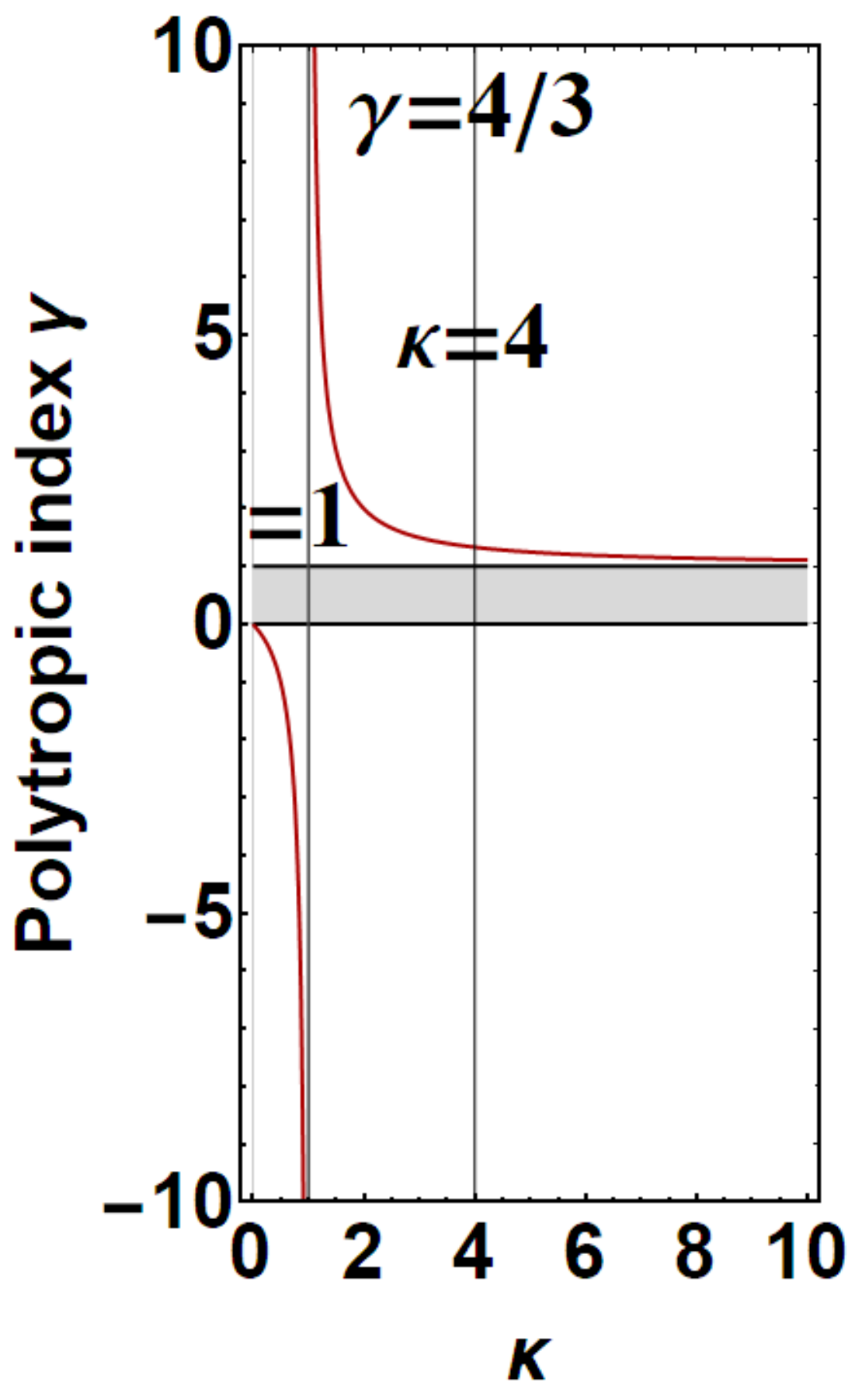}
     \includegraphics[width=8cm]{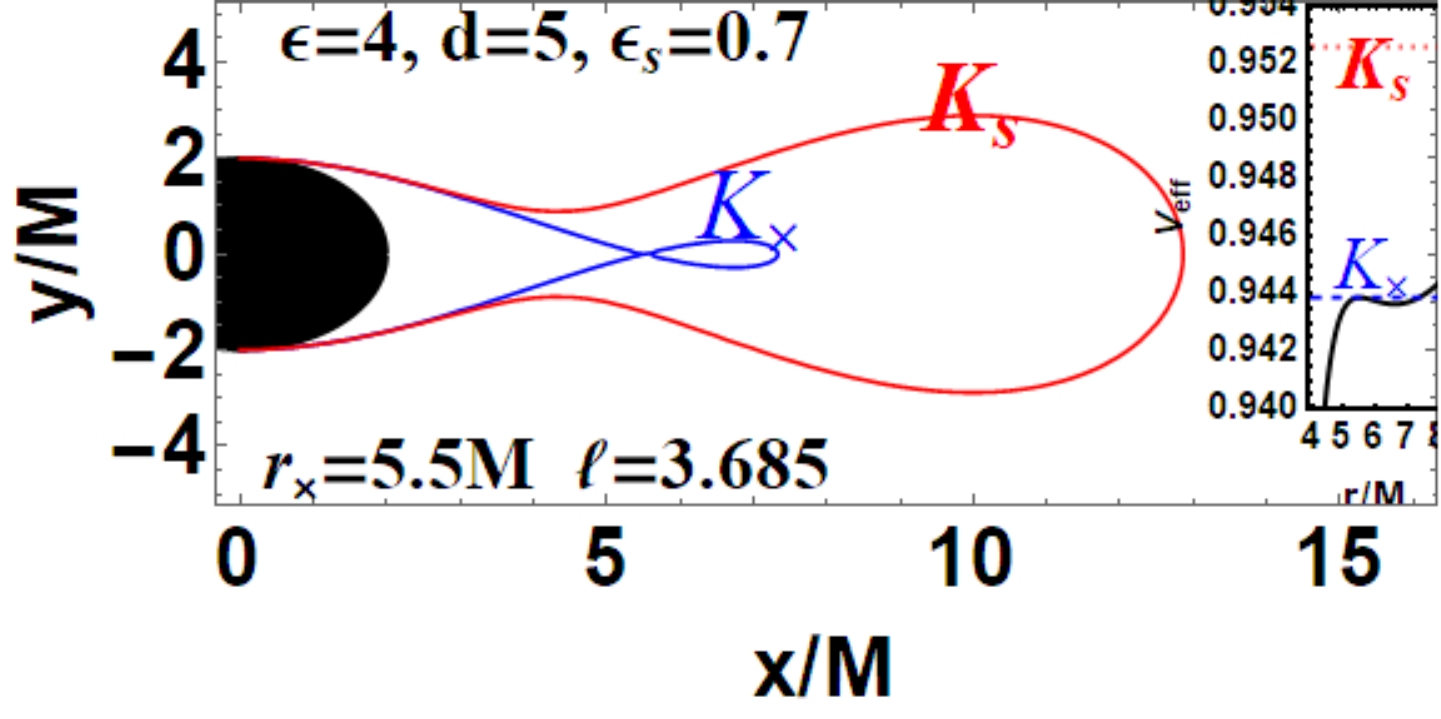}
  \caption{Upper left panel:  fluid specific angular   momentum $\ell$  and $K$ function (inside panel) evaluated in  quantities in Eqs\il(\ref{Eq:arbi-tra}) as functions of $\epsilon$. Upper right panel: $r_{\times}(\epsilon)$ and $r_s(\epsilon)$  defined  in Eqs\il(\ref{Eq:arbi-tra})  as functions of $\epsilon$.  Bottom left  panel: polytropic index $\gamma=\frac{1}{{k}-1}+1$ function of $k$ for the models considered in  Figs\il\ref{Fig:RegofreurOa}. Bottom right panel: tori profile for model specified in figure. Inside panel is the associated effective potential with $(K_{\times},K_s)$.}\label{Fig:RegofreurO}
\end{figure}
{Figs\il(\ref{Fig:RegofreurO}) shows the parameters  relevant for the analysis of the energetics of the RAD; here we relate the parameter $\ell$  to the polytropic index. We should note that although we selected specific functional relations (curves in the figure) one has to consider  the possibility to fit the observational data for a wide range of deducted parameter value, radial  distances are relatively small in terms of mass $M$ of the central object and finally we are concerned here with the behavior of the curve with the variation of the parameters which provides a relative  comparison with tori components.}
In the evaluation of these quantities  considered in  Figs\il\ref{Fig:RegofreurOa} we considered  $r_s$ chosen so that  there is  $W(r_{\times})<W(r_s)$ to ensure a non-zero flow thickness.
The limiting quantities, as $\mathcal{N}_{\lim}$, have been  evaluated assuming $r_{\times}=r_{mso}$ and $r_{s}=r_{mbo}$, as $(r_{\times}-r_s)$
is the limiting (maximum) distance,   and we have assumed that $r_s$ may approach the horizon--Figs\il\ref{Fig:RegofreurO}.
\begin{figure}
   \includegraphics[width=7cm]{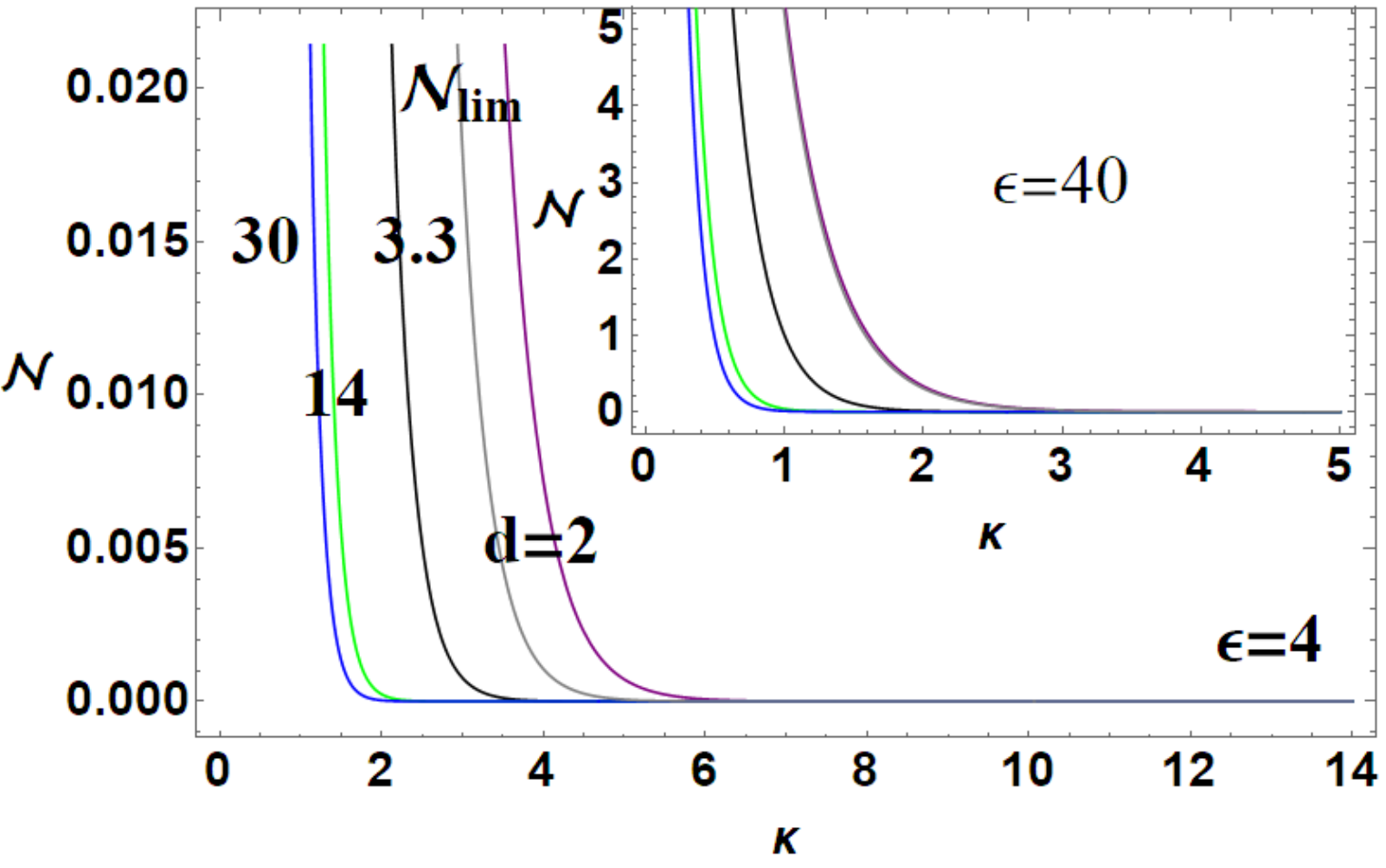}
  \includegraphics[width=7cm]{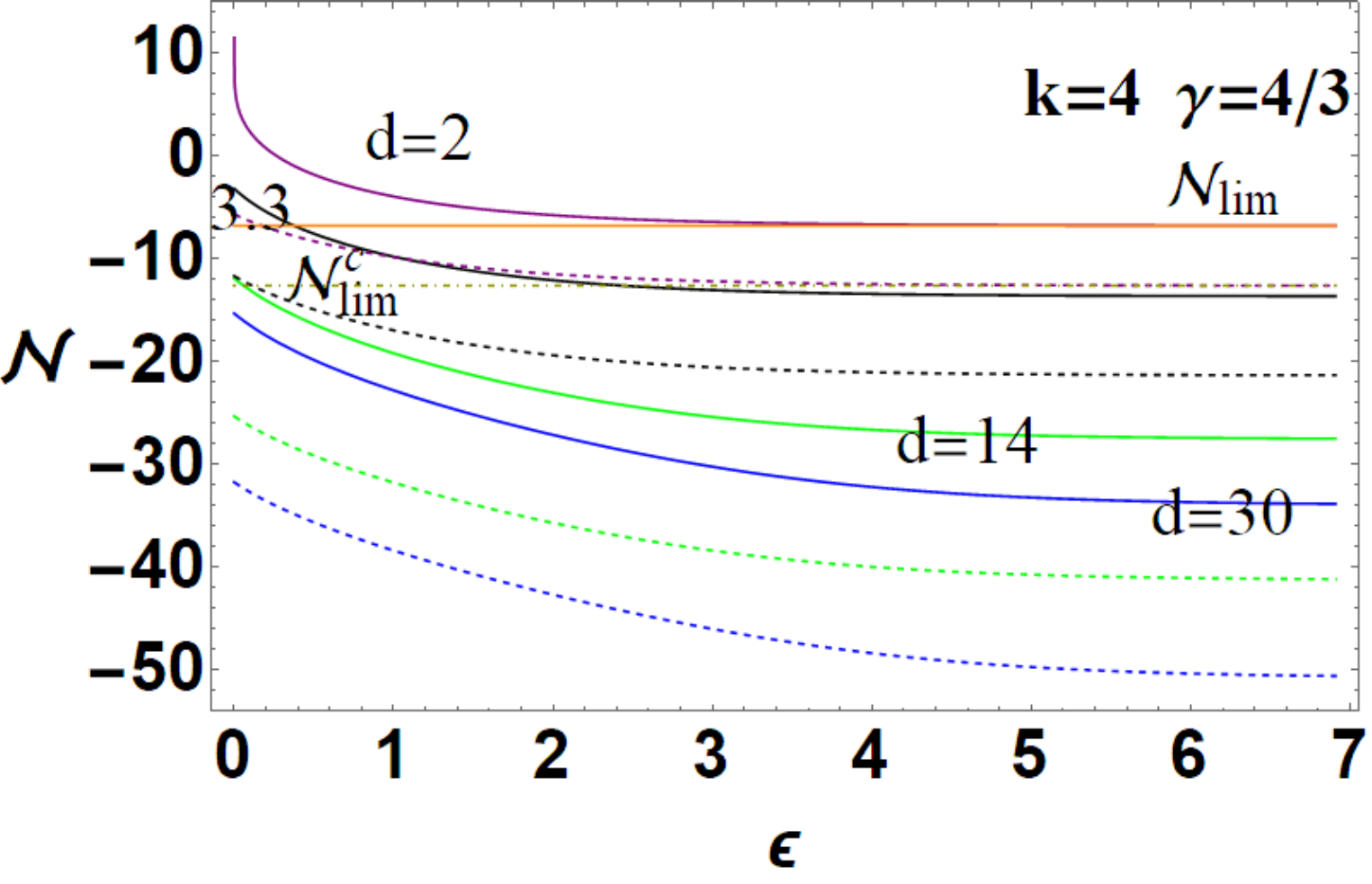} \\
  \includegraphics[width=7cm]{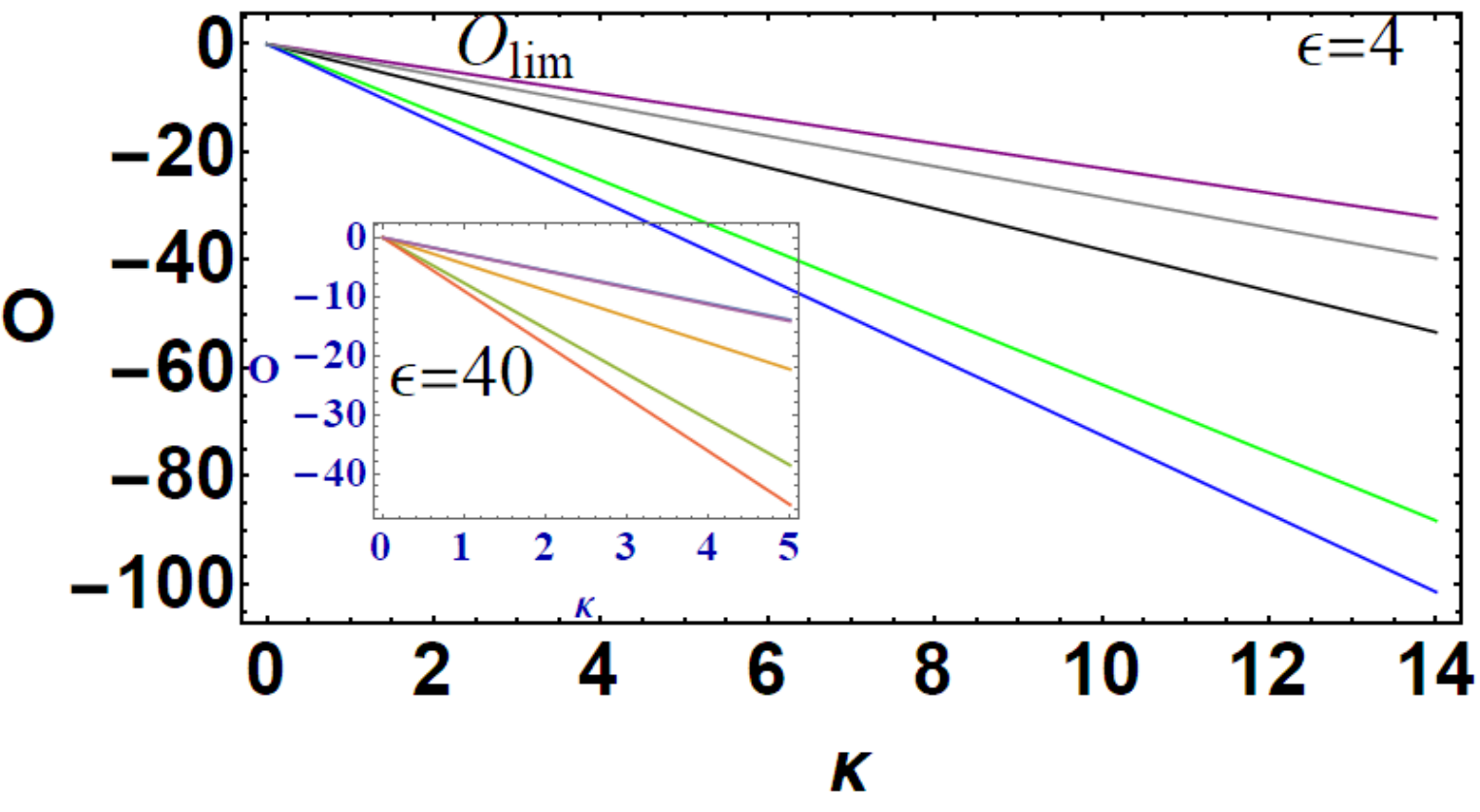}
 \includegraphics[width=7cm]{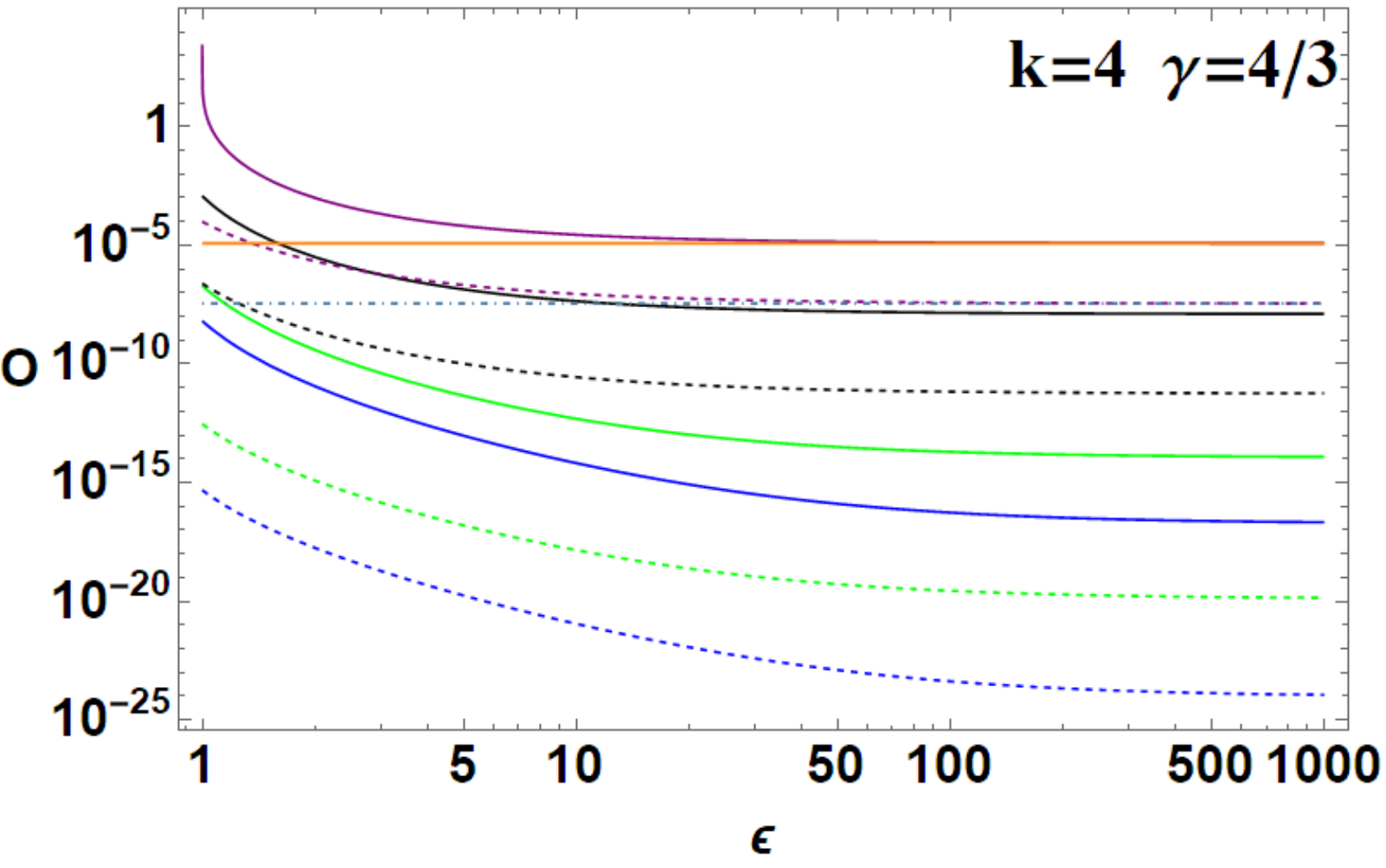}
  \caption{Quantities $\mathcal{N}$ (upper panels) and $\mathcal{O}$ (bottom panels)--defined in  Table\il(\ref{Table:Q-POs}) and evaluated on the models presented  in Figs\il\ref{Fig:RegofreurO}. Left panels Quantities $\mathcal{N}$  and $\mathcal{O}$ as functions of $\kappa$ for $\epsilon=4$ and $\epsilon=40$ (inside panel) for different values of $d$, limiting functions $\mathcal{N}_{\lim}$  and $\mathcal{O}_{\lim}$ are also shown. Models $(\epsilon,\kappa,d)$ refer to Figs\il\ref{Fig:RegofreurO} and  Eqs\il(\ref{Eq:arbi-tra}). Right panels: Quantities $\mathcal{N}$  and $\mathcal{O}$ as functions of $\epsilon$ for fixed polytropic  at different $d$.}\label{Fig:RegofreurOa}
\end{figure}
 {Figs\il(\ref{Fig:RegofreurOa}) focus on the dependence of the energetics quantities $\mathbf{E}$ on the polytropic equation of state  and location of inner edge. This analysis therefore has to be  compared  with results of  Sec.\il(\ref{Sec:morph}), and these considerations turn in terms of parameters analysis $\textbf{P}$  in  Figs\il(\ref{Fig:KKNOK},\ref{Fig:solidy}), to   the correspondent  tori as objects of the curves in figures, characterized by a selected values \textbf{E}.}
This comparative analysis  of  Figs\il\ref{Fig:RegofreurO}  and Figs\il\ref{Fig:RegofreurOa} allows the study of quantities  in dependence on   the distance of the toroids from the central black hole and  with variations of  $\epsilon\geq1$ and $d\geq 1$.
\section{Discussion and Conclusions}\label{Sec:conclu}
The main topic of this paper  is the  analysis of the  set of accreting tilted tori  around the  central \textbf{BH}.
We studied  clusters of  misaligned (inclined) tori  (\textbf{RAD}) orbiting one central Schwarzschild attractor.
For this purpose we have defined the toroidal components  of the gravitating system as an aggregate of  misaligned perfect fluid tori orbiting the central Schwarzschild black hole using the approach  developed  in   \cite{ringed,open,dsystem}  from the \textbf{eRAD} model construction.
We stress that  by considering a generic tilt angle  we  provide indication on the characteristics and observational properties   for the clusters of orbiting  tilted tori  rather then an analysis of the  single  toroidal component of the aggregate.
Constrains on existence of such configurations are discussed in  Sec.(\ref{Sec:mirpj}),   and particularly against  tori collision in Sec.\il(\ref{Sec:limiting})  and Paczynski instabilities-- Sec.\il(\ref{Sec:RADloq}). We studied the \textbf{BH-RAD}  energetics,  Sec.(\ref{Sec:energ-RAD-poli}), and an  overview of possible phenomena  associated to the  \textbf{RAD} structures, as  possible proto-jets emerging  in the  \textbf{RAD} structure. Many morphological and stability properties (presence of the cusp) of the torus  discussed Sec.\il(\ref{Sec:doc-ready}) are constrained by the distance of the  torus from the central attractor and regulated by the centrifugal component of the force balance inside the torus, therefore the  disk rotational law.
 From methodological view point we take advantage of spherical symmetry providing  the results for perfect fluids.
 Our analysis places constraints on the existence and stability of misaligned tori which can be used in  dynamical (time-dependent)  analysis of a similar system with these (evolving) initial configurations.
 With respect to the current literature considering similar  objects   within a numerical approach we frame the analysis in a fully analytical \textbf{RAD} models identifying one  leading function to describe  the distribution of tori. We  singled out definition of  \textbf{RAD} leading function $\ell(r)$ of Eq.\il(\ref{Eq:lqkp}), the \textbf{RAD}  energy function $ K(r)$ of Eq.\il(\ref{Eq:sincer-Spee}).
 The outcomes of this  approach results in      constraining  the inner structure of the \textbf{RAD}  and  tori  morphological  characteristics.

More in details:   in section (\ref{Sec:Misal}) we give
the basic equations for the description of this system providing the  {\textbf{RAD} leading function} governing the  tori  distribution in  the orbiting macro-structure.
We have therefore studied the model characteristics  on the basis of the  parameters determining  the particular configurations according to their stability as related to the toroids morphology. We  integrated the force balance equation for the  equi-density (equi-potential)
surfaces. The evaluation of the toroids geometrical thickness  has an essential role  in the evaluation of the effects of disc-seismology as clarified in  \cite{next}.
Then geometry of \textbf{RAD} accreting tori, stability and collision emergence are  focused   in Sec.\il(\ref{Sec:doc-ready}).
This analysis  also led  to determine the misaligned tori collisional emergence. Here we provide the  tori distributions in the  \textbf{RAD} considering the characteristic of the geometric thickness $\Sa$ and an evaluation of the geometric thickness of the disks considered  in the \textbf{RAD} frame establishing conditions under  which disks are geometrically thick according to the model parameters  and the limit value  $\Sa=1$.
This analysis identifies  the    sets of  \textbf{RAD} inclined  toroids having  equal $\mathbf{R}\in\{r_{out},r_{cent},r_{inner},\lambda\}$ or other characteristics as the torus thickness. The relevance of this analysis lies in the fact that the classes of toroids and therefore of \textbf{RADs}  so classified  might  correspond to  observational effects dependent on the characteristics \textbf{R} or $\Sa$ which are  similar (as we have seen  in some cases  in this work), thus it is important to model this type of toroids.
We consider particularly  the  torus  elongation  $\lambda(\ell,K)$, the location of the inner edge, the
location of the torus  center  $r_{cent}$,  corresponding to  the point of maximum density and hydrostatic pressure, the location of the maximum $r_{\max}\equiv x_{\max}$, giving    \textbf{RAD} tori maximum density and  the  torus thickness $S\equiv2h_{\max}/\lambda$.
 These conditions are significant  to  establish the stability  of the \textbf{RAD}  including misaligned tori and also possibly the time scale of the involved processes that are potentially observable and which depend on the distance (spacing) between the toroids and their thickness, on the presence or absence of a cusp.
More generally, the  toroidal geometrical  thickness is also  an indication of the  accretion disk model  and \textbf{BH} accretion rates correlation.
Then the presence of multi orbiting structures could influence the accretion rate of the central \textbf{BH}.  Therefore,
in Sec.\il(\ref{Sec:energ-RAD-poli}) we  evaluated  quantities  related  to tori energetics such as the  mass-flux,  the  enthalpy-flux (evaluating also the temperature parameter),
and  the flux thickness depending on the model parameters for
polytropic fluids. We discussed also the variation of these quantities depending on the distance from the source.
The \textbf{RADs} are therefore characterized by a special and distinctive ringed structure that, as  pointed out  in \cite{ringed,long,dsystem},  could be evidenced in  the X-ray emission spectrum and as an imprint of the discrete inner \textbf{RAD}  composition, or in the combined  oscillatory phenomena  associated to the  tori model observable for example by the  X-ray observatory ATHENA\footnote{http://the-athena-x-ray-observatory.eu/}.
 As a sideline of the present analysis,
 we also discussed some aspects of   \textbf{RAD} models including  proto-jet
that  can offer  interesting scenarios as associated to misaligned tori, aligned along  the toroid rotational axis, and therefore inclined with respect to the other toroidal components of the aggregate.
Misaligned disks investigation faces aspects of the accretion disk formation after different processes, where  misalignment can provide description of the early phases of formation of the \textbf{RAD} tori and probable means to justify the origin of counterrotating tori of the \textbf{RAD}.
 More generally, the analysis presented here  has a natural extension in the study of the situation for the Kerr attractors, the co-evolution of central \textbf{BH} with the disks and influence of magnetic fields.
  In a broader   perspective of analysis  we should  note that  disk misalignment is expected to  depend significantly on the central \textbf{BH} spin and particularly the dragging of frames has a determinant influence on the morphology and equilibrium of an accretion disk and this has also a notable importance in the misalignment. Magnetic fields  then have a different role in determining the initial tori inclination of the disk,   possibly supported by the flow dynamical pressure   from the  Lagrangian points, we will also focus in \cite{next}  on the case when the
 {leading \textbf{RAD} function}, defining the distribution of tori in the \textbf{RAD} with  misaligned disks   has changed to an alternative definition to include the effects of toroidal magnetic field.
Part of our analysis was also  dedicated to an evaluation of the disk geometry (specifically its thickness) which is  crucially  significant in many aspects of the accretion disk physics and phenomenology. In particular in \cite{next}
we  focus on a more specific analysis of the role of geometrical thickness in relation to the  disco-seismology effects for each toroid. In Sec.\il(\ref{Sec:sfer.J}) we  evaluated the conditions for which these can be considered geometrically thick.

In this frame we note here that
this model envisages also  the possibility  of  a static \textbf{BH}  ``embedded'' in a set of orbiting \textbf{RAD} tori--as \emph{accreting globules}--with the \textbf{BH}  horizon  ``covered'' to an observer at   infinity, with  the \textbf{RAD} representing a matter   covering  the central \textbf{BH}--see for example Figs\il\ref{Fig:sepa},\ref{Fig:Previosue},\ref{Fig:Forse1ce}.
 However, in this situation  it  remains to  discuss  the stability of these  static \textbf{BHs} immersed in the  set of accreting \textbf{RAD} tori.
In the development of such  possibility  there  are further  aspects to be considered as, for example,  the tori self-gravity\footnote{A related  interesting possibility  of this model  might consist  in  having  a  micro-\textbf{BH} embedded in a \textbf{RAD}. Micro-\textbf{BHs} are (hypothetical)  black holes, possibly with a cosmological origin  (primordial micro-\textbf{BHs}), which are essentially regulated  by an important  quantum mechanics role.   An essential aspect of these objects is related on the possibility that they could evaporate through Hawking radiation process implying elementary particles with a radiation rate that is as larger as the minimum is the size of \textbf{BH} (eventually leading to \textbf{BH} explosion).  The proposed model  may naturally   enter  in the family of different models  foreseing  for example micro-\textbf{BHs} immersed in neutron stars.  In our case we do not consider the tori self-gravity. Moreover it is essential  to establish a mechanism for the  formation of such tori orbiting  a micro-\textbf{BH}. A further interesting aspect of this possibility is the combination of matter dynamics surrounding the central \textbf{BH} and the \textbf{BH} dynamics  itself as for example the radiation emission from the \textbf{BH} together from jet emission from accreting configurations or the combination of the oscillation modes of the tori. These \textbf{BHs} where the accretion is absent would be surrounded by a empty region  of minimum radius  $M$, located between   between  $ r=2M$ and $r=3M$, with the perimeter of the accreting inner edge in  $]8\pi M,12\pi M[=]25.1327M,37.6991M[$.
This system could provide  an interesting combination of classical and quantum effects. (Note the primordial black holes of initial mass around the  $10^{15}$ grams would already be evaporated. Larger masses  \textbf{BHs} could now be observed through the emission of  $\gamma$ rays associated to their activities).
We note, however, that such  \textbf{BHs} in the final stages of evolution may no longer be described as as a classic black hole.
}.
An interesting aspect of the eventual presence of these spherical globules (on static attractors) would be the  presence of the differently oriented spins  orbiting tori. The tori instability can lead  then to a collapse with very high energy release.
The globulous assumption  clearly implies  a frozen (not dynamical) situations for at least a  significant  time scale in the attractor life, where the central black hole  can be still considered spherically symmetric.
The possibility that such globulous exists in a frozen state  has to likely feature small or zero \textbf{BH} spin, and tori orbiting at large  distance from the attractor (depending on  $r/a$,  $r/(a \sin\theta)$ and  $\ell/(a \sin\theta)$, where $a$ is the Kerr \textbf{BH} spin and $(\theta,r)$ are Boyer--Lindquist coordinates \cite{pugtot})  to reduce the dragging effects from the center spinning attractor  that would imply the evolution in configurations with torsion of matter due to the Lense--Thirring effect.
The unstable phases  of the globules  would be associated   to great energy release, with \textbf{BH} accretion rates  being approximately the sum  $\leqslant\sum_{i=1}^{n_{\times}}\dot{M}_{i}$
of the accreting tori  rates considering however the possible collisional effects, that could also lead to phases of interrupted accretion--see Sec.\il(\ref{Sec:energ-RAD-poli}).
Among other  possible observational effects that could  be associated with these structures, we mention the  possibility to observe tangled  {luminous anuli}, shaping the accreting \textbf{RAD} inner structure essentially related to the  \textbf{RAD}  inner edge,  supposed to be  the most active part of the  accreting disk.
(For a more accurate analysis of optical effects see for example \cite{KS10,S11etal,Schee:2008fc,Schee:2013bya}).
On the other hand, the zones of tori collisions with wide angle separation (in particular orthogonal tori) are the  surfaces areas $\la_i$.
These  regions, two for each pair of  colliding tori,  could appear as active  (unstable)  ``bright knots''. The knot formation phase could  be totally or partially overlapped  with  the formation phase of one or both \textbf{RAD}  tori of the configurations.
 Such bright (active) knots  have a radial separation (linear distance) and  can be evaluated as maximal at $2 r_1$, (for a maximum crossing) being $r_1$ the location of the inner edge of the torus of minimal area.
Moreover, more generally the velocity difference between the fluids streams in the knots and between these and the torus may produces a shear effect triggering some kind of instability. These instabilities eventually may grow into  material that (periodically) accretes on the configuration.
The possible origin and evolution of such knotty structure is however closely related to the timescale of the processes related to knots. The dynamical timescales characterizing the knots   should be considered in a dynamical model,   considering explicitly the collisional effects.

\medskip

 We  conclude summarizing   the main  aspects of the model setup and  discussing the  possible \textbf{RAD} associated  phenomenology. {We also mention  some topics  arising from the analysis developed in this investigation that would be planned  for future  work, for example the analysis of the proto-jets colliding with  tori in \textbf{RAD} and the introduction of the  \textbf{BH} spin in this \textbf{RAD} set-up.}

\begin{itemize}
\item[-]\textbf{The model setup}

This analysis is grounded  upon some assumptions.
\textbf{(I)} Firstly we considered a  central static \textbf{BH}, which is a convenient approximation as a first approach to the exploration of the misaligned tori in the \textbf{RAD} frame.  \textbf{(II)} We then assumed a GRHD perfect fluid, which is particulary adapted to  obtain  constraints  for the set of tori, especially  for GRMHD initial configurations. On the validity of this assumption with  respect to the more realistic case for example where magnetic fields are expected to play  a relevant role or in dynamic model, is clear considering that   the  torus  morphology     well adapts to the configurations provided even at later time in the  analysis made in the numerical GRMHD dynamical case.
  Many aspects here considered are therefore  well described  by HD approximation.
\textbf{(III)} The third assumption of  the model sup-up concerns the  disk rotational law.  This issue is a complex topic of  the  physics of an accretion disk, especially in the GRMHD-MRI frame,  where the angular momentum  distribution and angular momentum  transport in the
disk, is entangled to the mechanism of accretion and  related to turbulence  and the  viscosity inside the disk. This  is in fact  a controversial aspects in accretion process, particularly for geometrically thin accretion disk.
  Here  we faced this issue considering two main assumptions: \textbf{\emph{(1)}}  we adopted a Keplerian rotational law for the \textbf{RAD}  leading function,  $\ell(r)$ in Eq.\il(\ref{Eq:lqkp}), that represents the angular momentum distribution for the \textbf{RAD} tori set; \emph{\textbf{(2)}} For  each torus of the cluster we have assumed $\ell=$constant.
 The choice of  $\ell(r)$ has an immediate geometric sense  detailed  in Sec.\il(\ref{Sec:RADloq})   being directly connected  to  gravitational part of the fluid force balance equation (the Euler equation) and the system symmetries. The rotational law is per-se  arbitrary,  but this assumption is supported moreover by the  consistence with the von Zeipel results, and it is used particularly in the comparative analysis with the so called  Bondi spherical accretion  characterized by a disk slow rotation (sub Keplerian i.e. $<\ell(r)$)--\cite{abrafra}.  The choice $ \ell =$constant for  each torus and therefore the torus  parametrization with the $\ell$ value  is a very convenient choice in the  HD \textbf{RAD} macrostructure  scenario, since  here  we are more concerned with the global issue the \textbf{RAD} structure, as the tori location in the cluster, the location of the  maximum and minimum pressure points and other morphological characteristics, rather then the details of the physics of  each torus.
 This assumption on the other hand is very well known and widely adopted.  Indeed, more generally   for geometrically thick configurations it is  assumed that the tori  are essentially regulated by  effects of strong gravitational fields, considered dominant with respect to the  dissipative factors   and therefore having the major role  in the  unstable phases of the system. This assumption consequentially translates in  the assumption on  time scales of dynamical, viscouse  and thermal processes characterizing the torus and translating in the form of fluid energy  momentum tensor, the equation of state (and eventually the constitutive equations)  having  consequences clearly on the definition of the disk rotational law (centrifugal forces).
This in turn grounded the assumption of  perfect fluid energy-momentum tensor too. Therefore   during the evolution of dynamical processes, the functional form of the angular
momentum  together with the  entropy distribution depend  on the initial state of the system only rather then on
 the  less influential dissipative processes, leading essentially to an
 ad hoc momentum  distribution  inside the disk  \cite{Abramowicz:2008bk}. This feature  constitutes a great advantage of these models  which in fact result    extremely useful and predictive .
On the other hand, in these models the entropy is constant along the flow and according to the von Zeipel condition, the surfaces of constant angular velocity $\Omega$ and of constant specific angular momentum $\ell$ coincide and  the rotation law $\ell=\ell(\Omega)$ is independent of  the details of the equation of state. We include further considerations on these surfaces with regard to  the \textbf{RAD} fluid characterizations in Sec.\il(\ref{Sec:zeipel}).
\textbf{(IV)} A further assumption concerns the system  symmetries. A static spherically symmetric central Schwarzschild  \textbf{BH} is assumed for this  first analysis of misaligned tori cluster,   secondly  we assumed the  fluid flow be toroidal (we have use the analogy with a   magnetic-like orbiting  multipole),   obtaining   \textbf{RAD}  steady (or stationary) states  which is the usual assumption to construct these toroids--see also \cite{pugtot}.
\item[-]\textbf{RAD associated phenomenology and observable characteristics}
\begin{description}
\item[--]
\textbf{Recognizing a RAD from an emission spectrum}

From phenomenological viewpoint, the \textbf{RAD} unstable phases
turn  significant for the high energy phenomena, as especially related to accretion onto super-massive
black holes in \textbf{AGNs}, as well as  the extremely energetic phenomena in quasars observable in their X-ray emission, as  the X-ray obscuration and absorption by one of the \textbf{RAD} torus.
In this respect we  should emphasize two principal observational aspects  related to two typical  accretion disks dynamics that, framed within the \textbf{RAD} scenario, provide  interesting implications and  patters to recognize these structures: first we mention  the tori  proper modes of oscillation for the   misaligned tori  clustered in the \textbf{RAD}.  The radially oscillating tori of the ringed disk could be related to the high-frequency quasi periodic oscillations
which are now observed in non-thermal X-ray emission from compact objects (QPOs). The second  consequential element   of the \textbf{RAD} structure consists in the jet emission from the tilted toroidal components of the \textbf{RAD}.
Concerning the possible correlation with QPOs emission, which is considered in \cite{next}  the oscillations  of each component are added to others and are pulsations of the \textbf{RAD}, and  possibly the  globule creating   a rather distinct detectable emission spectra.
This is a still unclear feature of the X-ray astronomy which has been  largely
related to the physics of accretion and more specifically to the  inner parts of the disk.
More generally \textbf{RAD} inner structure may be  revealed by  X-ray spectroscopy, as relatively indistinct excesses
on top of the relativistically broadened spectral line  profile \cite{S11etal,KS10,Schee:2008fc}, where the predicted relatively indistinct excesses
 of the relativistically broadened  emission-line components,   arise in a well-confined
radial distance in each    toroid and the \textbf{RAD}.
\item[--]
\textbf{Recognizing a  "globulus" structure}

 Globules are interesting from the observational point of view constituting in fact a  possible \textbf{BH} embedding. The orbiting  matter  is a    multipole-system characterized by the presence of different tori.  It is  in fact possible that  these systems may be originated from different accreting periods of the central attractor life interacting with its galactic environment therefore being composed by  matter with diversified characteristic. A globulus can be pictured as a configuration of
 very thick,   Keplerian or super-Keplerian tori  having  angular momentum distribution equal or superior to $\ell(r)$ in Eq.\il(\ref{Eq:lqkp}). This structure would cover  almost completely the  central \emph{embedded} black hole. The \textbf{BH} horizon is then covered almost completely  from the observer-view at  different angles,  as  an \emph{embedding} of gravitating orbiting matter around the \textbf{BH}. Hence  the model  would picture a  \textbf{EBH} system, leading in fact  to consider  the  \textbf{BH}-disk  as one entire  single object. (In this model we  exclude  the  tori  self-gravity). The dynamics  of  \textbf{BH}-disk system is in fact often considered in this unified picture   for example in different approaches foreseing \textbf{BH}  energy  extraction  from interaction with the surrounding orbiting tori or  measurement of the \textbf{BH} spin by the analysis of the inner edge of the disk.  A globulus would be  recognizable by   a series of main  typical characteristics: \textbf{\emph{(i)}} the first element typical of a globulous structure   is  the presence of an internal-globulus  vacuum zone,   the region in the range $[r_+,r_{inner}^i]$, where  $r_{inner}^i$ is the inner edge of the inner quiescent  torus of the globulus and  radial distance $r$  (in Boyer--Lindquist frame) is  with respect to the central singularity. Of course  if  on the other had the inner torus is accreting, then this region would be characterized by the  presence of matter accretion onto the attractor.   The occurrence of this condition is thoroughly analyzed   in this analysis considering  $r_{inner}^{\times}$, cusp location and conditions related  such configurations. \textbf{\emph{(ii)}} The system is characterized by a complex, specific "multipoles structure" (with  "multipoles number"  equal to  $2n$ for $n$ \textbf{RAD} toroidal components, so that it would be a dypole in the Schwarzschild \textbf{BH}  case with one  torus)  characterized  by the different (\emph{satellite}) spins due to the  different mislaigned  tori rotation orientations and  the
central (\emph{primary}) \textbf{BH}  spin which in the case considered in this article is null. (In this article we consider more specifically a magnetic like multipole, the  fluid having only a toroidal flow). \textbf{\emph{(iii)}} The globulus is  subjected to  different  constraints due to the emergence of tori collision. Such constrains reduce to conditions  on the tori angular momentum, which is the value of model  parameter  $\ell$ characterizing each torus, a value  of the \textbf{RAD} rotational curve $\ell(r)$, and tori  morphology, here considered determined by  a set of tori morphological characteristic $\mathbf{M}$ as the torus geometrical thickness, height and  elongation on the torus symmetric  plane, emergence of cusp instability here largely  considered Sec.\il(\ref{Sec:Misal}) and Sec.\il(\ref{Sec:doc-ready}),  and other characteristics related to \textbf{RAD-BH } system energetics  $\mathbf{E}$  as the accretion rates or the cusp luminosity, considered in Sec.\il(\ref{Sec:energ-RAD-poli}). (Obviously in the case more specifically of globulus   it is necessary to consider the constraints imposed by the regions  in Figs\il(\ref{Fig:torithick},\ref{Fig:SMGermMs}) to identify each torus based on its geometric thickness,
 and Figs\il(\ref{Fig:woersigna}), thus deducing the conditions on tori parameters  and tracing them through these constraints on their morphology). \textbf{\emph{(iv)}} A peculiar feature relevant  for the observation and recognition of the \textbf{RAD} and globuli particularly is its extension and location of the outmost torus of the orbiting aggregate. This issue has been addressed evaluating the constraints on the $r_{out}^{o}$ outer edge of the outer torus of the \textbf{RAD}, and more generally the limiting on the locations of  maximum density point in the farthest torus , we mention particularly Sec.\il(\ref{Sec:ount}) and Sec.\il(\ref{Sec:zeipel}).
\textbf{ \emph{(v)}} Finally the globulus would be recognizable by an  articulated internal life triggered by each torus dynamics and empowered by the   entangled, ringed inner structure, made up by  a sequel of typical associated phenomena as tori collisions,  tori oscillation modes (eventually related to QPOs emission as we have considered in \cite{next}), and possibly internal accretions and jet emission.
\item[--]\textbf{On the jet emission in the RAD}

 A further interesting aspect of the \textbf{RAD} multipole structure is the possibility that this may be reflected in the associated jet emission.
  Regarding the  internal \textbf{RAD}  jet emission, the  possibility of an extended jet launching region has been widely considered  for the  \textbf{eRAD} model  in \cite{open,proto-jet,dsystem,long}.  Jet emission in \textbf{BH} accretion disks system  in fact  can be equally considered autonomously  from the occurrence of the open solutions with cusp, predicted in the perfect fluid tori considered here  under special   conditions on momentum and identified with "proto-jet".

  The  (multi-polar globules with) shells of jets in accreting balls,
can have huge consequences for  the \textbf{RAD} stability and observation and, possibly, having impact  also for  the galactic environment enrichment process.
In the cluster frame presented here and more generally in the \textbf{RAD} frame,  there are several points to be addressed concerning this intriguing aspects of the orbiting agglomerate.

Jet emission has been in fact largely  directly connected to accretion and more specifically  accretion disks and the inner regions of accreting disk.
Although the issue of jet  launch still needs to be  resolved,  a large part of  the current analysis  relates, at least geometrically, the inner edge of accreting disk to jet emission.  Consequently,  in this  frame of   the torus-jet emission correlation,  we  could  consider the occurrence for each toroidal component of the globule,   predicting accordingly  the presence of inner \textbf{RAD} globulus jets, whose detection would be an indication on the presence of the \textbf{RAD} structure.

   It should be pointed out that the orientation of the jet in the case of static attractor  can be  supposed to be orthogonal to the (torus) plane of symmetry resulting in the rotation (the dipole) direction, in the case of attractor with spin. We also know  to be ascertained  that the spin of the central attractor  has indeed a role in the launch of jets, in the collimation and in the determination of the direction of the jet, together with the presence of the  magnetic fields. The jets presence  is a  complex scenario that,  if observed, would open an important observational  window  both on the determination of the physics of \textbf{BH} and on the understanding of process of  energy emission.
 There are important  aspects connected  the HD proto-jets  structures, associated to the  energetics in  the \textbf{RAD} tori environment. The analysis in   \cite{ringed,open,dsystem,long},  introducing  the  \textbf{eRAD} models,  details the HD matter, multi-proto-jets in the Kerr \textbf{SMBH} spacetime having, therefore,  also an initial (in the vicinity of the launch point $r_j$)  counter-rotating component, and mixed corotating/counter-rotating funnels of material.
In \cite{proto-jet} we focused specifically  on  proto-jets configurations in \textbf{eRADs} orbiting a Kerr \textbf{SMBH} considering  thoroughly the  symmetries and limiting surfaces and  highlighting  the
 boundary limiting
surfaces  connected to  the emergency of the jet-like instabilities with the black hole dimensionless
spin. The energetics is addressed more specifically  in   \cite{Multy} by concentrating on the energy  released during  collision  and  time-dependent accretion   examined  in the  more simple
example of situations where deterministic chaos might
emerge, considering a modified accretion rate law. Then  jet emission  was considered in  \cite{Letter}   with emerging tori collisions
around super-massive Kerr black holes,
providing an evaluation of center-of-mass energy for two
colliding particles from the two interacting tori.

In future work we are planning to consider more specifically  the impacts  of the  proto-jets   with accreting tori  in the \textbf{RAD} configurations as globules.
In the case of Kerr \textbf{BH}, there can be the occurrence of   a  more external shell, "breaking" the internal accretion disk. This limiting occurrence, regulated by the background geometry and precisely by the  Kerr \textbf{BH}  dimensionless spin distinguishes also  the torus direction of  rotation--\cite{open,proto-jet,long}. In the case of a spherically symmetric spacetime,  this occurrence does not exist.  Considering the shells in Fig.\il(\ref{Fig:SIGNS}) and analysis of constraints in Sec.\il(\ref{Sec:doc-ready}), there are completely separate and  disconnected regions where these configurations can emerge, and yet in some of these it is still possible that   interaction between structures with maximum density in different regions  give rise to collision or possibly replenishment  and exchange of matter. The outer region is   associated  to generally  very thick tori with a large centrifugal component (with respect to the other components in  force balance equation of  Eq.\il(\ref{Eq:Eulerif0})) and with the characteristics  of energetics  considered in Sec.\il(\ref{Sec:er}). This region is followed by the intermediate region of quiescent tori and proto-jet emission and the inner region with quiescent or accreting tori having lower density and  momentum.
This is obviously because in the case of a static central attractor the direction of rotation of the fluid is not diversified against the gravitational background.
Noting that proto-jets  are open, cusped solutions  associated to geometrically thick tori as described by the model in Sec.\il(\ref{Sec:doc-ready}) and \cite{open,Lasota:2015bii},  we can expect that the associated toroidal fluid configurations provide eventually  also the source material,  and  determining the   initial   prompting  centrifugal component, especially in the HD model orbiting  a static central attract.  

Considering   different "companion disk"  models, this enlarged ground of elements affecting the proto-jets funnels can be enriched by the  viscosity and resistivity terms  which, as we have largely discussed in this analysis, in the context settled here  for the  constraints investigation, can be neglected in first approximation. Here we consider the limiting conditions on  fluid centrifugal components and the general relativistic effects of a Schwarzschild background, using therefore an adapted HD model as base for the enlarged situation with a  richer, more complex, embedding of accretion. On the other hand  in many GRMHD analyzes the only HD has proved to be a good comparative pattern and certain a good initial condition.

   However,  in \cite{proto-jet} we have  the dragging  due to the Lense--Thirring effect around  a Kerr \textbf{BH}, affecting possibly both  launching and collimation  of  the corotating  and counterrotating component of  the fluid and, at the same time, in the \textbf{eRAD} components the relative directions of proto-jet rotation is  parallel or antiparallel in all jet shells.  By imposing a static background, we have maintained the mutual co-rotation or  counter-rotation of the funnels, but this possibility is now expanded including  different relative angles of the rotation axes: consequently the proto-jets are no longer "parallel" or "anti-parallel", but they are  always "orthogonal" to the rotation plane of the  orbiting toroidal embedding  material  which eventually  constitutes both the replenishment  material and collision and centrifugal support.

Although proto-jets are expected to be somehow
 transient structures,  the presence of the tori of \textbf{RAD} may constitute a reservoir materials and momentum
to reform or enact possibly also associated to occurrence of  runaway  or  runaway--runaway instability,  constituting eventually   interrupted  sequences of complex outbursts with different shells  and a structured orbiting  disk. These solutions have  been differently associated to jet emission empowered by initial unstable fluid  centrifugal component associated to  a  minimum of pressure correspondent to the  surface  cusp.
In these structures we proved that the centrifugal  component  should be not "too large" 
that  the disk is stabilized against the formation of the cusp\footnote{We proved that proto-jets  are associated to fluids having initial specific angular momentum $\ell\in ]\ell_{mbo},\ell_{\gamma}[$  with  $r_{j}\in\mathbf{ R_1}=]r_{mbo},r_{\gamma}[=]3M,4M[$,  therefore the cusp is  located in an instability shell of radius $\Delta r=M$  located at $r=3M$ from the attractor and separated by the photon sphere  with $r_j<r_{\times}\in \mathbf{R_2}$. The center of maximum pressure is located in a shell of radius $\approx11M$  ($r_{center}\in \mathbf{R_4}=]r^b_{mbo},r^b_{\gamma}[ = ]10.4721M,22.3923M[$).} and are  in fact  "geometrically not correlated" directly with accretion  where the accreting  fluid has initial specific angular momentum lower then  the fluid supporting proto-jet, whose cusp is  closer to the \textbf{BH} then the accreting  point. The proto-jet structure is a shell englobing the accreting configuration as shown Fig.\il(\ref{Fig:SIGNS})  distinguishing  these  configurations from other open structures.
 It is  possible that tori eventually formed in the condition  for proto-jet as described here could more or less rapidly undergo a phase of angular momentum decreasing  bringing the  torus  in the condition for accretion.
The presence of proto-jet cusp is also regulated  by the $K$ parameter,   corresponding  to mass and density term parameter,  related  to very large tori, hence the reservoir of mater and funnels. There could be the concomitant formation of internal proto-jet associated to an outer  toroid and related to a disk  between them, which is   in accretion, replenishing also the cusp of the proto-jet.
The fluid of the inner shell has a higher specific  momentum of the intermediate shell where there is an accretion point and maximum pressure of the accretion disk, which then replenishes the jet with a  fluid with initial lower moment.
It is in fact  clear that the replenishment fluid  can have then a very  diversified (with respect to the proto-jet funnel) initial  rotation direction  and matter properties component.    We firstly propose  that  there are multiple, and  simultaneously tilted, proto-jets with very different materials, and secondly we foresee the possibility of impact   in the  \textbf{RAD}  of jet on disk which we plan to analyze with more details  elsewhere. Therefore for a forthcoming paper we plan to extend this set up to consider  in a clear and systematic way the jets and tori in schwarzschild geometry as well as  tilted tori in an approximated Lense--Thirring geometry and  the Kerr geometry.
\item[--]
\textbf{Outbursts and  geometry perturbation  triggered by RAD instabilities}

 The occurrence of the principal \textbf{RAD} instabilities initiates an interesting set of   associated \textbf{RAD} phenomenology. The consequences of the \textbf{RAD} instability processes  have to be considered  in the analysis of  energetic involving a \textbf{SMBH} interacting with orbiting tori.
There are four main typical \textbf{RAD} instabilities which can lead  to some kind of disruption or even catastrophic destruction  of the \textbf{RAD} internal structure characterized by large and relatively fast release of matter and energy outburst triggered from the occurrence of  tori accretion,  tori collision or a combination of these two effects.  We studied the energetics of \textbf{BH} accreting disk related to accretion phase  in Sec.\il(\ref{Sec:energ-RAD-poli}) in functions  of  the  model parameters, location of the cusps and   distance from the central \textbf{BH} $r/M$ scaled with the \textbf{BH} mass $M$. We include, in the discussion of \textbf{RAD} phenomenology, some further notes on the fourth  \textbf{RAD} instability consisting in the \emph{runaway-runaway } instability.  The runaway instability has also been considered playing a part in the   energy  extraction from the central \textbf{BH} engine,  the extraction of  the rotational energy of the  spinning  black hole  via the Blandford-Znajek mechanism,  and  in the \textbf{GRB} production. The  \emph{runaway-runaway }  process, triggered by the occurrence of runaway instability prompted by accretion from the inner   torus of the \textbf{RAD}  induces also a change in the \textbf{BH} parameters, therefore changing the spacetime geometry and  consequently changing the \textbf{RAD} structure. While the accretion from the inner  torus may establish the runaway instability, this can be  accompanied by a further process of torus-torus interaction, inducing  the other instabilities processes.
    Such tori collision may go  through a positive or negative feedback reaction, since
    the runaway mechanism is  essentially   an interactive process   involving both the attractor and  the accreting torus initially   due to the interaction of the  \textbf{BH}  with the  inner accreting torus via accretion of matter flowing  into the \textbf{BHs}----\cite{dsystem}. If  a runaway of the inner  \textbf{RAD} torus is established, then also the  outer torus of the \textbf{RAD}  is affected.  The  runaway instability is regulated  by
several  parameters, particularly the  dependence of the mass accretion rate-- here considered in Sec.\il(\ref{Sec:energ-RAD-poli}), the initial  the cusp location--Sec.\il(\ref{Sec:doc-ready}), the flow thickness--Sec.\il(\ref{Sec:sfer.J}).
Because of the change in the mass \textbf{BH} parameter, the accreting torus cusp can  move deeply  towards the torus, and
   increasing the mass lost rate $\dot{M}$
  due to  the cusp will bend inside the torus, \textbf{BH} mass increases, according to  the  mass accretion rate. If the non-zero \textbf{BH} spin is constant during this process then   this implies a  decreasing the spin-to-mass ratio. A similar  mechanism  of \textbf{BH}-accreting torus interaction can   arise  also after a  spin-shift for energy extraction for example, eventually  the question if there  is a phase of accretion process where  the central (static or spinning) \textbf{BH} may acquire an  electrical  charge and the  role  of this phase in the subsequence phases of \textbf{BH} matter interaction is in fact still a wide open  (not irrelevant) issue. A possible consequence of the entire process is however  that   the
accretion tori and  consequently  the \textbf{RAD}   never reach a  steady state. Note that otherwise the accreting inner  torus could even   be completely destroyed by such instability.
We should also note that, from the point of \textbf{RAD} morphology structure,  the accreting torus cusp may also move inwardly i.e.
towards the  central black hole, with the consequent   decreasing of  the mass transfer, ending eventually in  stabilizing
the entire  process. The runaway instability when  the cusp   moves outwardly,
penetrating the torus, results   in an increases of mass transfer rate. We considered  the dependence of the accretion  rate on the cusp location and other parameters in the analysis of Sec.\il(\ref{Sec:energ-RAD-poli}).
It is worth to point out however how
the study of this  situation  has been    carried out by considering  stationary
models \cite{Font02}, in a non-dynamical framework and therefore particularly adapted to the \textbf{RAD} scenario constructed here.
 In this approach the evolution of
the central black hole is  described  as a sequence of exact  black
holes solutions with a   different  mass, function of the  mass
accretion rate of the former state.
Pictures of  the systems during stages of its dynamics were studied, this method was carefully analyzed and discussed for geometrically thick torus around Kerr \textbf{BHs}    in \cite{pugtot}.
Finally we note that a relevant factor of this process especially in the \textbf{RAD} frame, where there is  the combination several processes and dynamics, consists in the typical   time-scales   that should be compared with the time-scale of the dynamical processes of the accreting torus  and  \textbf{RAD} dynamics.
\end{description}

\end{itemize}

{Misaligned disks are usually located, by observational evidences,   in
\textbf{AGNs} at
 relatively large distances from the central \textbf{SMBHs} \cite{2006MNRAS.368.1196L,Herrnstein1996,Greenhill2003}.  A warped inner accretion
can explain
the relation between radio-jets in \textbf{AGN} and the galaxy disk.  Evidences of misalignment and  details  on observational expectations of  the structure of  tilted disks and proto-jets  are discussed  for example in \cite{Miller-Jones:2019zla}. These investigations are focused on
  relativistic jets in a  stellar-mass
black holes
 launched  and "redirected" from the accretion
 and subjected to the  frame dragging effect.
Others are more focused on  the   images of
accreting black holes in presence of the
disk and jet misalignment \cite{Chatterjee:2020eqc}, developed by  3D
relativistic MHD simulations \cite{2011ApJ...730...36D,Fragile:2008sv}.
Many of these investigations address  particularly   the  flow structure and  the inner region of the
tilted disk
\cite{KKK,LLL,Teixeira:2014una}.}

{Therefore, observational data concern more accretion periods of  \textbf{SMBHs} located in \textbf{AGNs}, leaving  traces in counterrotating and even misaligned structures orbiting around the \textbf{SMBHs}.
\textbf{RADs}  might represent   the  episodic   accretion phases   with different evolutive patterns \cite{apite1,apite2,apite3,Li:2012ts,Oka2017,Kawa,Allen:2006mh},   grounding  the  process of accretion for  \textbf{SMBHs}      from    intermediate   or low mass \textbf{BH} seeds  ($10^4-10^2 M_{\odot}$), especially at cosmological distances with redshift  $\approx 6$.
These structures would be recognizable from the
      long and continuous accretion episodes  due to  tori   merging, and  eventually    involving also a relevant spin-shift and a  succession of accretion episodes  from misaligned disks, characterized by
 a sequence of   turning-on and turning-off of  super-Eddington accretion phases  to  sub-Eddington phases. The implications and potential observational consequences  of the new results connected to \textbf{RAD} analysis developed here  reside  on two different sides. Firstly,  the results  base the analysis of the peculiar limiting effects developed  in \cite{next} where the  implications of the \textbf{RAD} structures in  \textbf{QPOs} emission are analyzed. Further  peculiar object associated to the \textbf{RAD} morphology  is the orbiting globulis covering the central \textbf{BH}, a limiting situation that could give rise to a huge release of energy due to the unstable modes,  with a complex structure injecting matter   collapsing  onto the central attractor. In the simulation and observation of more complex systems in evolution, we would recognize  the constraints derived by quantities easy reducible to the $(\ell,K)$ parameter set   considered here. In these complex systems constraints would appear,  resulting from   the analysis of  Sec.\il(\ref{Sec:mirpj}) and Sec.\il(\ref{Sec:doc-ready})  evaluating the  \textbf{GRHD} effects.}

{As previously stated, one of  the future targets of this analysis is the extrapolation of the results and analysis developed here for the rotating background case represented by a central  Kerr \textbf{SMBH}. In Sec.\il(\ref{Sec:intro}) we discussed the role of Schwarzschild limit  as a first-step analysis for the spinning case. The exploration of the static geometry case is indeed useful to discern the Lense--Thirring   effects on the disks with respect to other factors contributing to the disks dynamics. As  noted in \cite{dsystem} for the \textbf{eRAD} in Kerr spacetime,  the  inner ringed structure is different,  in relation to the classes of different spin-mass ratios of the central attractor, for the set of corotating and counterrotating tori as related to the attractor, and for the  $\ell$corotating and  $\ell$counterrotating sequences.
For all these reasons the analysis of the Schwarzschild backgroundg serves as comparative test and first analysis step for more complex situations, for the extension to Kerr background and comparison in different situations determined also  by   different characteristics of the disk. }
\begin{acknowledgements}
D. P.  and Z. S. acknowledge the financial and institutional support of  the Institute of Physics, Silesian University in Opava.
Z. S. acknowledges the support of the Czech Science Foundation grant 19-039505.
 \end{acknowledgements}
\appendix
\section{Further notes on  tori construction and limiting configurations}\label{Sec:app}
\subsection{On   $\ell(r)$ and $K(r)$ functions}\label{Sec:four-b}
  In this section we briefly discuss  some geometric and dynamic  aspects  underlying the definition of the specific angular momentum adopted in  Eq.\il(\ref{Eq:lqkp}) and energy function $K(r)$ of Eq.\il(\ref{Eq:sincer-Spee}), focusing on their geometric origin related to the symmetries of the Kerr background.  The significance of quantity  (\ref{Eq:lqkp})  for  the extended matter in orbit in the \textbf{RAD}  context is then explored.
For this purpose we start  by considering  the Schwarzschild metric
\begin{equation}\label{11metrica}
ds^2=-e^{\nu(r)}dt^2+e^{-\nu(r)}(r)dr^2
+r^2\left(d\theta^2+\sin^2\theta d\phi^2\right),\quad  e^{\nu(r)}\equiv\left(1-2/r\right),
\end{equation}
 written in standard spherical coordinates
$(t,r,\theta,\phi)$.
 It is convenient to list here      the following   quantities
\bea\label{Eq:pro-conserva}
&&
\mathbf{(a)}\quad\{\Lambda\equiv u^r,\quad\Sigma\equiv u^t,\quad\Phi\equiv u^{\phi},\quad\Theta\equiv u^{\theta}\};
\quad\mathbf{(b)}\quad\{{E} \equiv -g_{ab}\xi_{t}^{a} p^{b}=-g_{tt} \Sigma ,\quad L \equiv
g_{ab}\xi_{\phi}^{a}p^{b}=g_{\phi \phi}\Phi\}, 
\eea
 the $\mathbf{(a)}$-quantities are the four  fluid velocity components and  $\mathbf{(b)}$-quantities are the  constants of motion for test particle geodesics with four-momentum $p^a$,   which are related to the  Kerr geometry Killing vectors $\xi_t$ and $\xi_\phi$.
From  $\mathbf{(a)}$ and  $\mathbf{(b)}$ we define the following
 quantities $\mathbf{(c)}=\{T, V_{eff}\}$,  together with the  relativistic angular frequency  $\Omega$ and the fluid specific angular momentum $\ell$ for Eq.\il(\ref{Eq:lqkp}):
\bea\label{Eq:poll-delh}
&&\mathbf{(c)}\quad\left\{\Lambda \equiv\sqrt{E^2-V_{eff}^2},\quad
\Theta\equiv \frac{T}{r^2}\right\};
\quad\mathbf{(d)}\quad\left\{\Omega \equiv\frac{\Phi }{\Sigma }=-\frac{g_{tt} L}{{E} g_{\phi \phi}}= -\frac{g_{tt} \ell}{g_{\phi\phi}},\quad
\ell\equiv\frac{L}{{E}}=-\frac{g_{\phi\phi} \Omega }{g_{tt} }=\frac{g_{\phi\phi}}{g_{tt}}\frac{\Phi}{\Sigma}
\right\}.
\eea
Assuming $\Lambda=0$, for the circular configurations,  there is from  Eq.\il(\ref{Eq:poll-delh}), $V_{eff}=E$ and we obtain explicitly
\bea\label{Eq:remai}
&&\mathbf{(\Lambda=0)}\quad u_t=V_{eff}=\sqrt{e^{\nu(r)}\left(1+\frac{L^2}{r^2\sin^2\theta}+\frac{T^2}{r^2}\right)},
\\\label{Eq:cin-cont-Hong}
&&\mathbf{(\Lambda=0\quad T=0)}\quad
V_{eff}=\sqrt{\frac{-g_{tt} g_{\phi \phi}}{g_{\phi \phi} +\ell^2g_{tt}}}=\sqrt{-\frac{g_{tt} ^2\Sigma^2}{g_{tt} \Sigma ^2 +g_{\phi\phi}\Phi^2}}=\sqrt{-\frac{{E}^2 g_{tt} g_{\phi\phi}}{{E}^2 g_{\phi\phi}+g_{tt}  L(\ell)^2}}.
\eea
 In Eq.\il(\ref{Eq:remai})  is  the \emph{effective potential} expressed in terms of $L$, constant of motion   for  test particles in circular motion,  and we included explicit dependence on the poloidal component of the fluid velocity $\Theta$, while in Eq.\il(\ref{Eq:cin-cont-Hong}),
$L(\ell)$ indicated  $L$ as function of  $\ell$ as in Eq.\il(\ref{Eq:poll-delh}). In the clusterized   set of misaligned tori, using the background and the toroidal disk symmetries, we can  consider, without loss of generality,  the condition $T=0$ (equivalent to $\Theta=0$) for each toroidal component of the cluster reducing, for each toroid   Eq.\il(\ref{Eq:remai}) to  Eq.\il(\ref{Eq:poll-delh}).
Within our assumptions $(\Lambda=0, \Theta=0,  \partial_tp=\partial_{\phi}p=0)$, from
the Euler equation (\ref{Eq:Eulerif0}) we obtain  the expression for the \emph{radial
pressure gradient},  regulated by the radial gradient of the effective potential which provides the integral form  Eq.\il(\ref{Eq:sha-conf}).
It is then convenient also to define the following angular momenta
\bea&&\label{Eq:wEo}
\mathbf{(e)}\;\{ L_{{K}}=\pm\sqrt{\frac{(\sin\theta)^2  r^2}{(r-3)}},\; L(\ell)=\sqrt{\frac{r^2(\sin\theta)^2}{e^{-\nu}r^2 \sigma^2\ell^{-2}-1}}\quad (\ell\neq0)\}; \mathbf{(f)}\;  \{\ell_r\equiv\sqrt{\frac{(\sin\theta)^2 r^3}{(r-2)}}=\sqrt{\frac{\ell}{\Omega}},\quad \ell_K=\sqrt{\frac{(\sin\theta)^2 r^3}{ (r-2)^2}}\},
\eea
where
 $L=L_{{K}}$  is the angular momentum defined in Eq.\il(\ref{Eq:pro-conserva}) as constant of for \emph{test particle circular $(T=0)$ motion},   regulated by the  effective potential in  Eq.\il(\ref{Eq:remai}). In Eq.\il(\ref{Eq:wEo})  $L(\ell)$ expresses  $L$ as function of the  fluid specific angular  momentum $\ell$ as in Eq.\il(\ref{Eq:poll-delh}) (clearly for very large radius  $r$ there is $L\approx \ell$), we note that $L$ exists for
$0\leq \ell<\ell_{r}$. In Eq.\il(\ref{Eq:wEo})-$\mathbf{(f)}$ we give
the fluid specific  angular momentum $\ell$   of Eq.\il(\ref{Eq:lqkp}) with the explicit  dependence on the  \textbf{BH} equatorial  plane. In figure (\ref{FIG:for-cre}) and Appendix\il(\ref{Sec:zeipel}) we   discuss further properties of these quantities.
 %
%
\subsection{Toroidal surfaces}\label{Sec:fasc-eur}
The analysis of the tori equatorial sections,  and the morphological characteristics of the associated  toroidal constant pressure and density surfaces given by Eq.\il(\ref{Eq:Eulerif0}), allows  to deduce  important properties  for the constraints on the orbiting  tori cluster.
From Eqs\il(\ref{Eq:Eulerif0}) we obtain the following expression for the toroidal surfaces on each plane $\theta$:
\bea\label{Eq:condizioni}
\forall \theta:\quad\left(\frac{2 \left(\mathcal{B}^2+K^2 Q\right)}{K^2 \left(Q-\mathcal{B}^2\right)+\mathcal{B}^2}\right)^2-\mathcal{B}^2-\mathcal{Z}^2=0,
\eea
where $Q\equiv \ell^2$, $\mathcal{B}(x,y)$ and $\mathcal{Z}(x,y)$ are functions of cartesian coordinates $x,y$ on an equatorial plane, for example we have fixed  $(\mathcal{B}=x \cos (\theta )+y \sin (\theta );) (\mathcal{Z}=y \cos (\theta )-x \sin (\theta ))$  in Figs\il\ref{Fig:sepa},\ref{Fig:DopoDra}--where $(x,y)$ are Cartesian coordinates.

\begin{figure}
\includegraphics[width=7.6cm]{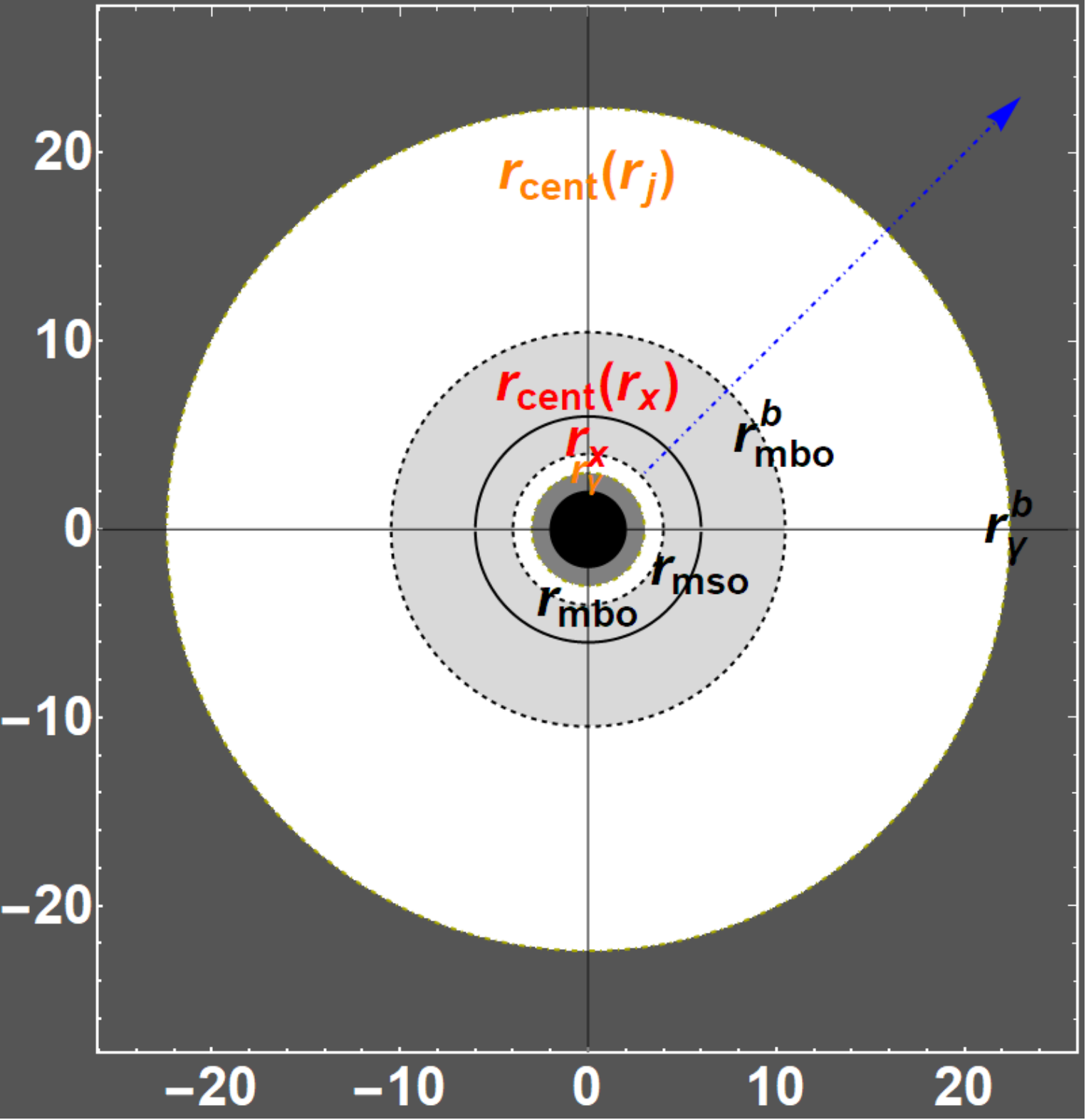}\hspace{0.2cm}
\includegraphics[width=9.09cm]{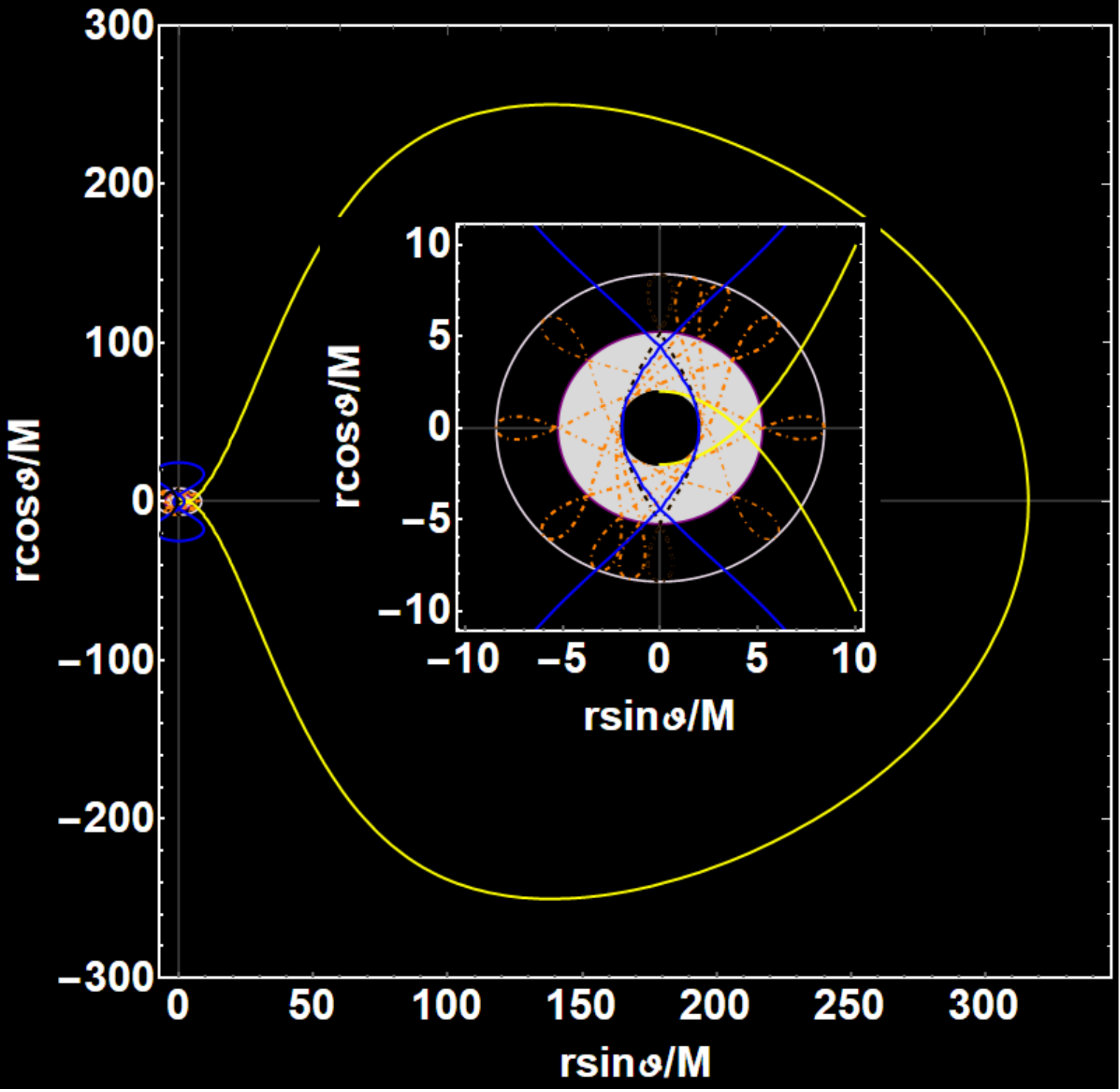}
\includegraphics[width=9.4cm]{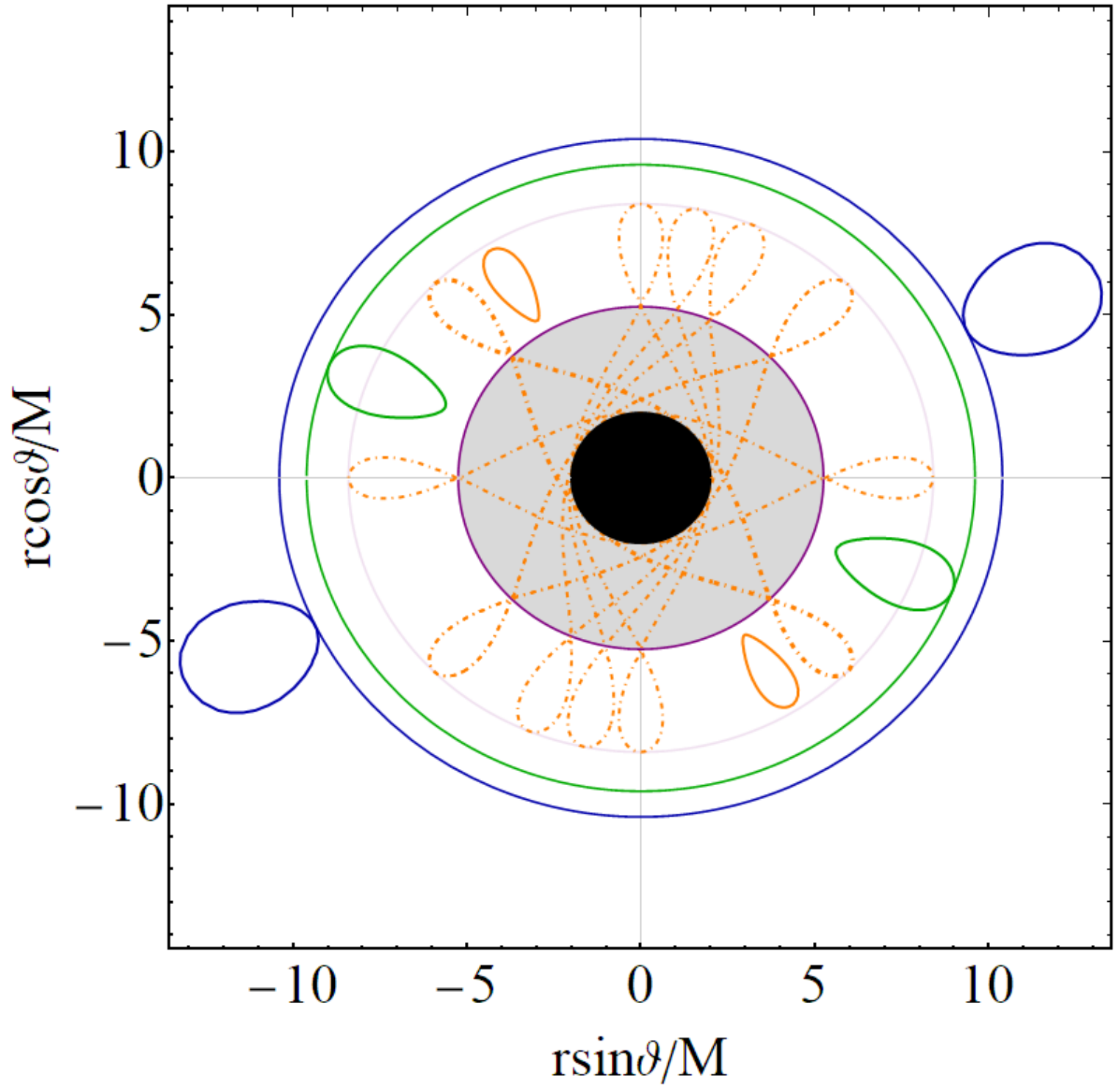}
\includegraphics[width=7.4cm]{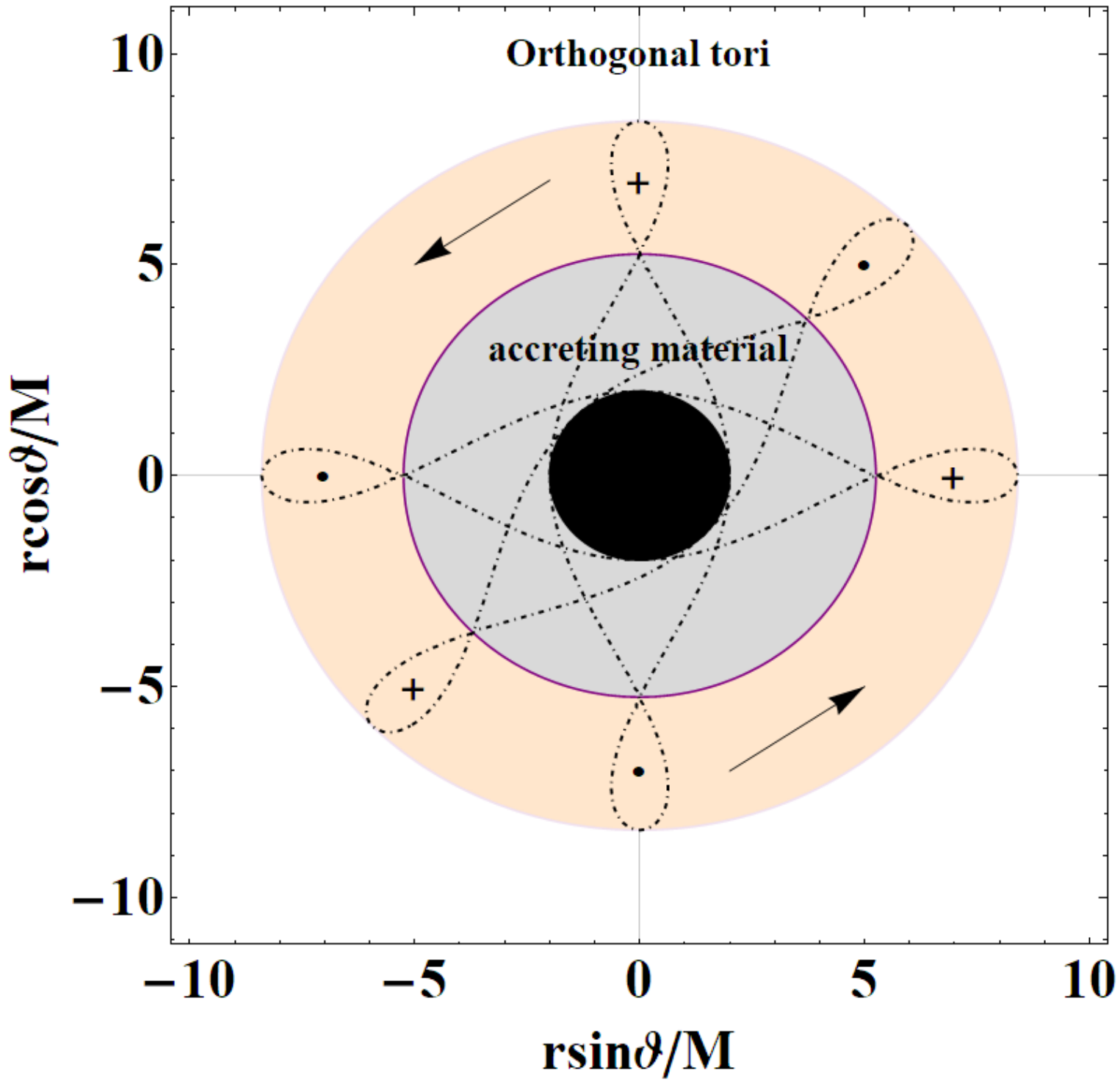}
\caption{Upper left panel: Cusps of accreting tori are in $r_{\times}\in]r_{mbo},r_{mso}[=]4M,6M[$, the center in $r_{cent}(r_{\times})\in[r_{mso},r_{mbo}^b[$.
Cusps of open cusped proto-jets configurations are in $r_j\in]r_{\gamma},r_{mbo}[=]3M,4M[$, the center in $r_{cent}(r_j)\in[r_{mbo}^b,r_{\gamma}^b[$. Configurations in $r>r_{\gamma}^b$ are quiescent--see Figs\il\ref{Fig:SIGNS}. Inner black region is the central \textbf{BH}.
Right upper panel and below panels: the tori equatorial sections as  equipressure surfaces (cross sections of the rigid Boyer surfaces) of Eq.\il(\ref{Eq:condizioni}). Central black region is the Schwarzschild \textbf{BH}. Boundary spheres $\Sa_{in}$ for the inner edge and $\Sa_{out}$ for the outer edge of tori  are also shown (colored circles centered on the central \textbf{BH} region)-- see also Figs\il\ref{Fig:sepa}.
Black arrows indicate the  velocity directions. In the orthogonal tori panel the velocity fields enter or come out orthogonally from the paper,    (dot $\bullet$)  are for ingoing  fluid  and outgoing (plus $ \mathbf{+}$) from the figure panel.
\label{Fig:DopoDra} }
\end{figure}
We now introduce the following limiting surfaces:
\bea\label{Eq:wendll}
y_s=\sqrt{\frac{4 \left[K^2 \left| x\right|  x^*+x (\left| x\right| -2)^2\right]^2}{\left[\left(K^2-1\right) x (\left| x\right| -2)^2-K^2 \left| x\right|  x^*\right]^2}-x^2},\quad
x_{k}^{\pm}\equiv\frac{3 K^2-4}{2 \left(K^2-1\right)}\pm\frac{1}{2} \sqrt{\frac{9 K^2 (K^2-8)}{\left(K^2-1\right)^2}}
\eea
($\left| x\right|$ is the absolute value of $x$, and $x^*$ gives the complex conjugate of $x$)
obtained from Eq.\il(\ref{Eq:condizioni}) where $\Qa=\ell(r)^2$ of Eq.\il(\ref{Eq:lqkp}).
These limiting configurations have a significant number of symmetries. We describe these properties by referring to the Fig\il(\ref{Fig:Foolelon}), the solutions for $K\in[K_{mso},1]$ and   $K>1$.  Surfaces  $y_s$ in Eq\il(\ref{Eq:wendll})  represent limiting  solutions for the existence of the toroidal configurations, providing the maximum and minimum limits on the inner edge of the associated rigid toroidal  surfaces.
\begin{figure}
\includegraphics[width=8.69cm]{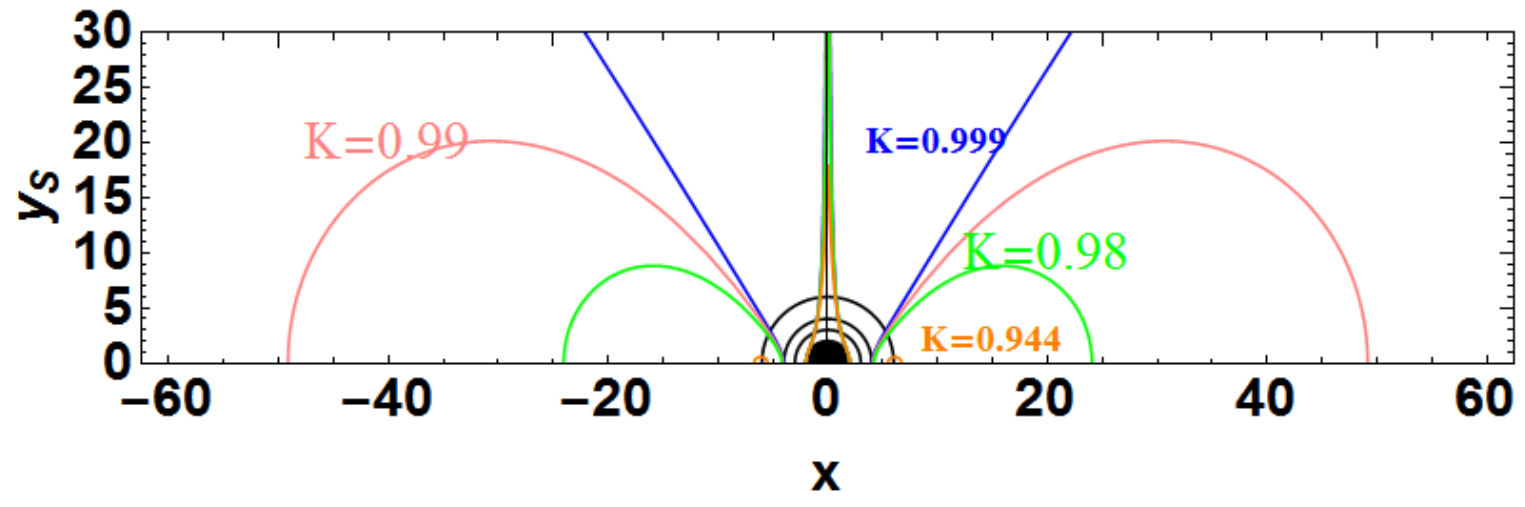}
\includegraphics[width=8.69cm]{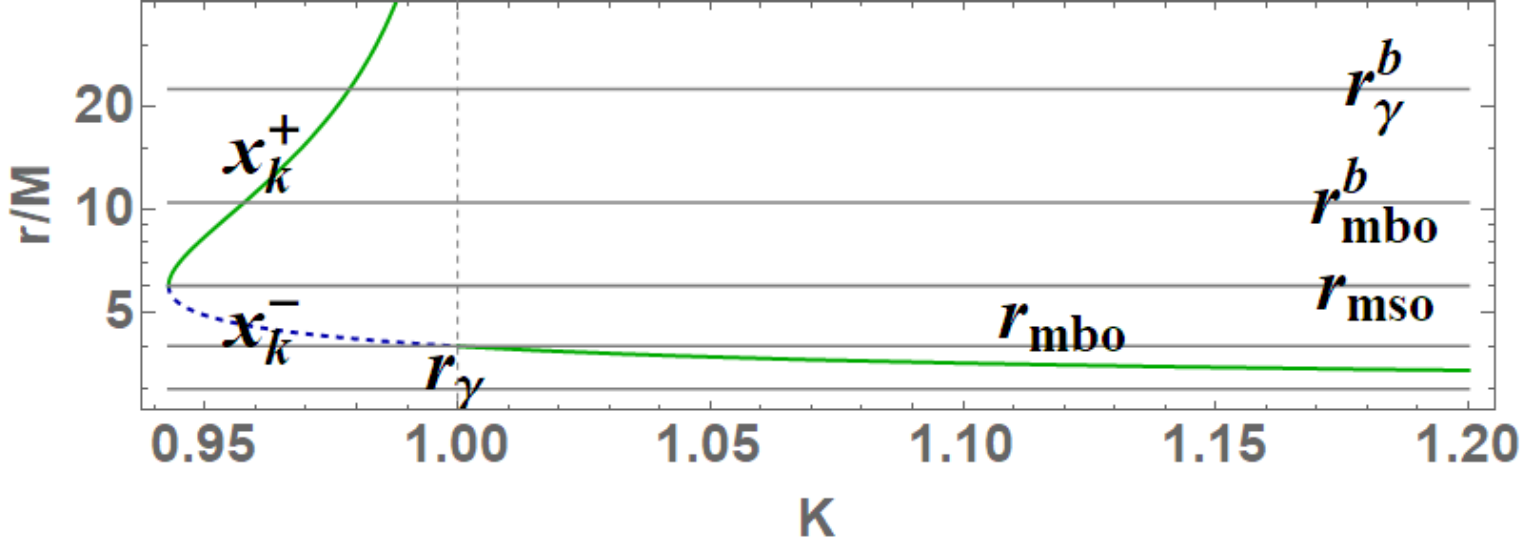}
\includegraphics[width=8.69cm]{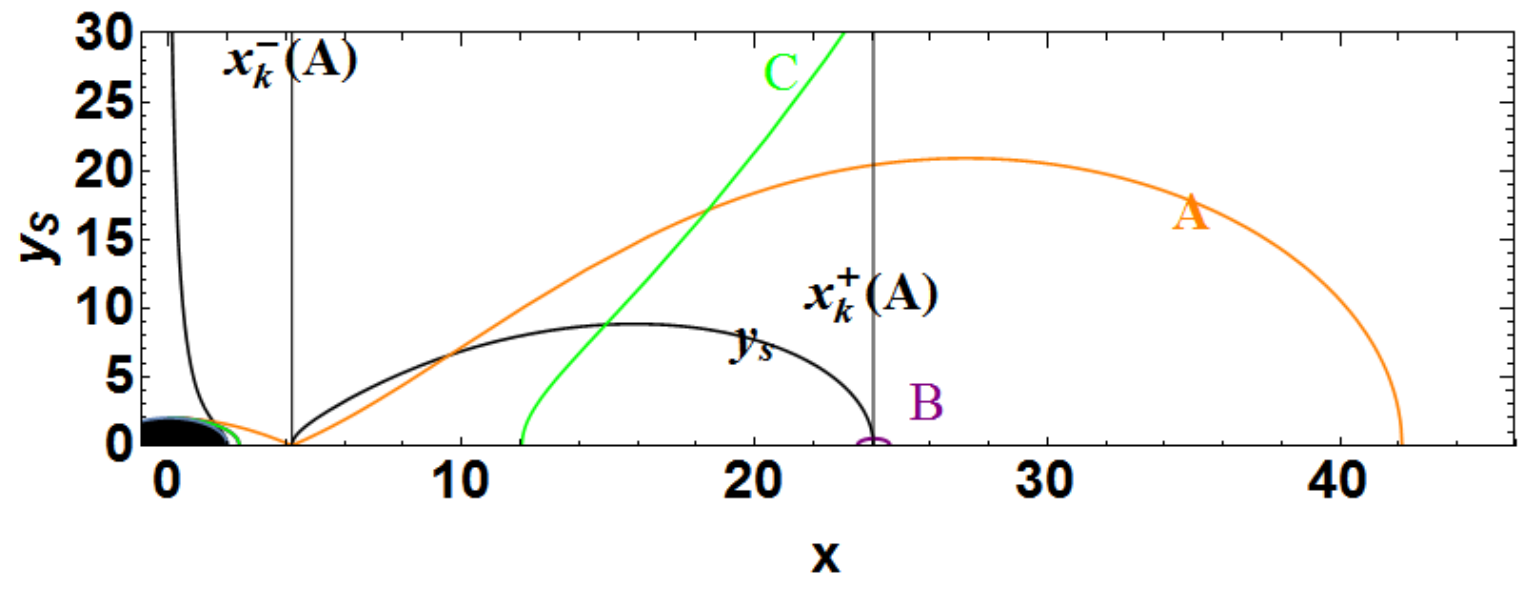}
\includegraphics[width=8.69cm]{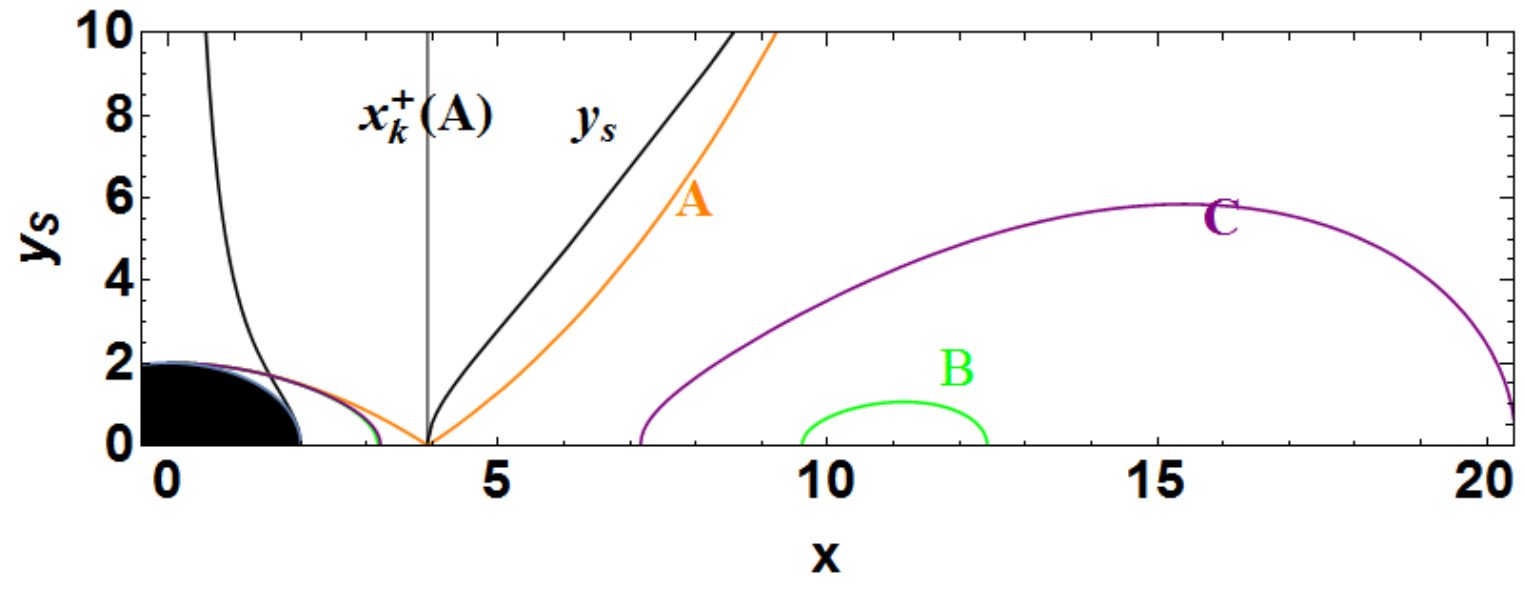}
\caption{Upper left panel: Limiting surfaces $y_s(x)$ in Eq.\il(\ref{Eq:wendll}) as functions of $x$ for different values of the $K$-parameter. Upper right panel: plots of
$x_k^{\pm}$ of  Eq.\il(\ref{Eq:wendll}) as functions of the $K$ parameter. The limit $K=1$ is also signed.   Bottom left panel: case $K=0.98$, plot of limiting surface $y_s$ (black curve): with   $r_1<r_2$ solutions of  $K(r)=0.98$,  models  (\textbf{A},\textbf{B},\textbf{C})
\textbf{A}:  $(\ell(r_1), K=0.98)$ and   $\mathbf{B}:$ $(\ell(r_2), K=0.98001)$ and
\textbf{C}: $(\ell(r_2), K=0.999)$, $x_k^{\pm}(K=0.98)$ is also shown. We can read the limit solution $y_s$ as providing the minimum  and maximum $(x_k^\pm)$ inner edge  location of the correspondent toroids $(\textbf{A},\textbf{B})$. Bottom right  panel:  case $K=1.01$, plot of limiting surface $y_s$ (black curve): with   $r_1$ solutions of  $K(r)=1.01$,  models  (\textbf{A},\textbf{B},\textbf{C}) are as follows
\textbf{A}:  $(\ell(r_1), K=1.01)$ and   $\mathbf{B}:$ $(\ell(r_1), K=0.9594)$ and
\textbf{C}: $(\ell(r_1), K=0.967)$, $x_k^{+}(K=1.01)$.
\label{Fig:Foolelon} }
\end{figure}
\subsection{On the upper limit on the \textbf{RAD} (globule) radius}\label{Sec:zeipel}
The issue of  the \textbf{RAD} (globulus) extension  is  correlated to  two important consequential aspects of the \textbf{RAD} structure, the  existence of the limit on the \textbf{RAD} mass and radius, as well as on  the  spin  of  the orbiting extended multipole object.  Eventually one can consider a cluster of  \textbf{RAD} where the  outer  tori has a preeminent role in the determination of  a possible \textbf{RAD} collision with another element of the cluster. However,
constrains on the existence of such  limit can be  provided by    factors  which are actually external to the model setup considered here, for example  due to the    other factors regulating  accretion disks and accretion processes as  presence of  magnetic fields or also  from the   \textbf{BH} interaction  with  its host environment  usually a galactic embedding. This last aspect requires clearly a focus on the characteristics of the  typical \textbf{RAD} environment.
It is possible that clusters of orbiting globules and clusters of clusters could be formed in some eras of \textbf{BH}  formation and especially in low activity periods characterized by  minor   interaction of the central \textbf{BH} with its host for example in the  (rather rarefied) galactic environment. The gravitating shells  of tori could constitute a frozen situation in which the center \textbf{BH}  would be inert and isolated in particular circumstances (cold \textbf{RAD}). Otherwise  the presence of small spins, would induce \textbf{BH} globuli collision followed by very violent destabilization effects.
Firstly  a \textbf{RAD} radius definition may be well established as the outer radius  $ r_ {out}^o$   of the outer torus of the \textbf{RAD}, this radius has been studied and  constrained in Sec.\il(\ref{Sec:doc-ready}). Particularly we refer to
 Eqs\il(\ref{Eq:over-top},\ref{Eq:mer-panto-ex-resul}) and
Figs\il(\ref{Fig:VClose},\ref{Fig:conicap}),
for accreting configurations in  Eq.\il(\ref{Eq:over-top})--Figs\il(\ref{Fig:woersigna}) and
Eqs\il(\ref{Eq:dan-aga-mich},\ref{Eq:vast-maj})-- Figs\il(\ref{Fig:tookP},\ref{Fig:lom-b-emi},\ref{Fig:SMGerm}).
Considering also the torus elongations (which sets the problem of inner edge location)  there is Eqs\il(\ref{Eq:lam})--Figs\il(\ref{Fig:SMGerm},\ref{Fig:VClose},\ref{Fig:conicap},\ref{Fig:lom-b-emi},\ref{Fig:woersigna})
 and  Eq.\il(\ref{Eq:polaplot})--Sec.\il(\ref{Sec:ount}),
  Eqs\il(\ref{Eq:vast-maj},\ref{Eq:rac.tes.tec})-- Figs\il(\ref{Fig:tookP},\ref{Fig:lom-b-emi}).
For start we can set the most external  torus in the region $r>r_{\gamma}^b$, and therefore with   $\ell>\ell_\gamma$.
(Note this implies a  constrain on the maximum density point $r_{cent}$ and the outer edge $r_{out}$ but not on  the inner edge of the outer disk, which can in fact be also in $r<r_{\gamma}^b$, the detailed analysis of this issue for \textbf{eRAD }Kerr attractors and its Schwarzschild limit  for  corotating or  the more probably outer counterrotating tori  can be found in \cite{open,long}).
For these  tori,  for  large $K$, the prevalence of self-gravity of the thick configuration can be indeed a consistent factor regulating the structure.
A further limit may originate  by constrains  on the  magnitude  of the relativistic angular  velocity $\Omega$, which is a  further relevant point for the fluid and disk,  and toroidal fluid velocity, $u^{\phi}$, which can be found here  from   $\ell(r)$, as in Eq.\il(\ref{Eq:poll-delh}),  and related to  the  (constant) von Zeipel   surfaces.
These quantities are related  in Eqs\il(\ref{Eq:wEo}), we refer also \cite{mnras} for a careful study of these related quantities in the Schwarzschild spacetime, however  below  we provide for convenience some further notes on these elements.

Therefore we consider here again  $L $, $ u^{\phi} $ and  $ \Omega $,  obtaining
\bea\label{Eq:it-impo-ger1}
\ell(r)=\frac{r^{3/2} \sigma }{r-2},\quad\Omega(\ell(r))=\frac{1}{(r-2) r^{3/2} \sigma },\quad
\partial_\ell\Omega(\ell(r))=\frac{\Omega(\ell(r))}{\ell(r)}=\frac{r-2}{r^3 \sigma ^2}=s
\eea
$\Omega (\ell (r)) $ is the curve of the  relativistic velocity evaluated on the  \textbf{RAD} rotation curve, $ s $ defines the surface of von Zeipel  which we consider in  Figs\il(\ref{FIG:for-cre}). We analyzed the asymptotic regime for  large  $r$ 
\bea\label{Eq:it-impo-ger}
&(\bullet)\quad \frac{g_{\phi\phi}}{g_{tt}}=-r^2 \sigma ^2-2 r \sigma ^2-4 \sigma ^2-\frac{8 \sigma ^2}{r}+\mathrm{O}\left[\frac{1}{r^2}\right],\quad&(\bullet)\quad\Omega=\frac{\left(\frac{1}{r}\right)^{3/2}}{\sigma }+\mathrm{O}\left[\frac{1}{r^{7/2}}\right],
\\\label{Eq:it-impo-ger2}
&(\bullet)\quad\ell=\sqrt{r} \sigma +2 \sqrt{\frac{1}{r}} \sigma +\mathrm{O}\left[\frac{1}{r^{3/2}}\right],&\quad (\bullet)\quad \frac{\Omega}{\ell}=\frac{1}{r^2 \sigma ^2}+\mathrm{O}\left[\frac{1}{r^3}\right].
\eea
The form of the  rotational law $\ell(r)$ for large $r$ evidences the  existence of the maximum density point $r_{\mathcal{M}}$. However the presence of a central \textbf{BH} spin, a condition which is to be considered  highly probable, modifies the location of this radius with respect to the corresponding spheres of stability and therefore has important consequences from the point of view of the formation of these globular tori  at large distances. In the case of Kerr attractor besides, there is the further issue  of the role of counter-rotating discs (in the \textbf{eRAD} case this are obviously well defined, in the \textbf{RAD} we intend  the counterrotating component of the torus spin with respect to the Kerr \textbf{BH}) that would be the most likely to be formed in the outer regions-- \cite{dsystem,long,Letter,Multy}
\begin{figure}
     \includegraphics[width=9cm]{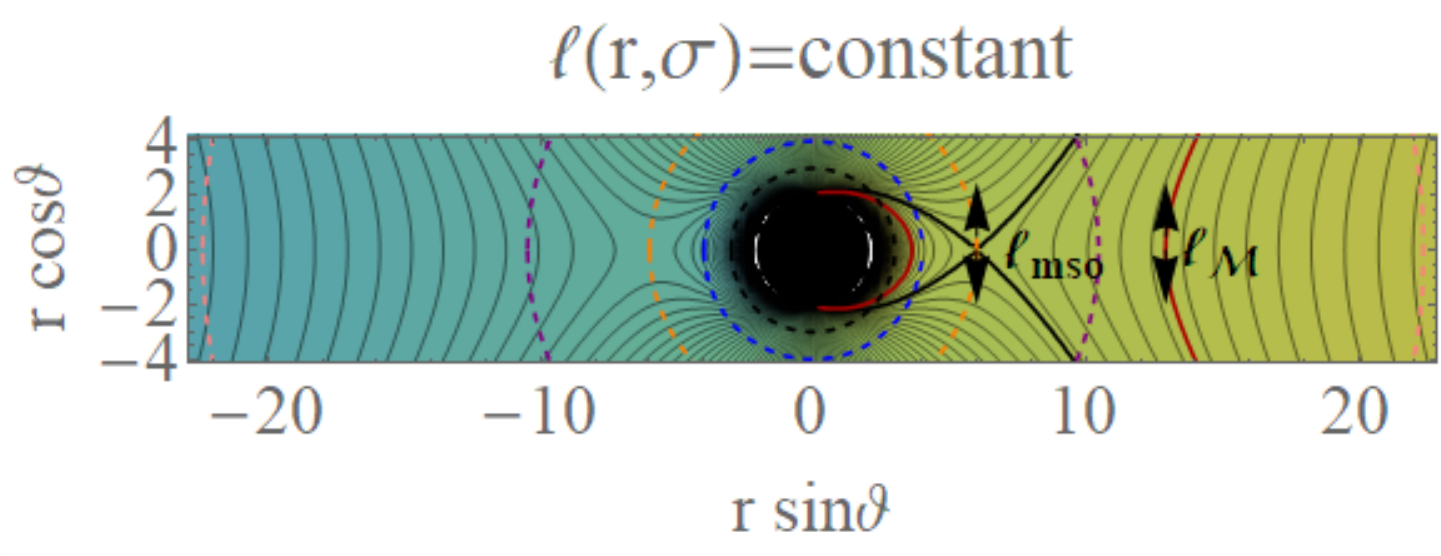}
      \includegraphics[width=7cm]{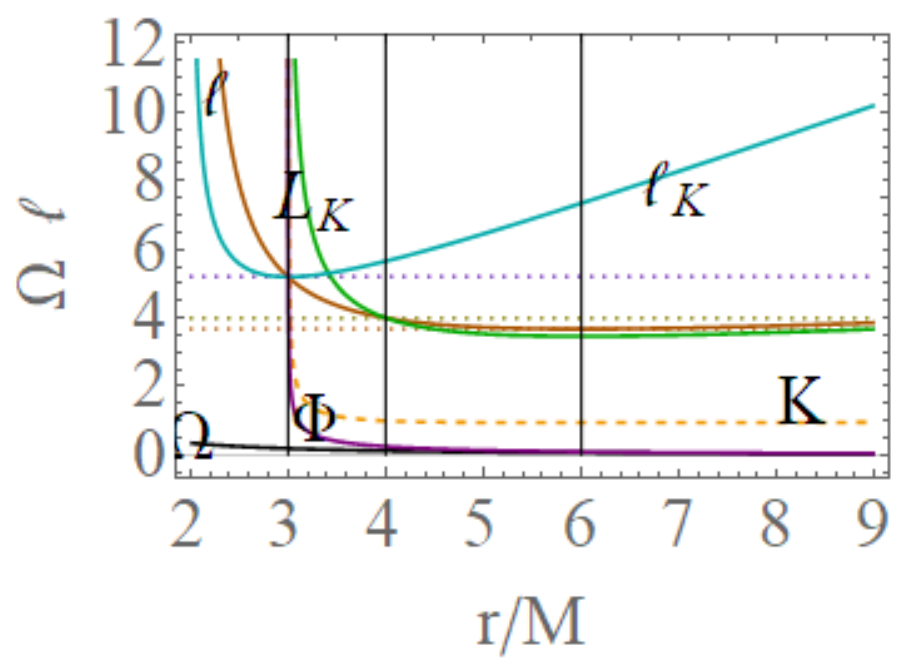}
             \includegraphics[width=7cm]{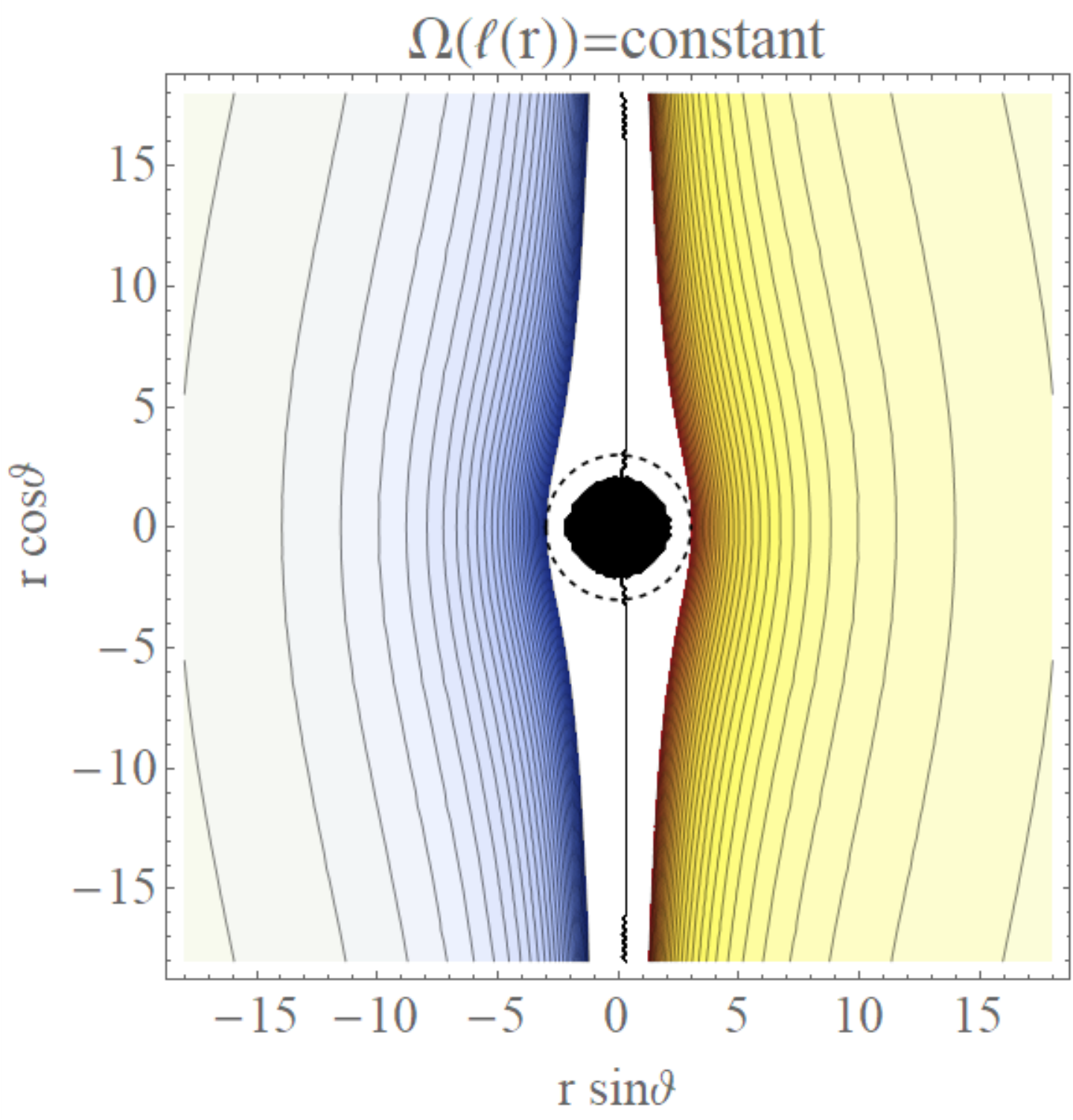}
          \includegraphics[width=7cm]{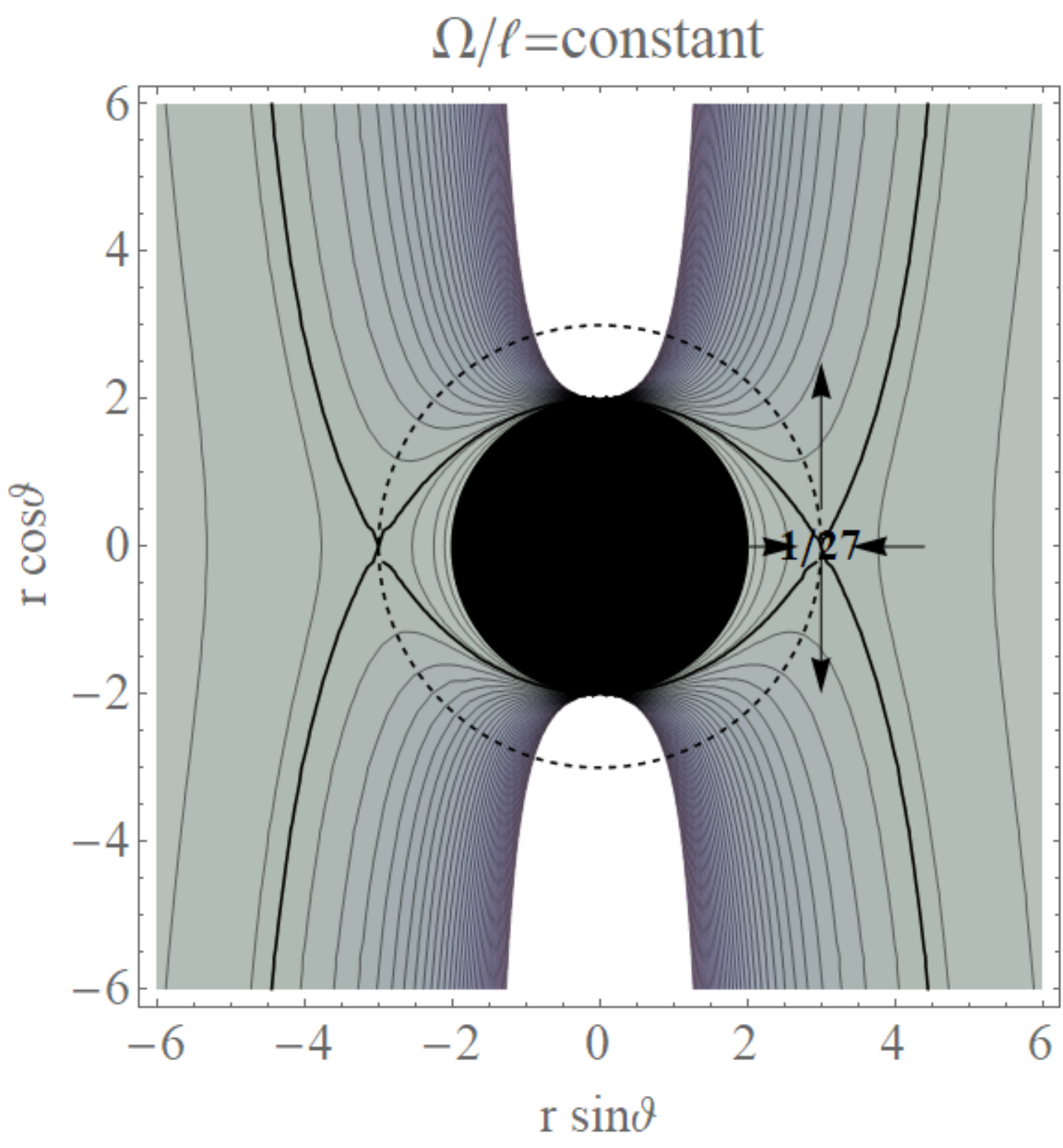}
  \caption{The analysis showed in figures follows discussion in Sec.\il(\ref{Sec:four-b}) and Sec.\il(\ref{Sec:zeipel}), figures refer  to the quantities defined in Eqs\il(\ref{Eq:it-impo-ger}), (\ref{Eq:it-impo-ger1}),(\ref{Eq:it-impo-ger2}), where $\Omega$ is the fluid relativistic angular velocity, $\ell$ is the fluid specific angular momentum, $\Omega/\ell$ represent the von Zeipel surfaces, black central region in the figures is the central Schwarzschild \textbf{BH}. Here $\sigma=\sin \theta$.  Circles $\ell_{mso}$ and $\ell_{\mathcal{M}}$,  $\ell_{\gamma}$ and $\ell_{mbo}$are represented--see Table (\ref{Table:pol-cy}). $\Phi=u^{\phi}$ is the toroidal fluid velocity. $K$ is the energy function of the RAD defined in Eq.\il(\ref{Eq:sincer-Spee}). $L_K$ is the conserved angular momentum for a test particle in circular motion (geodesic) associated to the Schwarzschild geometry Killing field $\xi_{\phi}$, for definition of  $ \ell_K$  and other quantities see Eq.\il(\ref{Eq:wEo}).}\label{FIG:for-cre}
\end{figure}

\end{document}